\documentclass[12pt,a4paper]{article}
\pdfoutput=1
\usepackage{jheppub}

\usepackage{amsmath,amsfonts,amsbsy,amssymb,array,accents,dsfont}
\usepackage{enumerate,array,latexsym,graphicx,mathrsfs,verbatim,psfrag}
\usepackage{enumerate}
\usepackage{graphicx}
\usepackage{caption} 
\usepackage{subcaption} 

\usepackage{bm}
\usepackage{amssymb}
\usepackage{amsmath}
\usepackage{cancel}
\usepackage{epigraph}

\newcommand{\captionfonts}{\small}

\makeatletter  % Allow the use of @ in command names
\long\def\@makecaption#1#2{%
  \vskip\abovecaptionskip
  \sbox\@tempboxa{{\captionfonts #1: #2}}%
 \ifdim \wd\@tempboxa >\hsize
    {\captionfonts #1: #2\par}
  \else
    \hbox to\hsize{\hfil\box\@tempboxa\hfil}%
  \fi
  \vskip\belowcaptionskip}
\makeatother   % Cancel the effect of \makeatletter

\usepackage{fancyhdr}
\usepackage{datetime}

\usepackage{mciteplus}

\usepackage[all]{hypcap}     %should be loaded AFTER hyperref, which should otherwise be loaded last.

\makeatletter
\renewcommand\section{\@startsection {section}{1}{\z@}%
                                   {-3.5ex \@plus -1ex \@minus -.2ex}%nn
                                   {2.3ex \@plus.2ex}%
                                   {\normalfont\Large\bfseries}}
\renewcommand\subsection{\@startsection{subsection}{2}{\z@}%
                                     {-3.25ex\@plus -1ex \@minus -.2ex}%
                                     {1.5ex \@plus .2ex}%
                                     {\normalfont\bfseries}}
\renewcommand\subsubsection{\@startsection{subsubsection}{3}{\z@}%
                                     {-2.5ex\@plus -1ex \@minus -.2ex}%
                                     {1.25ex \@plus .2ex}%
                                     {\normalfont\textit}}
\makeatother
\newcommand{\WZ}{{\text{\tiny WZ}}}
\newcommand{\WZW}{{\text{\tiny WZW}}}

\newcommand{\pa}{\ensuremath{\partial}}

\newcommand{\rhoo}{\ensuremath{\! \rho \:\! }}

\newcommand{\Ab}{\ensuremath{\bar{A}}}
\newcommand{\cAb}{\ensuremath{\bar{\mathcal{A}}}}

\newcommand{\longhookrightarrow}{\lhook\joinrel\longrightarrow}

\DeclareMathSymbol{\medhatsym}{\mathord}{largesymbols}{"62} % basic symbol
\newcommand\lowermedhatsym{%  adjust height
  \text{\smash{\raisebox{-1.28ex}{%
    $\medhatsym$}}}}
\newcommand\medhat[1]{% command to be used
  \mathchoice
    {\accentset{\displaystyle\lowermedhatsym}{#1}}
    {\accentset{\textstyle\lowermedhatsym}{#1}}
    {\accentset{\scriptstyle\lowermedhatsym}{#1}}
    {\accentset{\scriptscriptstyle\lowermedhatsym}{#1}}
}

\DeclareMathSymbol{\medtildesym}{\mathord}{largesymbols}{"65}% basic symbol

\DeclareMathAlphabet{\mathpzc}{OT1}{pzc}{m}{it}

\def\sl{\text{sl}}
\def \su{\text{su}}
\def \uo{\text{u(1)}}

\def\eps{\epsilon}
\def\vareps{\varepsilon}
\def\xone{$x^1$}
\def\xtwo{$x^2$}
\def\xthree{$x^3$}
\def\xfour{$x^4$}

\def\ellone{\ell_1}
\def\elltwo{\ell_2}
\def\mhat{\mathsf{m}_2}
\def\nhat{\mathsf{m}_1}

\newcommand{\eq}[1]{\eqref{#1}}

\def\tight#1{\! #1 \!}

\def\({\left(}
\def\){\right)}
\def\[{\left[}
\def\]{\right]}

\def\sltwo{{SL(2,\IR)}}

\def\sutwo{{SU(2)}}
\def\uone{U(1)}
\def\Uint{\cI}

\def\ie{{i.e.}}
\def\eg{{e.g.}}
\def\cf{{c.f.}}
\def\etc{{etc}}
\def\etal{{\it et.al.}}

\def\eff{{\rm eff}}
\def\lstr{\ell_{\textit{str}}}
\def\apl{\alpha'_{\textit{little}}}
\def\gstr{g_{\textit s}}

\def\rads{R_{\textit AdS}}

\def\suplabel{m}

\def\nfive{{n_5}}

\def\k{\ensuremath{\mathsf{k}}}
\def\vv{\ensuremath{\mathsf{v}}}

\def\sst#1{\scriptscriptstyle{#1}}

\def\x1x2{$x^1$-$x^2$}

\def\ytil{{\tilde y}}

\def\Ry{R_y}
\def\Rytil{R_\ytil}
\def\ktil{{\tilde k}}
\def\alphab{{\boldsymbol\nu}}

\def\varthetab{{\boldsymbol\vartheta}}
\def\etab{{\boldsymbol\eta}}
\def\mub{{\boldsymbol\mu}}

\def\omhat{{\boldsymbol\omega}}

\def\sst{\scriptscriptstyle}
\def\half{\frac12}
\def\coeff#1#2{{\textstyle \frac{#1}{#2}}}
\def\hf{\coeff12}
\def\tr{{\rm Tr}}
\def\One{{\hbox{1\kern-1mm l}}}

\def\barray{\begin{array}}
\def\earray{\end{array}}
\def\be{\begin{equation}}
\def\ee{\end{equation}}
\def\bea{\begin{eqnarray}}
\def\eea{\end{eqnarray}}
\def\bal{\begin{align}}
\def\eal{\end{align}}
\def\nn{\nonumber}

\newcommand{\bH}{{\mathbb H}}

\newcommand{\bR}{{\mathbb R}}
\newcommand{\bS}{{\mathbb S}}
\newcommand{\bT}{{\mathbb T}}
\newcommand{\bZ}{{\mathbb Z}}

\def\IR{\mathbb{R}}

\def\IZ{\mathbb{Z}}

\definecolor{cardinal}{rgb}{0.6,0,0}
\definecolor{darkgreen}{rgb}{0,0.4,0}
\definecolor{golden}{rgb}{0.92, 0.7, 0}
\definecolor{midnight}{rgb}{0, 0, 0.5}
\definecolor{darkblue}{rgb}{0, 0, 0.7}

\numberwithin{equation}{section}

\mathchardef\mhyphen="2D

\def\cA{\mathcal {A}} \def\cB{\mathcal {B}} \def\cC{\mathcal {C}}
\def\cD{\mathcal {D}} \def\cE{\mathcal {E}} \def\cF{\mathcal {F}}
\def\cG{\mathcal {G}} \def\cH{\mathcal {H}} \def\cI{\mathcal {I}}
\def\cJ{\mathcal {J}} \def\cK{\mathcal {K}} \def\cL{\mathcal {L}}
\def\cM{\mathcal {M}} \def\cN{\mathcal {N}} 
  
\def\cS{\mathcal {S}}  \def\cU{\mathcal {U}}
\def\cV{\mathcal {V}}

\def\cAbar{\bar{\mathcal{A}}}
\def\cChat{\medhat{\cal{C}}}

\def\one{{\hbox{\kern+.5mm 1\kern-.8mm l}}}
\def\zero{{\hbox{0\kern-1.5mm 0}}}

\def\id{\textrm{id}}

\newcommand{\Gg}{{\scriptscriptstyle \mathcal{G}}}
\newcommand{\Hh}{{\scriptscriptstyle \mathcal{H}}}
\newcommand{\Ii}{{\scriptscriptstyle \mathcal{I}}}

\newcommand{\omegatwo}{\ensuremath{\omega_2}}
\newcommand{\lhat}{\ensuremath{\hat{\ell}}}

\begin{document}

%\title{D-branes in Supertube Backgrounds} 
\title{Little Strings, Long Strings, and Fuzzballs} 

\author{Emil J. Martinec$^a$, Stefano Massai$^a$ {\it and}\,
  David Turton$^b$}

\vspace{0.85 cm}

\affiliation[a]{
Enrico Fermi Institute and Dept. of Physics \\
5640 S. Ellis Ave.,
Chicago, IL 60637-1433, USA 
}

\affiliation[b]{
Mathematical Sciences and STAG Research Centre, University of Southampton, \\
Highfield, Southampton, SO17 1BJ, UK
}

 \emailAdd{ejmartin@uchicago.edu}
 \emailAdd{massai@uchicago.edu}
 \emailAdd{d.j.turton@soton.ac.uk}

\vspace{0.7cm} 

\abstract{
At high energy densities, fivebranes are populated by a Hagedorn phase of so-called {\it little strings}, whose statistical mechanics underlies black fivebrane thermodynamics.  A particular limit of this phase yields BTZ black holes in $AdS_3$, leading us to the idea that in this context fuzzballs and highly excited little strings are one and the same.
We explore these ideas through an analysis of D-brane probes of fivebrane supertube backgrounds.
String theory dynamics on these backgrounds is described by an exactly solvable null-gauged WZW model. 
We develop the formalism of null gauging on worldsheets with boundaries, and find that D-branes wrapping topology at the bottom of the supertube throat are avatars of the ``long string'' structure that dominates the thermodynamics of the black hole regime, appearing here as excitations of supertubes lying near but slightly outside the black hole regime.
}

%%%%%%%%%%%%%%%%%%%%%%%%%%%%%%%%%%%%

\setcounter{tocdepth}{2}   % suppresses sub-sub-sections in toc
\maketitle

%%%%%%%%%%%%%%%%%%%%%%%%%%%%%%%%%%%%
%%%%%%%%%%%%%%%%%%%%%%%%%%%%%%%%%%%%

\baselineskip=15pt
\parskip=3pt

%\vskip 1cm
%\tableofcontents
%\newpage

\setcounter{footnote}{0}

%%%%%%%%%%%%%%%%%%%%%%%%%%%%%%%%%%%%
%%%%%%%%%%%%%%%%%%%%%%%%%%%%%%%%%%%%

\section{Introduction}

%%%%%%%%%%%%%%%%%%%%%%%%%%%%%%%%%%%%
%\subsection{Little strings, long strings, and fuzzballs}
\subsection{Fivebrane dynamics}

The dynamics of coincident fivebranes in string theory is governed by {\it little string theory}, a somewhat mysterious non-gravitational, nonlocal theory in six spacetime dimensions~\cite{Seiberg:1997zk,Aharony:1998ub} (for reviews, see~\cite{Aharony:1999ks,Kutasov:2001uf}).  We understand the outlines of little string theory, but little more.  For instance, it is nonlocal on the scale $\nfive\alpha'\equiv\apl$, where $\nfive$ is the number of fivebranes and $\alpha'$ is the inverse tension scale of the fundamental (F1) string.  Sufficiently supersymmetric backgrounds exhibit T-duality symmetry.  Fivebrane thermodynamics at sufficiently high energy density is dominated by a Hagedorn gas of little strings~\cite{Maldacena:1996ya}.  Yet much more remains obscure.

\begin{figure}[ht]
\centering
    \includegraphics[width=.52\textwidth]{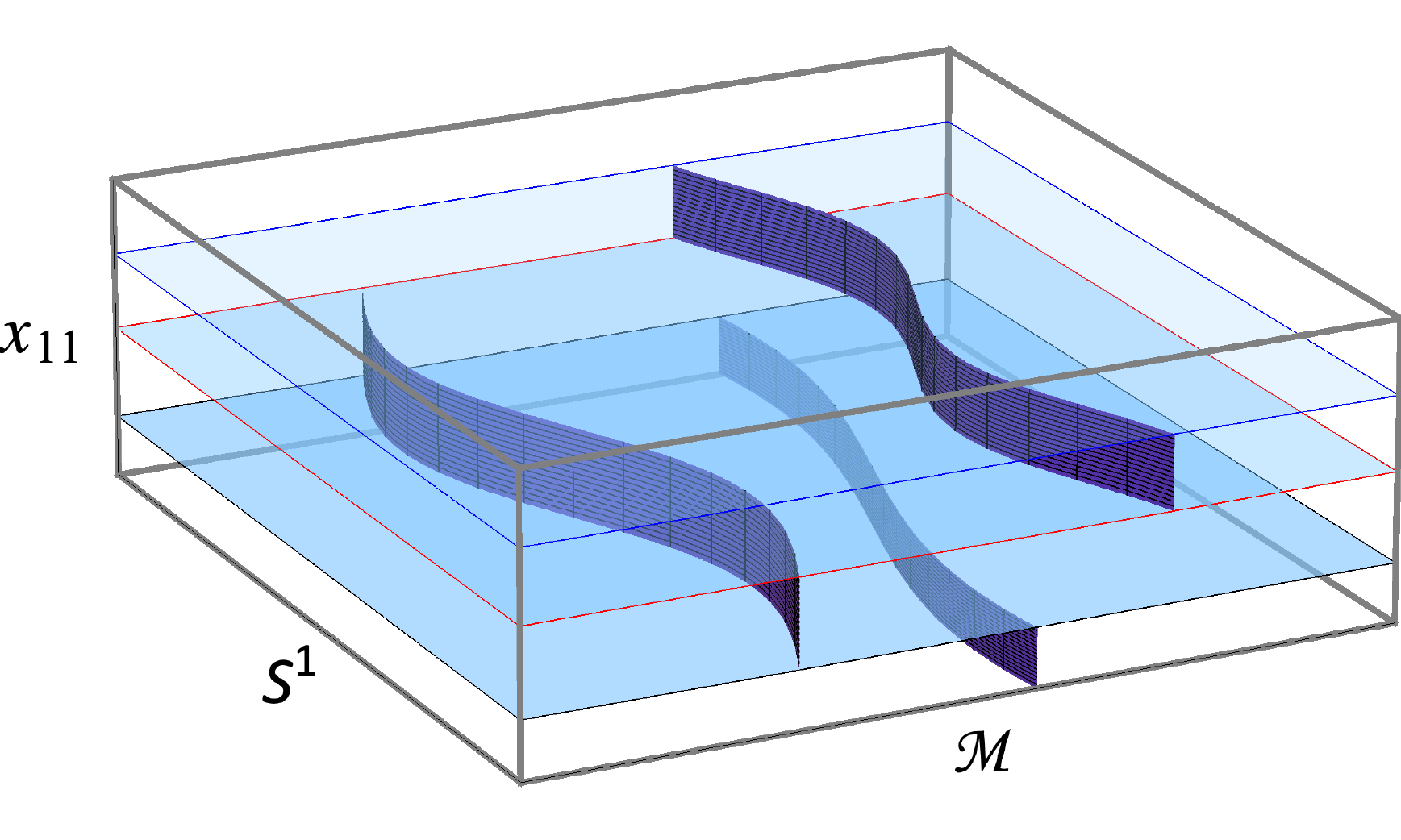}
\caption{\it 
Open M2-branes stretched between a stack of M5-branes.  If the latter are separated in their transverse space (not depicted here), in the type IIA limit this bound state reduces to D2-branes stretched between NS5-branes.
}
\label{fig:OpenMembrane}
\end{figure}

The presence of fivebranes fractionates fundamental string charge and tension. One can see this in the M-theory lift of type IIA, where the fundamental string is an M2-brane wrapped around the circular 11th dimension.  Upon encountering a stack of $\nfive$ coincident M5-branes (transverse to the circle), the wrapped membrane can split into $\nfive$ strips stretching between successive M5's around the circle, see Figure~\ref{fig:OpenMembrane}.  The charge fractionates, and so does the tension of the effective ``W-strings'', providing a heuristic picture of the origin of the little string's tension scale.

%%%%%%%%%%%%%%%%%%
%%%%%%%%%%%%%%%%%%

When the stack of fivebranes is wrapped around $\cM\times\bS^1$ (where $\cM=\bT^4$ or $K3$), sufficiently excited states are are microstates of black holes in the effective five-dimensional supergravity, whose entropy matches the Hagedorn entropy of the little string~\cite{Maldacena:1996ya},
\be
\label{Shag}
S = 2\pi\sqrt{ N_L} + 2\pi\sqrt{N_R} ~,
\ee 
where $N_{L,R}$ are the excitation levels of the little string.

If one binds $n_1$ strings to $\nfive$ fivebranes (\ie\ F1-NS5 or D1-D5 bound states), they will fractionate into little strings in a superselection sector of total little string winding number $n_1n_5$.  Momenta on the little string can be fractionated by amounts up to $n_1n_5$ (if the fivebranes have a suitable $\bZ_{\nfive}$ twisted boundary condition around the $\bS^1$), and entropy is enhanced by a factor $n_5$ over the entropy of Hagedorn fundamental strings in isolation.  This effect plays an important role in the infrared scaling limit $R_{\bS^1}\to\infty$ with the energy above the ground state $R_{\bS^1} E\equiv \varepsilon$ held fixed, which leads to an effective geometry $AdS_3\times \bS^3\times\cM$; the associated BTZ black holes have entropy 
\be
\label{BTZentropy}
S = 2\pi\sqrt{n_5n_1 (\varepsilon+n_p)/2-J_L^2} + 2\pi\sqrt{n_5n_1(\varepsilon-n_p)/2-J_R^2} ~.
\ee
This expression is the specialization of the little string's Hagedorn entropy~\eqref{Shag} to this scaling limit, using the Virasoro constraints on the little string~\cite{Maldacena:1996ya}
\be
M^2 
= \Bigl(\frac {n_p}R - \frac{mR}{\apl} \Bigr)^2 + \frac{4}{\apl} (N_L +J_L^2)
= \Bigl(\frac {n_p}R + \frac{mR}{\apl} \Bigr)^2 + \frac{4}{\apl} (N_R +J_R^2)
\ee
with $M\sim (mR/\apl)+\varepsilon/R$ in the limit, and we work in the superselection sector where the little string has $m=n_1n_5$ units of winding and $n_p$ units of momentum on $\bS^1$, and left/right angular momenta $J_{L,R}$ in the space transverse to the fivebranes.

\begin{figure}[ht]
\centering
  \begin{subfigure}[b]{0.4\textwidth}
  \hskip .5cm
    \includegraphics[width=\textwidth]{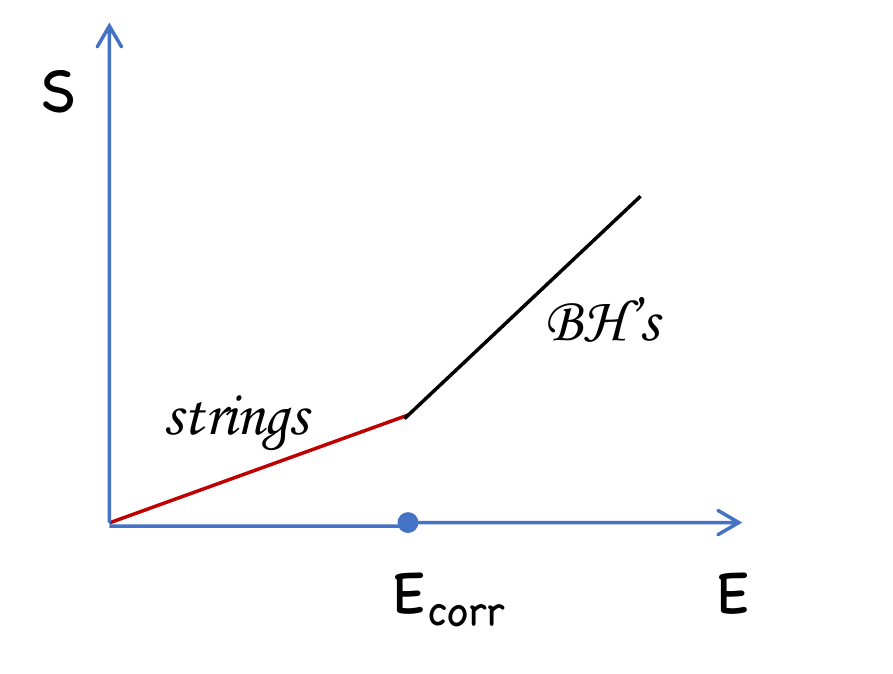}
    \caption{ }
    \label{fig:correspondence}
  \end{subfigure}
\qquad\qquad
  \begin{subfigure}[b]{0.45\textwidth}
      \hskip -.5cm
    \includegraphics[width=\textwidth]{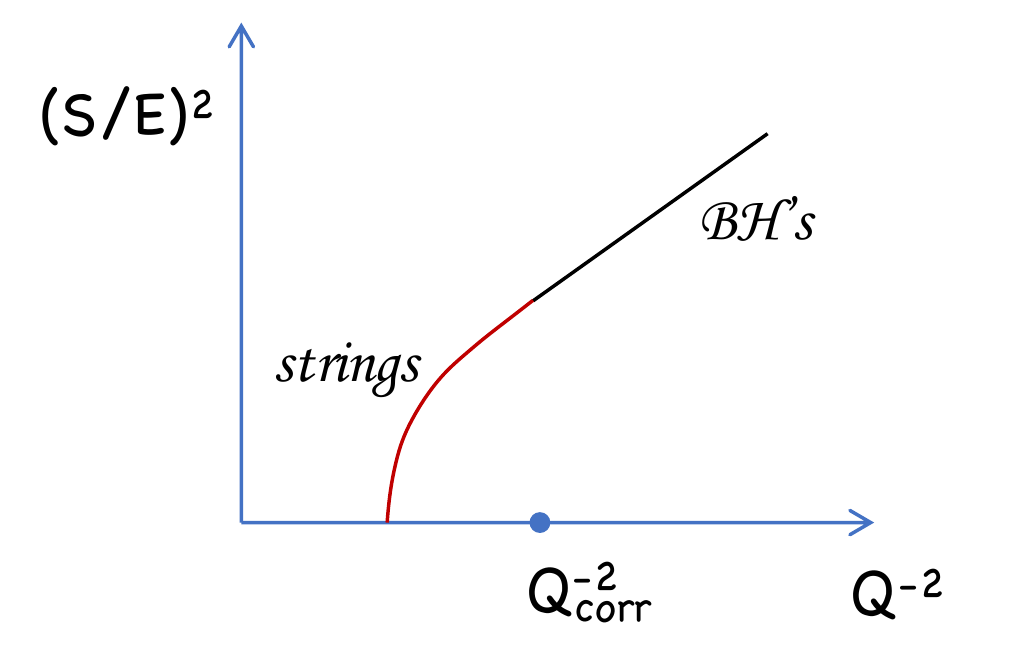}
    \caption{ }
    \label{fig:NS5-correspondence}
  \end{subfigure}
%
%\centerline{\includegraphics[width=5in]{AdSorbs-alt.pdf}}
%\setlength{\unitlength}{0.1\columnwidth}
\caption{\it 
The string/black hole correspondence principle: 
(a) In flat spacetime, the density of states is dominated by excited strings at low energies; above the correspondence point, black hole thermodynamics dominates.
(b) In NS5 throats, black hole states are not normalizable if the slope $Q$ of the linear dilaton (which governs the effective string tension~$\alpha'_\eff$) is too large, and perturbative strings govern the asymptotic density of states.
}
\label{fig:correspondence-transitions}
\end{figure}

The little string is a highly quantum object living down at the bottom of the throat of the $\nfive$ coincident fivebranes, with an effective coupling of order one, and so it is difficult to translate the above heuristic picture into a systematic, quantitative computational strategy.  However, there may be properties that are robust against interactions from which to glean further insights.  Consider for instance the correspondence transition~\cite{Horowitz:1996nw}, first considered in the context of bound states of fundamental strings and D-branes in asymptotically flat spacetime.  At low energies, the density of states is well-approximated by a gas of weakly interacting strings on the D-brane (sometimes this is a Hagedorn gas of the fundamental string, sometimes it is a gas of short open strings).  At the correspondence point, the string gas entropy matches the entropy of a black hole or black brane carrying the corresponding charges, and above this point black holes dominate the density of states, see Figure~\ref{fig:correspondence-transitions}.

This behavior is a somewhat more sophisticated version of the dynamics of quantum-mechanical particles interacting with gravity.  One doesn't treat an elementary particle as a small black hole because its Compton wavelength is much larger than its Schwarzschild radius; near the massive source, classical dynamics (and in particular, classical general relativity) does not apply because the quantum wavefunction of the particle is spread over a region much larger than any possible horizon scale -- the particle is not sufficiently localized to be a black hole.  Similarly, in a situation where string theory is below the correspondence point, string wavefunctions extend well beyond what would be the classical Schwarzschild radius, and string $\alpha'$ effects dominate over classical GR.  For instance, consider a large circular string let go to collapse toward its center of mass; in classical general relativity coupled to a classical string, a horizon would form and the final state would be a black hole, but at sufficiently weak string coupling, the final state will be a highly excited fundamental string~-- a horizon never forms.

%%%%%%%%%%%%%%%%%%
%%%%%%%%%%%%%%%%%%

The correspondence transition is somewhat different in the linear dilaton throat of NS5-branes, and its $AdS_3$ limit~\cite{Giveon:2005mi}.  In the linear dilaton case, the transition point is a function of $Q\lstr $ instead of $E\lstr$, where $Q$ is the slope of the linear dilaton, $\Phi = Q\rho$ with $\rho$ the appropriate radial coordinate in the fivebrane throat.  Similarly, in the AdS limit $R_{\bS^1}\to\infty$, the transition point is a function of $\rads/\lstr\:\!$, where $\rads$ is the $AdS_3$ radius of curvature.  As one approaches the correspondence point in the fivebrane throat, the wavefunctions of fundamental strings and D-branes near the bottom of the throat start to delocalize~\cite{Kutasov:2005rr,Giveon:2005mi}.  At the correspondence point, the asymptotic spectrum of fundamental strings and black objects matches; beyond the correspondence point, $\rads/\lstr<1$ (for $AdS_3$ throats; $Q\lstr>1$ for linear dilaton throats) and the density of states up to arbitrarily high energy is dominated by the Hagedorn density of states of fundamental strings rather than the Bekenstein-Hawking entropy of black objects. In fact black holes are thought to be absent from the spectrum, having non-normalizable wavefunctions.

It is tempting to believe that this same dynamics of the correspondence point is at work in little string theory.  Hagedorn thermodynamics is largely kinematic in nature, characterized by a statistical equilibrium between kinetic and stretching energy of the string gas.  One therefore might expect that the dominant effect of the large rate of joining/splitting interactions of the little string is to ensure ergodicity and a rapid exploration of the phase space, rather than to dramatically alter the equation of state.
The key distinction between the correspondence transition dynamics of fundamental strings and that of little string theory is that the little string is {\it always at its correspondence point} -- {\it the black fivebrane entropy~\eqref{Shag} equals the Hagedorn entropy of the little string}.  
Indeed, the little string correspondence point in the linear dilaton throat is~\eqref{Shag}, and in the $AdS_3$ limit is~\eqref{BTZentropy}; comparing to the fundamental string correspondence points $Q^2=1/\alpha'$ for the linear dilaton throat and $\rads^2=\alpha'$ in the $AdS_3$ limit~\cite{Giveon:2005mi}, one finds that the little string correspondence points are precisely the same as the fundamental string correspondence points, with the fundamental string tension replaced by the little string tension $\apl=\nfive\alpha'$.  If the little string behaves in the same way as the fundamental string, one expects the little string wavefunction to be delocalized in the fivebrane throat, at least out to the horizon scale of the relevant black hole.

Of course, little string holography is the statement that the {\it entirety} of the decoupled fivebrane throat is dual to the non-gravitational little string theory, so in a sense the little string wavefunction indeed extends over the entire throat and not just the horizon region.  However, outside the black hole horizon, the little string degrees of freedom are confined in the same way that the nonabelian degrees of freedom of $\cN=4$ SYM theory are confined in $AdS_5\times \bS^5$ outside the horizon of $AdS_5$ black holes.  One expects that the bulk gravity description of the wavefunction of these nonabelian degrees of freedom has its dominant support persisting out to the horizon scale, with the exterior of the black hole being well-described by the collective field theory of the singlet degrees of freedom, \ie\ supergravity.

In this scenario, the highly excited little string is the embodiment of the ``fuzzball'' in the context of linear dilaton and $AdS_3$ black holes.  The fuzzball paradigm~\cite{Mathur:2005zp} posits that the black hole interior is supplanted by some nonsingular quantum structure, whose underlying dynamics does not have a causal horizon.
The horizon in the low-energy {\it effective} theory is thought to arise from an inappropriate integrating out of the light IR degrees of freedom that carry the entropy.
Some discussions of the fuzzball proposal in the literature have emphasized the importance of microstates described by supergravity solutions that cap off smoothly without a horizon (see for instance~\cite{Bena:2007kg}); 
however it has been argued%
~\cite{deBoer:2009un,Bena:2012hf,Martinec:2015pfa,Eperon:2016cdd,Raju:2018xue}
that such capped geometries are highly coherent states which are quite non-generic in the ensemble of microstates.  Indeed, while there has been much recent progress in constructing and studying three-charge ``superstrata''%
 ~\cite{Bena:2015bea,Bena:2016agb,Bena:2016ypk,Bena:2017geu,Bena:2017upb,Bena:2017xbt,Bena:2018bbd,Bakhshaei:2018vux,Bena:2018mpb,Ceplak:2018pws,Heidmann:2019zws,Bombini:2019vnc,Bena:2019azk} and related solutions~\cite{Mathur:2011gz,Mathur:2012tj,Lunin:2012gp,Giusto:2013bda}, it seems unlikely that the set of solutions that are realizable solely in terms of geometry can account for the typical black hole microstate. However a more expansive characterization of the fuzzball paradigm (\cf~\cite{Mathur:2005zp,Skenderis:2008qn,Mathur:2008nj,Bena:2013dka,Martinec:2014gka,Mathur:2018tib}) allows for the possibility that stringy and quantum ingredients are essential, and it is this possibility which seems to be realized in the context of fivebranes.
The suggestion here is that in fivebrane throats and their $AdS_3$ limits, the interior structure of black holes consists of a little string condensate.  The role of smooth, capped geometries is to allow us a window into the black hole regime, as we now explain.

%%%%%%%%%%%%%%%%%%%%%%%%%%%%%%%%%%%%
\subsection{Emergence of long string structure ``in the bulk''}

The notion that fivebrane black holes consist of a ``deconfined'' phase of little strings places this example of holography squarely in line with examples of gauge/gravity duality wherein the black hole phase involves liberation of nonabelian modes of a strongly coupled Yang-Mills gauge theory~\cite{Itzhaki:1998dd}.  If black hole formation involves such a deconfinement phase transition, one should see the nonabelian degrees of freedom as virtual excitations which are more and more easily excited as one approaches the threshold of black hole formation.  For instance, one can imagine keeping the branes apart, and then letting them approach one another.  From the effective field theory point of view, a horizon forms when the branes are close enough that the nonabelian degrees of freedom start to become thermally excited, see for example%
~\cite{Kraus:1998hv,Giddings:1999zu,Brandhuber:1999jr,Danielsson:1999zt,Horowitz:2006mr} 
(though in the full theory, this horizon of the low-energy {\it effective} theory is not a {\it fundamental} barrier to information transport).

Much of this picture is based on intuitions derived from the weak-coupling regime of the gauge theory, where the gravitational field (and in particular the effective horizon) sourced by the branes is not part of the description; or from purely gravitational analyses of horizon formation and dynamics, where the branes are strongly coupled and hidden from view.  One would like to fill in the gap.  An important but hard problem is that of extracting gravitational physics from strong coupling dynamics of the gauge theory dual.  Approaching from the opposite direction, one might look for the W-particles or W-strings of the brane dynamics on the gravitational side of the duality, and to see what happens to them as one approaches the black hole threshold from below by bringing the background branes together.

Little strings have a number of avatars, depending on the duality frame.  In a type IIB frame, fractional instantons~\cite{tHooft:1981nnx,Guralnik:1997sy,Dijkgraaf:1997ku} in the $U(\nfive)$ gauge theory on fivebranes (either D5 or NS5) are 1+1 dimensional string-like objects whose tension is that of the little string.  In M-theory, M2-branes stretching between M5-branes behave as effective strings; in the reduction to type IIA, these become (when the M5's are suitably separated in their transverse space) D2-brane strips stretching between NS5-branes, again a string-like object when the branes are nearly coincident.  The process of separating the branes transversely and then reducing to type IIA has inverted the tension hierarchy between fundamental strings and little W-strings, and allows the latter to be studied in perturbative string theory, which is predicated on fundamental strings being the objects with the lowest tension.  This latter description, and ones related to it by perturbative dualities, will be our focus here.

%%%%%%%%%%%%%%%%%%
\begin{figure}[ht]
\centering
    \includegraphics[width=.5\textwidth]{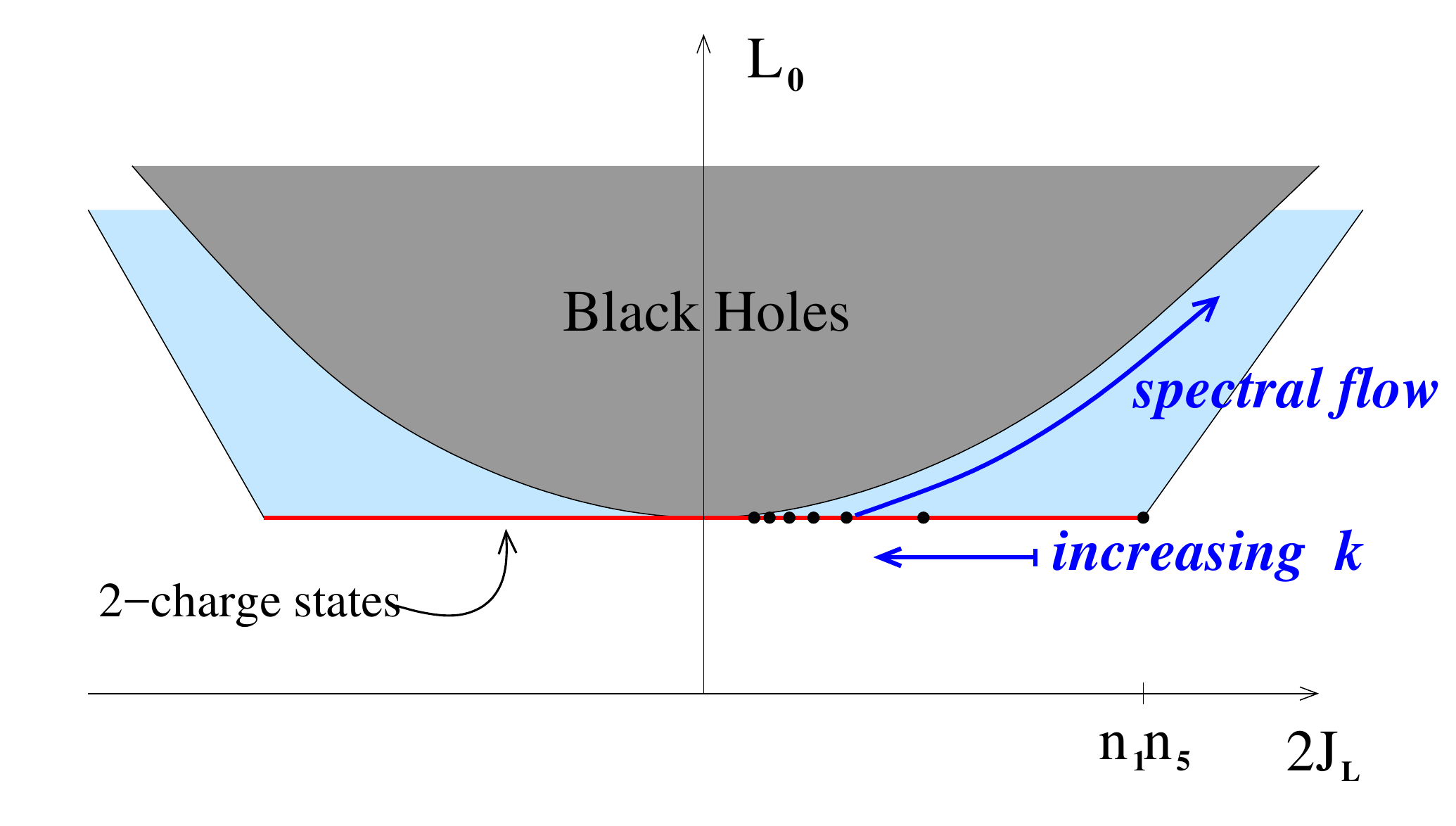}
\caption{\it 
Two-charge supertubes occupy the (red) BPS line in the spectrum of states with left-moving energy $L_0$ and angular momentum $J_L$.  Spectral flow of these states moves them parallel to the cosmic censorship bound where the BTZ black hole entropy formula~\eqref{BTZentropy} degenerates.
}
\label{fig:Spectrum-orb2}
\end{figure}
%%%%%%%%%%%%%%%%%%

Our route to W-branes near the black hole threshold begins with backgrounds having slightly separated fivebranes.  The $AdS_3$ limit, $R_{\bS^1}\to\infty$, of the string-fivebrane system has the benefit of a large collection of half-BPS states, variously known as {\it supertubes} or {\it two-charge BPS fuzzballs}, living on the BPS bound in the phase diagram of Figure~\ref{fig:Spectrum-orb2}.  From the point of view of the BTZ solution, generically the geometries are spinning too rapidly to be BTZ black holes; the angular momentum pries apart the underlying fivebranes, but as one dials down the angular momentum one can approach the threshold of black hole formation.
The geometry sourced by the branes can be completely worked out in the supergravity approximation~\cite{Lunin:2001fv,Lunin:2001jy,Lunin:2002iz,Kanitscheider:2007wq}; there is a family of nonsingular geometries that cap off in a structure of topological bubbles threaded by flux. The map between supergravity solutions and coherent microstates of the branes is known explicitly~\cite{Lunin:2001fv,Lunin:2001jy,Kanitscheider:2007wq,Giusto:2019qig}, and the geometric quantization of the phase space of classical solutions reproduces the microstate entropy~\cite{Rychkov:2005ji,Krishnan:2015vha}. The topological bubbles arise because the string/fivebrane system with angular momentum sources KK dipole charge; the local KK monopole (KKM) geometry is nonsingular, up to orbifold loci where monopole cores coincide.  As one descends the fivebrane throat, the geometry ``caps off'' smoothly before a horizon forms.  It might seem that the underlying fivebranes have completely disappeared into geometry and fluxes; that the notion of ``where the fivebranes are'' and how much they are separated has no precise answer; and that therefore the notion of what happened to the nonabelian degrees of freedom cannot be answered.  However, by carefully tracing through the duality structure one can see that these nonabelian degrees of freedom are in fact branes wrapping the KKM topology at the bottom of the throat; for a related example, see~\cite{Martinec:2015pfa}, and for related earlier work, see~\cite{Das:2005za}.

There has been some debate as to how one should interpret the two-charge solutions, in particular how the entropy of the system arises in different duality frames (see~\cite{Mathur:2018tib} for a recent discussion and further references).
The fuzzball paradigm was to some extent motivated by the idea that these two-charge solutions and the fact that they cap off without a horizon might be a good model for what happens when one adds a third charge to obtain a large black hole.  
However there are reasons to be cautious when asking how much of this physics might carry over to large black holes.
First of all, the geometry of the two-charge solutions is typically quite stringy.  For instance, in the NS5-F1 frame the KKM structures shrink and develop orbifold singularities as the angular momentum is reduced.  In fact, for the typical two-charge fuzzball solution with angular momentum less than of order $\sqrt{\vphantom{t}n_1n_5}$, the geometry in the vicinity of the cap of the geometry has curvatures of order the string scale or more~\cite{Martinec:1999sa,Mathur:2005ai,Chen:2014loa} in the local duality frame appropriate for the physics of the cap.  A supergravity analysis is thus not valid everywhere in the throat, and has significant corrections in the region of interest near the cap. 
In light of the correspondence principle discussion above, the two-charge BPS system is at or below its correspondence point; regarding it as an ensemble of black hole microstates may not be the most useful interpretation.

The semiclassical quantization of the BPS supertube moduli space outlined above mirrors a similar quantization of the moduli space of multi-center brane bound states using quiver quantum mechanics%
~\cite{Denef:2002ru,Denef:2007vg,deBoer:2008fk,deBoer:2008zn,deBoer:2009un,Sen:2009vz,Dabholkar:2010rm,Bena:2012hf,Lee:2012sc,Manschot:2012rx}.
There, vector multiplets in the quantum mechanics describe the locations of fiber degenerations in the geometry which cap it off; their expectation values characterize the depth of the throat and again relate it to the angular momenta of the constituents.  One has a similar structure in the onebrane-fivebrane supertube, but with the degenerations happening along a one-dimensional submanifold in five spatial dimensions rather than at discrete points in four spatial dimensions.  Nonzero angular momentum of the supertube is directly related to the formation of the topological structures that cap off the geometry at a finite redshift.

In the quiver QM models, the capped geometries lie on the Coulomb branch side of a Coulomb-Higgs phase boundary, with single-center black holes lying in the Higgs phase~\cite{Denef:2007vg,Bena:2012hf,Lee:2012sc,Manschot:2012rx}.  More precisely, upon integrating out the vector multiplets the effective hypermultiplet QM on the Higgs branch captures all the BPS states, and the Coulomb branch states can be described in either of the Coulomb or Higgs branch effective theories.  However, upon integrating out the hypermultiplets, the effective vector multiplet QM on the Coulomb branch does not contain zero angular momentum states that are intrinsic to the Higgs branch.  Similarly, in the $AdS_3$ limit of the onebrane-fivebrane system there is a dual spacetime CFT in terms of Higgs branch hypermultiplets that captures all the BPS states. The quantization of the BPS supertube moduli space described above is an analogue of the effective Coulomb branch QM.%
\footnote{Indeed, upon T-duality along $\bS^1$ the half-BPS NS5-F1 bound states become half-BPS momentum excitations of the NS5 branes.  Excitations of the scalars describing the transverse location of the fivebranes carry angular momentum and pry the fivebranes apart slightly onto their Coulomb branch.}
In what follows, we will use this Coulomb-Higgs language to describe the states at and near the BPS bound in the onebrane-fivebrane system.  Of particular interest are the additional degrees of freedom that are essential for a complete characterization of the state space, beyond the collective modes of the Coulomb branch.

Recently, new tools have become available~\cite{Martinec:2017ztd,Martinec:2018nco} that provide an exact worldsheet description of a special class of two-charge BPS configurations where the fivebranes are at the same time bound together, and slightly separated on their Coulomb branch, namely the round NS5-P and NS5-F1 supertubes studied in~\cite{Lunin:2001fv,Lunin:2001jy} (as well as three charge NS5-F1-P bound states obtained from these supertubes by solution-generating transformations known as {\it spectral flow}~\cite{Lunin:2004uu,Giusto:2004id,Giusto:2004ip,Jejjala:2005yu,Giusto:2012yz}).  On the one hand, the gravitational effects of the fivebranes are under control at the exact level in $\alpha'$, and perturbatively in $\gstr$.  Being solitonic objects, the NS5-branes' configuration is part of the classical background, with gravitational back-reaction fully taken into account.  On the other hand, this class of supertubes is rich enough that one can dial discrete parameters of the background to approach the black hole threshold and analyze the fluctuation spectrum.  The spectrum of closed strings was analyzed in~\cite{Martinec:2018nco}; our purpose here is to study in detail the D-brane spectrum.  The latter is of considerable interest in that, as discussed above, D2-branes stretching between NS5-branes in type IIA are the Coulomb branch avatars of the little string.  In the NS5-P supertube, this structure will be readily apparent; and T-duality will convert that structure to that of a D3-brane wrapping the bubbled geometry of the NS5-F1 supertube.%
\footnote{In this context, it is interesting to note that in the BPS states intrinsic to the Higgs branch in quiver QM models, the hypermultiplets being turned on are U-dual to D-branes wrapping the topology of bubbled solutions~\cite{Martinec:2015pfa}, and are thus similar in spirit to the W-branes being analyzed here.}  
All of this structure is under precise control since we have access to an exactly solvable worldsheet CFT.

%%%%%%%%%%%%%%%%%%
\begin{figure}[ht]
\centering
  \begin{subfigure}[b]{0.4\textwidth}
    \includegraphics[width=\textwidth]{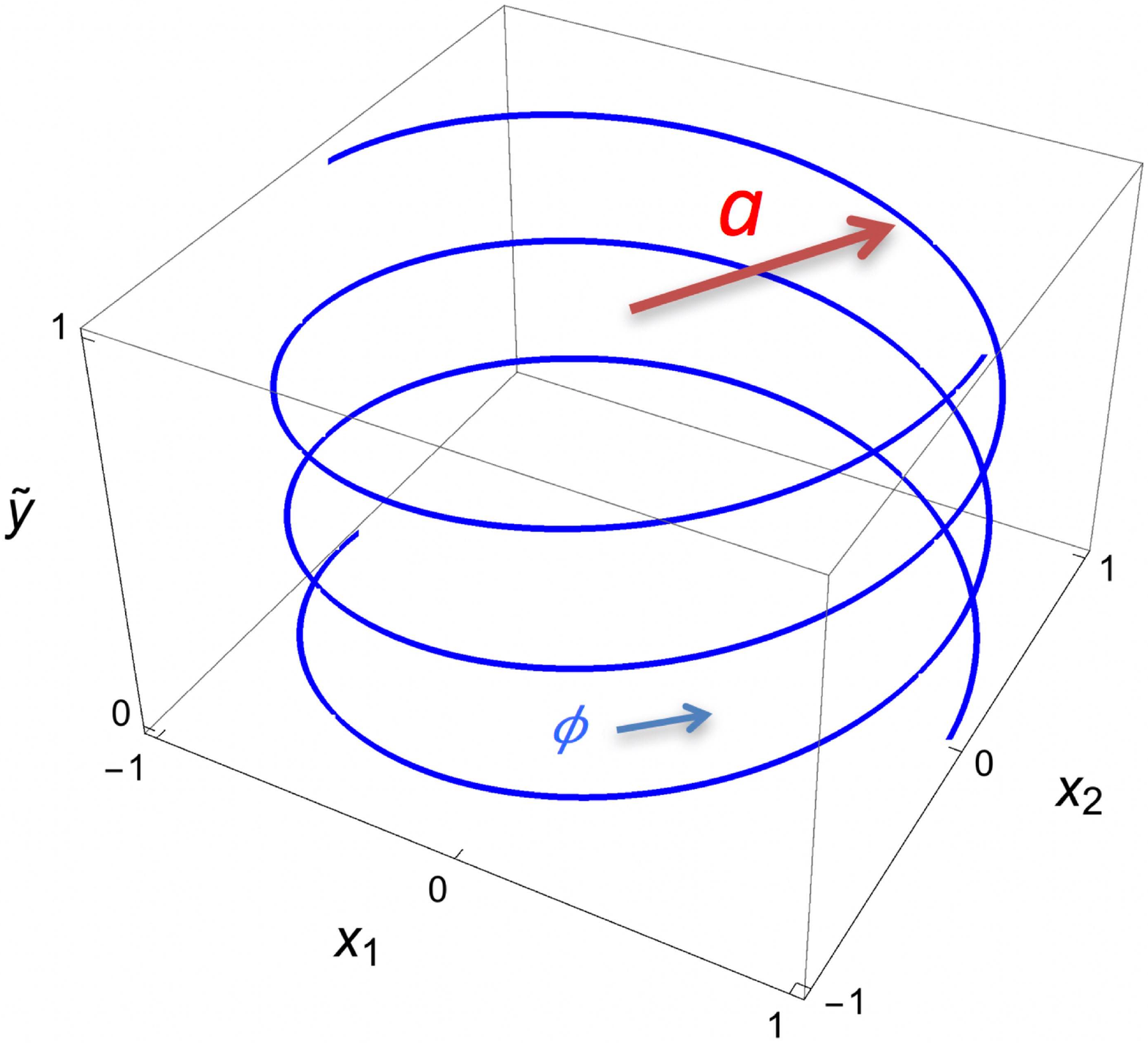}
    \caption{ }
    \label{fig:AdSchiralprimary3-1}
  \end{subfigure}
\qquad\qquad
  \begin{subfigure}[b]{0.4\textwidth}
    \includegraphics[width=\textwidth]{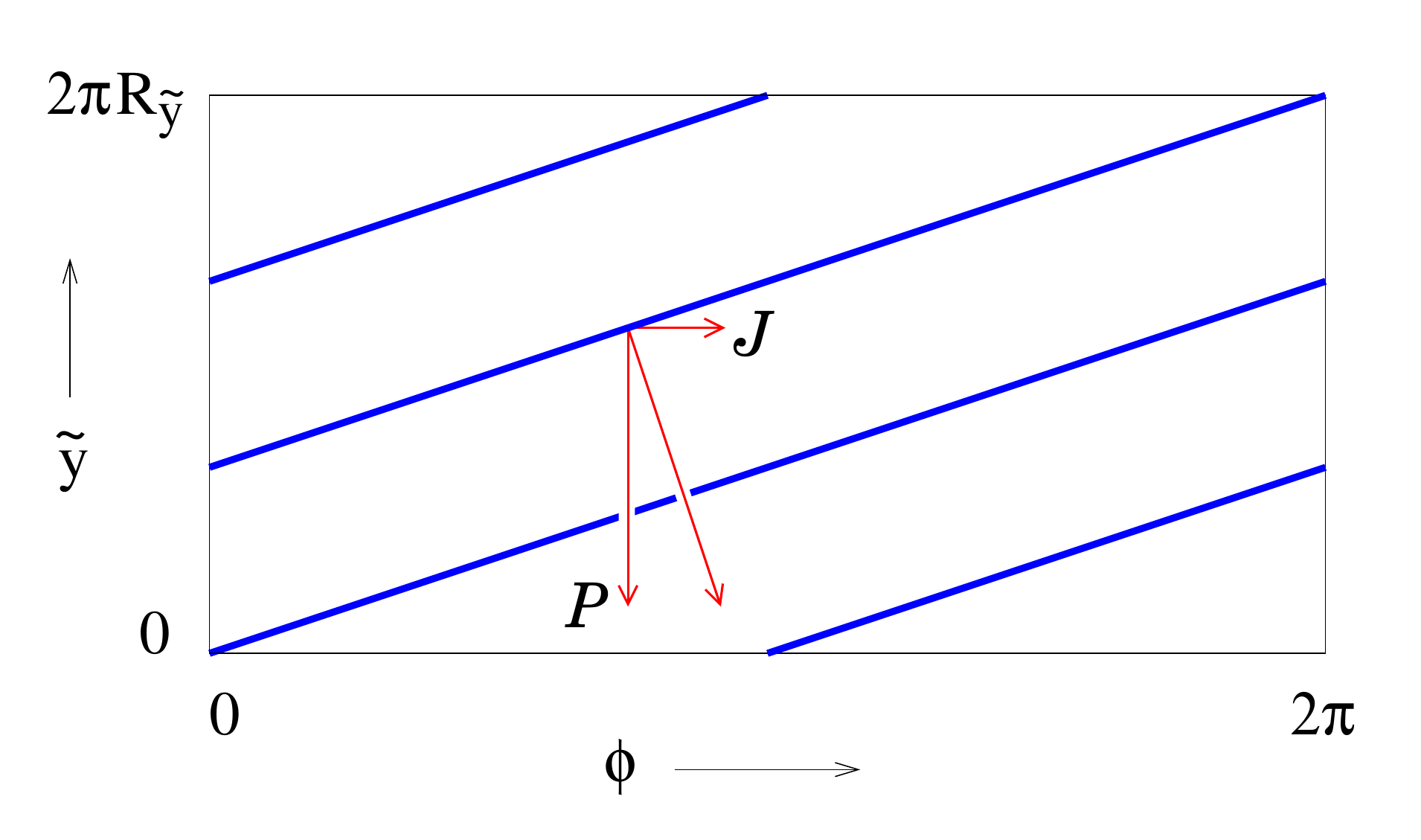}
    \caption{ }
    \label{fig:Supertube-alt3}
  \end{subfigure}
\caption{\it 
Source for the round NS5-P supertube.  (a) A single BPS source with $n_5=2$ and $k=3$; (b) Unrolling the circle of radius $\boldsymbol a$ reveals a fivebrane source moving transversely to its worldvolume on the $\ytil$-$\phi$ torus.  The fivebrane winds along the $(\nfive,k)$ cycle of this torus.
}
\label{fig:Supertube}
\end{figure}
%%%%%%%%%%%%%%%%%%

The round NS5-P supertube is obtained by macroscopically exciting a single chiral mode of the scalars $X^i$ describing the embedding of the $m^{\rm th}$ fivebrane worldvolume:
\be
\label{profile1}
X^1+iX^2
\equiv |X|\, e^{i\phi}
 = a \, \exp\Bigl[ i\frac{k}{\nfive} \frac{(t-\tilde y )}{\Rytil} + \frac{2\pi i \suplabel}{\nfive}\Bigr] 
\ee
where $m=1\dots\nfive$ labels the fivebranes, and $\ytil$ parametrizes the $\bS^1$ of radius $\Rytil$ in the compactification.  The monodromy of the solution winds together the fivebrane worldvolumes, into a single unit if $k$ and $\nfive$ are relatively prime.  The angular momentum of the branes in the $x^1$-$x^2$ plane supports the branes at finite separation, preventing their collapse to the origin and thus dynamically stabilizing the mass of W-branes at a finite value determined by the radius $a$ of the supertube,
\be
a = \frac{\sqrt{Q_5 Q_p}\,\Rytil}{k\lstr^2} 
~~,~~~~
Q_5 =  n_5\lstr^2
~~,~~~~
Q_p = \frac{n_p\gstr^2\lstr^8}{V_4\Rytil^2} ~,
\ee
where $V_4$ is the volume of $\cM$ and $n_p$ is the number of momentum quanta.%
\footnote{For the momentum charge to be part of the classical supergravity background, 
one must have $Q_p \gg \lstr^2$, and so typically $n_p\propto \gstr^{-2}$.}
The supertube with $\nfive=2$ and $k=3$ is depicted in Figure~\ref{fig:Supertube}.

In the T-dual NS5-F1 frame, $R_y=\lstr^2/\Rytil$ and $Q_p\to Q_1$; the quantum numbers of the family of round two-charge supertubes in the $AdS_3$ limit are indicated by the blue dots in Figure~\ref{fig:Spectrum-orb2}, together with the effect of spacetime spectral flow which produces three-charge supertubes.  Note that as the mode number $k$ increases, the supertube shrinks and coils more and more; as $k$ becomes macroscopic (bounded by $n_1n_5$), the state approaches the black hole threshold from below.  The W-brane tension becomes lighter and lighter as $k$ increases, due to the increasing redshift to the bottom of the throat where the supertube source is located.

%%%%%%%%%%%%%%%%%%
\begin{figure}[ht]
\centering
    \includegraphics[width=.5\textwidth]{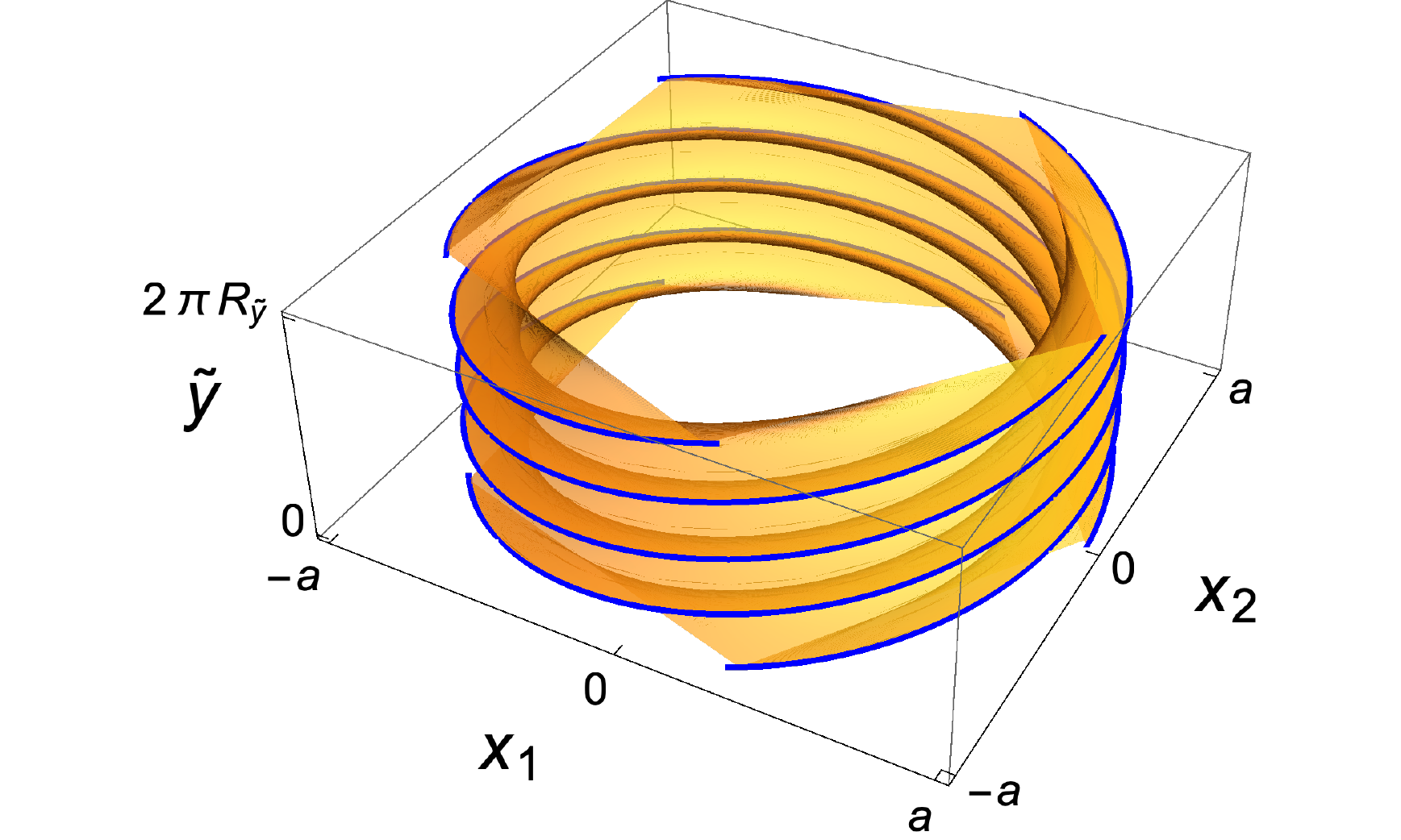}
\caption{\it 
An open D2-brane (gold) stretches between a spiraling stack of NS5-branes (blue).  The membrane only closes on itself after winding around the $\phi$ circle $k$ times; here $k=4$, $\nfive=5$.
}
\label{fig:WbraneFig}
\end{figure}
%%%%%%%%%%%%%%%%%%

Figure~\ref{fig:WbraneFig} depicts a W-string stretched between neighboring strands of an NS5-P supertube, and extending along the fivebrane worldvolume until it wraps around enough times to close on itself.  In the process, it winds $k$ times around the angular circle of the supertube source in the transverse $x^1$-$x^2$ plane, and $\nfive$ times around the circle $\bS^1_\ytil$ wrapped by the fivebranes. 
T-duality along $\bS^1_\ytil$ relates the NS5-P and NS5-F1 supertubes.  From Figure~\ref{fig:Supertube-alt3}, we see that the fivebrane worldvolume lies partly along and partly transverse to the $\ytil$ circle.  This affects the result of the T-duality, which for a longitudinal circle preserves NS5 charge, while for a transverse circle it transforms NS5 branes to Kaluza-Klein monopoles.  The coiled ring of NS5-branes thus becomes a coiled ring of KK monopoles under T-duality.  The local structure of nearly coincident KKM's is a slightly resolved $A_{k-1}$ singularity transverse to the ring.  The monodromy on the NS5-P side, that winds all the fivebranes together into a single strand, becomes on the NS5-F1 side a monodromy that cyclically permutes the $k$ two-cycles of the ring of $A_{k-1}$ singularities as one passes once around the ring.  The global topology of the KKM ring is thus $\bS^2\times\bS^1$ (with the $\bS^1$ being a $k$-fold cover of the $\phi$ circle of~\eqref{profile1}) rather than the $(\bS^2)^{k-1}\times\bS^1$ that one might have naively guessed from the effective local $(AdS_3\times\bS^3)/\bZ_k$ geometry at the bottom of the throat.

The D2 W-brane stretching between strands of the NS5-P supertube helix transforms under T-duality to a
D3 W-brane wrapping this coiled $\bS^2\times\bS^1$.  Under the T-duality, the fivebranes seem to have totally disappeared into geometrical flux, however they are not completely gone -- the underlying source structure is diagnosed by stringy probes.

%%%%%%%%%%%%%%%%%%%%%%%%%%%%%%%%%%%%
\subsection{Expanding the toolkit for fivebrane dynamics}

The tool that allows us to analyze these D-branes and the substringy structure they probe begins with the Wess-Zumino-Witten (WZW) model for the 10+2 dimensional group manifold%
\footnote{Here we choose the compactification $\cM=\bT^4$; for $\cM=K3$ one can consider a point in moduli space where the worldsheet theory is solvable, such as a torus orbifold or Landau-Ginsburg orbifold.  Of course, current algebra CFTs underly these constructions as well.}
\be
\label{Gupstairs}
\mathcal{G} =
\bigl(SL(2,\mathbb{R}) \times SU(2)\bigr) \times \bigl(\bR_t\times\bS^1_y\times \bT^4\bigl) ~,
\ee
and gauges a pair of null isometries so that the physical target spacetime geometry is 9+1 dimensional~\cite{Martinec:2017ztd}.  Roughly, the first factor in $\cG$ lies largely transverse to the fivebranes but has 5+1 rather than 4+0 dimensions; the role of the gauging is to eliminate the unwanted directions.  However, one has a choice to involve the second factor in the gauge current; the freedom in the choice of this admixture comprises a set of discrete parameters which determine the supertube shape, such as the parameter $k$ in~\eqref{profile1}.

It will turn out that the well-understood spectrum of D-branes in the component factors of $\cG$%
~\cite{Alekseev:1998mc,
Alekseev:1999bs,
Stanciu:1999id,
Alekseev:2000fd,
Bachas:2000fr,
Pawelczyk:2000hy,
Giveon:2001uq,
Israel:2005ek}
and in gauged WZW models%
~\cite{Maldacena:2001ky,
Gawedzki:2001ye,
Elitzur:2001qd,
Fredenhagen:2001kw,
Sarkissian:2002ie,
Walton:2002db,
Sarkissian:2002bg,
Sarkissian:2002nq,
Quella:2002ns,
Quella:2002fk,
Quella:2003kd}
allows us to describe W-branes in these special supertube backgrounds, in a manner closely related to the work of~\cite{Israel:2005fn}.
We begin in Section~\ref{sec:sugrasolns} with a summary of the relevant supergravity solutions:
\vspace{-1mm}
\begin{itemize}
\item
NS5-branes on the Coulomb branch, distributed along a circle~\cite{Sfetsos:1998xd,Giveon:1999px,Giveon:1999tq}
\vspace{-1mm}
\item
NS5-P supertubes~\cite{Mateos:2001qs,Lunin:2001fv}
\vspace{-1mm}
\item
NS5-F1 supertubes~\cite{Lunin:2001fv}
\vspace{-1mm}
\item
BPS fractional spectral flows of these supertubes~\cite{Lunin:2004uu,Giusto:2004id,Giusto:2004ip,Jejjala:2005yu,Giusto:2012yz}.
\end{itemize} 
\vspace{-0.5mm}
The original construction of string dynamics in the background of NS5-branes on the Coulomb branch~\cite{Giveon:1999px,Giveon:1999tq} used a noncompact version~\cite{Ooguri:1995wj} of the Calabi-Yau/Landau-Ginsburg correspondence~\cite{Martinec:1988zu,Vafa:1988uu} to describe the transverse space of the fivebranes in terms of WZW coset models
\be
\left(\frac{\sltwo}{\uone}\times\frac{\sutwo}{\uone}\right)/\bZ_{\nfive} ~.
\ee 
The reorganization of the gauge and orbifold groups into the gauging of a pair of left- and right-moving null isometries (introduced in~\cite{Israel:2004ir,Itzhaki:2005zr}),
\be
\label{GmodH}
\frac{\cG}{\cH} =
\frac{\bigl(SL(2,\mathbb{R}) \times SU(2)\bigr) \times \bigl(\bR_t\times\bS^1_y\times \bT^4\bigl)}
{\uone_L\times \uone_R} \;,
\ee
provides the freedom necessary to describe the remaining backgrounds by generalizing the embedding of $\cH$ from lying strictly in the first factor of $\cG$ to involving a mixture of both $\sltwo\tight\times\sutwo$ and $\bR_t\tight\times\bS^1_y$.

In Section~\ref{sec:gnlsm}, we review the gauged nonlinear sigma model, following the general formalism of~\cite{Hull:1989jk,Figueroa-OFarrill:2005vws}.  The presence of antisymmetric tensor flux complicates matters, in particular there is an intricate interplay between gauge invariance and the Wess-Zumino term, especially in the presence of worldsheet boundaries.  We then specialize the discussion to group manifolds $\cG$, using the symmetry analysis of%
~\cite{Alekseev:1998mc,
Bachas:2000fr,
Elitzur:2001qd,
Gawedzki:2001ye,
Walton:2002db,
Sarkissian:2002ie,
Sarkissian:2002bg,
Sarkissian:2002nq}
and especially the work of Quella and Schomerus~\cite{Quella:2002ns,Quella:2002fk,Quella:2003kd}  
to determine both the shape of the brane in simple examples as well as the two-form $\omega_2=B+\cF$ that solves the DBI equations of motion.

Section~\ref{sec:ST CFT} reviews the results of~\cite{Martinec:2017ztd}, showing how the choice of
gauged null isometries in~\eqref{GmodH} yields the fivebrane backgrounds of interest.  
Various D-branes in the $\sltwo$ and $\sutwo$ WZW models are then reviewed in Section~\ref{sec:GroupBranes}.  The canonical examples of D-branes on group manifolds preserve the maximum group symmetry, lying along a (twisted) conjugacy class of the group $\cG$; in addition, there are branes that preserve only a subgroup $\cH\subset\cG$ of the full symmetry%
~\cite{Maldacena:2001ky,
Gawedzki:2001ye,
Elitzur:2001qd,
Fredenhagen:2001kw,
Sarkissian:2002ie,
Walton:2002db,
Quella:2002ns,
Sarkissian:2002bg,
Sarkissian:2002nq,
Quella:2002fk,
Quella:2003kd};
their ``symmetry breaking'' worldvolumes lie along products of conjugacy classes of $\cG$ and $\cH$.  Since we will be gauging $\cH=\left(\uone_L\times\uone_R\right)\subset\cG$, we only need such a subgroup to be preserved, and the symmetry-breaking branes are indeed an essential ingredient of our construction.
We then assemble these component D-branes into W-branes in various situations:
NS5-branes separated onto their Coulomb branch in Section~\ref{sec:NS5Coul};
round NS5-P and NS5-F1 supertubes in Section~\ref{sec:NS5P-NS5F1};
and spectral flows of these supertubes carrying all three charges in Section~\ref{sec:specflowST}.

The analysis of Section~\ref{sec:NS5Coul} reproduces within the formalism of null gauging the results of Israel \etal~\cite{Israel:2005fn}, which used the coset orbifold description.  The W-branes of interest are constructed by starting with D-branes in $\cG$, whose worldvolumes are specified by conjugacy classes of the various group factors $\sltwo$, $\sutwo$, \etc; this D-brane core is then smeared along the orbits of $\cH$ to obtain a brane invariant under the gauge symmetry.
In Sections~\ref{sec:NS5P-NS5F1} and~\ref{sec:specflowST} we apply this method to the more general gaugings that yield supertubes as the effective geometry.  A key aspect of the construction is the non-factorized nature of the smearing procedure.  The gauge orbits combine motion in the various factors of $\cG$, and so the resulting brane after smearing is not a factorized product of D-branes in $\sltwo$, $\sutwo$, \etc.
We will also use the same smearing procedure to generate the spiraling W-brane geometry of Figure~\ref{fig:WbraneFig}, whose worldvolume lies along a diagonal combination of the physical (\ie\ gauge invariant) directions $\phi$ and $\ytil$ in~\eqref{profile1} (see Figure~\ref{fig:Supertube}).  This spiral trajectory, multiply covering a circle in the effective geometry, allows the W-brane to capture aspects of the long string structure of the black hole phase in a regime amenable to analysis in perturbative string theory.  
We conclude with a discussion of our results in Section~\ref{sec:Discussion}.

%%%%%%%%%%%%%%%%%%%%%%%%%%%%%%%%%%%%
%%%%%%%%%%%%%%%%%%%%%%%%%%%%%%%%%%%%
\section{Review of supergravity solutions}
\label{sec:sugrasolns}

The simplest background of interest here is that of $\nfive$ nearly coincident NS5-branes wrapped around $\bS^1\times\bT^4$.  In the decoupling limit $\gstr\to 0$, the geometry of coincident fivebranes can be written as
(choosing conventions where $\lstr=1$)
\begin{align}
\label{CHSsoln}
ds^2 &= \Bigl(-du\,dv +ds_{\scriptscriptstyle\mathbf T^4}^2 \Bigr)
+ n_5\Bigl[d\rho^2 + d\theta^2 + 
\sin^2\!\theta \,d\phi^2 + \cos^2\!\theta \,d\psi^2\Bigr]
\nn\\[8pt]
B   &= {n_5 \cos^2\theta } \, d\phi \wedge d\psi
\quad,\qquad
\Phi = -\rho ~.
\end{align}
where $u=t+y$, $v=t-y$.
The nonlinear sigma model on this background is exactly solvable~\cite{Callan:1991at} -- the directions along the brane are described by free fields, while the radial direction in the transverse space is a free field with linear dilaton, and the angular directions in the transverse space yield an SU(2) Wess-Zumino-Witten (WZW) model whose $\sutwo_L\times\sutwo_R$ current algebra symmetry has level $\nfive$.  While it is nice that the free string dynamics is exactly solvable, the S-matrix has no perturbative expansion~-- string wave packets sent down the throat inevitably reach the region of arbitrarily large string coupling near the fivebrane source at $\rho\to -\infty$.

%%%%%%%%%%%%%%%%%%%%%%%%%%%%%%%%%%%%

\paragraph{Coulomb branch NS5's:} 
The cure for this problem is to slightly separate the fivebranes onto their Coulomb branch moduli space~\cite{Sfetsos:1998xd,Giveon:1999px,Giveon:1999tq}.  
Neveu-Schwarz fivebranes separated in a $\bZ_{\nfive}$ symmetric array on their Coulomb branch source a background
\begin{align}
\label{NS5 coulomb}
ds^2 &= \Bigl(-du\,dv +ds_{\scriptscriptstyle\mathbf T^4}^2 \Bigr)
+ n_5\Bigl[d\rho^2 + d\theta^2 + \frac{1}{\Sigma_0}\Bigl(
{\cosh}^2\!\rho\sin^2\!\theta \,d\phi^2 + {\sinh}^2\!\rho\cos^2\!\theta \,d\psi^2\Bigr)\Bigr],
\nn\\[8pt]
B   &= \frac{n_5 \cos^2\theta \cosh^2\rho}{\Sigma_0} \, d\phi \wedge d\psi
\,,\qquad
e^{-2\Phi} = \frac{\Sigma_0}{\gstr^2\nfive}  \,, \qquad 
\Sigma_0 \;\equiv\; {\sinh^2\!\rho + \cos^2\!\theta} \,.
\end{align}
String theory on this background remains exactly solvable -- it is a non-compact version of the Calabi-Yau/Landau-Ginsburg (CY/LG) correspondence~\cite{Martinec:1988zu,Vafa:1988uu}, in this case given by the coset orbifold%
~\cite{Sfetsos:1998xd,Giveon:1999px,Giveon:1999tq}
\be
\Bigl(\frac{\sltwo}{\uone} \times \frac{\sutwo}{\uone}\Bigr)/\bZ_{\nfive}  ~,
\ee
whose low-energy S-matrix is perturbatively well-defined.  While it may appear that the geometry still has a strong coupling singularity at $\rho=0,\theta=\pi/2$, this is an artifact of the classical approximation to the sigma model; at the full quantum level, the coset sigma models are entirely well-behaved.

The gauge orbit in $\sltwo/\uone$ is a timelike circle of size $\sqrt\nfive\,\lstr$; likewise the gauge orbit in $\sutwo/\uone$ is a spacelike circle of the same size.  The effect of the $\bZ_\nfive$ orbifold is to rearrange the gauge group into $\cH=\uone_L\times\uone_R$ where the tangents to the gauge orbits are null directions in $\mathfrak{sl}_2\oplus\mathfrak{su}_2$.%
\footnote{More precisely, as discussed in~\cite{Martinec:2018nco} the global structure of the gauge group is non-compact in the timelike direction and thus $\bR_{L+R}\times U(1)_{L-R}$.  The distinction, while important, will not affect our considerations here.}

\paragraph{NS5-P supertubes:}
The realization that there is a more direct presentation of the coset orbifold in terms of null gauging leads immediately to generalizations describing supertubes.  A boost-like transformation on the fivebranes imparts momentum and angular momentum to the fivebranes, resulting in the NS5-P supertube.  In the null-gauged WZW description above, this amounts to tilting the orientation of the null vector so that it points partly along the $\bR_t\times\bS^1_\ytil$ directions~\cite{Martinec:2017ztd}.  Doing this symmetrically on left and right leads to the NS5-P supertube
\begin{align}
\label{smearedNS5Pmetric}
ds^2 &= \Bigl( -du\, dv + ds_{\scriptscriptstyle\mathbf T^4}^2 \Bigr)
+ \nfive\Bigl[d\rho^2+d\theta^2 +  \frac{1}{\Sigma_0} \Bigl( {\cosh}^2\!\rho\sin^2\!\theta \,d\phi^2 + {\sinh}^2\!\rho\cos^2\!\theta \,d\psi^2 \Bigr)\Bigr]
\nn\\[.1cm]
& \hskip 2cm 
+\frac{1}{\Sigma_0} \Bigl[ \frac{2 k }{ \Rytil} \sin^2\!\theta \,dv\, d\phi + \frac{k^2}{\nfive\Rytil^2} dv^2 \Bigr],
\\[8pt]
B  
&= \frac{n_5 \cos^2\theta \cosh^2\rho}{\Sigma_0} d\phi \wedge d\psi +
\frac{ k  \cos^2\theta}{\Rytil\,\Sigma_0}  dv \wedge d\psi
~,\qquad~~~
e^{-2\Phi}  = \frac{n_p\Sigma_0}{n_5 k^2 V_4} \;.
\nn
\end{align}
Again, although it might look as though the string is propagating in a geometry with a strong-coupling singularity, low-energy string dynamics is perturbatively well-behaved and consistent.

%%%%%%%%%%%%%%%%%%%%%%%%%%%%%%%%%%%%

\paragraph{NS5-F1 supertubes:}
T-duality of the NS5-P supertube along $\bS^1_\ytil$ leads to the NS5-F1 supertube. Introducing the notation  $\alphab\equiv k\Ry$, the NS5-F1 solution is
\begin{align}
\label{smearedNS5F1metric}
ds^2 &= \Bigl( -du\:\! dv + ds_{\scriptscriptstyle\mathbf T^4}^2  \Bigr)
+ {n_5}\Bigl[ d\rho^2+ d\theta^2+\frac1\Sigma \Bigl( {\cosh}^2\!\rho\sin^2\!\theta \,d\phi^2 + {\sinh}^2\!\rho\cos^2\!\theta \,d\psi^2 \Bigr)\Bigr] 
\nonumber\\[.1cm]
& \hskip .6cm 
+ \frac{2\alphab}{\Sigma} \Bigl(  {\sin^2\!\theta \, dt\,d\phi + \cos^2\!\theta \, dy\,d\psi}  \Bigr)
+\frac{\alphab^2}{\nfive\Sigma} \Bigl[ \nfive \sin^2\!\theta \, d\phi^2 +  \nfive\cos^2\!\theta \, d\psi^2  
+ du\:\! dv \Bigr],
\nonumber\\[8pt]
B  &= \frac{ \cos^2\!\theta (\alphab^2+\nfive\cosh^2\!\rho)}{\Sigma}  { d\phi\wedge d\psi - \frac{\alphab^2}{n_5\Sigma} \, dt\wedge dy } \nonumber\\
& \hskip .6cm 
{}+\frac{\alphab  \cos^2\!\theta}{\Sigma}  dt\wedge d\psi
+\frac{\alphab  \sin^2\!\theta}{\Sigma}  dy\wedge d\phi~,
\nn\\[8pt]
e^{-2\Phi} & = \frac{n_1\Sigma}{k^2\Ry^2\,V_4} ~,\qquad~~~  
\Sigma = \frac{\alphab^2}{\nfive} + \Sigma_0 \;.
\end{align} 
where now $u,v$ are defined in terms of the T-dual coordinate $y$, i.e.~$u,v=t\pm y$, and where we have divided some terms into two parts for later convenience.
This geometry has a local $\bZ_k$ orbifold singularity at $\rho=0$, $\theta=\pi/2$ that identifies the angles according to
\be
(y/\Ry,\psi) \sim (y/\Ry,\psi) + \frac{2\pi}{k} (1,-1) \;.
\ee

%%%%%%%%%%%%%%%%%%%%%%%%%%%%%%%%%%%%

\paragraph{NS5-F1-P supertubes:}
Fractional spectral flow of the above two-charge supertubes (\ie\ a particular large diffeomorphism of the angular coordinates) yields a larger set of backgrounds carrying three charges -- NS5 along $\bS^1\times \cM$ as well as both string winding and momentum along $\bS^1$%
~\cite{Giusto:2004id,Giusto:2004ip,Jejjala:2005yu,Giusto:2012yz}. In
the fivebrane decoupling limit, the solutions take the form:
\begin{align}
\label{GLMTmetric}
ds^2 &= {}- \frac{f_0}\Sigma \;\! du\;\! dv
+ \frac{\Delta_p}{\Sigma} \;\! dv^2
+ n_5 \bigl(d\rho^2 + d\theta^2 \bigr)
\nn\\
&\hskip 1cm
+ \frac{n_5}{\Sigma} \Bigl( \sinh^2\!\rho - s + \Delta_1 \Bigr) \cos^2\!\theta\, d\psi^2
\nn\\ 
&\hskip 1cm
+ \frac{n_5}{\Sigma} \Bigl( \sinh^2\!\rho+(s+1) + \Delta_1 \Bigr) \sin^2\!\theta\, d\phi^2
\\
&\hskip 1cm
- \frac{2{n_5\Delta_1}}{ k\Ry\Sigma}\Bigl({ s\cos^2\!\theta\,d\psi } - (s+1)\sin^2\!\theta\,d\phi\Bigr) dv
\nn\\
&\hskip 1cm
+ \frac{2{n_5\Delta_1}\etab}{k\Ry\Sigma}\Bigl({\cos^2\!\theta\,d\psi}+\sin^2\theta\,d\phi\Bigr) dy + dz_a\,dz^a~,
\nn\\[15pt]
%\end{align}
%
%\begin{align}
\label{GLMT_Bfield}
B_2 &= -\frac{\Delta_1}{\Sigma} \, dt\wedge dy
+\frac{n_5\cos^2\!\theta}{\Sigma}\Bigl( \sinh^2\!\rho + (s+1) + \Delta_1 \Bigr) { d\phi\wedge d\psi}
\\\
&\hskip 1cm
 - \frac{ {n_5\Delta_p}\etab}{k \Ry \Sigma} dy\wedge\bigl({\cos^2\!\theta\,d\psi} + \sin^2\!\theta\, d\phi\bigr) 
\nn\\
&\hskip 1cm
{ + \frac{{n_5\Delta_1}\cos^2\!\theta}{k\Ry\Sigma}\bigl((s\tight+1)\,dt -s\,dy\bigr)\wedge d\psi }
\nn\\
&\hskip 1cm
- \frac{{n_5\Delta_1}\sin^2\!\theta}{k\Ry\Sigma}\bigl(s\,dt -(s\tight+1)\,dy\bigr)\wedge d\phi~,
\nn\\[15pt]
e^{-2\Phi} &= \frac{n_1\etab\,\Sigma }{k^2\Ry^2 V_4}, 
\label{GLMT_Phi}
\end{align}
%%%%%%%%%%%%%%%%%%%%%%%%%%
%
where
\begin{align}
\label{GLMTparams}
\Delta_1&=\frac{k^2R_y^2}{\nfive\lstr^2} + s(s+1)
~,~~~~
\Delta_p=\frac{s(s+1) \nfive \lstr^2}{k^2\Ry^2} \,\Delta_1
~,~~~~
\etab = \frac{k^2R_y^2}{k^2R_y^2 + s(s+1)\nfive\lstr^2}~,
\nn\\[8pt]
f_0 &= \Bigl[ \sinh^2\!\rho -s\,\sin^2\!\theta + (s\tight+1)\cos^2\!\theta \Bigr]
~,~~~~
\Sigma = f_0 + \Delta_1
~.
\end{align}

%%%%%%%%%%%%%%%%%%%%%%%%%%%%%%%%%%%%

All of these backgrounds can be obtained~\cite{Martinec:2017ztd} by gauging null isometries in the WZW model on the group manifold
\be
\label{Gupstairs1}
\mathcal{G} =
\bigl(SL(2,\mathbb{R}) \times SU(2)\bigr) \times \bigl(\bR_t\times\bS^1_y\times \cM\bigl) ~;
\ee
the motion along these isometries is generated by left- and right-moving null currents 
\be
\label{generalcurrent}
\cJ=\ell^i J_i
~~,~~~~
\bar\cJ= r^i \bar J_i \;,
\ee 
where the index $i$ runs over the Lie algebra of $\cG$, and the null conditions are
\be 
\label{nullcond}
\langle {\boldsymbol{\ell}},\boldsymbol{\ell} \rangle = \langle \boldsymbol{r},\boldsymbol{r} \rangle = 0 \;.
\ee
Starting with a 10+2 dimensional group manifold $\cG$ and generating physical (9+1)-d spacetime as the set of gauge equivalence classes under the orbits of $\cH$, all the above backgrounds are obtained by varying the embedding $\cH\hookrightarrow \cG$.  Thus we turn now to a discussion of gauged nonlinear sigma models.

%%%%%%%%%%%%%%%%%%%%%%%%%%%%%%%%%%%%
%%%%%%%%%%%%%%%%%%%%%%%%%%%%%%%%%%%%

\section{The gauged nonlinear sigma model}
\label{sec:gnlsm}

We now discuss gauged nonlinear sigma models, following \cite{Hull:1989jk,Figueroa-OFarrill:2005vws}, and then specialize the analysis to group manifolds. We will mostly follow the presentation in \cite{Figueroa-OFarrill:2005vws}, and we will generalize the considerations of that paper to include more general boundary conditions as considered in%
~\cite{Maldacena:2001ky,
Gawedzki:2001ye,
Elitzur:2001qd,
Fredenhagen:2001kw,
Sarkissian:2002ie,
Walton:2002db,
Quella:2002ns,
Sarkissian:2002bg,
Sarkissian:2002nq,
Quella:2002fk,
Quella:2003kd}.

%%%%%%%%%%%%%%%%%%%%%%%%%%%%%%%%%%%%
\subsection{Gauging target space isometries}

The 2d nonlinear sigma model on a worldsheet $\Sigma$ with target manifold $\cM$ with metric $G$ and three-form flux $H$ has an action consisting of a kinetic term and a Wess-Zumino (WZ) term, as follows. (To reduce clutter in equations, we suppress some overall normalization factors in this section; we shall give the precise normalizations in \eq{eq:S-kin}--\eq{eq:S-WZ}.)
\be
\label{SWZ}
\cS = \cS_{K} + \cS_{WZ} = \half\int_\Sigma G_{ij}(\varphi)\, d\varphi^i \wedge \star d\varphi^j + \int_\cB H  
 ~,
\ee
where the three-manifold $\cB$ has boundary $\Sigma$.  Suppose $\cM$ admits Killing vectors $\xi_a$ under which $H$ is invariant, $d( \imath_a H)=0$ where $\imath_a$ denotes contraction along $\xi_a$; then one can try to gauge translations along $\xi_a$.  The kinetic term gains a minimal coupling to the gauge field
\be
\label{covderiv}
d\varphi^i  \longrightarrow  \cD\varphi^i = d\varphi^i - A^a \xi_a^i(\varphi)
\ee
while the WZ term can be gauged via
\be
\label{gWZ term}
\cS_{gWZ} = \int_\cB H + \int_{\varphi(\Sigma)} \bigl(A^a \!\wedge \theta_a + \half \imath_a \theta_b\, A^a\!\wedge\! A^b \bigr)
\ee
where the target space one-forms $\theta^a$ satisfy
\be
\imath_a H = d\theta_a 
~~,~~~~
\imath_a\theta_b = -\imath_b\theta_a ~.
\ee
Consistency of gauge transformations along the $\xi_a$ requires that the Lie derivative of $\theta_b$ along $\xi_a$ satisfy
\be
\label{integrability}
\mathcal{L}_a \theta_b = f_{ab}^{~~c} \:\! \theta_c
\ee
where $f_{ab}^{~~c}$ are the structure constants of some Lie algebra $\mathfrak{h}$.

%%%%%%%%%%%%%%%%%%%%%%%%%%%%%%%%%%%%
\subsection*{Including worldsheet boundaries}

The WZ term on a worldsheet $\Sigma$ with boundary must be defined with care, since the WZ term itself asks for a  three-manifold $\cB$ whose boundary is $\Sigma$, so naively the latter cannot have a boundary.  We consider for simplicity a single boundary component lying along a D-brane worldvolume $\cC$. We let $i:\cC \to \cG$ denote the canonical embedding, so $i^*$ is the pull-back to the brane worldvolume. The general sigma-model action is written
\be
\label{bdywzw}
\cS =  \int_\Sigma \cL_{\sst \bf kin} + \int_\cB H - \int_\cD \omegatwo
\ee
where $\cD\subset\cC$ is a disk in spacetime whose boundary coincides with the worldsheet boundary $\partial\Sigma$, $\cB$ is a three-dimensional submanifold of spacetime with boundary $\partial \cB = \Sigma\cup \cD$, 
$H$ is the standard Wess-Zumino term, and $\omegatwo$ is a two-form on the D-brane worldvolume $\cC$ in spacetime satisfying 
\be 
\label{dBeqH}
d\omegatwo  = i^* H  ~.
\ee
The idea is to ``fill in the hole'' in $\Sigma$ with a disk $\cD\subset\cC$ so that there is a proper closed surface that bounds $\cB$.
In string theory one identifies $H$ as the field strength of the NS two-form potential $B$, and $\omegatwo= B+\cF$, the combination of $B$ and the field strength $\cF$ of the gauge field on the D-brane worldvolume $\cC$ which is invariant under antisymmetric tensor gauge transformations $\delta B=d\Lambda$, $\delta \cF = -d\Lambda$. 

Gauging the Wess-Zumino term in the presence of a boundary involves an extension of these forms in the formal tensor product of forms on $\cM$ and on $\Sigma$ (for details, see~\cite{Figueroa-OFarrill:2005vws}):
\be
\Omega_3^{\sst\rm WZ} =  H + \theta^a F_a
~~,~~~~
\Omega_2 = \omegatwo + \imath_a \omegatwo\, A^a
- \half \imath_a\imath_b \omegatwo\; A^a\wedge A^b + h_a F^a
\ee
where $F$ is the field strength of $A$, and in addition to~\eqref{dBeqH} one imposes the constraint that $i^*\theta_a + \imath_a \omegatwo$ is exact, i.e.~\cite{Figueroa-OFarrill:2005vws}
\be
\label{hconstraint}
i^*\theta_a + \imath_a \omegatwo =  dh_a ~.
\ee 
The resulting modification of the gauged WZ term is
\be
\cS_{\rm WZ}  =  \int_\cB H - \int_\cD \omega_2^{\sst\cC}
+ \int_{\varphi(\Sigma)} \bigl( A^a \theta_a + \half\, \imath_a\theta_b\, A^a\wedge A^b \bigr)
+ \int_{\varphi(\partial\Sigma)} h_a A^a  ~.
\ee
The first and third terms comprise the gauged WZ term without boundary \eqref{gWZ term}, while the second term is the boundary term in~\eqref{bdywzw}; the last term is a boundary gauge interaction that ensures gauge invariance as a consequence of the property~\eqref{hconstraint}.

%%%%%%%%%%%%%%%%%%%%%%%%%%
\subsection{Specialization to group manifolds}

In the following, we will be interested in the situation where $\cM$ is a Lie group~$\cG$. We will ultimately be interested in matrix Lie groups, and so we will record expressions for matrix groups along the way.

We thus now apply this general formalism of gauged nonlinear sigma models to the specific case of the Wess-Zumino-Witten model on a group $\cG$, with the metric on $G$ given by the Cartan-Killing metric and the $H$ flux given by the three-form in group cohomology.  The constraint~\eqref{integrability} means that we are gauging a subgroup $\cH\subset\cG_L\times\cG_R$ of the isometries of $\cG$.  

We begin by setting up some notation and conventions. 
For a Lie group $\cG$ and corresponding Lie algebra $\mathfrak{g}$, identified with the tangent space $T_e(\cG)$ at the identity $e$, we define: 
\vspace{-1.5mm}
\begin{itemize}
\item 
$\lambda_g$ and $\rho_g$ are the left- and right-multiplication maps: $\lambda_g(g_0) \tight= g\:\! g_0 \,,$ \hfill  $\rho_g(g_0) \tight= g_0\:\! g \,. $
\item 
$\theta_L$ and $\theta_R$ are the left and right Maurer-Cartan one-forms, $\theta_L\big|_g = \lambda^*_{g^{-1}} \mathrm{id}$\,, \\
$\theta_R\big|_g = -\rho^*_{g^{-1}} \mathrm{id}$\,.
For matrix Lie groups, one can write
\be \label{eq:mc-1-forms}
\theta_L\big|_g \;=\; g^{-1} dg \,, \qquad\quad \theta_R\big|_g \;=\;  - dg \;\! g^{-1} \,.
\ee
\end{itemize}
Note that $\theta_L$ and $\theta_R$ are maps from $T_g(\cG) \to T_e(\cG)$ -- they are one-forms on $\cG$ with values in $\mathfrak{g}$. 
In general for any vector $v$, we have by definition
\be
\label{theta-on-vec}
\theta_L\big|_g(v|_g) \equiv (\lambda_{g^{-1}})_* \:\! (v |_g) \,
\ee
which is an element of $\mathfrak{g}\equiv T_e(\cG)$. 
The minus sign in the definition of $\theta_R$ in and above Eq.\;\eq{eq:mc-1-forms} follows the conventions of \cite{Figueroa-OFarrill:2005vws} and are chosen as such since the group action we will consider will be of the form\footnote{Since near the identity, $(e^{X})^{-1} = e^{-X}$ becomes $(1+X)^{-1} = (1-X)$, the push-forward of the map
%\be
$\cI(g) = g^{-1}\,,$
%\ee
at the identity $e$, is simply minus the identity map $\id$, that is $\;\cI_*\big|_e = - \id $.}
\be
g_0 ~\mapsto ~  g_\ell \, g_0 \, g_r^{-1} \,.
\ee

We denote the left-invariant vector field corresponding to $X\in\mathfrak{g}$ by $X^L$, similarly $X^R$ for the right-invariant vector field.  Note that the action of $\theta_{L}$ on $X^{L}$ ($\theta_{R}$ on $X^{R}$) is simply
\be
\theta_L(X^L) \;=\; X \,, \qquad \quad \theta_R(X^R) \;=\; X \,.
\ee
The Maurer-Cartan equation, for matrix groups, is
\be
d \theta_L = - \theta_L \wedge \theta_L = - \tr\left[ (g^{-1} dg)^2 \right] \,, 
\ee
where matrix multiplication is implied in the wedge. Similarly the standard bi-invariant metric is
\be \label{eq:metric0}
ds^2 ~=~  \frac{1}{2} d \theta_L ~=~ -\frac{1}{2} \tr \left[ (g^{-1}dg)^2 \right] \,
\ee
and the standard bi-invariant three-form is\footnote{The minus signs in Eqs.\;\eq{eq:metric0} and \eq{eq:threeformH} are calibrated for SU(2); for SL(2) we will have a relative minus sign once we introduce all appropriate normalizations in \eq{eq:S-kin}--\eq{eq:S-WZ}.}
\be \label{eq:threeformH}
H ~=~ -\frac{1}{3}\tr(\theta_L \wedge \theta_L\wedge \theta_L )  \;=\;-\frac{1}{3}\tr\left[ (g^{-1} dg)^3 \right] \,. 
\ee

In general, one can gauge any subgroup $\cH$ of the $\cG_L\times\cG_R$ isometries of $\cG$, subject to the constraints of anomaly cancellation. The action of $\cH$ is specified by left and right embedding homomorphisms, which we denote by $\ell: \cH \hookrightarrow \cG$ and $r: \cH \hookrightarrow \cG$ respectively, such that the action to be gauged is 
\be
\label{eq:Haction}
g ~\mapsto~ \ell(x) \,g\, r(x)^{-1} \,, \qquad x\in \cH \,.
\ee
The group embeddings $\ell$ and $r$ induce corresponding Lie algebra homomorphisms, which we also denote by $\ell$ and $r$.

We now review the constraints for a consistent gauging, following~\cite{Figueroa-OFarrill:2005vws}.
Let $X_a$ be a basis of $\cH$. For each $X_a$ there is a corresponding Killing vector field given by 
\be
\label{Killingvec}
\xi_a ~\equiv~ - \ell(X_a)^R-  r(X_a)^L   \,.
\ee
For matrix groups, for each Killing vector field $\xi_a$, there corresponds a tangent matrix field $\xi_a g$, given by
\be \label{eq:xi-g}
\xi_a g ~=~ \ell(X_a) \;\! g - g \;\! r(X_a)   \,.
\ee
For instance, given a coordinate $\psi$, if $\xi_a$ is the vector field $\frac{\partial}{\partial \psi}$, then $\xi_a g$ is the matrix field $\frac{\partial g}{\partial \psi}$.
One then has $\imath_a H = d\theta_a$, where
\be
\label{grouptheta}
\theta_a \;\equiv\;  \langle \ell(X_a),\theta_R \rangle - \langle r(X_a), \theta_L \rangle  \;,
\ee
where $\langle \cdot,\cdot\rangle$ is the inner product given by the Killing form on $\mathfrak{g}$, taking into account the normalization of the inner product given by the level $\k$ of the current algebra. For matrix groups, we take for now the canonical normalization $\langle A,B\rangle = \tr(AB)$; at the beginning of the next section we will be more specific about conventions for $\sltwo$ and $\sutwo$, which will involve a relative minus sign between the two groups. 
The constraints of anomaly cancellation $\imath_a\theta_b+\imath_b\theta_a=0$ then evaluate to
\be
\label{anomalyfree}
\bigl\langle \ell(X_a),\ell(X_b)\bigr\rangle  \;=\;  \bigl\langle r(X_a),r(X_b)\bigr\rangle ~.
\ee

Consider a D-brane with worldvolume $\cC\subset\cG$ with associated two-form $\omegatwo$ satisfying~\eqref{dBeqH}; such a D-brane descends to a brane in the coset theory $\cG/\cH$ if in addition the constraint \eqref{hconstraint} is satisfied.  Worldvolumes $\cC$ associated to (products of) twisted conjugacy classes of $\cG$ satisfy these properties, with the added bonus that the equations of motion derived from the DBI effective action are satisfied; and if enough of the chiral algebra of the WZW model is preserved by the worldsheet boundary conditions, one may be able to construct an exact CFT boundary state for the D-brane%
~\cite{Behrend:1999bn,Maldacena:2001ky,Gawedzki:2001ye,Gaberdiel:2001xm,Fredenhagen:2001kw}.  A general method for constructing such branes is laid out in~\cite{Quella:2002ns,Quella:2003kd}; we will now review some of this technology, beginning with the simplest branes that preserve the maximal chiral algebra symmetry.

%%%%%%%%%%%%%%%%%%%%%%%%%%%%%%%%%%%%
\subsection{Symmetry-preserving branes}
\label{sec:symmetrypreservingbranes}

%%%%%%%%%%%%%%%%%%%%%%%%%%%%%%%%%%%%

So-called {\it symmetry-preserving} branes set $J_a = \Omega_\Gg(-\bar J_a)$ on the worldsheet boundary, for all the currents of $\cG$; here $\Omega_\Gg$ is a group automorphism (which induces a corresponding automorphism of $\mathfrak{g}$ that we also denote by $\Omega_\Gg$).  The $\cG_L\times \cG_R$ symmetry 
\be
\label{WZWsym}
g(z,\bar z) \to \gamma_L(z) g(z,\bar z) \gamma_R(\bar z)
\ee
is then broken to the subgroup $\gamma_R=\Omega_\Gg(\gamma_L^{-1})$ on the boundary; a subgroup of $\cG_L\times\cG_R$ isomorphic to $\cG$ is the maximum amount of symmetry that can be preserved by the boundary conditions.  Thus if $f_\Gg\in\cG$ is an allowed boundary value for the sigma model fields, so is $g f_\Gg\Omega_\Gg(g^{-1})$ for any $g\in\cG$; the allowed boundary values thus lie in a twisted conjugacy class of $\cG$  
\begin{equation}
\label{conjclass}
\mathcal{C}_\cG^{(f_\Gg,\Omega_\Gg)} \;\equiv\; \left\{ g f_\Gg\, \Omega_\Gg (g^{-1})~ ,~~ g \in
  \cG \right\}
\end{equation}
where $f_\Gg$ is a fixed group element.
The worldvolume flux is given by the formula~\cite{Alekseev:1998mc}
\begin{equation}
\label{omegaf}
\omegatwo \;=\;  \tr\Bigl[ \Omega_\Gg (g^{-1}dg) \,f_\Gg^{-1} (g^{-1}dg) \,f_\Gg\,  \Bigr] \, .
\end{equation}
One can show that the property~\eqref{dBeqH} is satisfied.

Let us now consider gauging the $\cH$ action $g ~\mapsto~ \ell(x) \,g\, r(x)^{-1}$, with left and right embeddings $(\ell,r)$ satisfying~\eqref{anomalyfree}.
If the automorphism $\Omega_\Gg$ is such that $r=\Omega_\Gg\circ \ell$ for a subgroup $\cH$ then one can gauge $\cH$; the constraint \eqref{hconstraint} is satisfied, and the symmetry-preserving brane descends to a brane\footnote{Denoted an ``A-brane'' in~\cite{Maldacena:2001ky}.} on $\cG/\cH$.

We shall not review the details of these facts here; the steps can be
found in \cite{Quella:2003kd,Quella:2002ns,Quella:2002fk,Elitzur:2001qd,Walton:2002db,Figueroa-OFarrill:2005vws} and are parallel to those in the following subsection which treats in more detail the more involved example of symmetry-breaking branes.

%%%%%%%%%%%%%%%%%%%%%%%%%%%%%%%%%%%%
\subsection{Symmetry-breaking branes}
\label{sec:symmetrybreakingbranes}

Symmetry-breaking branes are constructed by taking the D-brane worldvolume to lie along a product
of ``generalized twisted conjugacy classes'', following the terminology of~\cite{Quella:2003kd}.  Symmetry-breaking branes are valid D-branes regardless of whether we choose to gauge $\cH$; however, they allow that possibility, or for that matter the gauging of any subgroup of $\cH$.  We will of course be interested in branes that preserve the chosen null gauging~\eqref{Gupstairs1}--\eqref{nullcond}.

Suppose we want to preserve only a subgroup $\cH$ of $\cG$;
in the simplest case, such a symmetry breaking brane worldvolume is given by the following product. Let 
\bea 
\label{eq:cg-ch-param-0}
 \cC_\cG ~=~ \mathcal{C}_\cG^{(f_\Gg,\Omega_\Gg)} &\,=\,& \bigl\{ \, g \;\! f_\Gg \; \Omega_\Gg (g^{-1})~ ,~~~~ g \in
  \cG \bigr\} \,, \nn\\[5pt]
	\cC_\cH &\,=\,& \bigl\{ \;\! \lhat(h) \:\! f_\Hh \, r(h^{-1})~ ,~~ h \in
  \cH \bigr\} \,,
\eea
where 
\be
\label{eq:lhatomegal-0}
\lhat ~\equiv \Omega_\Gg \circ \ell \,.
\ee
then the boundary is%
\footnote{Note that while at first glance it may seem as if we are smearing $\cC_\cG$ by the right group action of $\cH$, the action of $\cH$ on $\cG$ is specified in \eq{eq:Haction}, and the ordering in \eq{classproduct-gen-0} is simply a matter of convention: one could equally choose the opposite ordering and adjust \eq{eq:cg-ch-param-0}--\eq{eq:lhatomegal-0} appropriately.}
\be
\label{classproduct-gen-0} 
\cC  ~=~  \cC_\cG \cdot \cC_\cH ~=~ \bigl\{ \; c_\Gg \, c_\Hh ~~ \bigl|~~   c_\Gg \in \cC_\cG \,,~ c_\Hh \in \cC_\cH \bigr\} \,.
\ee

An important special case sets $\cC_\cH$ to be the embedding of a conjugacy class of $\cH$; here one relates the left and right embeddings via 
\be \label{embed-special-conj-0}
r = \lhat \circ \Omega_\Hh 
\ee
where $\Omega_\Hh$ is an automorphism of $\cH$. 
Writing $f_\Hh = \lhat(\hat{f}_\Hh)$ we then have
\be
\label{classproduct-conjH}
\cC_\cH ~=~ \lhat \Big( h \:\! \hat{f}_\Hh \, \Omega_\Hh (h^{-1}) \Big)
~=~ \lhat \Big(\mathcal{C}_\cH^{(\hat{f}_\Hh,\Omega_\Hh)}\Big) \,.
\ee
For the moment however, we will work with the more general boundary condition \eq{classproduct-gen-0}, and we will return to this point later.

The boundary condition \eq{classproduct-gen-0} breaks the symmetry preserved from $\cG$ to the $\cH$ action \eq{eq:Haction}.  More precisely, if $f\in\cG$ is an allowed boundary value of the sigma model fields, then so also is
$\ell(h) \, f \,r(h)$ for any $h\in\cH$.  Loosely speaking, one has taken a symmetry-preserving brane and smeared it along a generalized conjugacy class of $\cH $ embedded in $\cG$.  We now write down the flux on the branes specified by the symmetry-breaking boundary condition, and then demonstrate the gauge invariance, as done in \cite{Quella:2003kd}, generalizing the presentation of \cite{Figueroa-OFarrill:2005vws} to this boundary condition.

A general method for computing the two-form $\omega_2$ has been formulated in~\cite{Quella:2002ns,Quella:2003kd}. To write the flux, it is convenient to introduce the notation (in what follows $h \in \cH$ and $g \in \cG$)
\bea
\vartheta_\Gg \;\equiv\; dg \:\! g^{-1} \,, \qquad ~~ 
\vartheta_\Hh &\,\equiv\,& dh \:\! h^{-1},  \nn\\[1pt]
\alpha_\Gg ~\equiv~ \mathrm{Ad}_{c_{\Gg}^{-1}} \,,~~
\qquad \alpha_\Hh &\,\equiv\,& \mathrm{Ad}_{c_{\Hh}^{-1}} \!\; \lhat \;,  \\[1pt]
 \bar{\alpha}_\Gg ~\equiv~ \mathrm{Ad}_{c^{\phantom{-1}}_{\Gg}} \!\!\!\;\Omega_\Gg \,, \, 
\qquad \bar{\alpha}_\Hh  &\,\equiv\,&\mathrm{Ad}_{c^{\phantom{-1}}_{\Hh}} \!\!\!\; r \,, \nn
\eea
so that for example $\alpha_\Hh \vartheta_\Hh = c_\Hh^{-1} \lhat (\vartheta_\Hh) c_\Hh$.

The worldvolume flux for the product of these generalized conjugacy classes is given by
\begin{equation}
\label{SBomega}
  \omegatwo = \omegatwo(\cG) + \omegatwo(\cH)  + \omegatwo(\cH,\cG) 
\end{equation}
where 
\begin{align}
\label{omegas}
\omegatwo(\cG)  &\;=\;  \bigl\langle \Omega_\Gg \vartheta_\Gg \,,\, \alpha_\Gg \vartheta_\Gg \bigr\rangle \,,\cr
  \omegatwo(\cH)  &\;=\;  \bigl\langle r \vartheta_\Hh \,,\, \alpha_\Hh \vartheta_\Hh \bigr\rangle \,, \\
  \omegatwo(\cH,\cG) & \;=\;  \bigl\langle c_\Gg^{-1} dc_\Gg \,, \, dc_\Hh c_\Hh^{-1}\bigr\rangle ~=~ 
	 \bigl\langle (\alpha_\Gg-\Omega_\Gg) \vartheta_\Gg \,,\, (\lhat - \bar{\alpha}_\Hh )\vartheta_\Hh \bigr\rangle\,.\nn
\end{align}
One can directly verify that $i^*H = d\omegatwo$: $\,i^*H$ is computed by simply evaluating the three-form H in \eq{eq:threeformH} on the boundary $\cC$ in \eq{classproduct-gen}, and one uses \eq{anomalyfree} and \eq{eq:lhatomegal-0}.

For matrix groups, the flux evaluates to
\begin{align}
\label{omegas-2}
\omegatwo(\cG)  &\;=\;  \tr\Bigl[ \;\!  \Omega_\Gg \bigl(g^{-1} dg\bigr)\:\! f_\Gg^{-1} \bigl(g^{-1} dg\bigr)\:\! f_\Gg \:\! \Bigr] 
		\,, \cr
  \omegatwo(\cH)  &\;=\; \tr\Bigl[ \;\!  r \bigl(h^{-1} dh\bigr)\:\!  f_\Hh^{-1} \;\! \lhat \bigl( h^{-1} dh \bigr) \;\!  f_\Hh \:\! \Bigr] \,,  \\
  \omegatwo(\cH,\cG) & \;=\;  \tr  \Bigl[ \bigl( c_\Gg^{-1} dc_\Gg \bigr)\bigl( dc_\Hh c_\Hh^{-1}  \bigr) \Bigr]  \,. \nn
\end{align}
Before gauging we note that $\omegatwo$, and therefore the action \eq{bdywzw}, is invariant under the global $\cH$-action~\eq{eq:Haction}. To see this, it is convenient to note that the $\cH$-action~\eq{eq:Haction} corresponds to the following action at the level of $\cC_\cG$ and $\cC_\cH$ (here \eq{eq:lhatomegal-0} is important):
\be \label{eq:gauge-action-ch-cg}
c_\Gg ~\mapsto~ \tilde{c}_\Gg ~=~ c_\Gg \big|_{g \,\mapsto\, \ell(x) g} \;, \qquad
c_\Hh ~\mapsto~ \tilde{c}_\Hh ~=~ c_\Hh \big|_{h \,\mapsto\, x h} \;.
\ee

One can then proceed to gauge this symmetry, whereupon one must ensure that the constraint \eq{hconstraint} is solved. 
This can be done as follows, generalizing the
calculation performed in~\cite{Figueroa-OFarrill:2005vws} for the
symmetry-preserving boundary condition. We have 
\bea
i^* \theta_L \;=\; g^{-1}dg\big|_{g=c_\Gg c_\Hh} &\,=\,& \mathrm{Ad}_{c_{\Hh}^{-1}}(c_\Gg^{-1}dc_\Gg) + c_\Hh^{-1} dc_{\Hh} \,, \cr
&=&   \mathrm{Ad}_{c_\Hh^{-1}}(\alpha_\Gg-\Omega_\Gg)\vartheta_\Gg + (\alpha_\Hh-r)\vartheta_\Hh\,, 
\eea
\vspace{-7mm}
\bea
i^* \theta_R \;=\; - dg \;\! g^{-1}\big|_{g=c_\Gg c_\Hh} &\,=\,& -
dc_\Gg c_\Gg^{-1} - \mathrm{Ad}_{c_{\Gg}}(dc_\Hh c_\Hh^{-1})\,, \cr
&=& - (\mathrm{id} - \bar{\alpha}_\Gg)\vartheta_\Gg - \mathrm{Ad}_{c_\Gg}(\lhat-\bar{\alpha}_\Hh)\vartheta_\Hh \,.
\eea
Then from \eq{grouptheta} we have 
\be
\label{istargrouptheta}
i^*\theta_a =  \langle \ell(X_a),i^*\theta_R \rangle - \langle r(X_a), i^* \theta_L \rangle  \;.
\ee
Next, to compute $\imath_a \omegatwo$, we employ the method used in~\cite{Figueroa-OFarrill:2005vws} and apply this to the gauge action expressed as a simultaneous action on $c_\Gg$ and $c_\Hh$ in Eq.\;\eq{eq:gauge-action-ch-cg}. The Killing vector field corresponding to the gauge action is the sum of the Killing vector fields for the individual actions,
\be
\xi_a ~=~ \xi_a^{\Gg} + \xi_a^{\Hh} \,.
\ee
Here $\xi_a^{H}$ is the Killing vector corresponding to the action $h\mapsto x\:\! h $ in $\cH$. So $\xi_a^{\Hh}$ generates
\be
h ~\mapsto ~ e^{-t X_a} h \,,
\ee
so $\xi_a^{\Hh}$ corresponds to the right-invariant vector field $X_a^R$ on $\cH$. Furthermore, $\vartheta_{\Hh}$ is equal to minus the right Maurer-Cartan one-form on $\cH$, i.e.~$\vartheta_{\Hh}=-\theta_R^{\Hh}$. Since the interior product is linear, and $\xi_a^{\Gg}$ acts only on $c_{\Gg}$, we have
\be
\imath_a \vartheta_\Hh ~=~ \imath_{\xi_a^{\Hh}}\,\vartheta_\Hh ~=~ 
-  \imath_{\xi_a^{\Hh}}\,\theta_R^{\Hh} ~=~  \theta_R^{\Hh} (X_a^R) ~=~ - X_a \,.
\ee
Similarly, we have
\be
\imath_a \vartheta_\Gg ~=~ \imath_{\xi_a^{\Gg}}\,\vartheta_\Gg ~=~ 
-  \imath_{\xi_a^{\Gg}}\,\theta_R^{\Gg} ~=~  \theta_R^{\Gg} (\ell(X_a)^R) ~=~ - \ell(X_a) \,.
\ee
Applying these expressions to the flux in the form \eq{omegas}, one can directly verify that 
\be
\label{hconstraint-2}
i^*\theta_a + \imath_a \omegatwo =  0 ~.
\ee 
This establishes the classical consistency of the gauging, given Eq.\;\eq{anomalyfree}. Note that to show this we did not need to use the special relation $r = \lhat \circ \Omega_\cH $ \eq{embed-special-conj-0}, we worked generally.  
Thus the action is classically gauge invariant without imposing this constraint~\cite{Quella:2002fk,Quella:2003kd}.  However, there can be additional requirements on the D-brane worldvolume in order that the quantum theory is consistent, and Eq.\;\eqref{embed-special-conj-0} is one such constraint.
We will return to this point in Section~\ref{sec:factorized}.

\paragraph{More general symmetry-breaking branes}

One can generalize the construction of symmetry-breaking branes to worldvolumes specified the product of multiple conjugacy classes, corresponding to a chain of embeddings~\cite{Quella:2002ns,Quella:2003kd}
\be
\cH\equiv \cU_N \hookrightarrow \cU_{N-1} \hookrightarrow \cdots \hookrightarrow \cU_0\equiv \cG \,.
\ee
The boundary condition is a product of $N+1$ conjugacy classes, generalizing \eq{eq:cg-ch-param}--\eq{classproduct-gen}, and the flux on the branes contains a contribution from each of the $N+1$ groups in the embedding chain as well as a contribution from each pair of groups, generalizing \eq{SBomega}--\eq{omegas-2}. 

In Section \ref{sec:specflowST} we will use an embedding chain of length three; for use there we record some expressions for such a chain. We denote the intermediate group by $\Uint\equiv \cU_1$.
A priori we could consider independent left and right embeddings of $\cH$ into $\Uint$ and $\Uint$ into $\cG$, generalizing \eq{eq:cg-ch-param}, however we shall restrict attention to the simpler case in which the left and right embeddings are related by a generalization of \eq{embed-special-conj-0}. Explicitly, we consider the embeddings
\be
\label{gen embed} 
\cH~\underset{\textstyle\vareps_\Hh}{\longhookrightarrow}~  \Uint ~\underset{\textstyle\vareps_\Ii}{\longhookrightarrow}~ \cG  \;,
\ee
together with automorphisms $\Omega_\Gg$, $\Omega_\Ii$, $\Omega_\Hh$ of the respective groups. 
The action to be gauged is as before, $g ~\mapsto~ \ell(x) \,g\, r(x)^{-1}\;$ for $x\in \cH\,$, \eq{eq:Haction}, and we have
\begin{align}
\label{GLMTembedsteps}
\ell &= \vareps_\Ii \circ \vareps_\Hh
~~,~~~~
r =  \Omega_\Gg \circ \vareps_\Ii \circ \Omega_\Ii \circ \vareps_\Hh \circ \Omega_\Hh \;.
%~~,~~~~
\end{align}
The generalized conjugacy classes are then embeddings of twisted conjugacy classes of the respective groups:
\bea \label{eq:cg-ch-param}
 \cC_\cG ~=~ \mathcal{C}_\cG^{(f_\Gg,\Omega)} &\,=\,& \bigl\{ \, g \;\! f_\Gg \; \Omega_\Gg (g^{-1})~ ,~~~~ g \in
  \cG \bigr\} \,, \nn\\
	\cC_\Uint &\,=\,& 
	\;\! \Omega_\Gg \!\!\; \circ \!\!\; \varepsilon_\Ii \left( \mathcal{C}_\Uint^{(f_\Ii,\Omega_\Ii)}  \right) 
\,, \cr
	\cC_\cH &\,=\,& 
	\Omega_\Gg \!\!\; \circ \!\!\; \varepsilon_\Ii \!\!\;\circ \Omega_\Ii \!\!\; \circ \!\!\; \varepsilon_\Hh
	\left( \mathcal{C}_\cH^{(f_\Hh,\Omega_\Hh)}  \right) 
	\,,
\eea
and the boundary is given by
\be
\label{classproduct-gen} 
\cC  ~=~  \cC_\cG \cdot \cC_{\Uint}\cdot \cC_\cH ~=~ \bigl\{ \; c_\Gg \, c_\Ii \, c_\Hh ~~ \bigl|~~   c_\Gg \in \cC_\cG \,,~c_\Ii \in \cC_{\Uint}\,, ~ c_\Hh \in \cC_\cH \bigr\} \,.
\ee
The flux on the brane is the appropriate generalization of \eq{SBomega}--\eq{omegas-2}, with six parts in total, three from each of the groups separately $\omegatwo(\cG)$, $\omegatwo(\Uint)$, $\omegatwo(\cH)$, and three from the pairs of groups, $\omegatwo(\cH,\cG)$, $\omegatwo(\cH,\Uint)$, $\omegatwo(\Uint,\cG)$.

%%%%%%%%%%%%%%%%%%%%%%%%%%%%%%%%%%%%
%%%%%%%%%%%%%%%%%%%%%%%%%%%%%%%%%%%%

\section{Gauged WZW models for supertubes}
\label{sec:ST CFT}

The supergravity backgrounds of Section~\ref{sec:sugrasolns} have an exact worldsheet description as gauged WZW models.  The construction of~\cite{Martinec:2017ztd} gauges left and right null isometries on the group manifold
\be
\label{Gupstairs2}
\mathcal{G} =
SL(2,\mathbb{R}) \times SU(2) \times \bR_t\times\bS^1_y\times \bT^4 ~;
\ee
in this way one builds, in incremental stages, worldsheet string theory for each of the backgrounds of Section~\ref{sec:sugrasolns}.
We thus specialize in the following to the WZW model on $\cG$, and specify the Killing vectors $\xi^a$ to be 
gauged in each case.

%%%%%%%%%%%%%%%%%%%%%%%%%%%%%%%%%%%%

We begin by specifying our conventions for the worldsheet sigma models on $\sutwo_\k$ and $\sltwo_\k$, which introduce some additional overall factors with respect to the general presentation above.
For the $\sltwo$ factor we will find it convenient to use the equivalent $SU(1,1)$ description, though we will still denote elements by $g_\sl$. The sigma models that we will consider, before gauging, will contain elements
\be \label{eq:su11}
(g_\sl, g_\su) \; \in \; \mathrm{SU}(1,1) \times \mathrm{SU}(2).
\ee
We use Euler angle group parameterizations as follows:
\be
\label{Eulerangles-1}
g_{\sl} \;=\; e^{\frac{i}{2}(\tau-\sigma)\sigma_3}e^{\rho \sigma_1}e^{\frac{i}{2}(\tau + \sigma)\sigma_3} 
 ~, ~~~~~~~ g_{\su} \;=\; e^{\frac{i}{2}(\psi-\phi)\sigma_3}e^{i \theta \sigma_1}e^{\frac{i}{2}(\psi + \phi)\sigma_3} \,.
\ee
In order to have one timelike and five spacelike directions, the metric involves a relative sign between the two group factors. To ease the notation we write the expressions in the absence of the worldsheet boundary, as this suffices to specify the overall normalizations. We then have
\bea 
\label{S_wzw}
\mathcal{S}_{\mathrm{kin}} \;=\;  \mathcal{S}_{\mathrm{kin}} + \mathcal{S}_{\WZ}(g) \,,
\eea
where
\begin{align}
\begin{aligned} \label{eq:S-kin}
\mathcal{S}_{\mathrm{kin}} &=\; \frac{1}{\pi}\int G_{ij}(\varphi)\, d\varphi^i \wedge \star d\varphi^j  \\
&=\; \frac{\k}{2\pi}\int \text{Tr}
\Bigl[(\partial g_\sl) g_\sl^{-1} (\bar{\partial} g_\sl) g_\sl^{-1}\Bigr]  \,-\, \frac{\k}{2\pi}\int \text{Tr}\Bigl[(\partial g_\su) g_\su^{-1} (\bar{\partial} g_\su) g_\su^{-1}
\Bigr]\,,
\end{aligned}
\end{align}
and where 
\bea \label{eq:S-WZ} 
\mathcal{S}_{\WZ}(g) ~=~ \frac{1}{\pi}\int\limits_M H  ~=~
\frac{\k}{2\pi}\int\limits_M \frac13 \tr \left[ (g_\sl^{-1} dg_\sl)^3 \right] -  \frac{\k}{2\pi}\int\limits_M \frac13 \tr \left[ (g_\su^{-1} dg_\su)^3 \right] . 
\eea
We work in the large $n_5$ limit, in which to leading order $\k=n_5$, giving the line element
\be \label{eq:metric}
ds^2 \,=\, G_{ij} d\varphi^i d\varphi^j \,=\,  
n_5 \big(  - \cosh^2 \rhoo d\tau^2 + d\rho^2 +  \sinh^2 \rhoo d\sigma^2
+d\theta^2 + \sin^2\!\theta \,d\phi^2 +  \cos^2\!\theta d\psi^2 \big)\,,
\ee
and the $H$-flux
\be \label{eq:Hslsu}
H ~=~ n_5 \big( \sinh 2\rho \,d\rho \wedge d\tau \wedge d\sigma + \sin 2\theta \,d\theta \wedge d\psi \wedge d\phi \big)\,.
\ee
Correspondingly, the expressions for the fluxes (and related quantities such as the one-forms $\theta_a$) in the previous section should be scaled by a factor of $\nfive/2$ in our explicit applications below.

%%%%%%%%%%%%%%%%%%%%%%%%%%
\subsection{Fivebranes on the Coulomb branch}
\label{sec:coulbranch}

As mentioned above, the original description of NS5-branes on the Coulomb branch in a circular $\mathbb{Z}_{n_5}$-symmetric configuration used the Landau-Ginsburg orbifold 
\be
\left(\frac{\sltwo}{\uone}\times\frac{\sutwo}{\uone}\right)/\bZ_{\nfive} ~,
\ee
stressing their relation to non-compact Calabi-Yau manifolds~\cite{Giveon:1999px,Giveon:1999tq} near a singular point in their moduli space through the Calabi-Yau/Landau-Ginsburg correspondence~\cite{Martinec:1988zu,Vafa:1988uu}.

We shall work instead with an alternative description using the gauging of null isometries~\cite{Israel:2004ir}
in the $\sltwo\times\sutwo$ part of the 10+2 dimensional ``upstairs'' group $\cG$ in~\eqref{Gupstairs}, with parametrization as described in Eqs.\;\eq{eq:su11}-\eq{eq:Hslsu}.

The group we wish to gauge is $\uone_L\times\uone_R$,%
\footnote{More precisely, as discussed in~\cite{Martinec:2018nco}, the global structure of the gauge group is $\bR\times\uone$, where $\bR$ is generated by the (timelike) vector combination of the left and right null vectors, and \uone\ is generated by the (spacelike) axial combination.  Here we will be interested in the local structure of the gauge group, and will therefore ignore such subtleties.}
so a basis of generators of the Lie algebra $\mathfrak{u}(1)_L \oplus \mathfrak{u}(1)_R$ is simply given by a pair of real numbers,
\be
X_a \;=\; (\alpha, \beta).
\ee
Given $\alpha \in \mathfrak{u}(1)_L$, $\beta \in \mathfrak{u}(1)_R$, we gauge the action
\be
\label{Coulgaugetransfs}
(g_\sl, g_\su) ~\mapsto ~ \left(e^{{i} \alpha\sigma_3} \:\! g_\sl \, e^{{i} \beta \sigma_3} \,,\,
{ e^{{-i} \alpha \sigma_3} \:\! g_\su \, e^{{i} \beta\sigma_3}} \right) .
\ee
Let us translate this into the notation of \cite{Figueroa-OFarrill:2005vws}. We have homomorphisms $\ell$, $r$ describing the embedding of the above action -- we use the same notation for the group action and the induced Lie algebra action. 
We have 
\begin{align}
\begin{aligned} \label{eq:5BCB-embed}
\ell (X_1) = \ell (\alpha ) = \big(i \alpha \sigma_3 \,, \, {- i \alpha \sigma_3 } \big) 
\;\equiv\; \Big( \ell_{\sl}(\alpha)\,,\ell_{\su}(\alpha) \Big) 
~~&, ~~~~  r(X_1) = 0 ~, 
\\
r (X_2) = r(\beta) = \bigl( {} -i  \beta \sigma_3 \,, \,{- i \beta \sigma_3 } \bigr) 
\;\equiv\; \Big( r_{\sl}(\beta) , r_{\su}(\beta) \Big) 
~~&,~~~~ \ell(X_2) = 0 ~,
\end{aligned}
\end{align}
so that the separate actions to be gauged are
\be
(g_\sl, g_\su) ~\mapsto ~ \left( e^{\ell_{\sl}(\alpha)}\:\! g_\sl \,, e^{\ell_{\su}(\alpha)} \:\! g_\su \right) \,,
\qquad
(g_\sl, g_\su) ~\mapsto ~ \left(  g_\sl \, e^{-r_{\sl}(\beta)},  g_\su \, e^{-r_{\su}(\beta)} \right) .
\ee

The Killing vectors $\xi_a$ corresponding to the two basis elements $X_a$  are 
\be
\label{xione coulomb}
\xi_1 \;=\;  \bigl( {\partial_\tau}+{\partial_\phi} \bigr) 
-\bigl({\partial_\sigma}+{\partial_\psi} \bigr)  \,,
\ee
for the left action, and for the right action one has
\be
\label{xitwo coulomb}
\xi_2 \;=\;   \bigl( {\partial_\tau}+{\partial_\phi} \bigr) 
+\bigl({\partial_\sigma}+{\partial_\psi} \bigr)\,.
\ee
Note that if we were to set $\alpha = \beta$ we would be gauging away
\be \label{eq:ax-vec}
\xi_1 + \xi_2 \;=\; 2 \big( {\partial_\tau}+{\partial_\phi} \big) \,,
\ee
that is a (timelike) combination of axial gauging in $\sltwo$ and vector gauging in $\sutwo$. Similarly, if we set $\alpha = -\beta$ we would be gauging away
\be \label{eq:vec-ax}
\xi_2 - \xi_1 \;=\; 2 \big( {\partial_\sigma}+{\partial_\psi}  \big) \,,
\ee
that is a (spacelike) combination of vector gauging in $\sltwo$ and axial gauging in $\sutwo$.

The background $H$ before gauging is given in \eq{eq:Hslsu}; from \eq{grouptheta}, the $\theta_a$ are
\bea
\label{theta coulomb}
\theta_1 &=& \nfive \Big[ -\left(\cosh^2\rhoo \,  d\tau + \sinh^2\rhoo \,  d\sigma \right) - \left(\cos^2\theta \,d\psi- \sin^2\theta\, d\phi  \right) \Big] \,\nn\\[3pt]
\theta_2 &=& \nfive \Big[ \left(  \cosh^2\rhoo \,  d\tau -\sinh^2\rhoo \,  d\sigma \right) - \left( \cos^2\theta \,d\psi+ \sin^2\theta \,d\phi  \right) \Big] .
\eea
From \eq{eq:5BCB-embed} we see that the anomaly cancellation constraint \eq{anomalyfree} is satisfied.

The kinetic terms in the sigma model action involve the covariant derivative~\eqref{covderiv} with a gauge potential $A^a$ for gauging each Killing vector $\xi^a$.
We have two independent gauge fields $(A^1,\Ab^1)$ and $(A^2,\Ab^2)$; the kinetic terms involve 
\be
\label{gauged KE}
\mathcal{D} \varphi^i \, G_{ij} \, \overline{\mathcal{D}} \varphi^j
\;=\; (\partial \varphi^i - A^a \:\! \xi_a^i ) \, G_{ij} \,  (\bar\partial \varphi^j - \Ab^a \:\! \xi_a^j) \,.
\ee
Compared to the analysis of~\cite{Martinec:2017ztd}, this seems twice too many, however the fact that the currents being gauged are null results in the absence of the left component of the gauge field for the left null current in the action, and similarly for the right component of the right current. This happens as follows.

The kinetic term~\eqref{gauged KE} can be written in matrix notation as 
\be
-\frac{n_5}{2} \tr\Bigl[\bigl(g^{-1}\cD g\bigr)\bigl(g^{-1}\overline\cD g\bigr) \Bigr]  \,,
\ee
where the group element $g$ and the trace run over the various factors in $\cG$, and where there is a minus sign to be understood in the definition of the SL(2) trace, see \eq{S_wzw}--\eq{eq:metric}.
The terms quadratic in gauge fields are
\begin{align}
\frac{n_5}{2} \bigl(A^1\bar A^2+A^2\bar A^1\bigr)  \,\tr \Big[& \bigl(g^{-1}(\xi_1 g)\bigr)  \bigl(g^{-1}(\xi_2 g)\bigr)\Big] 
\\[3pt]
&=  \frac{n_5}{2} \bigl(A^1\bar A^2+A^2\bar A^1\bigr)  \,\tr \Big[ \ell(X_1) \:\! g \, r(X_2) \:\! g^{-1} \Big] 
\nn
\end{align}
where we have used \eq{eq:xi-g}; the terms involving $A^1\bar A^1$ and $A^2\bar A^2$ have vanished since the embeddings are chiral $(\ell(X_2)=r(X_1)=0)$, Eq.\;\eq{eq:5BCB-embed}.
The Wess-Zumino term involves $\imath_a \theta_b-\imath_b\theta_a$; using~\eqref{theta-on-vec}, \eqref{Killingvec}, \eqref{grouptheta},
one has for our chiral embeddings  
\be
{}  \frac{n_5}{2} \bigl(-A^1\bar A^2+A^2\bar A^1\bigr)  \,\tr \Big[ \ell(X_1) \:\! g \, r(X_2) \:\! g^{-1} \Big] \,.
\ee
%\be
%- n_5 \tr \Big[ \ell( A^1)  g \, r(\bar A^2) g^{-1}\Big] + n_5 \tr \Big[ \ell(\bar A^1) \, g \, r(A^2) g^{-1} \Big] \,.
%\ee
As a result, the sum of the gauge kinetic terms and Wess-Zumino terms that are quadratic in gauge fields depends only on $A^2,\bar A^1$, with the contributions from $A^1, \bar A^2$ cancelling between the two.   
The terms linear in the gauge fields reinforce/cancel similarly, so all together, the gauge kinetic terms and the WZ terms in~\eqref{SWZ}-\eqref{gWZ term}
combine in such a way that the gauge field components $A^1$,  $\Ab^2$ simply drop out completely and do not appear at all in the action.  The resulting action is that of the asymmetrically gauged models given in~\cite{Bars:1991pt,Quella:2002fk}.
Relabelling $\cA=A^2$, $\cAb=\Ab^1$, the full Lagrangian becomes 
\be \label{eq:NS5-P-lagrangian}
\cL ~=~ \cL_{\WZW} + 2 \cAb \left( J_3^{\sl} + J_3^{\su} \right)+ 2 \cA \left( \bar{J}_3^{\sl} - \bar{J}_3^{\su} \right) - 4 \nfive(\sinh^2\rho + \cos^2 \theta) \cA \cAb \, 
\ee
where $\cL_{\WZW}$ is the ungauged Lagrangian and where 
the conventions for the $J_3$ currents are given in the appendix, Eqs.\;\eq{eq:J3su}, \eq{eq:J3sl}.  Thus we recover the action for fivebranes on the Coulomb branch of~\cite{Israel:2004ir,Martinec:2017ztd}, which upon integrating out the gauge fields gives the background~\eqref{NS5 coulomb}. 

The absence of half the gauge field components is a direct consequence of the gauging of null isometries, and is not specific to this choice of group manifold.  In holomorphic worldsheet coordinates, the sigma model Lagrangian has the form
\be
(G_{ij}+B_{ij}) \partial \varphi^i \bar\partial\varphi^j  ~;
\ee
the left and right null Killing vectors mean that the matrix $G+B$ has left and right null vectors, and when these isometries are gauged, the gauge field components related to these null vectors are absent from the action.  When the worldsheet has a boundary, this property will extend to the matrix $\,G+B+\cF$, where $\cF$ is the field strength of the D-brane gauge field associated to the boundary.  This feature will have consequences for the DBI effective action, as we will see in the following.

We now proceed to the more general null gaugings that lead to supertubes and spectral flowed supertubes; we pause here to note that a potentially interesting extension of the present work could be to investigate connections with recent work on integrable deformations of asymmetrically gauged WZW models~\cite{Driezen:2019ykp} (see also \cite{Driezen:2018glg}).

%%%%%%%%%%%%%%%%%%%%%%%%%%
\subsection{NS5-P and NS5-F1 supertubes}
\label{sec:two-charge supertubes}

More general null embeddings of \uone$\times$\uone\ can be specified through the gauge currents
\begin{align}
\label{nullvectorparam}
U(1)_L : \quad \mathcal{J} &= l_1 J^{\sl}_3 + l_2 J^{\su}_3  + l_3 \hat{P}_{t,L} +
  l_4 \hat{P}_{y,L} \, , \\
U(1)_R: \quad \bar{\mathcal{J}}& = r_1 \bar J_3^{\sl} + r_2 \bar  J_3^{\su}  + r_3 \hat{P}_{t,R} +
  r_4 \hat{P}_{y,R} \,,  \nonumber
\end{align}
where 
\begin{align}
\hat{P}_{t,L}\equiv \partial t
~~,~~~~
\hat{P}_{t,R}\equiv \bar\partial t
%\nn\\
~~,~~~~
\hat{P}_{y,L}\equiv \partial y
~~,~~~~
\hat{P}_{y,R}\equiv \bar\partial y ~,
\end{align}
and where the $\sltwo$ and $\sutwo$ currents are given in~\eqref{eq:J3sl}, \eqref{eq:J3su} respectively.
The null conditions 
\begin{equation}
\label{eq:nullconstr}
0 = \langle \boldsymbol{\ell}, \boldsymbol{\ell} \rangle = n_5 (-l_1^2 +l_2^2)-l_3^2
+l_4^2  
~~,~~~~
0 = \langle \boldsymbol{r}, \boldsymbol{r} \rangle = n_5 (-r_1^2+r_2^2) -
r_3^2+r_4^2  
\end{equation}
ensure anomaly cancellation and independence of the left and right gaugings.

The gauged action is then 
\be
\cL ~=~ \cL_{\WZW} + { 2 \cAb \mathcal{J} + 2 \cA \bar{\mathcal{J}} - 4\nfive \:\! \Sigma\;\!  \cA \cAb \;, }
\ee
where
\be
\nfive \Sigma \;=\; \frac12 \Bigl[ n_5\bigl(l_1 r_1 \cosh 2\rho - l_2 r_2 \cos 2\theta \bigr)+l_3 r_3 - l_4 r_4  \Bigr] \,.		
\ee

The double-null choice $|l_1|=|l_2|$, $|l_3|=|l_4|$ (and similarly for the right coefficients $r_i$) tilts
 the null isometry into the $\bR_t\times \bS^1_y$ direction, leading to NS5-P and NS5-F1 supertube backgrounds~\cite{Martinec:2017ztd}.  Specifically, letting
\be \label{eq:NS5-P-current-params}
l_1=l_2=1 ~,~~~ l_3=-l_4 = -\frac{k}{\Rytil}
~~,~~~~
r_1=-r_2=1 ~,~~~ r_3=-r_4 = -\frac{k}{\Rytil}
\ee
leads to an NS5-P supertube that (for $\nfive,k$ relatively prime)
wraps together the $\nfive$ fivebranes into a single source that coils
$k$ times around the $\phi$ circle in the transverse angular $\bS^3$.
T-duality to the NS5-F1 supertube simply amounts to flipping the sign
of $l_4$, and relabelling the radius of the $\bS^1$,
$\Rytil=\lstr^2/R_y$, so that $l_4 =l_4^{\sst\rm(P)} = k/\Rytil$ becomes $l_4=l_4^{\sst\rm(F1)}= -kR_y$.  
For future reference, we can combine the gauge
transformations for both these possibilities into
\begin{align}
\label{STgaugetransfs}
\delta\tau & \;=\; l_1\alpha+r_1\beta \;=\; (\alpha+\beta) \;,\qquad\quad~~~~
\delta\sigma \;=\; -l_1\alpha+r_1\beta \;=\; -(\alpha-\beta)\;,
\nn\\
\delta\phi &\;=\; {l_2\alpha-r_2\beta \;=\; (\alpha+\beta) \;,\qquad\quad ~~~~
\delta\psi \;=\;  -l_2\alpha-r_2\beta \;=\; -(\alpha-\beta) }\;,
\nn\\
\delta t &\;=\; l_3\alpha+r_3\beta \;=\; -{kR_y}\,(\alpha+\beta)
\;,\quad~~
\delta y \;=\; { -l_4^{\sst\rm(F1)}\alpha } -r_4\beta \;=\;  {kR_y}\,(\alpha-\beta)\;,
\nn\\
\delta \ytil &\;=\; { -l_4^{\sst\rm(P)}\alpha } -r_4\beta  \;=\; -{kR_y}\,(\alpha+\beta)\;.
\end{align}
%where the gauge transformations have been written in the NS5-P frame; 
Note that T-duality, which interchanges the value of $l_4$ between $l_4^{\sst\rm(P)}$ and $l_4^{\sst\rm(F1)}$, 
 is equivalent to interchanging $y$ and $\ytil$ in this expression.

The form of the currents~\eqref{eq:J3sl} makes clear why the geometry of the NS5-F1 supertube is asymptotically that of the linear dilaton fivebrane throat, and in the cap locally $AdS_3\times \bS^3$.  
For large $\rho\gg \hf\log(kR_y/\nfive\lstr)$, the largest contribution to the current comes from motions along $\sltwo$, and so a good approximation to the physical spacetime comes from fixing a reference point along the gauge orbit $\tau=\sigma=0$ and examining the geometry along the other directions.  There is not so much difference between the tilted null gauging of the supertube and that of fivebranes on the Coulomb branch, or for that matter the linear dilaton throat~\eqref{CHSsoln} of coincident fivebranes.%
\footnote{The CHS geometry~\eqref{CHSsoln} fits within the null gauging framework -- it is obtained by gauging the null currents that generate the Borel subgroup of $\sltwo$, leaving the remaining factors in $\cG$ untouched.}  
On the other hand, in the cap region $\rho\ll \hf\log(kR_y/\nfive\lstr)$, the gauge current lies mostly along $t$ and $y$, thus a good approximation to the geometry in this region fixes these coordinates, largely leaving alone $\sltwo\times\sutwo$, and the geometry is thus well-approximated locally by $AdS_3\times \bS^3$.  

The $\bZ_k$ orbifold structure of the NS5-F1 supertube arises from a discrete residual gauge symmetry remaining after fixing the $y$ coordinate.  The factor of $k$ in the gauge transformation of $y$ in~\eqref{STgaugetransfs} means that while asymptotically a spatial gauge orbit $(\alpha-\beta)\in(0,2\pi)$ covers the range $\delta\sigma=2\pi$ of the $\sltwo$ spatial coordinate being fixed, in the cap the range $(\alpha-\beta)\in(0,2\pi/k)$ is sufficient to cover the entire range $\delta y=2\pi R_y$.  Thus in gauge fixing $y$ in the cap, one should decompose the axial gauge orbit as
\be
(\alpha-\beta) = 2\pi \Bigl(\frac{\eta}k + \frac mk \Bigr)
~~,~~~~
\eta\in(0,1)
~~,~~~~
m=0,1,\dots,k-1 
\ee
and use $\eta$ to fix a point in the $y$ circle;
the residual discrete $\bZ_k$ gauge group parametrized by $m$ keeps $y$ fixed and yields an orbifold identification of $AdS_3\times\bS^3$.

%%%%%%%%%%%%%%%%%%%%%%%%%%
\subsection{Three-charge NS5-F1-P supertubes}
\label{sec:GLMTnullgauging}

Further generalization to more generic null vectors yields worldsheet sigma models for the three-charge backgrounds of~\cite{Giusto:2004id,Giusto:2012yz} obtained by a spacetime spectral flow transformation of these supertubes.

The null current directions are given in the parametrization~\eqref{nullvectorparam} as 
\begin{align}
\label{GLMTcurrents}
l_1 = 1
~,~~~
l_2 = 2s+1
~~,~~~~
{l_3}   \,=\, -k R_y\, (1 + \varthetab )
~~&,~~~~ 
 {l_4}   \,=\, -kR_y (1-\varthetab) 
 ~, \nn\\
 r_1 = 1
 ~,~~~
 r_2 = -1
 ~~~~~,~~~~
{r_3}   \,=\, -k R_y\, (1 + \varthetab )
~~&,~~~~
{r_4}  \,=\, +kR_y (1+\varthetab)  ~ ,
\end{align}
where $s$ is the left-moving spectral flow parameter, and
\begin{equation}
\varthetab ~=~ \frac{1-\etab}{\etab} ~=~ \frac{s(s+1)\nfive\lstr^2}{k^2 R_y^2} ~.
\end{equation}
Note that for $s=0$, we recover the NS5-F1 supertube.
There is a further generalization to the non-supersymmetric ``JMaRT'' solutions with both left and right spectral flow parameters $s,\bar s$; however, since the closed string background is already unstable to rapid decay via perturbative string radiation when we couple it to asymptotically flat spacetime, we will not consider the D-brane spectrum here (most of its structure differs little from the supersymmetric backgrounds above).

The gauge orbits are now 
\begin{align}
\label{GLMTgaugetransfs}
\delta\tau &\,=\, 	(\alpha+\beta) \;,~~~~~\;
\delta\sigma \,=\, 	-(\alpha-\beta)
\;,~~~~~
\delta t \,=\, 		{(-kR_y - \ktil\Rytil)}\,(\alpha+\beta)\;,
\\
 \delta \phi &\,=\, {(s+1)\,(\alpha+\beta) 	+ s\,(\alpha-\beta)
\;,~~~~~~~~
\delta \psi \,=\, 	-s\,(\alpha+\beta) 	-(s+1)\,(\alpha-\beta)}\;,
\nn\\
\delta y &\,=\, 		-{\ktil\Rytil}\,(\alpha+\beta)	+{kR_y}\,(\alpha-\beta)
\;,~~~~~~
\delta \ytil \,=\, 	-{kR_y}\,(\alpha+\beta)+{\ktil\Rytil}\,(\alpha-\beta) \;,
\nn
\end{align}
where 
\be
\Rytil = \frac{\lstr^2}{R_y}
~~,~~~~
\ktil = \frac{s(s+1)}{k} \, \nfive  ~.
\ee
Note the manifest T-duality of~\eqref{GLMTgaugetransfs} under $y\leftrightarrow \ytil,\,$ $k\leftrightarrow \tilde k,\,$ $R_y\leftrightarrow\Rytil\:\!$.

The gauge orbit structure once again determines an orbifold identification in the cap when we use the gauge freedom to fix $y$ (or $\ytil$ in the T-dual geometry, where spectral flow has induced an F1 charge proportional to $s$ leading to a structure similar to the NS5-F1 cap).
We can parse the spatial gauge parameter as 
\be  
(\alpha-\beta) = 2\pi \Bigl( \frac{\eta}{q} + \frac pq \Bigr)  
\ee
where $\eta \in (0,1)$ and $p\in\{0,1,É,q-1\}$.  
Also, following the analysis in~\cite{Giusto:2012yz,Martinec:2018nco} let 
\be
\label{factorizations}
k=\ellone\elltwo
~~,~~~~
s=\nhat\ellone
~~,~~~~
s+1=\mhat\elltwo  
~~,~~~~
\ktil=\nhat \mhat \nfive
~.
\ee
There are now two canonical choices:
\begin{enumerate}
\item
We can use $\eta$ to gauge fix $y$ if we are working in the ``mostly NS5-F1'' frame (\ie\ the frame where $s=0$ leads to the NS5-F1 supertube).  Then $q=k$, since all we need is the (0,1/k) interval of the gauge parameter circle to fix a point on the $y$ circle.  There is then the residual discrete $Z_k$ part of the gauge group parametrized by $p$.  Now that we have gauge fixed $y$, the remaining spatial coordinates $\rho,\sigma,\theta,\phi,\psi$ are the spatial directions of $AdS_3\times\bS^3$.  There is the additional identification above.  Thus things look exactly like the discussion in section 2.4 of~\cite{Martinec:2018nco}, and we conclude that there is a $\bZ_{\ellone}$ orbifold singularity at $\theta=\pi/2$ and a $\bZ_{\elltwo}$ orbifold singularity at $\theta=0$.
\item
We can use $\eta$ to gauge fix $\ytil$ if we are working in the T-dual ``mostly NS5-P'' frame (\ie\ the frame that reduces to an NS5-P supertube when $s=0$).  Then $q = s(s+1)n_5/k = \nhat \mhat n_5\equiv \ktil$ since we only need a $1/q$ fraction of the gauge orbit to fix a point on the $\ytil$ circle.  We see that we have exactly the same structure, but with $k$ replaced by $\ktil=\nhat \mhat n_5$.  This is exactly what~\cite{Giusto:2004ip} found by performing T-duality on $y$, and we find it here rather directly through an analysis of the gauged WZW model.  There is a $\bZ_{\nhat}$ orbifold singularity at $\theta=\pi/2$ and a $\bZ_{\mhat}$ orbifold singularity at $\theta=0$.
\end{enumerate}
Note that the gauge orbits never degenerate in the target space $\cG$, because $y$ and $\ytil$ never pinch off, and the gauge group acts effectively on both for $s\ne 0$.  When present, such a degeneration causes the curvature and dilaton to diverge in the classical sigma model effective geometry (though of course such divergences are an artifact of the supergravity approximation and are absent in the exact tree-level string theory, as discussed above); but here, the geometry is regular apart from the orbifold singularities specified above.

For further details, we refer to~\cite{Martinec:2017ztd,Martinec:2018nco}.

%%%%%%%%%%%%%%%%%%%%%%%%%%%%%%%%%%%%
%%%%%%%%%%%%%%%%%%%%%%%%%%%%%%%%%%%%

\newpage

\section{Review of D-branes in SU(2) and SL(2,R)}
\label{sec:GroupBranes}

We now survey known results for D-branes in $\sltwo$ and $\sutwo$, as they will be useful ingredients in our analysis -- smearing them along the $\uone\times\uone$ gauge orbits will yield examples of D-branes in supertube backgrounds.
In this section, we suppress all factors of the level $\k=\nfive$ of the WZW models, to reduce clutter in formulae.  They may be restored easily, for instance all the fluxes are proportional to $\nfive$.

%%%%%%%%%%%%%%%%%%%%%%%%%%%%%%%%%%%%%%%%%

%\refstepcounter{subsection}
%\subsection*{\thesubsection \quad $SU(2)$ D-branes} \label{sec:SU2branes}

\subsection{$SU(2)$ D-branes}\label{sec:SU2branes}

%%%%%%%%%%%%%%%%%%%%%%%%%%%%%%%%%%%%%%%%%

\begin{figure}[ht]
\centering
  \begin{subfigure}[b]{0.4\textwidth}
    \includegraphics[width=\textwidth]{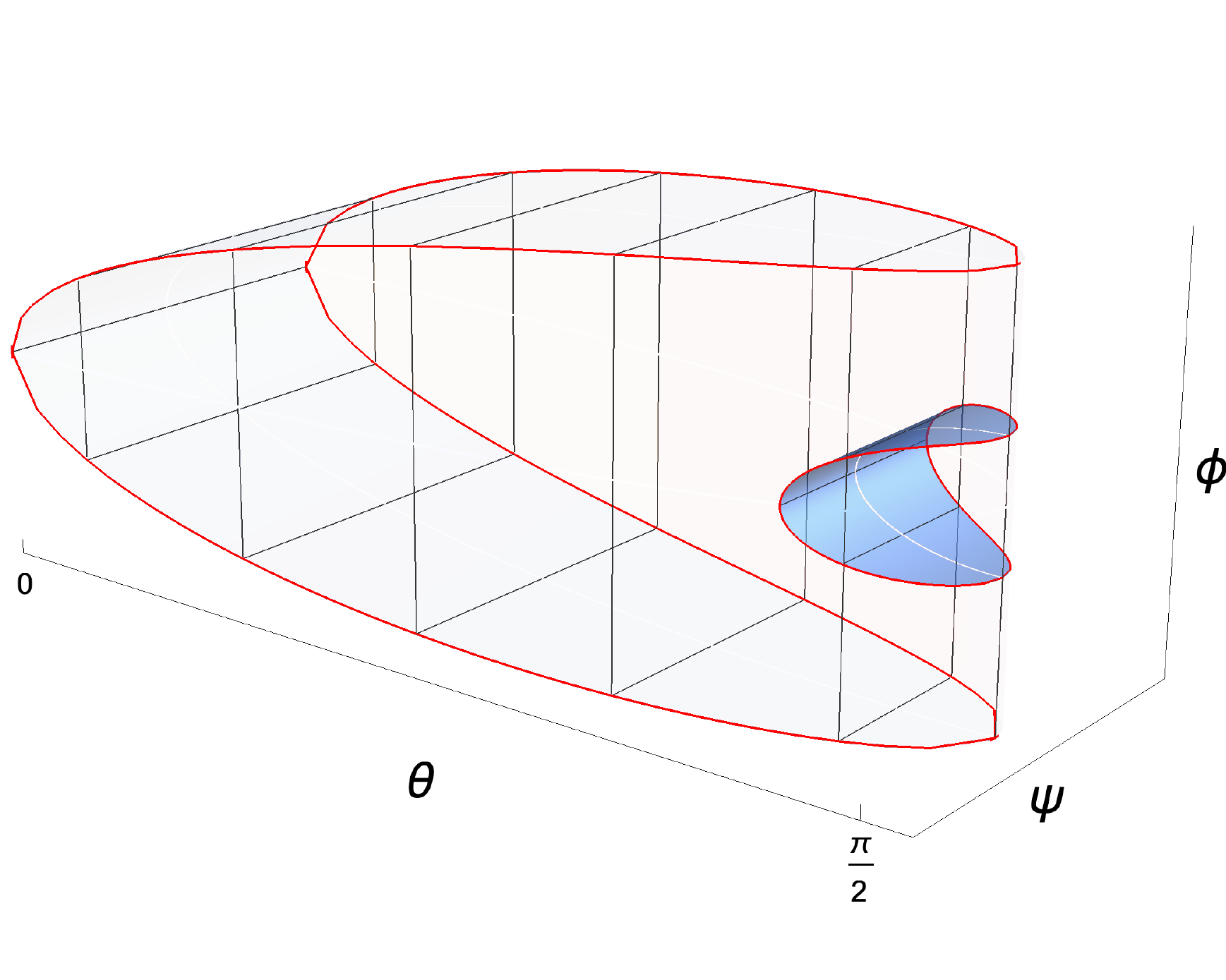}
    \caption{ }
    \label{fig:S2_brane}
  \end{subfigure}
\hskip 1.5cm
  \begin{subfigure}[b]{0.4\textwidth}
    \includegraphics[width=\textwidth]{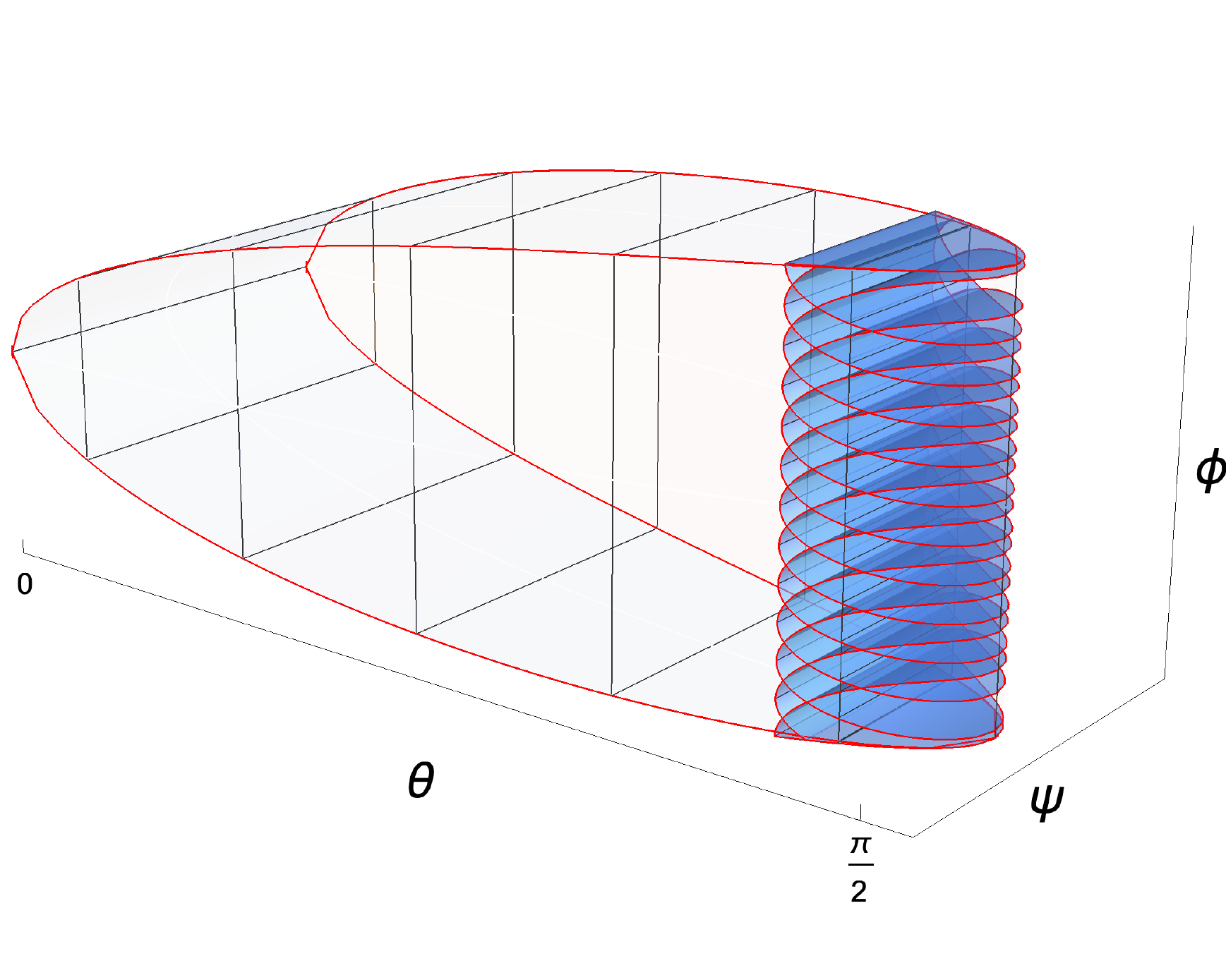}
    \caption{ }
    \label{fig:S2_brane_smeared}
  \end{subfigure}
\caption{\it 
$SU(2)$ branes. The $SU(2)$ group manifold is depicted in Euler angles as the $\phi$-$\psi$ torus
fibered over the polar $\theta$ interval (the torus at fixed $\theta$ has been cut open to a rectangle with opposite sides identified for visualization purposes).
(a) Symmetric $\bS^2$ brane (in blue); the torus identification makes the azimuthal circle, and the $\bS^2$ is this circle fibered over an embedded interval that begins and ends at $\theta=\pi/2$. 
(b) Symmetry-breaking brane obtained by smearing the $\bS^2$ brane along $\phi$; the brane fills a finite region of $\sutwo$ given by $\theta>\theta_0$.}
\label{fig:SU2branes}
\end{figure}
%%%%%%%%%%%%%%%%%%

We begin by describing D-branes in the $SU(2)$ group manifold,
following the geometric approach outlined in Section \ref{sec:gnlsm}.

\subsubsection{Symmetry-preserving branes}

Symmetry-preserving branes are described by the twisted conjugacy
classes \eqref{conjclass}
\begin{equation}
  \mathcal{C}_{\cG}=\mathcal{C}_{\sst \rm SU(2)}^{(f_{\cG},\Omega_{\Gg})}
  = \{ g f_{\su} \,\Omega_{\Gg} (g^{-1})\, ,
  \quad g \in SU(2) \} \, .
\end{equation}
If $f_{\su} = \pm \id$ these worldvolumes are just points,
while nontrivial $f_\su$ describes D-branes wrapping an $S^2\subset SU(2)$.  Nontrivial automorphisms $\Omega_{\Gg}$ in $\sutwo$ are always inner automorphisms, and correspond to a rotated orientation of the $\bS^2$ within $\sutwo$. 
For instance, in Section~\ref{sec:NS5Coul} we will be interested in a rotation automorphism $\Omega_{\Gg}^-$ that locates the N/S poles of the $\bS^2$ at $\theta=\pi/2$; this brane is depicted in Figure~\ref{fig:S2_brane}.  The untwisted brane with $\Omega_{\Gg}=\id$ is the same shape but with $\phi\leftrightarrow\psi, \theta\to\pi/2-\theta$, and so has its poles anchored at $\theta=0$.

For simplicity, we consider first the untwisted $S^2$ brane; setting 
$f_\su = e^{i \mu \sigma_3}$  
and taking the trace, we find the defining relation 
\begin{equation}\label{embeddingSU2brane}
\cos\theta \cos \psi = \cos \mu \, .
\end{equation}
These branes are puffed up by a worldvolume flux $\omega_2$ given by \eqref{omegaf}, with an additional prefactor $n_5/2$ (see comment below Eq.\;\eq{eq:Hslsu}). Since we are currently suppressing factors of $n_5$, we write
\begin{equation}
\omegatwo \;=\; \frac12\tr\Bigl[  (g^{-1}dg) \;\! f_\su^{-1}(g^{-1}dg) \;\! f_\su \:\! \Bigr] \, .
\end{equation}
In order to compute this form it is useful to use a parametrization
for $g$ such that the boundary locus takes the
form~\cite{Sarkissian:2002bg} 
\begin{equation} \label{eq:cg-param-su}
\mathcal{C}_{\cG} \;=\; 
\begin{pmatrix} \cos\mu + i X^3 & ~ i X^1 + X^2\\ 
i X^1 - X^2 & ~ \cos\mu - i X^3 \end{pmatrix}\,
\end{equation}
which is related to the Euler angle parametrization \eq{Eulerangles-1} by 
\begin{equation}
X^1 \,=\, \sin \theta \cos \phi \, ,\quad X^2 \,=\, \sin \theta \sin \phi \,
,\quad X^3 \,=\, \cos \theta \sin \psi \, .
\end{equation}
For example we can take
\begin{equation}\label{SU2braneparam}
g =\frac{1}{\sqrt{2\sin\mu (\sin \mu - X_3)}}
\begin{pmatrix} - X^1 + i  X^2 & ~~ - X^3+\sin\mu 
\\ X_3 -\sin \mu & ~~ -X_1- i X_2 \end{pmatrix} \,.
\end{equation}
The form $\omega_2$ is then given by the following expression
\begin{equation}\label{fluxsymmpresbrane}
\omega_2 \;=\; \frac{\cos \mu}{X^3} d X^1 \wedge d X^2 \, ,
\end{equation}
which can be expressed using the embedding equation 
\eqref{embeddingSU2brane} variously as
\begin{equation}\label{S2branefluxEuler}
\omega_2 \;=\; \cos^2 \theta \, d\phi \wedge d\psi \;=\;  \frac{\sin\theta \cos \mu}{\sin \psi} \, d\theta \wedge
d\phi \;=\;  \pm \frac12 \frac{\cos \mu \sin
  2\theta}{\sqrt{\cos^2\theta - \cos^2\mu}} \, d\theta \wedge d\phi   \, .
  \end{equation}
Note that in the hyperspherical parametrization  \eqref{SU2param2} $\omega_2$ has the
form 
\begin{equation}
\label{S2branefluxhyperspherical}
\omega_2 \;=\; \frac12 \sin 2\chi \:\!  \sin \vartheta \,d\vartheta \wedge d\varphi \, .
\end{equation}

It is straightforward to check that this result agrees with a DBI analysis.
 If we parametrize the worldvolume of the $S^2$ brane as follows:
\begin{equation}
\phi \,=\, \xi_0 \, , \quad \psi \,=\, \xi_1 \, ,\quad \theta \,=\, \theta(\xi_1)  \,,
\end{equation}
the DBI action is
\begin{equation}
\mathcal{L} \;=\; e^{-\Phi}\sqrt{\det (G + B + \cF)} \;=\;  \sqrt{(\dot{\theta}^2 +\cos^2\theta)\sin^2\theta + (B_{\phi
    \psi} + \cF_{\phi\psi})^2} \, ,
\end{equation}
where $\dot \theta = \partial_{\xi_1} \theta(\xi_1)$. 
From the first equality in \eq{S2branefluxEuler} we have
\be
(\omega_2)_{\phi \psi} \;=\;  B_{\phi \psi} + \cF_{\phi \psi} \;=\; \cos^2\theta\,,
\ee
so the DBI action becomes
\be
\label{pfdiskDBI}
\cL \;=\;  \sqrt{\cos^2\theta + \dot{\theta}^2 \sin^2\theta} \; .
\ee
The embedding equation \eqref{embeddingSU2brane} is a solution of
the resulting equations of motion.

\subsubsection{Symmetry breaking branes}

We now consider symmetry-breaking branes in $SU(2)$
obtained by smearing the D-branes described above along a twisted conjugacy 
class of a $U(1)$ subgroup $\cH$: 
\begin{equation}
  \mathcal{C}_\cH^{(f_{\cH},\Omega_{\pm})}
  = \{ h f_{\cH} \,\Omega_{\pm} (h^{-1})\, ,
  \quad h \in \cH \} \, , 
\end{equation}
where the automorphisms $\Omega_{\pm}$ act on $h = e^{i \lambda}$ as
\begin{equation}
\label{Omegapm}
\Omega_{\pm} (e^{i \lambda}) = e^{\pm i \lambda} \, .
\end{equation}
Note that $\mathcal{C}_\cH^{(f_{\cH},\Omega_{+})}$ reduces to a
point, so by using the automorphism $\Omega_{+}$ one recovers the
symmetry-preserving branes.
On the other hand, the inversion automorphism leads to $\cC_\cH$ 
isomorphic to $\cH=U(1)$, embedded in $\cG$.
The worldvolume of the symmetry breaking branes is then given by
\begin{equation}\label{prodconjclass}
 \mathcal{C}_{\mathcal{G}}  \cdot \epsilon (\mathcal{C}_\cH^{(f_{\cH},\Omega_{-})}) \;, \qquad\quad
\mathrm{with}\quad\mathcal{C}_{\mathcal{G}}  = \mathcal{C}_{\sst \rm SU(2)}^{(f_{\su})} \,,
\end{equation} 
and where  we take the embedding map $\epsilon$ to be
\begin{equation}
\epsilon (e^{i \alpha}) = e^{i  \alpha \sigma_3} ~.
\end{equation}
From \eqref{prodconjclass} we see that the brane is
described by the relation
\begin{equation}\label{SU2smearembed}
\cos \theta \cos (\psi - \alpha) = \cos \mu \, ,
\end{equation}
namely  $\cos \theta \geq \cos \mu$. The branes fill part of the
$SU(2)$ group (see Figure~\ref{fig:S2_brane_smeared}, where we have again depicted the brane twisted by the automorphism that sends $\theta\to\pi/2-\theta$, and $\phi\leftrightarrow\psi$; this twisted brane is relevant to the constructions in sections~\ref{sec:NS5Coul}-\ref{sec:specflowST}). 
The worldvolume flux $\omega_2$ is given by \eqref{SBomega}. Note that 
in the present case $\omega_2(\cH)$ vanishes. In order to evaluate the forms
$\omega_2(\cG)$ and $\omega_2(\cH, \cG)$ we can parameterize
$\cC_{\mathcal{G}}$ as in \eq{eq:cg-param-su}--\eqref{SU2braneparam},
with
\begin{equation}\label{SU2braneparamalpha}
X^1 = \sin \theta \cos (\phi-\alpha) \, ,\quad X^2 = \sin \theta \sin (\phi-\alpha )\,
,\quad X^3 = \cos \theta \sin (\psi -\alpha)\, .
\end{equation}
The form $\omega_2(\cG)$ is given by \eqref{fluxsymmpresbrane}, while
we find
\begin{equation}
\omegatwo(\cH,\cG)  = \left(X^2 \, d X^1 - X^1 \, d X^2 - \cos \mu \,d
  X^3\right) \wedge  d\alpha  . 
\end{equation}
By using \eqref{SU2braneparamalpha} and the embedding equation
\eqref{SU2smearembed} we find 
\begin{equation}
\label{omega2SU2branesmeared}
\omega_2  = \pm \frac{\cos\mu \tan \theta}{\sqrt{\cos^2\theta -
    \cos^2\mu}} \,d\theta \wedge d\phi - \sin^2\theta\, d\phi \wedge d\psi \,,
\end{equation}
where the $\pm$ arises from the sign of $X^3$,
  similarly to Eq.\;\eq{S2branefluxEuler}. We will see a similar
  structure in the following subsection.

The same result can be obtained from a DBI computation.
If we smear the $S^2$ brane along $\psi$ we can parametrize the
worldvolume by
\begin{equation}
\phi = \xi_0 \, ,\quad \psi = \xi_1 \, ,\quad \theta = \xi_2 \, .
\end{equation}
Turning on a non-zero flux $\mathcal{F} = \cF_{\theta \phi}\,  d\theta
\wedge d\phi$ we find that the matrix $G+ B+\mathcal{F} $ is
\begin{equation}
G + B + \mathcal{F} =
  \begin{pmatrix}
    \sin^2\theta & B_{\phi\psi} & -\mathcal{F}_{\theta \phi}\\
    -B_{\phi\psi} & \cos^2\theta & 0 \\
    \mathcal{F}_{\theta \phi} & 0 & 1 
    \end{pmatrix} \, ,
\end{equation}
  where we choose the gauge $B_{\phi\psi}  = -\sin^2\theta$ in order to agree with
\eqref{omega2SU2branesmeared}. The effective action is thus
\begin{equation}\label{S2smearedeffaction}
\mathcal{L} = \sqrt{\cF_{\theta\phi}^2\cos^2\theta +\sin^2\theta} \, .
\end{equation}
Demanding that
\begin{equation}
\frac{\partial \cL}{\partial \cF_{\theta\phi} } = \frac{\cF_{\theta
    \phi}\cos^2\theta }{\sqrt{\cF_{\theta\phi}^2\cos^2\theta
    +\sin^2\theta}} =\text{const} = \cos \mu \, ,
\end{equation}
we find
\begin{equation}
\cF_{\theta \phi} =\pm \frac{\cos\mu\tan\theta}{\sqrt{\cos^2\theta -\cos^2\mu}} \, .
\end{equation}
This solution agrees with \eqref{omega2SU2branesmeared}, taking into account
$\omega_2 = B + \cF$.

%%%%%%%%%%%%%%%%%%%%%%%%%%%%%%%%%%%%%%%%%

%\refstepcounter{subsection}
%\subsection*{\thesubsection \quad $SL(2,\mathbb{R})$ D-branes} \label{sec:SL2branes}

\subsection{$SL(2,\mathbb{R})$ D-branes}\label{SL2branes}

%%%%%%%%%%%%%%%%%%%%%%%%%%%%%%%%%%%%%%%%%

We now review both symmetry-preserving and symmetry-breaking
branes in $SL(2,\mathbb{R})$. As the discussion closely parallel the
one for $SU(2)$ we will be brief; see for example%
~\cite{Alekseev:1998mc,
Bachas:2000fr,
Elitzur:2001qd,
Gawedzki:2001ye,
Walton:2002db,
Sarkissian:2002ie,
Sarkissian:2002bg,
Sarkissian:2002nq,
Fredenhagen:2001kw,
Quella:2002ns,
Quella:2002fk,
Quella:2003kd}
for additional details.

%%%%%%%%%%%%%%%%%%%%%%%%%%%%%%%%%%%%
\subsubsection{Symmetry-preserving branes}

\begin{figure}[ht]
\centering
  \begin{subfigure}[b]{0.24\textwidth}
    \includegraphics[width=\textwidth]{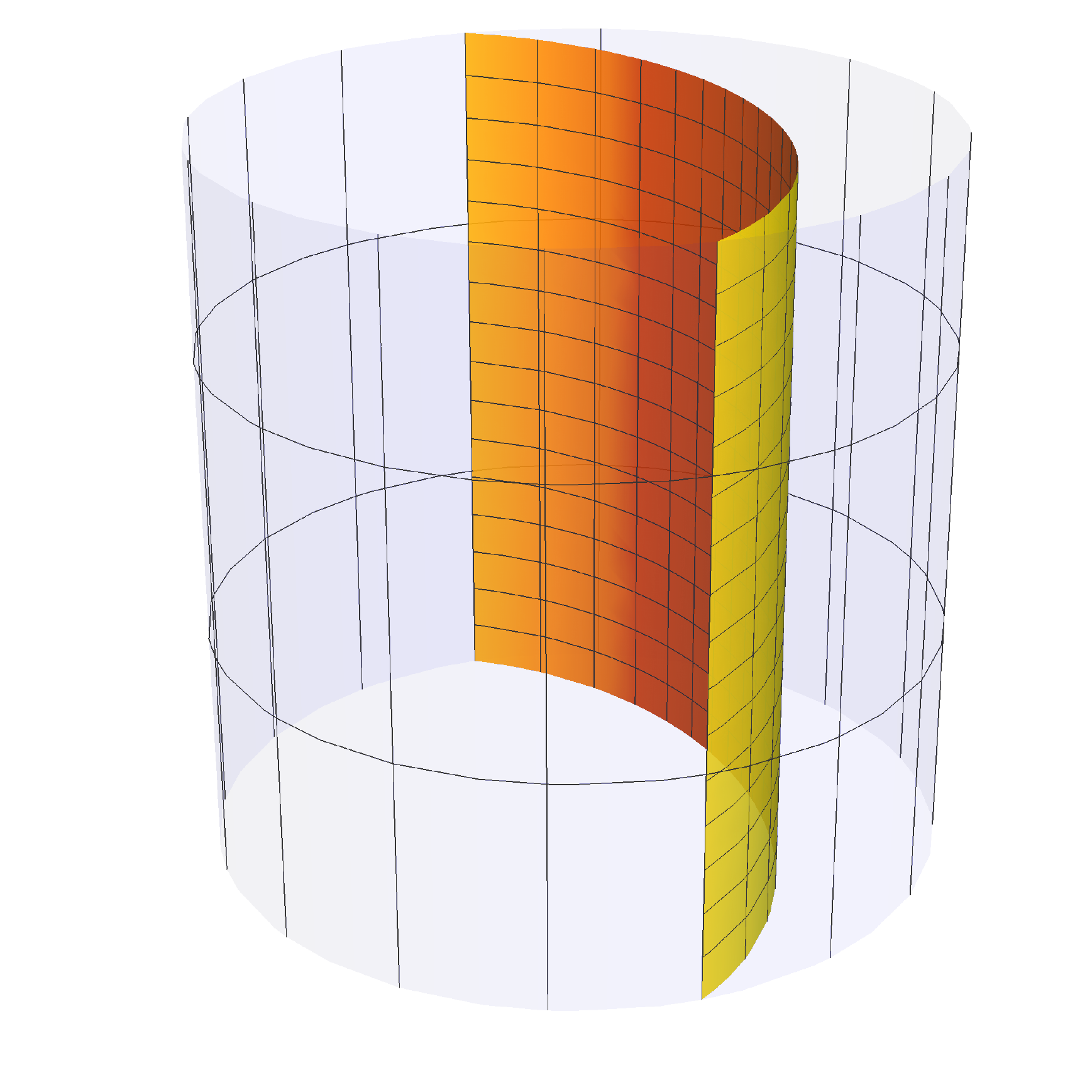}
    \caption{ }
    \label{fig:AdS2_brane}
  \end{subfigure}
  \begin{subfigure}[b]{0.24\textwidth}
    \includegraphics[width=\textwidth]{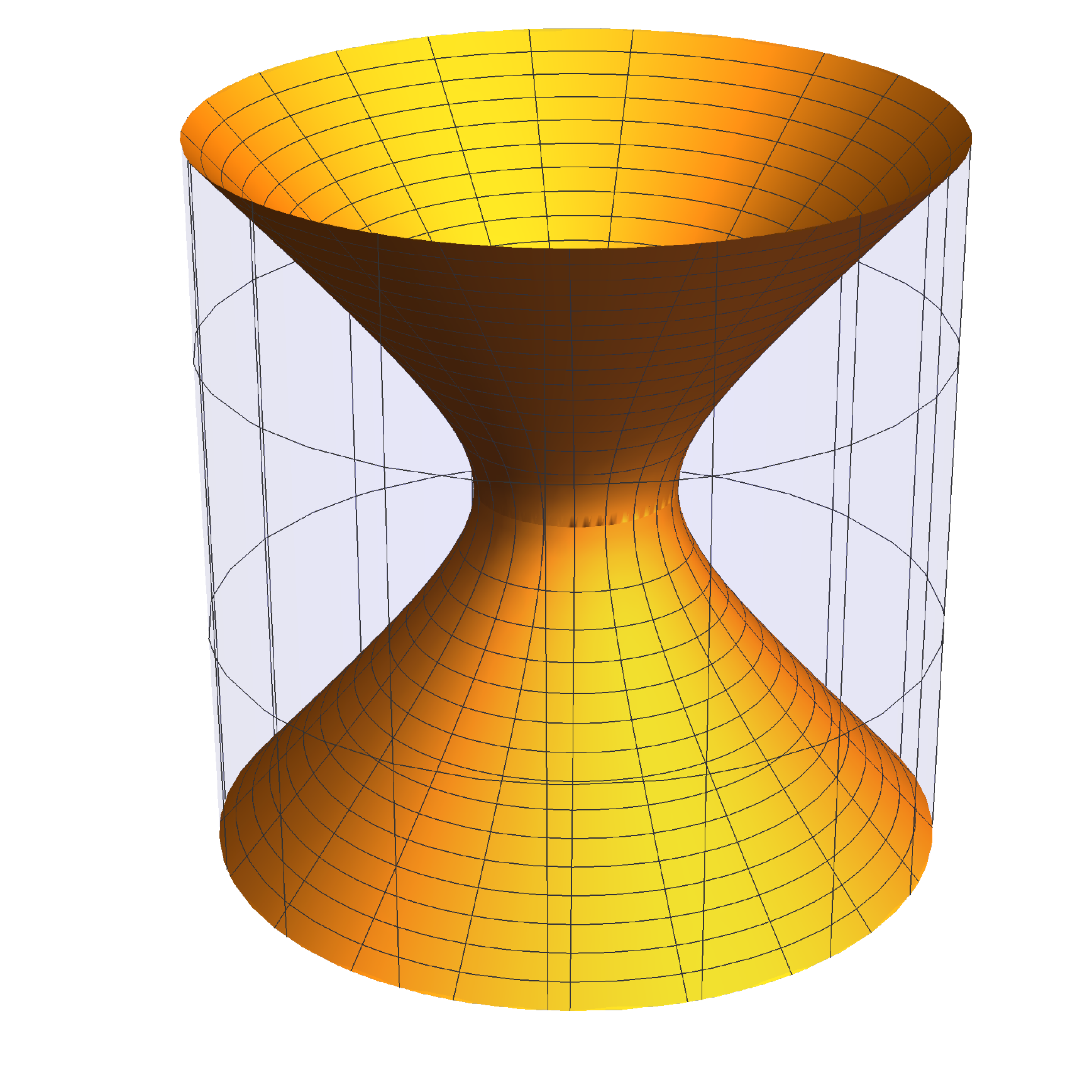}
    \caption{ }
    \label{fig:dS2_brane-0}
  \end{subfigure}
  \begin{subfigure}[b]{0.24\textwidth}
    \includegraphics[width=\textwidth]{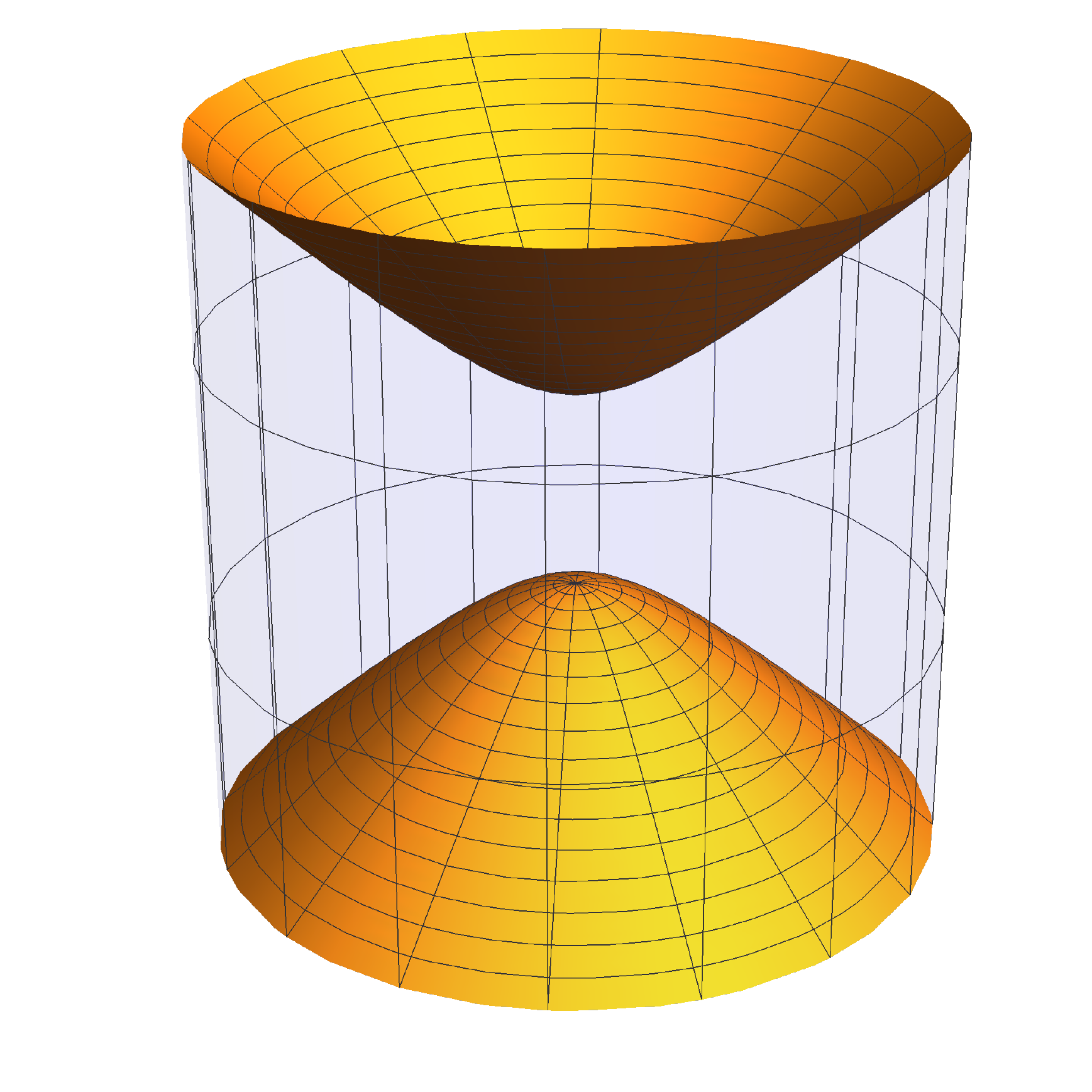}
    \caption{ }
    \label{fig:H2_brane}
  \end{subfigure}
\caption{\it 
Symmetry-preserving $SL(2,\mathbb{R})$ branes:
(a) $AdS_2$;
(b) $dS_2$;
(c) $\bH_2$.
}
\label{fig: SL2branes}
\end{figure}

The generic twisted conjugacy classes for $\sltwo$ are depicted in Figure~\ref{fig: SL2branes}; we now consider them in turn.

%%%%%%%%%%%%%%%%%%
%%%%%%%%%%%%%%%%%%

%%%%%%%%%%%%%%%%%%%
\paragraph{$AdS_2$ brane:}
If
$\Omega$ in \eqref{conjclass} is outer, we can take (up to group conjugation)
\begin{equation}
\Omega (g) = \sigma_1 g \sigma_1 \, .
\end{equation}
The defining relation for
this conjugacy class is $\tr (g\sigma_1) = \tr(f_{\sl}
\sigma_1)$. Taking $f_\sl= e^{ \mu \sigma_1}$ this reduces to 
\begin{equation}\label{eq:AdS2embedding}
  \sinh\rho \cos \sigma =
\sinh \mu \, .
\end{equation}
This defines $AdS_2$ sections of $SL(2,\mathbb{R})$. 
The worldvolume flux is given by~\eqref{omegaf}; this can be evaluated by using coordinates adapted to such $AdS_2$ slicing of $\sltwo$.  
Mapping to Euler angle coordinates one finds 
\begin{equation}\label{AdSfluxw2}
\omega_2 = \pm \frac12 \frac{\sinh \mu \sinh 2\rho}{\sqrt{\sinh^2\rho-
  \sinh^2\mu}} \, d\rho \wedge d\tau  \, .
\end{equation}
The two signs correspond to the two different branches of the
embedding \eqref{eq:AdS2embedding}. It is straightforward to show that the embedding equation
\eqref{eq:AdS2embedding}, together with the worldvolume flux
\eqref{AdSfluxw2}, provide a solution of the DBI equations.

\paragraph{$dS_2$ brane:}

Taking $\Omega = \id$ and $f_{\sl} = e^{\mu \sigma_3}$  we find the
brane defined by the embedding
\begin{equation}
\cosh \rho \cos\tau = \cosh \mu \, ,
\end{equation}
which defines a $dS_2$ world-volume. 
These branes have a super-critical worldvolume flux given by
\begin{equation}
\omega_2= \pm \frac12 \frac{\cosh \mu \sinh 2\rho}{\sqrt{\cosh^2\rho
    -\cosh^2 \mu}} \, d\rho \wedge d\sigma  \, .
\end{equation}

%%%%%%%%%%%%%%%%%%%
\paragraph{$\bH_2$ brane:}

Finally, for $\Omega = \id$ and $f = e^{i  \mu\sigma_3}$  we get 
\begin{equation}
  \cosh \rho \cos\tau = \cos \mu \, ,
\end{equation}
which defines a two-sheeted hyperboloid.
Such $\bH_2$ branes are formally a solution of the DBI equations
with a worldvolume density of D-instantons. We now find:
\begin{equation}
\omega_2 = \pm \frac12 \frac{\cos \mu \sinh 2\rho}{\sqrt{\cosh^2\rho
    -\cos^2 \mu}} \, d\rho \wedge d\sigma \, .
\end{equation}
Note that at $\mu$ = 0 the $\bH_2$ and $dS_2$ world-volumes degenerate to a
light-like brane.

\subsubsection{Symmetry-breaking branes}
\label{sec:symbreakbranes}

We now describe the symmetry-breaking branes
obtained by smearing the branes described above along 
a non-trivial conjugacy class of an abelian subgroup.

%%%%%%%%%%%%%%%%%%
\begin{figure}[t]
\centering
  \begin{subfigure}[b]{0.24\textwidth}
    \includegraphics[width=\textwidth]{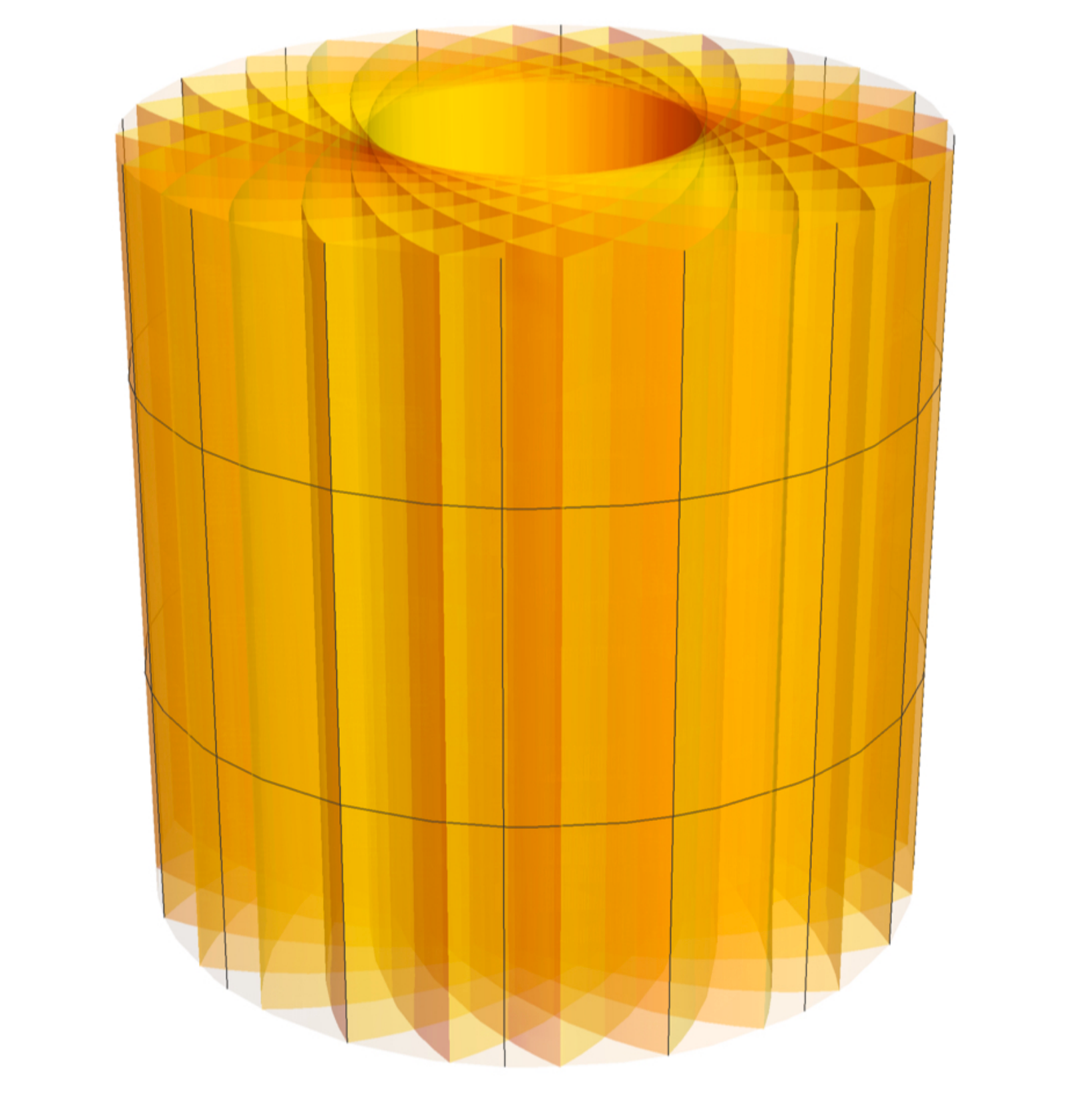}
    \caption{ }
    \label{fig:AdS2_brane_smeared}
  \end{subfigure}
  \begin{subfigure}[b]{0.24\textwidth}
    \includegraphics[width=\textwidth]{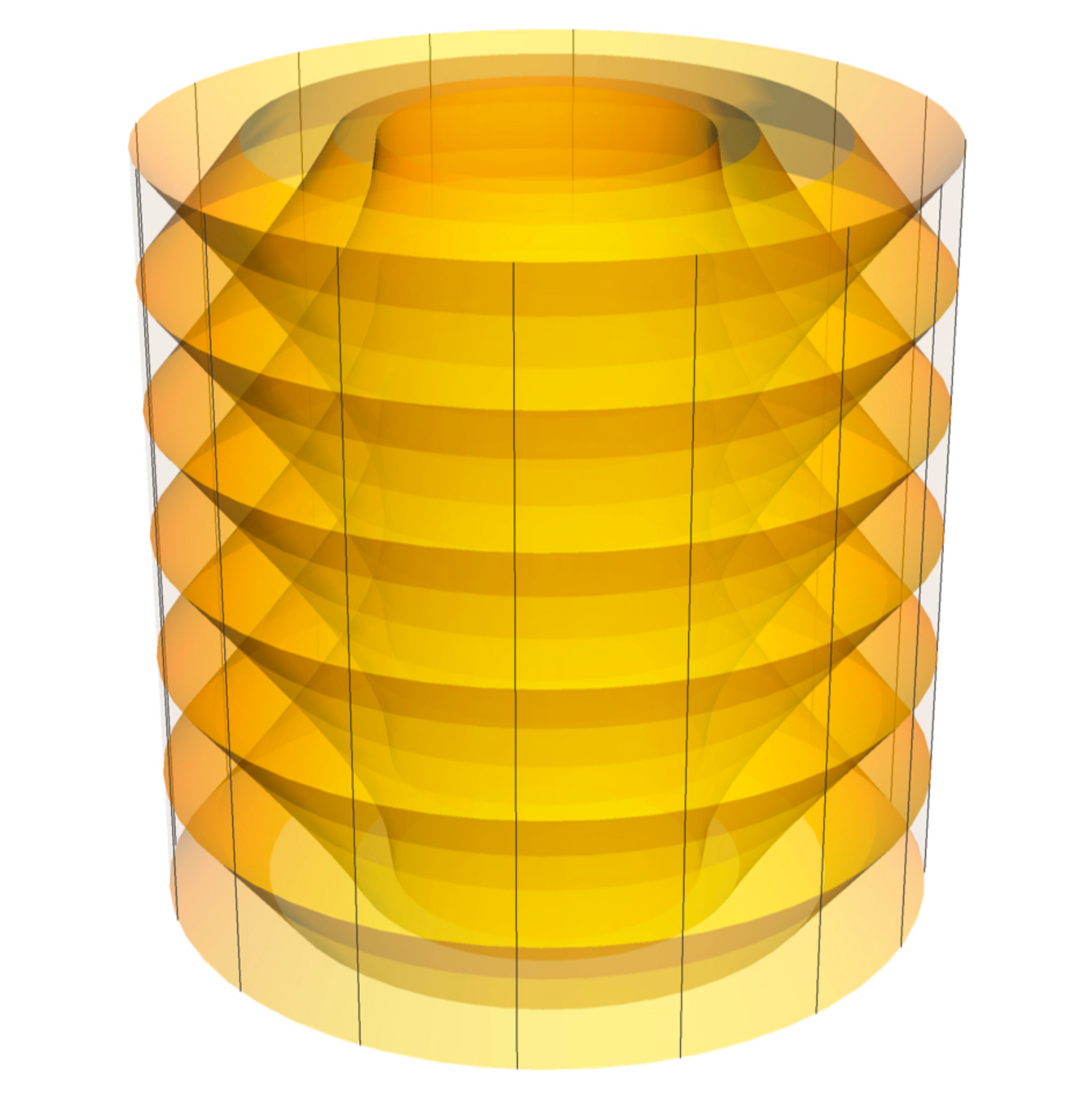}
    \caption{ }
    \label{fig:dS2_brane_smeared-2}
  \end{subfigure}
  \begin{subfigure}[b]{0.24\textwidth}
    \includegraphics[width=\textwidth]{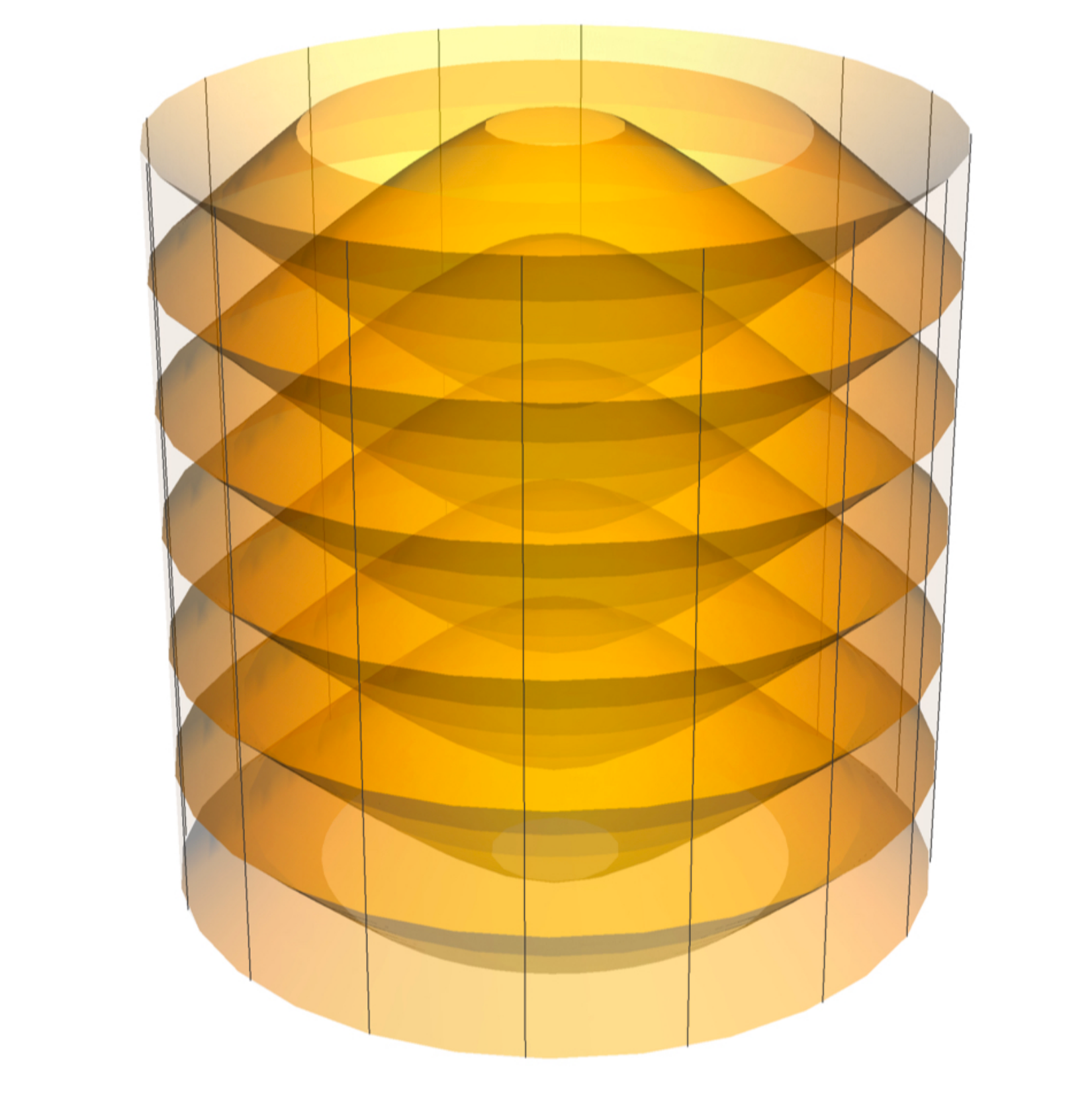}
    \caption{ }
    \label{fig:H2_brane_smeared}
  \end{subfigure}
\caption{\it 
Symmetry-breaking $SL(2,\mathbb{R})$ branes:
the worldvolume obtained by smearing
(a) the $AdS_2$ brane along $\sigma$;
(b) the $dS_2$ brane along $\tau$;
(c) the $\bH_2$ brane along $\tau$.
Note that the smeared $\bH^2$ brane fills $\sltwo$, and that although the worldvolumes for the smeared $AdS_2$ and $dS_2$ branes are similar, they are foliated differently and carry different two-form fluxes $\omega_2$.
}
\label{fig:smearedbranes}
\end{figure}
%%%%%%%%%%%%%%%%%%

%%%%%%%%%%%%%%%%%%%
\paragraph{Smeared $AdS_2$ brane}

Starting
from an $AdS_2$ brane, taking the trace we see that the condition
following from \eqref{classproduct-gen-0} is
\begin{equation}\label{smearedAdS2brane}
\sinh\rho \cos (\sigma+\alpha) = \sinh \mu \, ,
\end{equation}
namely $\sinh \rho \geq \sinh \mu$. The $AdS_2$ worldvolume has been
smeared along the $\sigma$ direction and the brane is filling the
$AdS_3$ space outside the radius $\rho_{\mu}=\mu$ (see Figure~\ref{fig:smearedbranes}). For $\mu = 0$ the whole
space is filled. Since $\alpha$ in \eqref{smearedAdS2brane} is
generically double valued, each element of the group is covered twice. 
The worldvolume flux can be determined  from~\eqref{SBomega}-\eqref{omegas}
\cite{Quella:2002ns,Sarkissian:2002bg}, following the same procedure
discussed for the $SU(2)$ branes.
The result is
\begin{equation}\label{eq:w2smearedads2}
\omega_2 = \pm \frac{\sinh\mu\coth\rho}{\sqrt{\sinh^2\rho
  -\sinh^2\mu}} \, d\rho\wedge d\tau -  \cosh^2 \rho\, d\sigma \wedge
d\tau \, .
\end{equation}

%%%%%%%%%%%%%%%%%%%
\paragraph{Smeared $dS_2$ brane}

Similarly, for the $dS_2$ brane we find 
\begin{equation}
\omega_2=\pm \frac{\cosh\mu\tanh\rho}{\sqrt{\cosh^2\rho
  -\cosh^2\mu}} \, d\rho\wedge d\sigma -  \sinh^2 \rho \, d\sigma \wedge
d\tau \, .
\end{equation}
Note that the smearing cures the large $\rho$ divergence of the flux
of the
symmetry-preserving branes.

%%%%%%%%%%%%%%%%%%%
\paragraph{Smeared $\bH_2$ brane}

Starting from an $\bH_2$ brane, one can construct a non-trivial
symmetry-breaking brane by smearing the worldvolume along the $\tau$
direction: 
\begin{equation}
\cosh \rho \cos (\tau - \alpha) = \cos \mu \, .
\end{equation}
The brane fills all the space. The worldvolume flux is now
\begin{equation}
\label{smearedH2flux}
\omega_2 =\pm \frac{\cos\mu\tanh\rho}{\sqrt{\cosh^2\rho
  -\cos^2\mu}} \, d\rho\wedge d\sigma -  \sinh^2 \rho \, d\sigma \wedge
d\tau \, .
\end{equation}

%%%%%%%%%%%%%%%%%%%
\paragraph{Smeared identity brane}

While we have not mentioned it so far, there is a special conjugacy class in $\sltwo$, namely the conjugacy class of the identity.  This describes a pointlike brane sitting at the origin $\rho=\tau=0$ in $\sltwo$.  For our applications, we then want to smear this brane along the orbits of the gauge group.  In particular we can smear along $\tau$ to arrive at a symmetry-breaking brane whose worldvolume is the worldline of a particle sitting at $\rho=0$ and extended along the timelike direction parametrized by $\tau$.  Because the orbit is one-dimensional, the two-form $\omega_2$ is trivial.

%%%%%%%%%%%%%%%%%%%%%%%%%%%%%%%%%%%%

%%%%%%%%%%%%%%%%%%%%%%%%%%%%%%%%%%%%
%%%%%%%%%%%%%%%%%%%%%%%%%%%%%%%%%%%%

%%%%%%%%%%%%%%%%%%%%%%%%%%%%%%%%%%%%
%%%%%%%%%%%%%%%%%%%%%%%%%%%%%%%%%%%%

\section{NS5-branes on the Coulomb branch}
\label{sec:NS5Coul}

In the null gauging formalism, a D-brane with a $p\!+\!1$ dimensional worldvolume downstairs in 9+1 dimensions gains another 1+1 dimensions in the group manifold upstairs in 10+2 dimensions, since the brane upstairs must be invariant under the $\uone\times\uone$ gauge translations.  One can accomplish this using the technology of Section~\ref{sec:gnlsm}, arbitrarily lifting the brane upstairs to 10+2 and then smearing it along the gauge orbits.

Let's first think about NS5-branes on the Coulomb branch.
In the supergravity approximation, the geometry of $n_5$ NS5 branes on their Coulomb branch is characterized by a single harmonic function
\begin{equation}\label{eq:NS5cartesian}
ds^2 = -du\, dv + ds^2_\cM + Z_5\, dx^i dx_i 
~~,~~~~ 
H_3 = -\epsilon^{l}_{\phantom{l}ijk}\pa_l Z_5 
~~ ,~~~~ 
e^{2\Phi} = g_s^2 Z_5 \, ,
\end{equation}
with
\begin{equation}
Z_5 = 1+\sum_{a=1}^{n_5} \frac{\lstr^2}{\left| x^i - x^i_a\right|^2}  ~.
\end{equation}
The decoupling limit scales $\gstr\to 0$ with $x^i/(\gstr\lstr)$ held fixed, and amounts to dropping the constant term in $Z_5$.

%%%%%%%%%%%%%%%%%%%%%%%%%%%%%%%%%%%%

Consider a D1-brane probe lying in the directions transverse to the NS5 worldvolume.  These are in fact trivial to describe downstairs in 9+1d at the level of the DBI effective action
\be
\cS = \int e^{-\Phi}\sqrt{det\bigl(G+B+\cF\bigr)}  ~;
\ee
for a static D1 in the transverse space, the warp factor $Z_5$ in the metric pulled back to the D1 worldvolume cancels exactly against the contribution $e^{-\Phi}$ of the dilaton.  As a result, the brane shape does not see the warp factor $Z_5$ and is thus a straight line in the transverse $\bR^4$.

We can characterize such a straight line in part via an equation $c_i x^i=C$.  
Let us use spherical bipolar coordinates on $\bR^4$, related to the Cartesian coordinates $x^i$ via
\be
\label{bipolars}
x^1+ix^2 = \cosh\rho \,\sin\theta\; e^{ i(\phi-\tau)}
~~,~~~~
x^3+ix^4 = \sinh\rho\, \cos\theta\; e^{ i(\psi-\sigma)}  ~~.
\ee
We have used the same coordinate labels as the Euler angles of $\sltwo\times\sutwo$ in order to facilitate the lift to 10+2 dimensions.
Note that these coordinates parametrize the physical transverse space to the fivebranes, in coordinates invariant under the gauge transformations generated by the Killing vectors~\eqref{eq:ax-vec}-\eqref{eq:vec-ax}.  With this embedding, the ring of fivebranes lies along the unit circle in the $x^1$-$x^2$ plane and at the origin in the $x^3$-$x^4$ plane, which is the locus $\rho=0,\theta=\pi/2$ (absorbing the factor $a$ in~\eqref{profile1} in a rescaling of coordinates).
We note that it is often easier to visualize the structure by taking the vector and axial combinations of the gauge parameters above, rather than the left/right parametrization of the gauge transformations, for the purpose of visualizing the shape of the brane upstairs at fixed time(s).  We will concentrate on the spatial shape of the brane, and thus the smearing along the axial gauge motion~\eqref{eq:vec-ax}.

The branes described in this section were considered
in~\cite{Israel:2005fn} using the coset orbifold description; here we recast their work in the formalism of null gauging, in preparation for the generalization to supertubes.

%%%%%%%%%%%%%%%%%%%%%%%%%%%%%%%%%%%%
\subsection{Factorized branes}
\label{sec:factorized}

Special cases of the straight-line D1-branes in the Coulomb branch NS5
background can be understood as coming from the gauging of branes that
start off as factorized boundary conditions in $\sltwo\times\sutwo$,
using the formalism of Section~\ref{sec:gnlsm}.  To this end, we want a brane in the ``upstairs'' group $\cG$ of equation~\eqref{Gupstairs2} that projects to the above 1+1d brane ``downstairs'' in 9+1d physical spacetime upon gauging of $\cH=U(1)_L\times U(1)_R$, where for the moment we restrict ourselves to $\cH$ embedding in $\sltwo\times\sutwo$.  As in Section~\ref{sec:gnlsm}, we denote the embeddings of the left and right null $\uone$'s into $\cG$ as $(\ell(h_L),r(h_R))$.  For NS5 branes on the Coulomb branch, the gauge motion is given in~\eqref{Coulgaugetransfs}, which shifts the Euler angles as in~\eqref{STgaugetransfs} (with the tilt parameter $k$ set to zero).  It will prove convenient to work with the linear combinations $\zeta= \alpha+\beta$ that parametrize temporal gauge transformations shifting the Euler angles $\tau$ and $\phi$, and $\eta= \alpha-\beta$ parametrizing spatial gauge transformations shifting $\sigma$ and $\psi$.  The left and right embeddings are then%
\footnote{We will ignore global issues involving the $\uone$ gauge groups; and notationally rewrite $\uone$ quantities in terms of the arguments of the phase circle, \eg\
$\Omega^\eps(e^{i\xi})=e^{i\eps\xi}$ will be written $\Omega^\eps(\xi)=\eps\xi$.}
\begin{align}
\bigl( g_\sl, g_\su \bigr) &\longrightarrow 
\ell(\zeta,\eta)  \bigl( g_\sl, g_\su \bigr)  r(-\zeta,-\eta)
  = 
\bigl( e^{ \frac i2(\zeta+\eta){\sigma_3}} g_\sl e^{ \frac i2(\zeta-\eta){\sigma_3}} \, , \, 
e^{ -\frac i2(\zeta+\eta){\sigma_3}} g_\su e^{ \frac i2(\zeta-\eta){\sigma_3}}  \bigr) ~.
\end{align}
To specify a brane in the formalism of~\cite{Quella:2002fk,Quella:2003kd} reviewed in Section~\ref{sec:gnlsm}, in addition to the embeddings $(\ell,r)$ of $\cH$ into $\cG$ one needs a pair of group automorphisms $\Omega_\cG$ and $\Omega_\cH$, with the constraint 
\be
\label{EmbeddingCondition}
r = \Omega_\cG \circ \ell \circ \Omega_\cH  ~~.
\ee
In~\cite{Quella:2002fk,Quella:2003kd}, this condition guarantees the preservation of an enlarged chiral algebra for the boundary states considered.  It seems to be necessary within the class of branes we are considering in order to preclude the appearance of manifestly unphysical branes, for instance a D1-brane that terminates at a point in space where there is no NS5-brane.%
\footnote{A generalization of this formalism is suggested in Appendix A of~\cite{Quella:2002fk} (see also Appendix D of~\cite{Quella:2003kd}) that drops this constraint, however while such boundary conditions may be allowed at the semiclassical level, there may be further constraints needed to ensure the absence of quantum anomalies.
}
We note the useful identities for the Euler angles~\eqref{Eulerangles-1}
\begin{align}
\Omega^-_\sl(g_\sl) &=\Gamma_\sl \,g_\sl(\rho,\tau,\sigma) \,\Gamma^{-1}_\sl = g_\sl(\rho,-\tau,-\sigma)
~~~~\, ,~~~~
\Gamma_\sl = \sigma_1
\nn\\
\Omega^-_\su(g_\su) &= \Gamma_\su \,g_\su(\theta,\phi,\psi) \, \Gamma^{-1}_\su = g_\su(\theta,-\phi,-\psi)   
~~,~~~~
\Gamma_\su = -i\sigma_1  ~~.
\end{align}
For $\sltwo$ this is a nontrivial outer automorphism, while for $\sutwo$ one has an inner automorphism.  In particular, for $\uone$ embeddings into $\sutwo$ and $SU(1,1)$ lying along the $\sigma_3$ direction, $\vareps(\eta)=\exp[i\eta{\sigma_3}]$, one has the properties
\begin{align}
\Omega^+_{\sl,\su}\circ\vareps\circ\Omega_\pm(\eta) &= \exp\Bigl[\pm i\eta{\sigma_3}\Bigr]
\nn\\
\Omega^-_{\sl,\su}\circ\vareps\circ\Omega_\pm(\eta) &= \exp\Bigl[\mp i\eta{\sigma_3}\Bigr]
\end{align}
where $\Omega^+_{\sl,\su}=\id$ are the corresponding identity automorphisms, and $\Omega_\pm$ are the identity and inversion automorphisms of $U(1)$ defined in~\eqref{Omegapm}.

Let us parametrize
\begin{align}
\label{conjclasses}
\Omega_\cH(\zeta,\eta) = (\eps \zeta, -\eps\eta)
~~,~~~~
\Omega_\cG(g_\sl,g_\su) = \bigl( \Omega^{\eps_\sl}(g_\sl) , \Omega^{\eps_\su}(g_\su) \bigr)
\end{align}
with $\eps,\eps_\sl,\eps_\su=\pm$.
The constraint~\eqref{EmbeddingCondition} is then satisfied provided
\be
\eps \:\! \eps_\sl = -1
~~,~~~~
\eps \:\! \eps_\su = +1 ~~.
\ee

We are considering branes that are factorized products of conjugacy classes $\cC_\cG=\cC_\sl\cdot \cC_\su$; the relation $\eps_\sl \eps_\su=-1$ correlates the choices, so that untwisted conjugacy classes in $\sltwo$ are associated to twisted conjugacy classes in $\sutwo$, and vice versa.
Consider first the choice $\eps_\sl=+$; then one has $\eps=-$ and therefore $\eps_\su=-$.  The $\sltwo$ conjugacy classes with $\eps_\sl=+$ are $dS_2$ and $\bH_2$, which are extended along both spatial directions $\rho$ and $\sigma$ of $\sltwo$; nontrivial conjugacy classes in $\sutwo$ are $\bS^2$ branes, also extended in two spatial dimensions.  With $\eps=-$, the conjugacy class $\cC_\cH$ smears $\cC_\cG$ along temporal gauge orbits parametrized gauge orbits and is pointlike along spatial gauge orbits.  All told, the product $\cC_\cG\cdot \cC_\cH$ has four spatial dimensions, and gauges down to a D3 brane downstairs in 9+1d.  While such branes are of interest, our focus here is on D1-brane probes.%
\footnote{The D3 branes are bound states of the D1's, puffed up in the $\sutwo$ directions by the Myers effect~\cite{Myers:1999ps}.}  
We can reduce the dimensionality by taking one of the two conjugacy classes $\cC_\sl$ or $\cC_\su$ to be trivial.  Thus we set
\begin{align}
\label{BraneAtConstantX1}
\half\tr\bigl[g_\su  \,\Gamma_\su\bigr] &= \sin\theta\,\cos \phi = C_\su 
~~,~~~~
\Gamma_\su = -i\sigma_1  ~~,
\nn\\[5pt]
\half\tr\bigl[g_\sl \bigr] &= \cosh\rho\, \cos \tau  = C_\sl 
\end{align}
with the choice $C_\su=1$ so that the $\sutwo$ component
collapses to a point brane at $\theta=\pi/2$ and $\phi=0$.  From the bipolar coordinates~\eqref{bipolars}, we see that the brane lies in the \xone-\xtwo\ plane.  The $\sltwo$ conjugacy class is extended in the radial direction $\rho$ and along $\sigma$, which is the only coordinate acted on nontrivially by spatial gauge transformations; all points along the $\sigma$ circle at fixed $\rho$ are identified under the gauge projection, and the brane upstairs descends to a D1 probe in the \xone-\xtwo\ plane that extends out to spatial infinity.

The temporal smearing implemented by $\cC_\cH$ is a simultaneous translation along $\phi$ and $\tau$, so that the brane locus $\cC_\cG\cdot\cC_\cH$ upstairs is
\be
\sin\theta\,\cos(\phi-\zeta) = 1
~~,~~~~
\cosh\rho\,\cos(\tau-\zeta) = C_\sl ~.
\ee
The choice of relative rotation has fixed $\phi-\tau=0$ and thus $x^1=C_\sl$; this can be adjusted to any desired angle by generalizing the choice of inner automorphism to $\Gamma_\su=\cos\nu(-i\sigma_1)+\sin\nu(-i\sigma_2)$.
For $C_\sl>1$, one has a $dS_2$ brane upstairs, which drops down from $\rho=\infty$, reaches a minimum radius $\cosh\rho=C_\sl$, and then runs back out to infinity.  Correspondingly, the D1 probe downstairs is a straight line at fixed \xone\ that passes outside the ring of fivebranes, which lie along the unit circle in this plane.  On the other hand, for $C_\sl<1$ the $\bH_2$ brane upstairs has two components, each describing a segment of D1-brane downstairs that ``ends'' on the unit circle (which in bipolar coordinates is $\rho=0,\theta=\pi/2$), at $x^1=C_\sl$, where the ring of fivebranes is located.  Upstairs, the brane geometry is perfectly smooth; the ``ending'' of the brane is simply a smooth degeneration of the spatial gauge orbit parametrized by $\sigma$.
Note that it is the embedding constraint~\eqref{EmbeddingCondition} that imposes the relation $\eps_\sl\eps_\su=-1$ that relates the choices of conjugacy classes in $\sltwo$ and $\sutwo$.  Without this relation, one could for instance choose an $\bH_2$ brane together with an untwisted brane in $\sutwo$; the D1-brane would end on the unit circle in the $x^3$-$x^4$ plane where there are no fivebranes, which is manifestly unphysical.

The two components of the $\bH_2$ brane describe the segments of a straight line D1 brane lying outside the unit circle at fixed $x^1=C_\sl<1$.  Define the ``impact parameter'' $b=C_\sl$ of this probe.  The complementary segment lying inside the unit circle at $x^1=b$ sits at $\rho=0$ and varying $\theta$, and describes a D1 brane stretching between fivebranes.  Upstairs it is described as a point brane in $\sltwo$ and a nontrivial $\bS^2$ brane in $\sutwo$ (again smeared along the timelike gauge orbit by $\cC_\cH$), \ie\ the locus~\eqref{BraneAtConstantX1} with $C_\sl=1$ and $C_\su=b$.  

\begin{figure}[ht]
\centerline{\includegraphics[width=.4\textwidth]{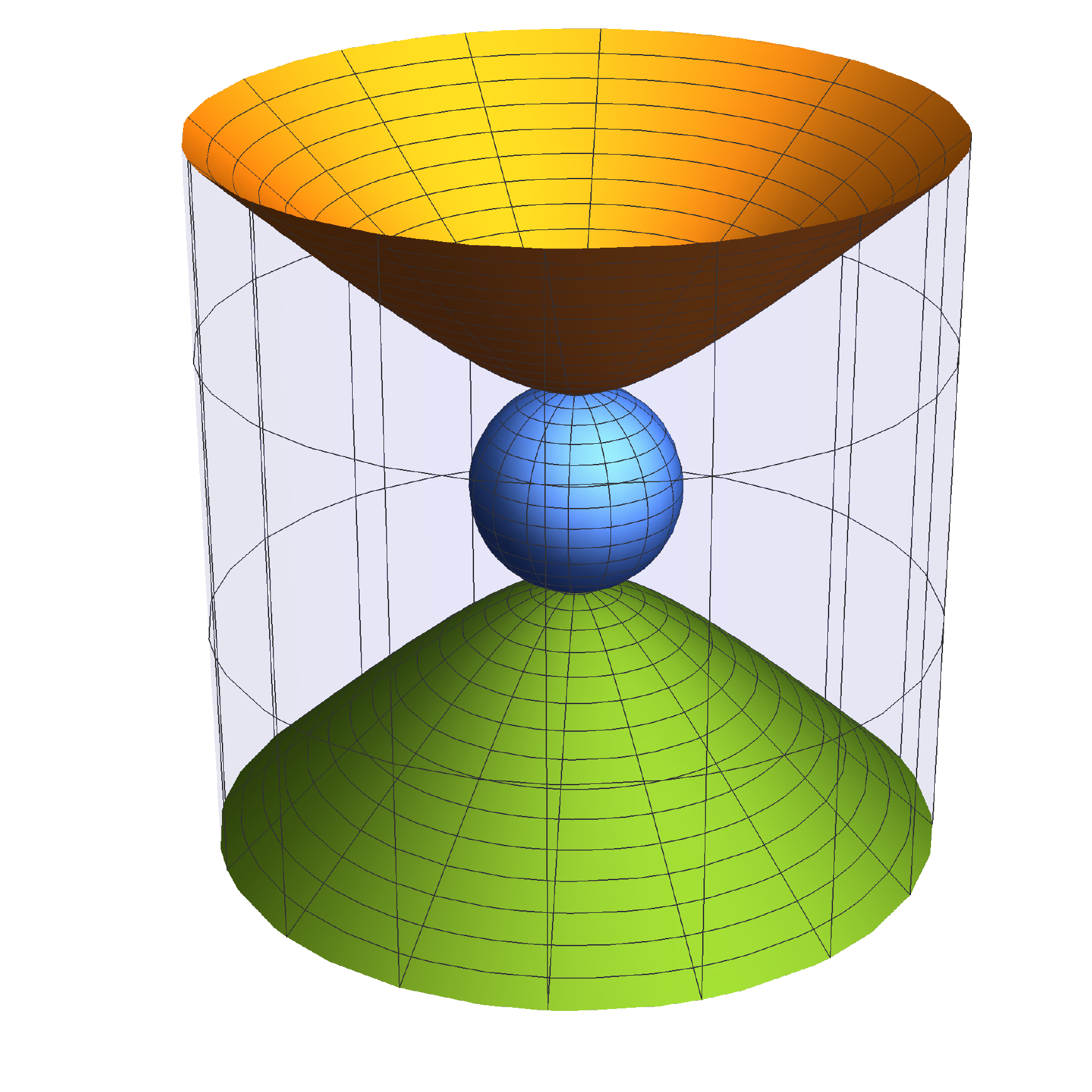}}
\setlength{\unitlength}{0.1\columnwidth}
\caption{\it 
An $\bH_2$ brane describes D1-brane segments stretching from the NS5-branes to infinity, while an $\bS^2$ brane describes the complementary segment stretching between NS5-branes, when lifted to 10+2 dimensions.  In both cases, the azimuthal direction is the orbit of spatial gauge transformations; all points on an orbit project down to the same point in physical spacetime.  The brane must also be smeared along the orbits of temporal gauge transformations.
}
\label{fig:H2-S2_brane}
\end{figure}

The various components upstairs of this fractionated D-string are depicted in Figure~\ref{fig:H2-S2_brane}.
The two component $\bH_2$ branes are depicted in yellow and green, with the $\bS^2$ component depicted as a blue sphere filling in the gap in between in the figure.
It should be emphasized that this sphere occupies a factor of the spacetime group manifold orthogonal to that occupied by the $\bH_2$ brane.  Sadly our world hasn't enough macroscopic dimensions to faithfully depict the plumbing-fixture structure at the juncture of the two components of the D-brane worldvolume, so we have simply inserted the $\bS^2\subset\sutwo$ brane into the gap left by the pair of $\bH_2$ branes in $\sltwo$ to indicate how the segments are joined together.
%%%%%%%%%%%%%

Thus factorized branes can describe an example of the process of probe D1-branes intersecting and breaking on NS5 branes through a topological transition in which the D-brane's 2d spatial worldvolume pinches off via a standard Riemann surface plumbing fixture; upstairs in 10+2d, there are no NS5-branes -- there is only a smooth 10+2d flux geometry.  The fivebrane locus arises through the degeneration of the gauge orbits; the D-brane worldvolume knows about the fivebranes because it must lie along the gauge orbits, and thus it must degenerate whenever it intersects the fivebranes.  In this way, the gauged WZW model can encode the phenomenon of D-brane worldvolumes ``ending'' on NS5-branes.   

We have considered the various options for factorized branes with $\eps=-$ in~\eqref{conjclasses}, finding straight line D1 probes lying in the \xone-\xtwo\ plane at $x^3=x^4=0$.  If instead we take $\eps=+$, we get straight line D1 probes in the \xthree-\xfour\ plane at $x^1=x^2=0$.  Now in order to satisfy~\eqref{epsconstraints} we require $\eps_\sl=-$ and $\eps_\su=+$.  Again to describe a D1 brane rather than a D3 brane, one of the conjugacy classes must be trivial, and so we consider a pointlike brane in $\sutwo$ at $\theta=0$, \ie\ at $x^1=x^2=0$, extended along $x^4$ at $x^3=\textit{const}$.  
This is the conjugacy class
\be
\label{epsconstraints}
\frac12 \tr[g_\su] = \cos\theta\,\cos\psi = C_\su
~~,~~~~
\frac12\tr[g_\sl\Gamma_\sl] = \sinh\rho\,\cos\sigma = C_\sl
\ee
with $C_\su=1$ to specify the pointlike brane at $\theta=0$ (and $\psi=0$), and $C_\sl$ specifies the impact parameter of the brane trajectory in the \xthree-\xfour\ plane.  The $\sltwo$ brane describes an $AdS_2$ conjugacy class; in this case, the combined brane is invariant under temporal gauge transformations shifting $\tau$ and $\phi$, and the effect of multiplying by $\cC_\cH$ is to smear along spatial gauge orbits.

We should note that the branes that are factorized between the various group factors in $\cG$ are very special.  A D1-brane downstairs with a general position and orientation will inextricably correlate its location in $\sltwo$ and $\sutwo$ along its worldvolume.  For instance, a brane along the straight line $x^1=x^4=0$, $x^2=c$ imposes the condition in bipolar coordinates
\be
\cosh\rho \sin\theta = c
\ee 
which does not appear to be a condition on class functions, or any other natural group-theoretic quantity.  Nevertheless, in representative examples such as this one we have been able to lift the D-brane worldvolume to $\cG$ by extending it along the gauge orbits of $\cH$, and find the two-form flux that solves the DBI equations of motion.  We omit the details of this example, which are a straightforward application of the above methods.

%%%%%%%%%%%%%%%%%%%%%%%%%%%%%%%%%%%%
\subsection{Flux quantization effects}
\label{sec:fluxquant}

Note that in the null gauging approach, the gauge action degenerates where the NS5's are located.  Naively this is the locus where the coefficient of $\cA\cAbar$ in the gauge action vanishes.   But this quantity vanishes along the entire circle $\rho=0$, $\theta=\pi/2$, and doesn't distinguish the locations of the NS5's specifically; in other words the naive sigma model does not exhibit the breaking of rotational symmetry in the $x^1$-$x^2$ plane of the ring of fivebrane sources to $\bZ_\nfive$, which is a nonperturbative effect in~$\alpha'$.  
However, when the brane intersects the ring of fivebranes twice, we can see some aspects of the discrete structure through the quantization of worldvolume $\cF$ flux carried by the brane in 10+2 dimensions.  For the $\bS^2\subset\sutwo$ branes that are pointlike in $\sltwo$, this flux is the usual induced lower-dimensional brane charge that keeps the brane puffed up due to the Myers effect~\cite{Bachas:2000ik}.
After smearing along the gauge orbits, the two-form $\omega_2$ is
given by the twisted version of~\eqref{omega2SU2branesmeared}
\begin{equation}
\omega_2  = \pm\frac{C_\su \cot \theta}{\sqrt{\sin^2\theta -
    C^2_\su}} \,d\theta \wedge d\psi + \cos^2\theta\, d\phi \wedge d\psi \, .
\end{equation}
If we integrate the flux on a fixed time(s) slice, the conjugacy classes are quantized as
\be
C_\su = \cos\mu
~~,~~~~
\mu = 2\pi j /\nfive
~~,~~~~
j=0,\frac12,1, ...,\frac12\nfive 
\ee 
due to the quantization of magnetic flux on the smeared $\bS^2$~\cite{Maldacena:2001ky,Walton:2002db,Sarkissian:2002bg,Israel:2005fn}, and this fixes the relative location of the N/S poles of the $\bS^2$ along the $\phi$ circle at $\theta=\pi/2$ where the fivebranes are located.

Similarly, for the component described by $\bH_2$ branes, one must smear the brane along the gauge orbit parametrized by $\tau$ (the brane lies at $\theta=\pi/2$, where $\psi$ is trivial); there is again a worldvolume magnetic flux~\cite{Israel:2005ek,Fotopoulos:2004ut}  
(see Eq.\;\eqref{smearedH2flux}),
\begin{equation}
\label{F2onH2}
\omega_2 =\pm \frac{C_\sl\tanh\rho}{\sqrt{\cosh^2\rho
  -C_\sl^2}} \, d\rho\wedge d\sigma -  \sinh^2 \rho \, d\sigma \wedge
d\tau \, .
\end{equation}
While the flux for the unsmeared brane grows exponentially in $\rho$ and is not normalizable, for the smeared brane the flux is integrable; quantization of the integrated flux once again yields $C_\sl=\cos\mu$ with $\mu=2\pi \frac n{\nfive}$, $n\in\IZ$.
Thus the $\cF$ flux is again quantized for smeared $\bH_2$ branes, and leads to the same discrete locations for the fivebranes.  

The lift of a D1-brane in 9+1d is a 2+2 brane in 10+2d; the spatial sections of probe 2+2 branes that drop down from spatial infinity are topologically cylindrical in the 10+2 lift, consisting of a 1+1-dimensional ``spine'' given by an embedding of the D1 worldline into 10+2, that is then spun around the orbits of the spatial and temporal gauge groups.
When the D1-brane encounters a fivebrane downstairs in 9+1 dimensions, upstairs in 10+2 the spatial circle of the cylinder pinches off.  The D-brane can then undergo a topology change which allows the two sides of the pinch to separate in the longitudinal directions of the NS5.  For a straight-line D1-brane in the $x^1$-$x^2$ plane and at the origin in $x^3$-$x^4$, we saw this happen when the brane intersects the unit circle in the \xone-\xtwo\ plane.

But if the brane pinches off into distinct topological components, the total integer magnetic $\cF$ flux carried by the brane must partition into integer-quantized pieces; however, the $\bS^2$ and $\bH_2$ branes only carry integer flux for particular quantized values of the corresponding conjugacy classes in $\sutwo$ and $\sltwo$ respectively, and a general offset from the origin $b<1$ for the straight-line D1-brane won't correspond to one of these quantized values.  The way the D1-brane accommodates this is by blowing up the intersection of the two brane components at $\rho=\theta=0$, into a small ``plumbing fixture'' that allows flux to leak from one component to the other.  Equivalently, there is a small condensate of bifundamental strings with one end on the $\bH_2$ and one on the $\bS^2$, which compensates the induced Myers flux such that the total flux is integral.  When this condensate is sufficiently large, it results in the blowing up of the intersection into a geometrical plumbing fixture.  In either description, the fractionated component segments of the D1-brane are stitched back together into a single D1-brane that can move away from the discrete fivebrane locations.

\begin{figure}[ht]
\centering
    \includegraphics[width=0.395\textwidth]{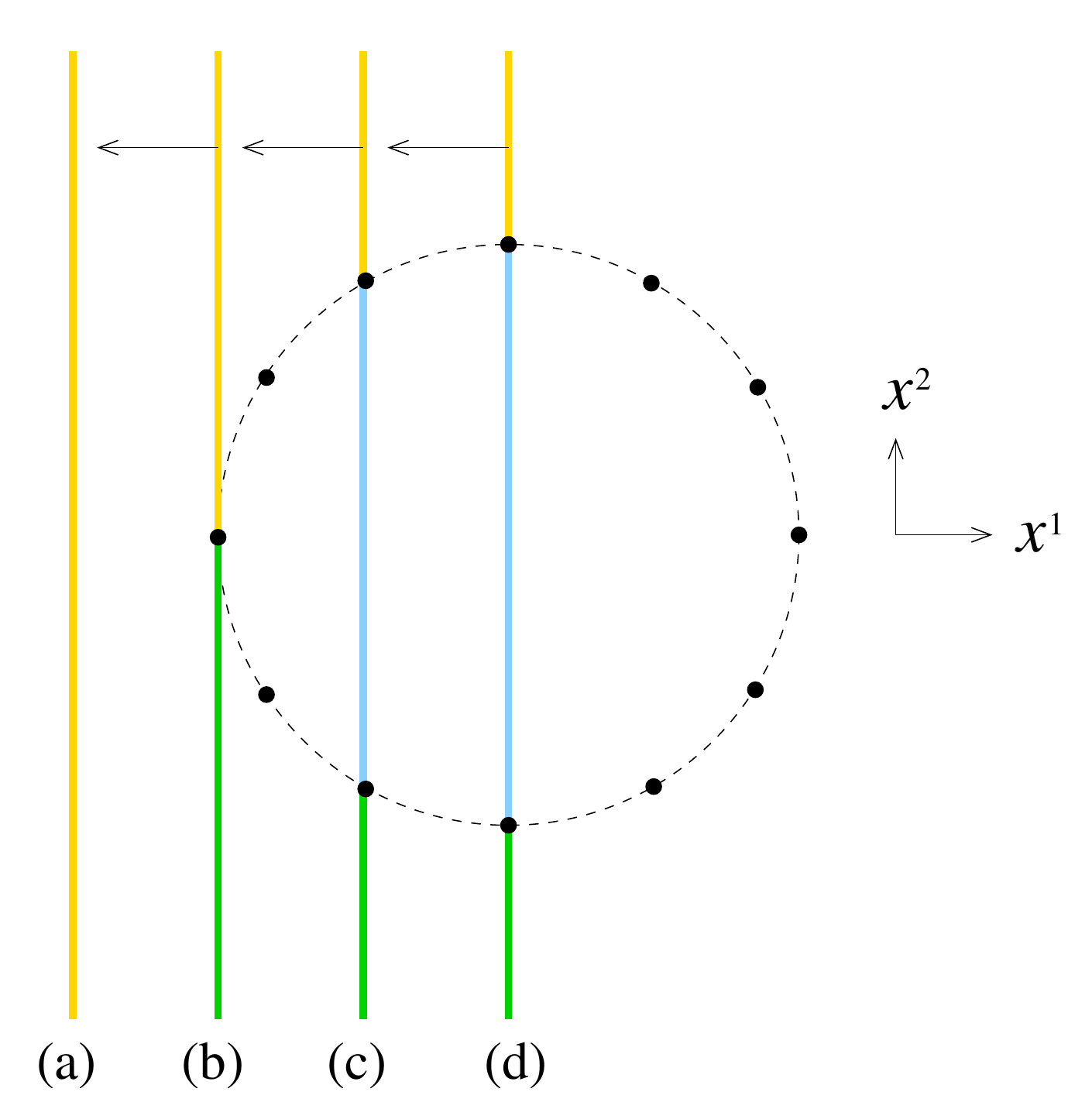}
\caption{\it 
A sequence of D1-branes probing the circular array of NS5-branes,
varying the displacement $x^1=b$ of branes lying in the $x^1$-$x^2$ plane.  The NS5's fractionate the D1 probe into segments indicated in yellow, blue, and green.
}
\label{fig:RingProbe}
\end{figure}

\begin{figure}[ht]
\centering
  \begin{subfigure}[b]{0.24\textwidth}
    \includegraphics[width=\textwidth]{dS2_brane.pdf}
    \caption{ }
    \label{fig:dS2_brane}
  \end{subfigure}
  \begin{subfigure}[b]{0.24\textwidth}
    \includegraphics[width=\textwidth]{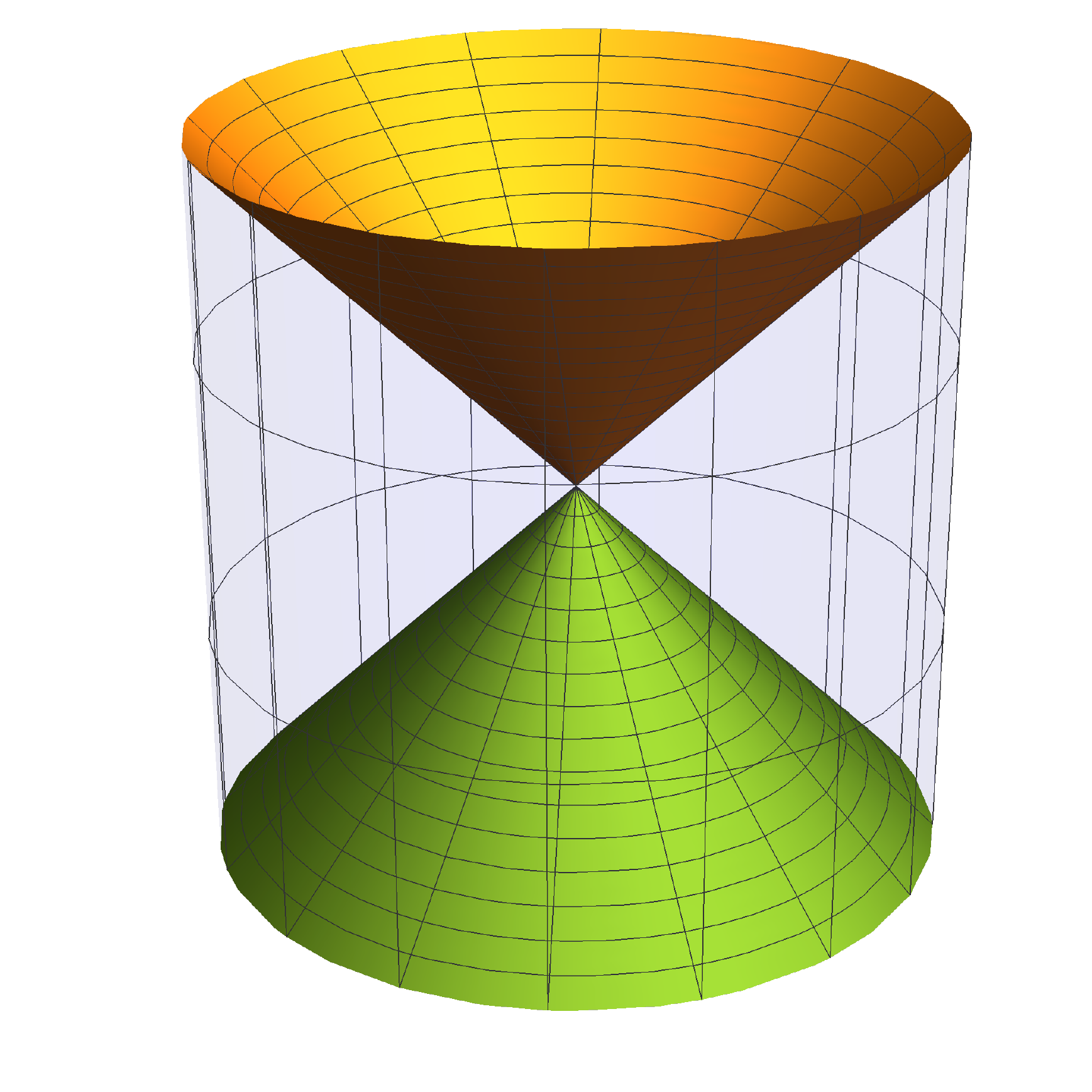}
    \caption{ }
    \label{fig:H2-S2_brane-3}
  \end{subfigure}
  \begin{subfigure}[b]{0.24\textwidth}
    \includegraphics[width=\textwidth]{H2-S2_brane-1.pdf}
    \caption{ }
    \label{fig:H2-S2_brane-1}
  \end{subfigure}
  \begin{subfigure}[b]{0.24\textwidth}
    \includegraphics[width=\textwidth]{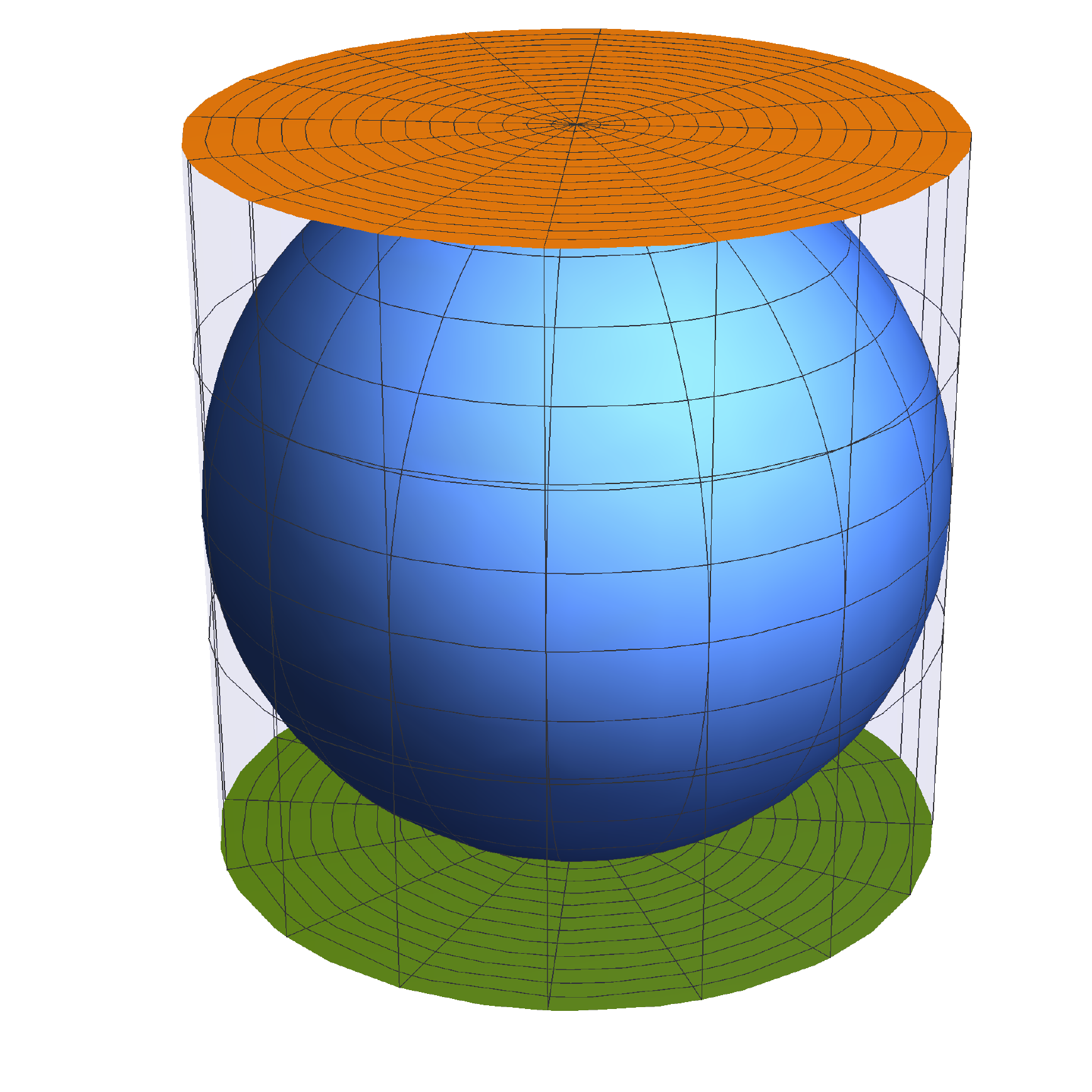}
    \caption{ }
    \label{fig:H2-S2_brane-2}
  \end{subfigure}
\caption{\it 
10+2d lift of D1-branes probing the circular array of NS5-branes.
(a) The $dS_2$ brane is the lift of a D1-brane passing outside the ring;
(b) Degeneration limit where the $dS_2$ brane splits into a pair of $\bH_2$ branes;
(c) $\bH_2$ brane segments are the lift of D1-branes that extend to infinity and end on NS5-branes (the latter being the 10+2d locations where the gauge group action degenerates), while an $\bS^2$ brane describes the D1 segment extending between NS5-branes;
(d) Maximum size $\bS^2$ segment extends between minimum $\bH_2$ branes.
}
\label{fig:RingProbe12d}
\end{figure}

An instructive thought experiment is to consider D1-branes in the $x^1$-$x^2$ plane at constant $x^1=b$, varying the impact parameter $b$ and considering the the brane as it sequentially passes across the fivebrane ring, see Figure~\ref{fig:RingProbe}.  Suppose $\nfive$ is a multiple of four, which simplifies the discussion; for instance in the figure, $\nfive=12$ and we can label the NS5-branes by the hours on a clock-face.  Put a D1-brane going along the vertical line through 12:00 and 6:00.  The lift of this brane to 10+2 is depicted in Figure~\ref{fig:H2-S2_brane-2}.
The part of the brane in the disk bounded by the source ring is an $\bS^2$-brane which is the largest size $\bS^2$ brane, that has $\nfive/2$ (\ie\ six) units of $\cF$ flux.  The parts of this D1-brane outside the source ring are $\bH_2$-branes that are flat in $AdS_3$; they do not bend up or down because they carry no $\cF$ flux.  Now start moving the D1-brane to the left, toward 9:00, keeping it vertical.  Every time one crosses an hour-point, \eg\ 11:00-7:00 (depicted in Figure~\ref{fig:H2-S2_brane-1}), or 10:00-8:00, the amount of flux on the $\bS^2$-brane component changes by two units; and the upper and lower $\bH_2$-branes each get a unit of $\cF$ flux, and as a result get more dimpled, \ie\ bend more towards one another.  The flux is transferred from the $\bS^2$-brane to the $\bH_2$-branes through the neck of the plumbing fixture described above, that opens up as the brane traverses the region between hour-points where the NS5's are located.  Locally at an NS5 pinch-point, the region near the origin $\rho=0$ of the $\bH_2$ brane intersects a north or south pole of the $\bS^2$-brane at $\theta=\pi/2$.  Condensing the strings with one end on the $\bS^2$-brane and the other end on the $\bH_2$-brane opens out a plumbing fixture that connects the two components, and allows $\cF$ flux to leak from one to the other (equivalently, these strings carry charge on both branes which transfers gauge flux from one to the other).

%%%%%%%%%%%%%%%%%%
%%%%%%%%%%%%%%%%%%

%%%%%%%%%%%%%%%%%%
%%%%%%%%%%%%%%%%%%

Eventually one gets to the 9:00 point (see Figure~\ref{fig:H2-S2_brane-3}), where the $\bS^2$-brane has become pointlike and carries no $\cF$ flux; the $\bH_2$-branes now touch at the center $\rho=0$ of $AdS_3$, and the two sides of the pinch each contain $\nfive/4$ (\ie\ three) units of $\cF$ flux.  If we continue to increase the impact parameter, the two $\bH_2$ branes join together into a single $dS_2$ brane as shown in Figure~\ref{fig:dS2_brane}.

If we instead move the branes to the right of the 12:00-6:00 line, the signs reverse, so that by the time one gets to the 3:00 point where one again has a pair of extreme $\bH_2$ branes in $AdS_3$ touching at their tips, each carrying $-\nfive/4$ units of flux, but also there is the point-brane in $\bS^3$ that carries $+\nfive$ units of $\cF$ flux when it is at the opposite pole to the one that carries zero flux, so that once again the total $\cF$ flux is $\nfive/2$.%
\footnote{Note that the flux is only defined modulo $\nfive$ due to the effects of large gauge transformations of the NS $B$-field~\cite{Maldacena:2001ss}.}
So there is a consistent picture where the lifted D1-brane carries a fixed amount of $\cF$ flux, and when the brane encounters an NS5, the cylindrical D2+2 brane in 10+2 pinches off in such a way that an integral amount of $\cF$ flux is carried by each side of the pinch.  In the in-between spaces where the D1-brane is crossing the NS5 source ring but is not quite intersecting the NS5's, the brane never quite reaches the ring $\rho=\theta=0$, instead opting to avoid this locus by blowing up the brane via a plumbing fixture that allows either side to carry non-integer $\cF$ flux.

%%%%%%%%%%%%%%%%%%%%%%%%%%%%%%%%%%%%
\paragraph{Aside on Page charges:}

One can ask, what is this quantized charge carried by the branes upstairs in 10+2 dimensions?  It is modeled after the induced D0-brane charge carried by D2-branes in the $\sutwo$ WZW model.  The situation has been studied in%
~\cite{Alekseev:2000jx,Marolf:2000cb,Figueroa-OFarrill:2000lcd};
the gauge invariant generalization of $\int \cF$ over the D-brane worldvolume is the {\it Page charge}
\be
\label{pagecharge}
\int_{\cB} H - \int_{\cD} (B+\cF)
\ee
where $\cB$ is a three-manifold whose boundary is the worldvolume $\cD$ of the D2-brane.  
The Page charge is invariant under small gauge transformations of $B$ and~$\cF$.  However, under large gauge transformations it shifts by a multiple of the level~$\nfive$.  One can see this from its definition~\eqref{pagecharge}; there are two choices for the three-manifold $\cB$ whose boundary is the D-brane worldvolume $\cD$, whose difference is $\bS^3$, and thus yield values for the Page charge differing by $\int_{\bS^3}H = 4\pi^2\nfive$.

%%%%%%%%%%%%%%%%%%%%%%%%%%%%%%%%%%%%
%%%%%%%%%%%%%%%%%%%%%%%%%%%%%%%%%%%%
\section{D-branes on the round supertube}
\label{sec:NS5P-NS5F1}

Having described the D-branes that stretch between static NS5-branes separated on their Coulomb branch using the formalism of null gauging, we are now ready to generalize the discussion to the modified null gauging that leads to two-charge NS5-P and NS5-F1 supertubes.  We begin with the former, as the geometrical structure is somewhat more intuitive; and then describe the effects of the T-duality that takes us to the NS5-F1 background.

%%%%%%%%%%%%%%%%%%%%%%%%%%%%%%%%%%%%
\subsection{NS5-P supertube}

We now include the factors $\bR_t \times \bS^1_\ytil$ in our considerations.  The gauge group acts via
\begin{align}
&\ell(\zeta,\eta)  \bigl( g_\sl, g_\su , e^{it}, e^{i\ytil} \bigr)   r(-\zeta,-\eta) 
\\
&\hskip .5cm = 
\bigl( 
e^{i\alpha{\sigma_3}} g_\sl e^{i\beta{\sigma_3}} \, , \,
e^{i\alpha{\sigma_3}} g_\su e^{-i\beta{\sigma_3}}  \, , \, 
e^{-i\alphab\alpha}e^{it}  e^{-i\alphab\beta}  \, , \,
e^{ -i\alphab\alpha}e^{i\ytil}  e^{ -i\alphab\beta} 
 \bigr) ~,
\nn
\end{align}
with $\alpha= \hf(\zeta+\eta)$, $\beta= \hf(\zeta-\eta)$, and $\alphab=k/\Rytil$.
As before, we define the group automorphisms
\be
\Omega_\cG \bigl( g_\sl, g_\su , e^{it}, e^{i\ytil} \bigr) = 
\bigl( \Omega^{\eps_\sl}(g_\sl) , \Omega^{\eps_\su}(g_\su) , \Omega^{\eps_t}( e^{it} ), \Omega^{\eps_\ytil}(e^{i\ytil}) \bigr)
\ee
and recall the definition of $\Omega_\cH$ of equation~\eqref{conjclasses}; once again, we seek factorized branes of the form $\cC_\cG\cdot\cC_\cH$, where the automorphisms that define the choices of twisted conjugacy classes $\cC_\cG$ and $\cC_\cH$ are related by the embedding constraint~\eqref{EmbeddingCondition}.

We are primarily interested in W-branes stretching between strands of the NS5 helix.  Thus we adapt the results of the Coulomb branch analysis of the previous section, and take
\be
\label{NS5Pepsilons}
\eps = -
~~,~~~~
\eps_\sl= +
~~,~~~~
\eps_\su= -
~~,~~~~
\eps_t = \eps_\ytil = +
\ee
so that the embedding constraint~\eqref{EmbeddingCondition} is satisfied.  The starting conjugacy class is thus an $\bS^2$ brane whose poles are anchored on the fivebrane ring at $\rho=0,\theta=\pi/2$, and pointlike in all the other group factors.  This brane is then smeared along $\cC_\cH$, which because $\eps=-$ is nontrivial in the timelike direction only.

Previously, the gauge group didn't act in the physical time direction $\bR_t$ of static fivebranes; we implicitly took the brane to be extended along this direction and didn't need to look further.  Now the embedding constraint has forced the starting brane locus $\cC_\cG$ to be pointlike in $\bR_t$, so we must explicitly smear it in the physical time direction transverse to the timelike gauge orbits.  The formalism is easily adapted to accomplish this task -- we simply enlarge the group $\cH$ to $\cH'=\cH\times\uone_\cK$ where $\uone_\cK$ is generated by timelike currents $\cK$, $\bar\cK$ satisfying
\be
\label{KJcorrelator}
\bigl\langle \cJ\, \cK \bigr\rangle = 0
~~,~~~~
\bigl\langle \bar\cJ\, \bar \cK \bigr\rangle = 0  
~~,~~~~
\bigl\langle \cK\, \cK \bigr\rangle = \bigl\langle \bar\cK\, \bar \cK \bigr\rangle
~~.
\ee
These conditions ensure that the starting conjugacy class $\cC_\cG$ is consistently smeared along $\uone_\cK$ as well as $\cH$; effectively, these are the conditions that we be able to gauge $\uone_\cK$ in addition to $\cH$ and so the brane is consistently extended along all of $\cH'$.  Note that we are not saying that we will gauge $\uone_\cK$ (we will not); we are simply saying that we could if we wanted to, and that is sufficient for our purposes.
Let
\begin{align}
\label{Kdefn}
\cK &= k_1 J_3^\sl + k_2 J_3^\su + k_3 \partial t + k_4 \partial \ytil
\nn\\
\bar \cK &= \bar k_1 \bar J_3^\sl + \bar k_2 \bar J_3^\su + \bar k_3 \bar \partial t + \bar k_4 \bar \partial \ytil  ~;
\end{align}
the condition for anomaly freedom amounts to
\be
\nfive(-k_1+k_2)+\alphab(k_3+k_4) = 0
~~,~~~~
\nfive(-\bar k_1 - \bar k_2) + \alphab(\bar k_3+\bar k_4) = 0  ~~,
\ee
(recall $\alphab=k/\Rytil$)
which we satisfy by setting
\begin{align}
\label{kchoice}
k_a &= \mu(\omhat,+\vv\omhat,\Rytil,-\vv\Rytil) 
\\
\bar{ k}_a &= \mu(\omhat,-\vv\omhat,\Rytil,-\vv\Rytil) 
~~,~~~~ \omhat \equiv k/\nfive  
\nn
\end{align}
where $\mu$ is a normalization.
The parameter $\vv$ describes the motion of the probe brane along the fivebrane strands it is attached to, a spiral motion simultaneously along $\phi$ and $\ytil$; note that the physical requirement that $\cK$ is timelike restricts $\vv^2<1$.  
Note that we could have also included in $\cK,\bar \cK$ a linear combination of the currents generating translations on the $\bT^4$ compactification which the fivebranes wrap~-- the moduli space of the W-particle is the entire fivebrane worldvolume, which is the tensor product of the supertube spiral with this $\bT^4$~-- but we have omitted it for simplicity.
Smearing along $\uone_\cK$ is accomplished by taking $\Omega_\cK=\Omega^-$.  The brane locus upstairs
\be
\label{NS5P conj class prod}
\cC_\cG\cdot \cC_{\cH'}
\ee
is thus 2+2 dimensional -- two spatial dimensions from $\cC_\cG$ and two timelike directions from $\cC_{\cH'}$.

To see that we have arrived at a physically sensible brane, consider a section of the supertube geometry~\eqref{smearedNS5Pmetric} defined by $\rho=0$, corresponding to setting $x_3 = x_4 = 0$. In this limit the B-field reduces to a constant and the metric simplifies to:
\begin{align}
ds^2 & = -du dv + n_5\Big[ d\theta^2 +\tan^2\theta \Bigl(d\phi + \frac{\omhat}{\Rytil} d v\Bigr)^2 + \frac{\omhat^2}{\Rytil^2} dv^2  \Big]
~ ,\nn \\
e^{-2\Phi} & = \frac{a^2\cos^2\theta}{\gstr^2n_5^2} ~.
\end{align}
Setting $z = \sin \theta$  and defining
\begin{equation}
   u' = u - \frac{n_5 \omhat^2}{\Rytil^2} v 
   ~~,~~~~ 
   v' = v 
   ~~,~~~~
   \phi' = \phi + \frac{\omhat}{\Rytil} v ~,
\end{equation}
we can write the metric as
  \begin{align}
ds^ 2 &=-d u' d v'+ \frac{n_5}{1-z^2}\left( dz^2 +
  z^2  (d\phi') ^2\right) \, , \nn\\
  e^{-2\Phi} & \sim 1-z^2 \, .
\end{align}
This geometry is $\bR^{1,1}$ times a parafermion disk (the geometry of the $\sutwo/\uone$ gauged WZW model%
~\cite{Witten:1991yr,Kiritsis:1991zt,Dijkgraaf:1991ba}
parametrized by $(z, \phi')$.  
Because once again the metric and dilaton are controlled by the same
warp factor, D-branes are straight lines in the $(z, \phi')$
plane~\cite{Maldacena:2001ky} at constant $\ytil'=\hf(u'-v')$.  Effectively, these coordinate transformations map the $\rho=0$ section of the geometry back to that of NS5-branes on the Coulomb branch, and so we can use results from the previous section to describe probe D1-branes stretching between the fivebrane strands of the supertube.  We can also slightly generalize by giving the probe brane a constant velocity $\vv'$ in the $\ytil'$ direction.

We thus take the probe brane worldvolume downstairs to be parametrized by
\begin{align}
\label{downparam}
u' &= e^{-\gamma} \,\xi''_0
\nn\\
v' &=  e^{+\gamma}\,\xi''_0
\nn\\
\phi'& = \xi_1
\\
\theta &= \theta(\xi_1)
\nn
\end{align}
where $t' = \hf(v'+u')$ and $\theta(\xi_1)$ is given by the solution of the embedding equation
\begin{equation}
\label{thetaxi1coulomb}
\sin \theta \cos (\xi_1) = C \, .
\end{equation}
Mapping this embedding back to the original unprimed variables yields
\begin{align}
t &= \Bigl(\cosh\gamma + \frac{\nfive\omhat^2}{2\Rytil^2}e^{\gamma}\Bigr)\xi''_0
\nn\\
\ytil &= -\Bigl(\sinh\gamma - \frac{\nfive\omhat^2}{2\Rytil^2}e^{\gamma}\Bigr)\xi''_0
\\
\phi& = \xi_1-\frac{\omhat}{\Rytil}\,e^{\gamma}\,\xi''_0
\nn
\end{align}
Note that regardless of its boost $\gamma$, the brane worldvolume always lives along~\cite{Martinec:2017ztd} 
\be
\label{helixpitch}
\frac{d\phi}{d(v/\Rytil)} = - \omhat = -\frac{k}{\nfive} ~.
\ee
We would like a brane in 10+2 dimensions whose gauging yields the above brane in 9+1d physical spacetime.  The fact that the spatial sections look like the $\sutwo/\uone$ projection of a symmetry preserving $\bS^2$ brane in $\sutwo$ suggests that we start with this brane and smear it to get something that is both gauge invariant and projects onto the above brane in 9+1d.  Indeed, let us set
\begin{align}
\label{STbraneupstairs}
\tau &= -\xi_3
\nn\\
t &=  (1 - \vv' )\xi'_0 + \frac{k}{\Rytil} \xi_3
\nn\\
\tilde y &=   -\vv' \xi'_0 + \frac{k}{\Rytil} \xi_3
\\[5pt]
\phi &= \xi_1 - \frac{\omhat}{\Rytil} \xi'_0  - \xi_3
\nn\\
\psi &= \xi_2
\nn
\end{align}
where $\xi_2\propto \eta$ parametrizes the spatial gauge motion~\eqref{STgaugetransfs}, $\xi_3\propto \zeta$ parametrizes the temporal gauge motion, and $\theta(\xi_1)$ solves~\eqref{thetaxi1coulomb}.  Comparing to~\eqref{downparam}, one has
\be
\label{vsoln}
\xi'_0 = e^{\gamma}\xi''_0  
~~,~~~~
\vv' = e^{-\gamma}\sinh\gamma - \frac{\nfive\omhat^2}{2\Rytil^2}  ~.
\ee
Note that for $\nfive\omhat^2>\Rytil^2$, one must have $\vv'\ne0$.

Note that the spatial $\sltwo$ Euler angle $\sigma$ is redundant at $\rho=0$ so we may omit it, and so the azimuthal direction $\psi$ of the $\bS^2$ brane lies along the spatial gauge orbit~-- the starting point is already invariant under the axial gauge group and needs no further smearing spatially.

We see that $\xi_1,\xi_2$ parametrize our starting $\bS^2$ brane.  Both the temporal gauge motion parametrized by $\xi_3$ and the physical timelike direction parametrized by $\xi'_0$ are $\uone$ isometries of $\cG$; the brane worldvolume is thus a product of conjugacy classes
\be
\cC_{\su}(\xi_1,\xi_2) \cdot \cC_{\uo\times\uo}(\xi'_0,\xi_3) 
\ee
as advertised (suppressing trivial pointlike conjugacy classes of factors in $\cG$ apart from $\sutwo$).  In order to match with our choice~\eqref{Kdefn}-\eqref{NS5P conj class prod}, we note that we have the freedom to shift $\cK$ by an arbitrary amount of $\cJ$, and similarly for $\bar\cK$; we can use this freedom to set $k_1=\bar k_1=0$ which was implicitly chosen in~\eqref{STbraneupstairs}.   The current $\cK'$ corresponding to this choice is related to the current $\cK$ of~\eqref{kchoice} via
\be
\cK' = - \frac{\cK/\mu - \omhat \cJ}{\Rytil(1-\vv)}
~~,~~~~
\vv' = - \frac{\nfive\omhat^2+\Rytil^2\vv}{\Rytil^2(1-\vv)}
\ee
so that the brane parametrization becomes
\begin{align}
\label{STbraneupstairs alt param}
\tau &= \mu\omhat \xi_0 -\xi_3
\nn\\
t &=  \mu\Rytil\,\xi_0  + \frac{k}{\Rytil} \xi_3
\nn\\
\tilde y &=  \mu\vv\Rytil\, \xi_0 + \frac{k}{\Rytil} \xi_3
\\
\phi &= \xi_1 + \mu\vv{\omhat}\xi_0 - \xi_3
\nn\\[5pt]
\psi &= \xi_2 ~~.
\nn
\end{align}

Thus with a simple modification of the setup describing D-branes stretching between fivebranes on the Coulomb branch, we can describe D-branes stretching between fivebrane strands of the NS5-P supertube.  One is free to slide the endpoints of the probe brane along the fivebranes, and so the D-brane has an $\bS^1$ moduli space that is the $k$-fold cover of the $\phi$ circle, \ie\ the supertube source ring.  Motion along this moduli space is specified by the velocity parameter $\vv$ in the current $\cK$ that generates translations in physical time, which as we have mentioned can be generalized to include motions along $\bT^4$ as well as along the supertube spiral.

%%%%%%%%%%%%%%%%%%%%%%%%%%%%%%%%%%%%
\subsection{NS5-F1 frame}

The NS5-F1 supertube is obtained by T-dualizing along the $\tilde y$
direction; we let $y$ parametrize the T-dual circle.  
In 9+1d, the T-dual of the D1-brane stretching between NS5-brane strands is a D2-brane wrapping a two-cycle created by slightly resolved KK monopoles.
In the gauged 10+2d worldsheet theory, T-duality amounts to simply flipping axial to vector gauging of this circle.
Following through the same steps as the previous section, one finds the same choices~\eqref{NS5Pepsilons} are required to describe W-branes wrapped at the cap of the geometry, except for a flip of $\eps_\ytil=+$ to $\eps_y=-$; in other words, the brane is wrapped around the $y$ circle since it was pointlike on the $\ytil$ circle.
A D1+1 brane in 9+1d lifts to a D2+2 brane upstairs in 10+2d; T-duality transforms this to a D3+2 brane.  If the brane upstairs had Dirichlet boundary conditions in $\ytil$, it now has Neumann boundary conditions in the coordinate $y$ parametrizing the T-dual circle; the location $\ytil_0$ of the brane in $\bS^1_\ytil$ arises as the value of the Wilson line of the gauge field $A_y$ on the brane,
\be
\ytil_0 = \oint dy \, A_y  ~.
\ee
The motion along the brane moduli space now consists of moving along $\phi$ while continuously changing the value of the Wilson line in the proportion $\omhat$.

\begin{figure}[ht]
\centering
    \includegraphics[width=0.7\textwidth]{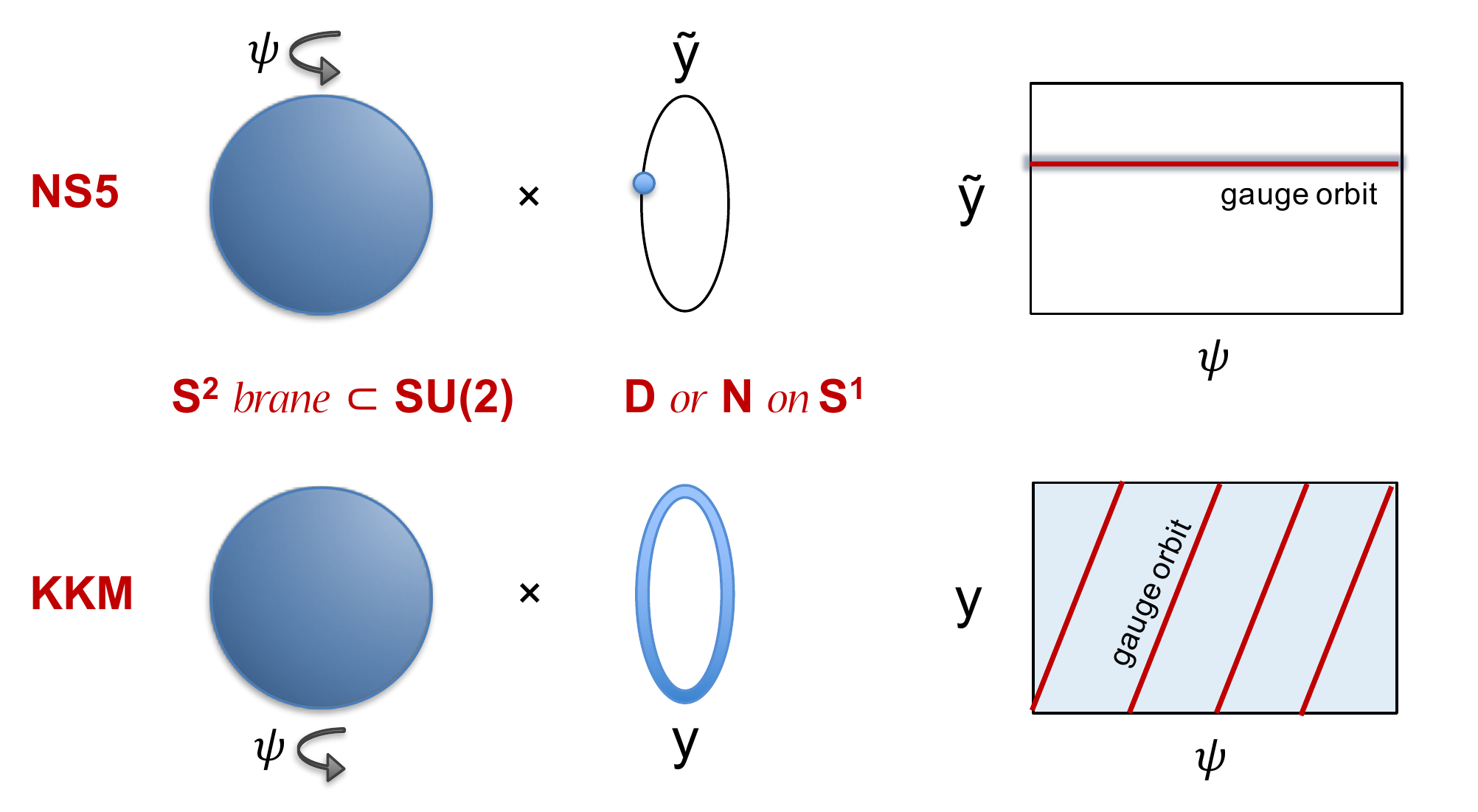}
\caption{\it 
T-dual pictures of the W-brane.  In the NS5-P frame in type IIB (top), D1-branes stretching between NS5's lift to $\bS^2$ branes smeared along the temporal gauge orbit and are pointlike on the $\ytil$ circle.  In the type IIA NS5-F1 frame (bottom), the NS5 source becomes a coiled KK monopole loop; the W-brane is now a D2 wrapping a vanishing cycle of coincident KKM's, and lifts to an $\bS^2\times\bS^1$ brane extended along the T-dual $y$ circle (and again smeared along gauge orbits).  On the right, the spatial gauge orbits on the respective $\ytil$-$\psi$ and $y$-$\psi$ tori are depicted to show how identification along gauge orbits upstairs recovers the picture of stretched/wrapped W-branes downstairs.
}
\label{fig:NS5-vs-KKM}
\end{figure}

The effect of gauge transformations~\eqref{STgaugetransfs} now shows that both $\psi$ and $y$ transform under spatial (axial) gauge transformations.  
The brane fills both of these directions, and the spatial gauge orbits are oblique lines in the $(\psi, y)$ torus, with slope determined by $k$, see Figure~\ref{fig:NS5-vs-KKM}.

%%%%%%%%%%%%%%%%%%
%%%%%%%%%%%%%%%%%%

The brane worldvolume (and Wilson line $\ytil$) can be parametrized as
\begin{align}
\tau &= \mu\omhat \xi_0-\xi_3
\nn\\
t &=  \mu\Rytil\, \xi_0 +\frac{k}{\Rytil} \xi_3
\nn\\[-3pt]
y &= \xi_4
\\[3pt]
\phi &= \xi_1 + \mu\vv\omhat \xi_0 - \xi_3
\nn\\[3pt]
\psi &= \xi_2   
\nn\\
\ytil &= \mu\vv\Rytil\, \xi_0 + \frac{k}{\Rytil} \xi_3 ~~,
\nn
\end{align}
with again $\theta(\xi_1)$ solving~\eqref{thetaxi1coulomb}. 
In other words, we start with a brane worldvolume $\cC_\cG$ that is a spatial $\bS^2$ brane in $\sutwo$ (parametrized by $\xi_1,\xi_2$) that is also extended along $\bS^1_y$ (parametrized by $\xi_4$), and smear it along the two time directions -- the physical time direction parametrized by $\xi_0$ in the same way as before, and the timelike gauge orbit parametrized by $\xi_3$ specified in~\eqref{STgaugetransfs}.  The starting brane $\cC_\cG$ is already invariant under spatial gauge transformations, which shift $\psi,y$ in the proportion
\be
\delta(y/\Ry) = -k \delta\psi~~.
\ee

The D3+2 brane upstairs has a spatial worldvolume $\bS^2\times\bS^1$, and from the figure we see that because the spatial gauge orbits involve motion along $y$, they never degenerate.  If as in Section~\ref{sec:two-charge supertubes} we fix $y$ as a gauge choice, we remove the $\bS^1$ factor, and the physical spatial D-brane worldvolume is topologically an $\bS^2$.  Note that as described there, gauge fixing $y$ leaves a residual discrete $\bZ_k$ gauge identification, but this does not change the topology of the brane -- $\bS^2/\bZ_k$ is still topologically a two-sphere.  This identification will, however, reduce the action of the brane by a factor $k$ -- a result similar to the action cost of fractional branes on orbifolds~\cite{Douglas:1996xg,Douglas:1997de}.

Thus the T-duality between NS5 branes and KK monopoles is reflected in the structure of their W-branes.  A D1-brane stretching between nearly coincident NS5-branes in type IIB becomes a D2-brane wrapping a vanishing (\ie\ stringy) cycle of coincident KK monopoles in type IIA.  The lift to 10+2 dimensions makes the relation quite transparent.

%%%%%%%%%%%%%%%%%%%%%%%%%%%%%%%%%%%%
\subsection{W-strings}

We have described the 10+2d lift of D1-branes stretching between fivebrane strands in the NS5-P supertube or D2-branes wrapping KK monopole topology in the NS5-F1 supertube; these objects are W-particles.  To get W-strings, we still need to extend the D-brane worldvolume along the fivebranes (in a direction which is {\it not} the spatial gauge orbit direction, rather it is a physical direction transverse to that).   This smearing along the fivebranes in the NS5-P frame gives us a D2+1 brane strip stretching between NS5's downstairs, lifting to a D3+2 brane upstairs; and in the NS5-F1 frame, a D3+1 brane wrapping the $\bS^2\times \bS^1$ topology in the cap lifts to a D4+2 brane upstairs.

This smearing can either be along $\bT^4$, or along the supertube spiral; in the latter case the brane worldvolume is a $k$-fold cover of the supertube source circle in the $x^1$-$x^2$ plane, in a harbinger of the long string structure expected in the black hole phase, see Figure~\ref{fig:WbraneFig}.
The extra spatial dimension of this coiled W-string lies along the moduli space of the above W-particle -- the spiral along the fivebranes with pitch $\nfive/k$ along $\ytil,\phi$.  We thus have an extra spatial brane direction parametrized by motion generated by
\begin{align}
\label{Ksupertube}
\cL &= \kappa_0\bigl[k J_3^\su - \nfive\, \partial (\ytil/\Rytil)\bigr] + \kappa_i\, \partial (x^i/R_i)
\nn\\
\bar\cL &= \kappa_0\bigl[k \bar J_3^\su - \nfive\, \bar\partial (\ytil/\Rytil)\bigr] + \kappa_i\, \bar\partial (x^i/R_i) ~~.
\end{align}
where $\kappa_0,\kappa_i\in\bZ$.
Here we have restored a possible contribution that gives the W-brane winding along the $\bT^4$ compactification parametrized by $x^i$, $i=6,7,8,9$.

To implement the smearing, we add $\cL,\bar\cL$ to the group $\cH$ along the lines of Eqs.\;\eqref{KJcorrelator}--\eqref{NS5P conj class prod}, so that the brane worldvolume is extended along the $U(1)$ orbits generated by $\cL,\bar\cL$.
For the NS5-P frame the brane is pointlike along $\ytil$ in $\cC_\cG$ but then gets smeared along its moduli space, which is a correlated motion in $\ytil$ and $\phi$; so now we have a brane extended along the $(k,-\nfive)$ cycle of the $\ytil$-$\phi$ torus.  T-duality converts this brane filling one dimension of this two-torus to one that fills the entire two-torus parametrized by $y$ and $\phi$, which now carries a flux $\cF$ determined by the data $(\nfive,k)$
(see for instance~\cite{Hashimoto:1997gm}).%
\footnote{Counting dimensions upstairs in the NS5-F1 frame, before smearing the brane is locally a spatial $\bS^2_{\vartheta,\psi}\times \bS^1_y$, where $\vartheta$ is related to $\theta,\phi$ via the analogue of~\eqref{Euler-hyperspherical relation} for twisted conjugacy classes:
$$ \sin\theta \cos\phi = \sin\mu  ~~,~~~~  \sin\theta\sin\phi = \cos\vartheta\cos\mu ~.  $$
The smearing along the moduli space fills the fourth spatial dimension of $\sutwo\times\bS^1_y$ in the region $\theta>\mu$.  Gauging then gets us down to 9+1d with a brane having three spatial dimensions, which comprise the topological cycle $\bS^2\times\bS^1$ at the tip of the geometry.}

%%%%%%%%%%%%%%%%%%%%%%%%%%%%%%%%%%%%
\subsection{The DBI effective action}

The effective action on D-branes at leading order in the derivative expansion is the DBI effective action
\be
\cS_{DBI} = \int e^{-\Phi}\sqrt{\det(G+B+\cF)}~.
\ee
We would like an expression for this effective action for the above D-branes, in terms of the branes upstairs in $\cG$.  There are a few wrinkles to straighten out.  First of all, as mentioned at the end of Section~\ref{sec:coulbranch}, null gauging by its very nature implies that the matrix $M=G\tight+B\tight+\cF$ evaluated on the brane worldvolume upstairs in $\cG$ has both a kernel and cokernel due to the null isometries of the background, and therefore its determinant vanishes.  Our prescription for computing the physical DBI determinant in the quotient theory on $\cG/\cH$ is to evaluate the DBI action slightly off-shell, and extract the coefficient of the vanishing as one takes the fields on-shell.  This coefficient is the determinant $\cE$ of the minor in the space transverse to the null isometries.
Since $M$ is not a Hermitian matrix, we consider instead
\be
\label{minorkey}
\cE^2\equiv\frac{d}{d\lambda} \det\bigl(M^\dagger M-\lambda\One \bigr)\Bigl|_{\lambda=0}
~~,~~~~
M=G+B+\cF ~~.
\ee
as a basis-independent definition of the determinant $\cE$.  Our prescription for the DBI action is then to consider
\be
\cS_{DBI} = \mub_0\int \sqrt{\cE} ~,
\ee
where $\mub_0$ is an overall constant.

Evaluating the induced metric and the two-form $\omega_2=B+\cF$ from
equation~\eqref{SBomega} in the brane
parametrization~\eqref{STbraneupstairs alt param}, we find
\begin{align}
\omega_2 &= \pm \frac{n_5 C \cot \theta}{\sqrt{\sin^2\theta - C^2}}
           \, d\theta \wedge d\psi + n_5 \cos^2 \theta\, d\phi \wedge d\psi  \nn \\
  & =
\cos^2\theta\,  (\mu k \vv \, d\xi_0 \wedge d\xi_2 + n_5 \, d\xi_2\wedge
d\xi_3) - n_5\sin^2\theta \, d\xi_1 \wedge d\xi_2
~,
\end{align}
where we used the embedding equation \eqref{thetaxi1coulomb}. The matrix $M$ evaluates to
\be
M = 
\left(\begin{matrix} 
\mu^2\Bigl(-\coeff{k^2(1-\vv^2\sin^2\theta)}{\nfive}-\coeff{1-\vv^2}{\Rytil^2}\Bigr) & ~\mu k\vv\sin^2\theta & ~\mu k\vv\cos^2\theta & ~\mu k\vv\cos^2\theta \\
\mu k\vv\sin^2\theta & \nfive(\sin^2\theta+\dot\theta^2) & ~-\nfive\sin^2\theta & ~-\nfive\sin^2\theta \\
-\mu k\vv\cos^2\theta & \nfive\sin^2\theta & ~\nfive\cos^2\theta & ~\nfive\cos^2\theta \\
\mu k\vv\cos^2\theta & -\nfive\sin^2\theta & ~-\nfive\cos^2\theta & ~-\nfive\cos^2\theta 
\end{matrix} \right) ~~.
\ee
The kernel of $M$ is then spanned by the direction parametrized by $\xi_3 - \xi_2$, while the cokernel is spanned by the direction parametrized by $\xi_3 + \xi_2$; these are of course the left and right gauge directions on the brane, as expected. 
Extracting the effective action from the linear term in the characteristic polynomial~\eqref{minorkey} we find
\begin{equation}
\label{NS5P DBI}
\mathcal{L} \sim  \sqrt{ n_5
(1-\vv^2)(\sin^2\theta +\dot{\theta}^2\cos^2\theta)} ~ ,
\end{equation}
using the normalization $\mu = (k^2 + n_5\Rytil^2)^{-1/2}$ for the physical time $\xi_0$ in~\eqref{STbraneupstairs alt param}.
The embedding equation \eqref{thetaxi1coulomb} solves the equations of motion derived from the above effective action; this part of the action is identical to that for D1-branes in $\sutwo$, equation~\eqref{pfdiskDBI} (up to the map $\theta\to\pi/2 - \theta$ appropriate for the twisted conjugacy class used here).

Note that the quantity $\cE$ already incorporates the spatial variation of the dilaton, \ie\ the factor $\Sigma_0$ in equation~\eqref{smearedNS5Pmetric}; the overall constant $\mub_0^{\,2}=n_p/(\nfive k^2V_4)$ coming from the dilaton is determined by physical considerations.  The dilaton in gauged WZW models arises from the coefficient of the term quadratic in gauge fields (see for example~\cite{Tseytlin:1992ri}).%
\footnote{The transformation of the dilaton under T-duality similarly picks up a factor of the volume of the torus being dualized, because T-duality can be realized in terms of gauging of U(1) isometries~\cite{Buscher:1987qj}.}
The spatial dependence $e^{-2\Phi}\propto\Sigma$ is the same as the varying size $\cV_\eta$ of the spatial gauge orbits in $\cG$, because
\be
\cV_\eta^2=\Bigl\langle \bigl(\ell(X_1)\tight-r(X_2)\bigr),\bigl(\ell(X_1)\tight-r(X_2)\bigr) \Bigr\rangle = 2\bigl\langle \ell(X_1),r(X_2) \bigr\rangle
\ee
is indeed the coefficient of the term quadratic in the gauge potentials in the null-gauged action, which yields the dilaton.  On the other hand, the contribution to the effective action~\eqref{NS5P DBI} from the $\xi^2$-$\xi^3$ directions is identical to that for twisted $\bS^2$ branes in $\sutwo$, which already incorporates the effects of the varying dilaton.  Gauge invariance requires that the brane upstairs lies along the orbits of $\cG$, and so its induced volume element in the DBI action will be proportional to the volume of the gauge orbits, times the volume of the brane worldvolume on the coset; the first factor is the dilaton, and the second factor is the DBI induced volume element.

%\newpage
%%%%%%%%%%%%%%%%%%%%%%%%%%%%%%%%%%%%
\section{Three charge supertubes}
\label{sec:specflowST}

We now turn to the three charge supertubes~\eqref{GLMTmetric}-\eqref{GLMTparams} described as a null gauged WZW model in Section~\ref{sec:GLMTnullgauging}.  The geometry has topology in the cap consisting of $\bZ_{\ellone}$ and $\bZ_{\elltwo}$ orbifold singularities at $\theta=0,\pi/2$, and so we expect to find D2-branes wrapping the orbifold vanishing cycles in a manner similar to the D2-branes in the NS5-F1 geometry, which lifted to $\bS^2\times\bS^1$ branes upstairs in $\cG$.  

%%%%%%%%%%%%%%%%%%%%%%%%%%%%%%%%%%%%
\subsection{Finding factorized branes}
\label{sec:GLMTfactorbranes}

The constraint~\eqref{EmbeddingCondition} can no longer be satisfied by embedding the gauge group $\cH$ into $\cG$ in a single step; in fact, one cannot generally satisfy this condition at all, even with a more general embedding chain of the form~\eqref{gen embed}.  We have, however, been able to find solutions with a two-step embedding chain, for particular choices of the circle radius $\Ry$.

The construction of D-branes in $\cG$ respecting the asymmetric gauge action~\eqref{GLMTcurrents} is similar to the construction of branes in the $T^{p,q}$ spaces $\bigl(\sutwo_{k_1}\tight\times\sutwo_{k_2}\bigr)/U(1)$ treated in%
~\cite{Sarkissian:2002nq,Quella:2002fk,Quella:2003kd}.  In that example, one is gauging the asymmetric $\uone$ embedding
\be
(g_1,g_2) \to \bigl(g_1 e^{ip\eta{\sigma_3}} , e^{iq\eta{\sigma_3}} g_2 \bigr)  ~,
\ee
which is non-anomalous if $k_1p^2=k_2q^2$.  The method for constructing branes adds a further stage in the embedding chain,
\be 
\Bigl(\cU_2=\uone\Bigr)\underset{\textstyle\vareps_2}{\longhookrightarrow} \Bigl( \cU_1=\uone\times\uone\Bigr) \underset{\textstyle\vareps_1}{\longhookrightarrow} \Bigl(\cU_0 = \sutwo\times\sutwo\Bigr) ~.
\ee
The left and right embeddings are related by~\eqref{GLMTembedsteps}
\begin{align}
\label{GLMTembedsteps-1}
\ell &= \vareps_1 \circ \vareps_2
~~,~~~~
r =  \Omega_0 \circ \vareps_1 \circ \Omega_1 \circ \vareps_2 \circ \Omega_2
%~~,~~~~
\end{align}
with the embeddings
\begin{align}
&\vareps_2(e^{i\eta}) = \bigl(1, e^{i\eta} \bigr)
~~,~~~~
\vareps_1(e^{i\eta_1},e^{i\eta_2}) = \bigl( e^{ip\eta_1{\sigma_3}},e^{iq\eta_2{\sigma_3}} \bigr)
~~.
\end{align}
and the exchange automorphism
\be
\Omega_1(h_1,h_2) = (h_2,h_1) ~~.
\ee
The modified embedding relation~\eqref{GLMTembedsteps} consistently generalizes the embedding constraint~\eqref{EmbeddingCondition} that relates left and right embeddings of the gauge group.
The construction of D-branes now follows as before, given a pair of automorphisms $\Omega_{0}\equiv\Omega_\cG$, $\Omega_2\equiv\Omega_\cH$.  The extended embedding chain leads to branes smeared along a product of conjugacy classes
\be
\cC_{\su\times\su}^{\Omega_0} \cdot \Bigl(\Omega_0 \circ \vareps_1 \bigl(\cC_{\uo\times\uo}^{\Omega_1}\bigr)\Bigr)
\cdot\Bigl(\Omega_0\circ\vareps_1\circ\Omega_1\circ\vareps_2\bigl(\cC_{\uo}^{\Omega_2}\bigr)\Bigr)  ~~.
\ee
A suitable generalization of the construction of the two-form $\omega_2$ leads to the general expression described at the end of Section~\ref{sec:symmetrybreakingbranes}, see~\cite{Quella:2002fk,Quella:2003kd} for details.

In the three-charge supertube, $\cH$ also has an asymmetric action on $\cG$.  We have not been able to find a solution to the constraint~\eqref{GLMTembedsteps} in general; however a special choice of radius yields a very similar structure to the above, namely for
\be
\Ry=R_\star=\sqrt{\nfive}\,s/k 
~~,~~~~
\Rytil =\sqrt{\nfive}\, (s+1)/\ktil
~~,
\ee 
so that  
\be
l_4 = -k\Ry+\ktil\Rytil = \sqrt{\nfive}
~~,~~~~
r_4 = (k\Ry+\ktil\Rytil) = (2s+1)\sqrt{\nfive}
\ee
and thus the gauge action along $\bS^1_y$ has the same structure as along $\sutwo$, with left and right interchanged just as in the example above.   Then the gauge action is
\begin{align}
&\ell(\zeta,\eta)  \bigl( g_\sl, g_\su , e^{it}, e^{iy} \bigr)   r(-\zeta,-\eta) 
\\
&\hskip .5cm = 
\bigl( 
e^{ i\alpha{\sigma_3}} g_\sl e^{ i\beta{\sigma_3}} \, , \,
e^{ - i(2s+1)\alpha{\sigma_3}} g_\su e^{ i\beta{\sigma_3}}  \, , \, 
e^{-i\kappa_+\alpha}e^{it}  e^{-i\kappa_+\beta}  \, , \,
e^{-i\kappa_-\alpha}e^{iy}  e^{ -i\kappa_+\beta} 
 \bigr) ~,
\nn
\end{align}
with again $\alpha= \hf(\zeta+\eta)$, $\beta= \hf(\zeta-\eta)$; and $\kappa_-=l_4$, $\kappa_+=r_4$.
We choose the intermediate group 
\be
\cU_1 = U(1)^3\equiv\Uint
\ee
(later we will enlarge this to $U(1)^4$ in order to smear along physical time), with the embedding chain
\be 
\cH~\underset{\textstyle\vareps_\cH}{\longhookrightarrow}~  \Uint ~\underset{\textstyle\vareps_\Uint}{\longhookrightarrow}~ \cG  ~~,
\ee
and set
\begin{align}
\vareps_\cH(\zeta,\eta) &= \bigl(\zeta+\eta\,,\,(2s\tight+1)(\zeta+\eta)\,,\, -(\zeta+\eta) \bigr)
\nn\\
\vareps_\Uint(\eta_1,\eta_2,\eta_3) &= \bigl( e^{ i\eta_1{\sigma_3}},e^{ - i\eta_2{\sigma_3}},e^{-i(2s+1)\eta_1\sqrt\nfive},e^{i\eta_3\sqrt\nfive} \bigr)
\nn\\
\Omega_{\cH}(\zeta,\eta) &= (\eps\zeta,-\eps\eta)
\\
\Omega_\Uint(\eta_1,\eta_2,\eta_3) &= (\eps'\eta_1,\eps''\eta_3,\eps'''\eta_2) 
\nn\\
\Omega_{\cG} \bigl( g_\sl,g_\su,g_t,g_y \bigr) &= \bigl(  \Omega^{\eps_\sl}(g_\sl), \Omega^{\eps_\su}(g_\su), \Omega^{\eps_t}(g_t), \Omega^{\eps_y}(g_y) \bigr) ~~.
\nn
\end{align}

With the goal of describing a brane localized in the cap of the
geometry, we choose $\eps_\sl=+$ so that the brane does not extend to
spatial infinity; one can then check that the embedding
conditions~\eqref{GLMTembedsteps} are satisfied provided 
\be
\label{epsrelations}
\eps_t=+
~~,~~~~ 
\eps' = -\eps
~~,~~~~
\eps'' = -\eps\eps_\su
~~,~~~~
\eps''' = { \eps \eps_y}
\ee
The extended embedding chain leads to branes smeared along a product of conjugacy classes
\be
\label{Cprod}
\cC_{\cG}^{\Omega_\cG} \cdot \Bigl(\Omega_\cG \circ \vareps_\Uint \bigl(\cC_{\Uint}^{\Omega_\Uint}\bigr)\Bigr)
\cdot\Bigl(\Omega_\cG\circ\vareps_\Uint\circ\Omega_\Uint\circ\vareps_\cH\bigl(\cC_{\cH}^{\Omega_\cH}\bigr)\Bigr)
\ee

We have the choice of a starting conjugacy class in $\cG$.  We choose the trivial conjugacy class in $\sltwo$ and $\bR_t$, an $\bS^2$ brane in $\sutwo$ at either $\theta=0$ (for $\eps_\su=+$) or $\theta=\pi/2$ (for $\eps_\su=-$).  In contrast to the two-charge case, this starting conjugacy class is not invariant under any subgroup of $\cH$, temporal or spatial, and so will have to be smeared along both directions by the last two factors in~\eqref{Cprod} (for which we adopt the shorthand notation $\cC_{\Uint}\cdot\cC_\cH$, leaving the embedding chain implicit).

Regardless of the sign of $\eps$, the product of the last two factors $\cC_{\Uint}\cdot\cC_\cH$ always smears $\cC_\cG$ in the temporal gauge direction, but not in the other timelike direction.  As in the two-charge case, one must construct independent timelike currents $\cK,\bar\cK$ transverse to the gauge currents $\cJ, \bar\cJ$ and adjoin them to $\cH$ in order to build a brane that is extended along the physical time direction.  We will attend to that issue after the analysis of smearing along gauge orbits, since the two are essentially independent.  We discuss the two choices in turn.

\paragraph{{{${\boldsymbol{\eps\tight=-}}$}~:}}
The analysis of the previous section suggests the choice $\eps\tight=-$ for the automorphism $\Omega_\cH$ defining $\cC_\cH$ for the localized W-brane.  Then $\cC_\cH$ is extended along the timelike gauge direction parametrized by $\zeta$.
We find that the product $\cC_{\Uint}\cdot\cC_\cH$ adds only one space
and one time direction (thus smearing only along the gauge orbits and
not adding further spatial dimensions) only for the choices 
\be
\eps = -
~~,~~~~
\eps_\su \eps_y = -
~~.
\ee
In terms of the parametrizations $(\zeta,\eta)\in\cH$ and $(\eta_1,\eta_2,\eta_3)\in\Uint$, one finds for these choices that the product of conjugacy classes $\cC_{\Uint}\cdot\cC_\cH$ can be parametrized by $(\zeta,\eta_3)$ with the other parameters being redundant; $\zeta$ parametrizes the right-hand gauge orbit, and $\eta_3$ a spacelike direction.  The product of conjugacy classes embeds in $\sltwo\times\sutwo\times\bR_t\times \bS^1_y$ via
\be
\label{CprodU1H}
\cC_{\Uint}\cdot\cC_\cH = \Bigl( e^{ i \zeta \sigma_3} \, , \, e^{ i (\zeta+\eta_3)\sigma_3} \, , \,
-(2s\tight+1)\sqrt{\nfive}\,\zeta  \, , \,  \eps_y\sqrt{\nfive}\,(\eta_3\tight+\mu_3) { - (2s\tight+1)\sqrt{\nfive}\,\zeta} \Bigr)  ~~,
\ee
where $\mu_3$ is a constant.
For $\eps_\su\tight=-$ (so $\theta\sim \pi/2$), one requires $\eps_y\tight=+$ and so $\cC_\cG$ consist only of $\bS^2\subset\sutwo$; the only additional spatial direction beyond this $\bS^2$ is the $\bS^1$ parametrized by $\eta_3$.  One obtains a brane upstairs with three spatial dimensions, projecting down to a D2 brane wrapping the $\bS^2$ of the $\bS^2\times\bS^1$ cycle at $\theta=\pi/2$ in the cap of the 9+1d geometry.  This brane is the direct analogue of the W-particle in the two-charge case, wrapping in that case a vanishing cycle of the local $\bZ_k$ singularity in the NS5-F1 supertube and free to move along the $\bS^1$ of the supertube coil.  Here one has much the same structure for the $\bZ_{\ellone}$ 
singularity at $\theta=\pi/2$ of the three-charge supertube, which has a moduli space coiled $\ellone$ times around the $\phi$ circle. 

For $\eps_\su\tight=+$ (so $\theta\sim 0$), one sets instead $\eps_y\tight=-$, and now the starting brane $\cC_\cG$ is $\bS^2\times\bS^1_y$.   
We seem to have landed on a T-dual description of the W-particle -- the starting brane $\cC_\cG$ is localized in $\ytil$ rather than $y$, and so repeating the analysis in that duality frame one expects to have a moduli space of motion that spirals in $\ytil$ and $\psi$.  One can verify this conclusion by modifying the last of the relations~\eqref{epsrelations} to $\eps''' = - \eps \eps_y$, which results in the T-dual description.
This starting brane is now smeared along the spatial direction parametrized by $\eta_3$ in the product of conjugacy classes~\eqref{CprodU1H}, and so all told the brane~\eqref{Cprod} has a four-dimensional spatial volume $\bS^2\times\bS^1$ in the $\ytil$ frame.

The asymmetry here between the descriptions of W-particles at $\theta=0$ and $\theta=\pi/2$ seems to be a limitation of the method of~\cite{Quella:2002fk,Quella:2003kd}, which forces a particular relation between left and right gauge actions that, while sufficient to guarantee the existence of a brane built out of factorized conjugacy classes, is by no means necessary.  It turns out that a different choice of special radius
\be
\Ry = R_*= \sqrt\nfive\, (s+1)/k
~~,~~~~
\Rytil = \sqrt{\nfive}\, s/\ktil
\ee
leads to much the same analysis, but with the feature that, only for the special choice
\be
\eps = -
~~,~~~~
\eps_\su \eps_y = + ~~,
\ee
will the product $\cC_{\Uint}\cdot\cC_\cH$ have only one spacelike and one timelike dimension.
At this new radius $R_*$, when the starting point $\cC_\cG$ is pointlike on $\bS^1_y$ (so that $\eps_y\tight=+$), it is now localized near $\theta=0$ (\ie\ $\eps_\su\tight=+$) rather than at $\theta=\pi/2$ as it was when $\Ry=R_\star$, due to the flip from twisted to untwisted $\sutwo$ conjugacy class in $\cC_\cG$.  For the opposite sign choices, $\eps_\su\tight=-$ and $\eps_y\tight=-$, one wants to use again the T-dual description by setting
$\eps''' = - \eps \eps_y$.
The starting brane $\cC_\cG$ is localized in $\ytil$, so that in the T-dual frame it again has topology $\bS^2\times\bS^1$.  Thus the structures at $\theta=0,\pi/2$ are exchanged relative to those found for $\Ry=R_\star$ when we set instead $\Ry=R_*$.

Clearly the existence of such branes is not dependent on the value of the radius $\Ry$ of $\bS^1_y$, and we expect that a relaxation of some of the requirements above should allow a construction of W-particles for general radius $\Ry$, at both $\theta=0,\pi/2$ -- that being able to construct W-particles at the orbifold loci only for two special radii $R_\star$ and $R_*$ is an artifact of the particular method and not a general limitation.

What seems to be happening is that for $\Ry=R_\star=\sqrt\nfive\, s/k$, we are in the happy situation that we can start with a brane that is a factorized product of conjugacy classes~-- an $\bS^2$ brane at $\theta=\pi/2$ times a point brane in $\bS^1_y$, or an $\bS^2$ brane at $\theta=0$ times a point brane in $\bS^1_\ytil$ -- which is then smeared over $\cC_{\Uint}\cdot\cC_\cH$, as well as a conjugacy class $\cC_\cK$ for the physical timelike direction generated by some current $\cK,\bar\cK$ (we will discuss the possibilities for this current below).  In order to implement this in the embedding chain for $\cH$ above, we can define an augmented group embedding for $\cH'=\cH\times\cK$, and a new intermediate group $\cI'=\cI\times\cK$ that simply passes through the embedding of physical time smearing generated by $\cK$.  The product 
\be
\cChat\equiv\cC_{\Uint'}\cdot\cC_{\cH'}
\ee
has two timelike dimensions, and for $\eps_\su\eps_y=-1$ has only one spacelike dimension.  Roughly speaking, the starting factorized brane gets smeared along the 1+1 dimensional gauge orbits, as well as the physical time direction.  For $\theta=\pi/2$ the starting brane $\cC_\cG$ is analogous to our description of the W-particle in the two-charge supertube, now localized in $y$, while for $\theta=0$ it is localized in $\ytil$.  
Conversely, for $\Ry=R_*=\sqrt\nfive (s+1)/k$, we impose $\eps_\su\eps_y=+1$ and the two descriptions are flipped~-- the starting $\cC_\cG$ brane at $\theta=0$ is analogous to the W-particle of the two-charge supertube, here localized in $y$, while the one at $\theta=\pi/2$ is localized in $\ytil$.

For general radius $\Ry$, neither of these $\theta=0,\pi/2$ W-particles can be described from a starting point where the brane locus $\cC_\cG$ is purely Neumann or purely Dirichlet in $y$, instead the brane has a mixed boundary condition that is correlated to what the brane is doing in $\sutwo$.  In other words, the starting point cannot be a brane that is factorized between $\sutwo$ and $\bS^1_y$.  But this is indeed the generic state of affairs; the spatial gauge group parametrized by $\eta=\hf(\alpha-\beta)$ in~\eqref{GLMTgaugetransfs} transforms each of $\phi,\psi,y,\ytil$ in a correlated way, and generically the cycle wrapped by a W-particle will not cleanly factorize between $\sutwo$ and $\bS^1_y$ (or $\bS^1_\ytil$).  From this perspective, it was rather remarkable that for the special radii $R_\star$ and $R_*$ one found success with a factorized starting point and using a bit of trickery.

\paragraph{{${\boldsymbol{\eps\tight=+}}$}~:}
There is also the alternative choice, $\eps\tight=+$.  Now $\cC_\cH$ is the spatial gauge orbit parametrized by $\eta$, and $\cC_{\Uint}\cdot\cC_\cH$ is three-dimensional.  This choice leads to branes wrapping a topology $\bS^2\times\bT^2$ upstairs in $\cG$, regardless of the choices of automorphism $\eps_\su,\eps_y$ in $\sutwo$ and $\bS^1_y$; and thus one always has a brane of topology $\bS^2\times \bS^1$ at either of the orbifold loci downstairs in $\cG/\cH$.  These branes are of interest as well -- they again appear to be particular orientations of the W-strings we have been looking for.  
An analysis of the various choices of the remaining independent signs $\eps_\su,\eps_y$ shows that the product $\cC_{\Uint}\cdot\cC_\cH$ has two physical spatial dimensions in addition to the gauge orbit parametrized by $\eta$; in the full product of conjugacy classes~\eqref{Cprod}, it smears $\cC_\cG$ along both the spatial gauge orbit, and in addition two physical spatial circles.  Thus, for an initial $\bS^2$ brane near $\theta=\pi/2$ whose azimuthal direction is $\psi$, the additional smearing is along both $\phi$ and $y$; for an $\bS^2$ brane near $\theta=0$ whose azimuthal direction is $\phi$, the additional smearing is along both $\psi$ and $y$.  There is no choice that reduces to the W-particle of the previous section when $s=0$. 
Thus after smearing, the brane upstairs always has four spatial directions with topology $\bS^2\times\bT^2$,%
\footnote{The starting brane $\cC_\cG$ can be either $\bS^2$ (for $\eps_y=+$) or $\bS^2\times\bS^1$ (for $\eps_y=-$); the smearing by $\cC_{\Uint}\cdot\cC_\cH$ is always along two spatial dimensions, one of which is redundant for $\eps_y=-$.  As a result the final brane upstairs always has spatial topology $\bS^2\times \bT^2$.}
and projects down to a localized D3 brane in 9+1d. 

We have not checked whether this $\eps\tight=+$ brane is the same as or different from what one gets from smearing the W-particle along its moduli space.  The different 4d branes upstairs could have different fluxes even though they occupy the same worldvolume.

%%%%%%%%%%%%%%%%%%%%%%%%%%%%%%%%%%%%
\subsection{Smearing along physical time}
\label{sec:GLMTtimesmear}

Independently, we must smear the above brane along the physical timelike direction.  Thus we seek timelike currents $\cK, \bar\cK$ having vanishing two-point function with the gauge currents $\cJ,\bar\cJ$.  We again define currents $\cK,\bar\cK$ as in~\eqref{Kdefn} and adopt the ansatz
\be
k_1=\bar k_1
~~,~~~~
k_2=-\delta_2\bar k_2
~~,~~~~ 
k_3=\bar k_3
~~,~~~~
k_4=-\delta_4\bar k_4
\ee
where $\delta_2,\delta_4=\pm$.  For the gauge
currents~\eqref{GLMTcurrents} of the three-charge background, the
constraints~\eqref{KJcorrelator} have various solutions depending on
$\delta_2,\delta_4$ (choosing the normalization $k_3=\bar k_3=1$ and
setting $k_2=\vv=-\delta_2\bar k_2$): 
\begin{align}
\delta_2\tight=+,\delta_4\tight=+
~&:~~~
k_4 = { -}\frac{k \Ry}{s\tight+1}\, \vv
~,~~
k_1 = \frac{k\Ry+\ktil\Rytil}{\nfive} + \vv\Bigl((s\tight+1) + \frac{sk\Ry^2}{\ktil}\Bigr)
\nn\\[3pt]
\delta_2\tight=-,\delta_4\tight=+
~&:~~~
k_4 ={ -} \frac{k\Ry}{s}\, \vv 
~~~,~~~~
k_1 = \frac{k\Ry+\ktil\Rytil}{\nfive} { +}  \vv\Bigl(s + \frac{(s\tight+1)k\Ry^2}{\ktil}\Bigr)  
\nn\\[3pt]
\delta_2\tight=+,\delta_4\tight=-
~&:~~~
k_4 = { +}  \frac{\ktil\Rytil}{s\tight+1}\, \vv 
~,~~
k_1 = \frac{k\Ry+\ktil\Rytil}{\nfive} + \vv\Bigl((s\tight+1) +\frac{s\ktil\Rytil^2}{k}\Bigr)
\nn\\[3pt]
\delta_2\tight=-,\delta_4\tight=-
~&:~~~
k_4 = { +}\frac{\ktil\Rytil}{s}\, \vv 
~~~,~~~~
k_1 = \frac{k\Ry+\ktil\Rytil}{\nfive} { +}  \vv\Bigl(s +\frac{(s\tight+1)\ktil\Rytil^2}{k}\Bigr)
~.
\end{align}
For the first two choices, the fact that $k_4,\bar k_4$ are proportional to $\Ry$ tells us that the moduli space is along $y$ rather than $\ytil$; similarly the last two choices having $k_4,\bar k_4$ proportional to $\Rytil$, describe brane motion along $\ytil$.  When $\delta_2\tight=-$, the moduli space involves motion along $\psi$, while for $\delta_2\tight=+$, it involves motion along $\phi$.

\begin{figure}[ht]
\centering
  \begin{subfigure}[b]{0.44\textwidth}
  \hskip -.5cm
    \includegraphics[width=\textwidth]{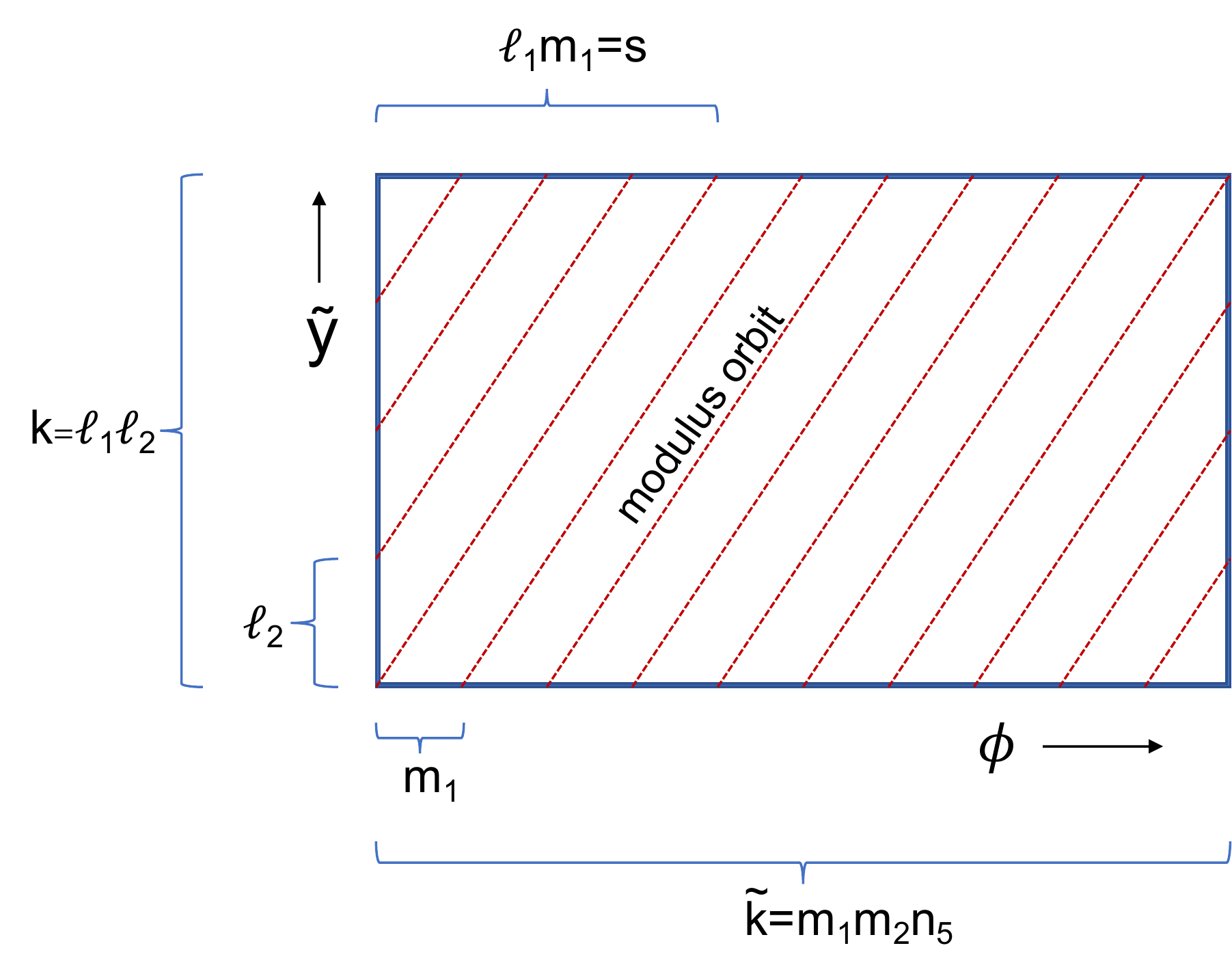}
    \caption{ }
    \label{fig:STtorus}
  \end{subfigure}
\qquad~
  \begin{subfigure}[b]{0.44\textwidth}
  \hskip -.5cm
    \includegraphics[width=\textwidth]{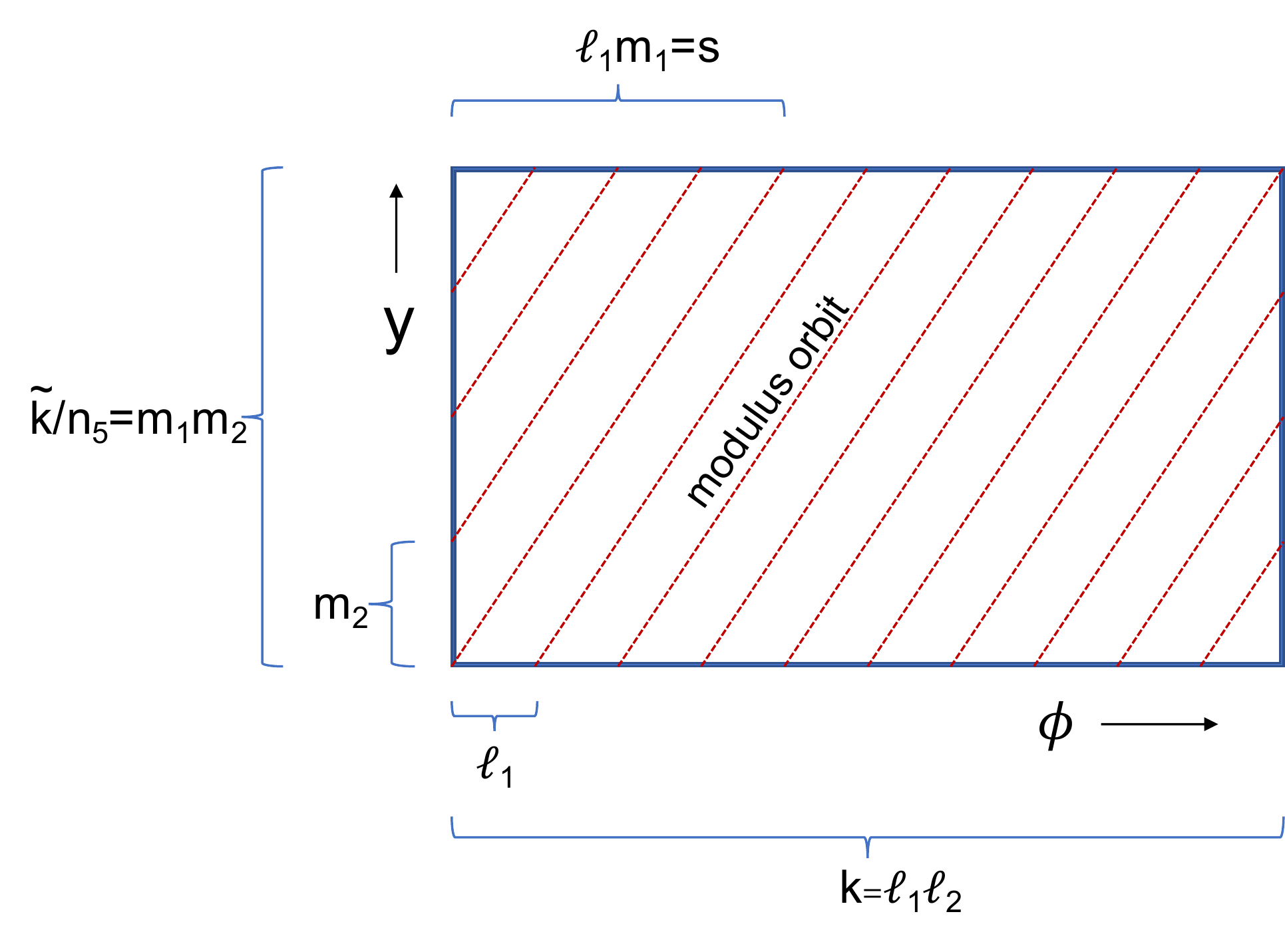}
    \caption{ }
    \label{fig:STdualtorus}
  \end{subfigure}
\caption{\it 
(a) Brane moduli space on the $\ytil$-$\phi$ torus.  There is an equivalent picture for the $\ytil$-$\psi$ torus, with $\ellone\leftrightarrow\elltwo$ and $\nhat\leftrightarrow \mhat$.
(b) The T-dual picture on the $y$-$\phi$ torus exchanges the roles of $k$ and $\ktil=\nhat \mhat\nfive$.
}
\label{fig:ST-torus}
\end{figure}

The four choices of sign correlate with the four branes found above for $\epsilon\tight=-$.  We found branes localized in $y$ and $\ytil$; and near $\theta\tight=0$ localized in $\psi$, and near $\theta\tight=\pi/2$ localized in $\phi$.  The currents $\cK,\bar\cK$ implement the following spiral motions, depending on the choice of $\delta_2,\delta_4$:
\begin{enumerate}
\item
$\delta_2\tight=+,\delta_4\tight=+$: Spiral motion in the $y$-$\phi$ torus with a pitch
\be
\label{pp spiral}
\frac{d(y/\Ry)}{d\phi} = { -}\frac{k}{s+1} = { -}\frac{\ellone}{\mhat } ~~.
\ee
There are $\mhat$ strands of the orbit over each point in $\phi$.
\item
$\delta_2\tight=-,\delta_4\tight=+$: Spiral motion in the $y$-$\psi$ torus with a pitch
\be
\label{mp spiral}
\frac{d(y/\Ry)}{d\psi} = { -}\frac{k}{s} = { -}\frac{\elltwo}{\nhat }~~.
\ee
There are $\nhat$ strands of the orbit over each point in $\phi$.
\item
$\delta_2\tight=+,\delta_4\tight=-$: Spiral motion in the $\ytil$-$\phi$ torus with a pitch
\be
\label{pm spiral}
\frac{d(\ytil/\Rytil)}{d\phi} = +\frac{\ktil}{s+1} = +\frac{\nhat \nfive}{\elltwo}~~.
\ee
There are $\elltwo$ strands of the orbit over each point in $\phi$.
\item
$\delta_2\tight=-,\delta_4\tight=-$: Spiral motion in the $\ytil$-$\psi$ torus with a pitch
\be
\label{mm spiral}
\frac{d(\ytil/\Rytil)}{d\psi} = { +}\frac{\ktil}{s} = { +}\frac{\mhat \nfive}{\ellone}~~.
\ee
There are $\ellone$ strands of the orbit over each point in $\phi$.
\end{enumerate}
These choices correspond to the coiling of the $\bS^1$ in the $\bS^2\times\bS^1$ topology at the tip of the geometry at $\theta=0,\pi/2$, for both the spectrally flowed supertube described in the $y$ coordinate (choice 1 at the end of Section~\ref{sec:GLMTnullgauging}), and the T-dual description in terms of $\ytil$ (choice 2 at the end of Section~\ref{sec:GLMTnullgauging}).  In the two-charge case, we saw that the coiling of this topology was revealed in the spiral of the T-dual coordinate -- the spiral of the $\bS^2$ vanishing cycle around the $\phi$ circle of the NS5-F1 supertube was seen in the trajectory of the fivebrane in the T-dual $\ytil$ coordinate.  The pitch of the spiral $d(\ytil/\Rytil)/d\phi=-\nfive/k$ indicated that there were $k$ locations of the fivebrane source in $\ytil$ for any given value of $\phi$, and so in the NS5-F1 frame one has a local $\bZ_k$ orbifold singularity.  The moduli space for the various choices~\eqref{pp spiral}-\eqref{mm spiral} is depicted in Figure~\ref{fig:ST-torus}.

%%%%%%%%%%%%%%%%%%
%%%%%%%%%%%%%%%%%%

%%%%%%%%%%%%

\refstepcounter{subsection}
\subsection*{\thesubsection \quad Aside on non-commutativity} \label{sec:noncommutativity}
%\subsection{Aside on non-commutativity}\label{sec:noncommutativity}

As a consequence of the asymmetric smearing, and more generally for branes at general radii which are necessarily non-factorized, the brane is generically not localized purely in $y$ nor in $\ytil$.  The situation is somewhat similar to D-branes wrapping a non-primitive cycle on a $\bT^2$ of the sort depicted in Figure~\ref{fig:Supertube-alt3} (see for instance~\cite{Hashimoto:1997gm}; and also~\cite{Recknagel:2013uja}, section 1A for a discussion of the relevant CFT boundary states).  Suppose the primitive cycles of the torus have coordinates \xone\ and \xtwo\ 
of radii $R_1$ and $R_2$.  
Neither primitive cycle has a purely Dirichlet or Neumann boundary condition; instead for a brane wrapping the $(p,q)$ cycle one has a brane making an angle $\chi$ with respect to these directions, where
\be
\tan\chi = \frac{q}{p} \frac{R_2}{R_1} ~~.
\ee
The D-brane boundary state for such a brane has the structure
\begin{align}
| \chi; \tilde x_0, x_0 \rangle\!\rangle &= \cN \sum_{n,w\in\bZ} 
\exp\Bigl[ \frac{ix_0 n}{R_2'} + 2i R_1' \tilde x_0 w \Bigr]
\bigl| n,w \bigr\rangle\!\bigr\rangle_\chi
\nn\\[5pt]
\bigl| n , w \bigr\rangle\!\bigr\rangle_\chi &=
\exp\Bigl(-\sum_{m>0}\frac1m {\bf a}^t_{-m} M_\chi \bar{\bf a}_{-m}\Bigr)\,
\bigl| n q,-np;wp,wn \bigr\rangle
\\[5pt]
M_\chi &= \left( {\cos 2\chi~~~~\sin 2\chi \atop \sin 2\chi~ -\cos2\chi}\right)
~~,~~~~
R_1'=\frac{pR_1}{\cos\chi}
~~,~~~~
R_2'=\frac{R_2\cos\chi}{p}
\nn
\end{align}
where ${\bf a}_{m},\bar{\bf a}_{m}$ are left/right mode operators for $x^1,x^2$.
Roughly speaking one has a Dirichlet boundary condition on a rotated combination of \xone, \xtwo, and a Neumann boundary condition on the orthogonal combination.  The D-brane boundary state will have a delta function of the zero mode $x_0$ in the orthogonal direction to the $(p,q)$ cycle; the Neumann nature of the $(p,q)$ cycle yields a delta function of the zero mode coordinate $\tilde x_0$ for the $(-q,p)$ cycle of the T-dual torus, which is the Wilson line of the brane.  But from the point of view of the individual free field CFT's for \xone\ and \xtwo, the brane is not fully localized or delocalized.%

One way to think about the D-brane boundary state in such a situation employs a doubling of the zero modes on both $\bS^1$ factors to include both a coordinate $x_0$ on the circle as well as a coordinate $\tilde x_0$ on the T-dual circle.  One may want to consider such a doubling on general grounds, for instance the vertex operator algebra of exponentials of a compact free scalar can shift both the winding and momentum quantum numbers $p,w$ of closed string states, and so one wants a Fourier conjugate for each.  But the zero modes $x_0,\tilde x_0$ conjugate to $p,w$ do not commute~\cite{Freidel:2017wst,Freidel:2017nhg}.  Boundary states for D-branes are localized in position space rather than momentum/winding space, and thus must select a ``polarization'' in the ``phase space'' of these coordinates, depending only on one linear combination.  In the example above of a D1-brane on a $(p,q)$ cycle of $\bT^2$, one has a four-dimensional ``phase space'' of the doubled zero modes for \xone, \xtwo; the boundary state chooses a polarization that is not diagonal in these coordinates, but rather diagonalizes the coordinates $x_0,\tilde x_0$ along the $(p,q)$ cycle and its dual, that commute with one another but not the remaining pair of the four coordinates on the doubled two-torus.

With asymmetric gauging it seems we again have the boundary conditions providing a brane that is neither purely Dirichlet nor Neumann on the coordinate axes $\phi,\psi,y,\ytil$.  The zero mode coordinates on $\sutwo$ lead to a non-commutative geometry on D-branes~\cite{Alekseev:1999bs}, the zero modes of $y,\ytil$ are also non-commutative~\cite{Freidel:2017wst,Freidel:2017nhg}; the boundary state is thus expected to lead to a somewhat non-commutative structure on the D-branes wrapping the topology at the bottom of the 3-charge supertube throat.

%%%%%%%%%%%%%%%%%%%%%%%%%%%%%%%%%%%%

%\newpage
%%%%%%%%%%%%%%%%%%%%%%%%%%%%%%%%%%%%
\section{Discussion}
\label{sec:Discussion}

We have seen that the sub-string scale structure of two-charge supertubes (and the associated three-charge backgrounds obtained by spacetime spectral flow) can be exposed through a quantitative analysis of D-brane probes, yielding a wealth of information about dynamics near the threshold of black hole formation; this analysis is made possible by the exact solvability of the round supertube background, through the gauging of a group of null isometries of a Wess-Zumino-Witten model. 

While the generic supertube D-brane has boundary conditions that inextricably correlate the various factors in the WZW group $\cG$ of equation~\eqref{Gupstairs}, particular D-branes are given by products of conjugacy classes of $\cG$ and of the subgroup $\cH$ being gauged.  In the two-charge NS5-P supertube, we found D-branes bound to the fivebranes; and in the NS5-F1 supertubes, we found their T-duals which wrap KK monopole structures in the cap of the geometrical background.  In the three-charge case, we found a suitable generalization (for special choices of the radius $\Ry$) involving an intermediate group $\cH\subset\Uint\subset\cG$.  These results led to a complete characterization of the corresponding D-brane worldvolumes and the fluxes that support them, at least at the level of the DBI action.  Furthermore, one expects that exact CFT boundary states for these branes can be constructed along the lines of~\cite{Quella:2002ns,Quella:2002fk,Quella:2003kd}.  Fortunately, this class of D-branes includes examples of the ``W-branes'' which expose the long string structure that arises near the threshold of black hole formation in linear dilaton and asymptotically $AdS_3$ spacetimes.

The $\sutwo$ factor in $\cG$ encodes the locations of the NS5-branes in the NS5-P supertube; a D1-brane stretching between NS5-branes in $\cG/\cH$ lifts in part to an $\bS^2$ brane in the $\sutwo$ factor of $\cG$, with the polar direction of the $\bS^2$ giving the path between fivebranes, and the azimuthal direction of the $\bS^2$ related to the axial gauge orbit.  The gauge orbit degenerates at the poles of the $\bS^2$ where the fivebranes are located.  Thus, while upstairs in $\cG$ the D-brane worldvolume is completely smooth, downstairs in $\cG/\cH$ it projects to a line segment that ends abruptly.  For D2-brane W-strings, this structure extends along a second, longitudinal direction of the fivebranes~-- either along the $\bT^4$ compactification, or along the supertube spiral along $\bS^1_\ytil$, or some combination of the two (see equation~\eqref{Ksupertube}), so that the spatial worldvolume upstairs in $\cG$ is $\bS^2\times\bS^1$.  The helical structure of the supertube leads to the characteristic ``long string'' feature that the $\bS^1$ is a multiple cover of the supertube source ring in the four physical dimensions transverse to the compactification $\bS^1_y\times\bT^4$.

Because we have an exact worldsheet CFT, T-duality from the NS5-P frame to the NS5-F1 frame is simply a relabelling of the CFT data.  D1-branes stretching between NS5's dualize to D2-branes wrapping vanishing cycles of nearly coincident KK monopoles.  
The T-dual description of the D-branes we constructed in the NS5-P frame yields the corresponding D-branes for the NS5-F1 frame in a straightforward manner.  Although the fivebranes have ``disappeared'' into flux, the CFT keeps track of the structure in the cap of the geometry that they generate.  That structure exhibits the expected T-dual W-branes, and illuminates the duality between NS5-branes and KK monopoles.

%%%%%%%%%%%%%%%%%%%%%%%%%%%%%%%%%%%%
\subsection{Is our lamp-post in a good location?}

\begin{figure}[ht]
\centering
    \includegraphics[width=.6\textwidth]{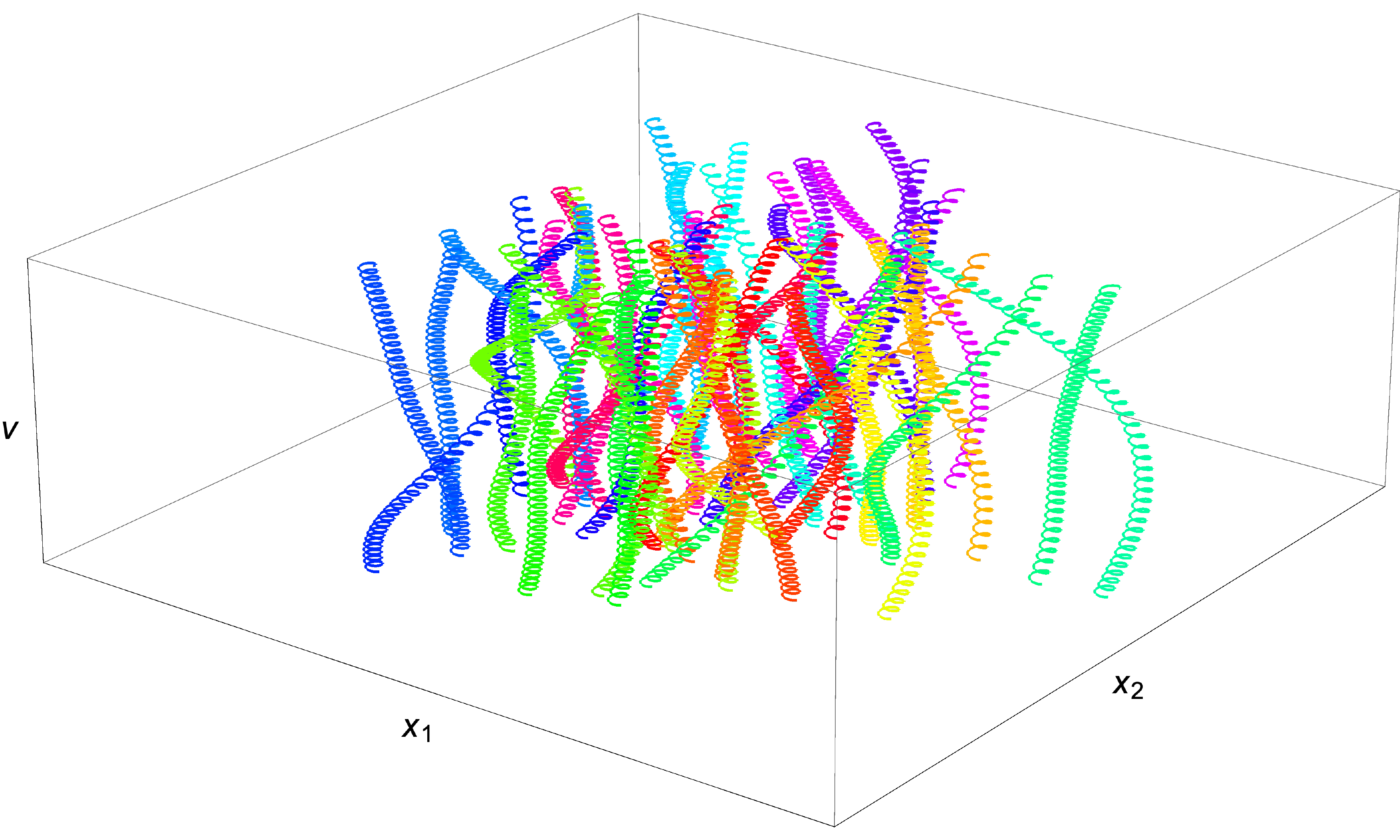}
\caption{\it 
Generic supertube profile for an NS5-P supertube at low angular momentum, consisting of a single mode of high wavenumber making a tight spiral, together with a generic sprinkling of low wavenumber modes causing a random walk of the profile on larger scales.  The $\nfive=50$ fivebrane strands have been wound into a single supertube; color adiabatically evolves along the profile to reveal the supertube's wandering in the transverse \xone-\xtwo\ plane.  
}
\label{fig:ThermalPrimary}
\end{figure}

One can ask to what extent our results are generic, and what features are special to the particular round supertubes we can study in detail.  The round supertube is of course a highly non-generic coherent state; for instance, in the NS5-P frame it is built by populating a single wavenumber and polarization mode of the fivebrane to a macroscopic level, leading to the characteristic spiral of Figure~\ref{fig:AdSchiralprimary3-1}.  A more generic low angular momentum supertube profile executes a random walk in the transverse space, and looks more like Figure~\ref{fig:ThermalPrimary}.
%
%%%%%%%%%%%%%%%%%%
%%%%%%%%%%%%%%%%%%

Clearly the fivebranes continue to be separated on their Coulomb branch, but the strands no longer neatly line up along a single trajectory in the \xone-\xtwo\ plane.  The supertube still winds $\nfive$ times around the $\ytil$ circle, and thus so also will the W-brane; however, the minimal W-brane shape will be much more complicated, and not simply a multiple cover of a single trajectory.  Rather, for $k\gg\nfive$ it looks more like $\nfive$ separate coils having of order $k/\nfive$ windings.  Because of the transverse spread of the profile, the more generic supertube may actually be farther from the threshold of black hole formation than the rather special configurations that we have studied here~-- the W-branes are heavier and exhibit less of the expected long string structure.  The geometry has many higher multipoles excited and is not locally $AdS_3\times\bS^3$.  In contrast, the round supertube {\it is} locally $AdS_3\times \bS^3$, a property it shares with the BTZ black hole.  

It may be that when excited, the supertube is driven toward more compact, coiled profiles like the round supertube, \ie\ toward the origin of the Coulomb branch where the long string structure becomes fully liberated~\cite{Marolf:2016nwu}.  In the process, the supertube must shed its angular momentum, or at least carry it in such a way that fivebrane strands can come together to make the W-branes light, as we will discuss below.  Decreasing the angular momentum by increasing the mode number $k$ of the dominant excitation makes the supertube more pointlike, 
and W-brane excitations will then push the system into a black hole phase;%
\footnote{There are also momentum modes on the type IIA NS5-P supertube that don't carry transverse angular momentum, such as the self-dual antisymmetric tensor modes and their scalar superpartner; exciting these rather than the transverse scalars also decreases the supertube radius and increases the depth of the throat sourced by the supertube (see for instance~\cite{Lunin:2002iz}).}
alternatively, the supertube may maintain a finite radius, with W-brane excitations forming a black ring.  So indeed, the special round supertubes are very non-generic configurations, but this may actually be a good thing if the goal is to study the black hole or black ring threshold.

%%%%%%%%%%%%%%%%%%%%%%%%%%%%%%%%%%%%
\subsection{Round supertubes and exotic phases}

In regimes where black holes dominate the spectrum, one can think of the black hole solution itself as a sort of ``ensemble geometry'' which captures the thermodynamics.  In particular, the classical action determines the constitutive relation $S(E)$ by connecting the ADM mass to the Wald entropy, both of which are Noether charges of the underlying diffeomorphism symmetry of gravity~\cite{Wald:1993nt,Iyer:1994ys}.%
\footnote{Usually in thermodynamics, this constitutive relation must be supplied from some analysis of the underlying microscopics; it is remarkable that gravity {\it knows} what the result must be, even though it doesn't know what the underlying microscopics is.  In the AdS/CFT context, and perhaps more broadly, gravity is a collective mode of the underlying microscopics, whose emergent diffeomorphism symmetry determines the outcome.}
If there are multiple black objects that might occur, the one with the most entropy dominates.
The asymptotics of the spectrum in $AdS_3$ is governed by the BTZ solution, connected to the Cardy formula for the density of states~\eqref{BTZentropy} of the spacetime CFT.  
In the regime where semiclassical gravity applies, namely large central charge and large supergravity charge radii $Q_i$, there are additional ``ensemble geometries'' which govern the thermodynamics of intermediate phases~\cite{Bena:2011zw}.  These phases fill in the region between the locus $S_{\rm\sst BTZ}=0$ where the BTZ solution ceases to dominate the ensemble, and various unitarity bounds (for instance, the requirement that $\varepsilon>0$).  

\begin{figure}[ht]
\centering
    \includegraphics[width=.6\textwidth]{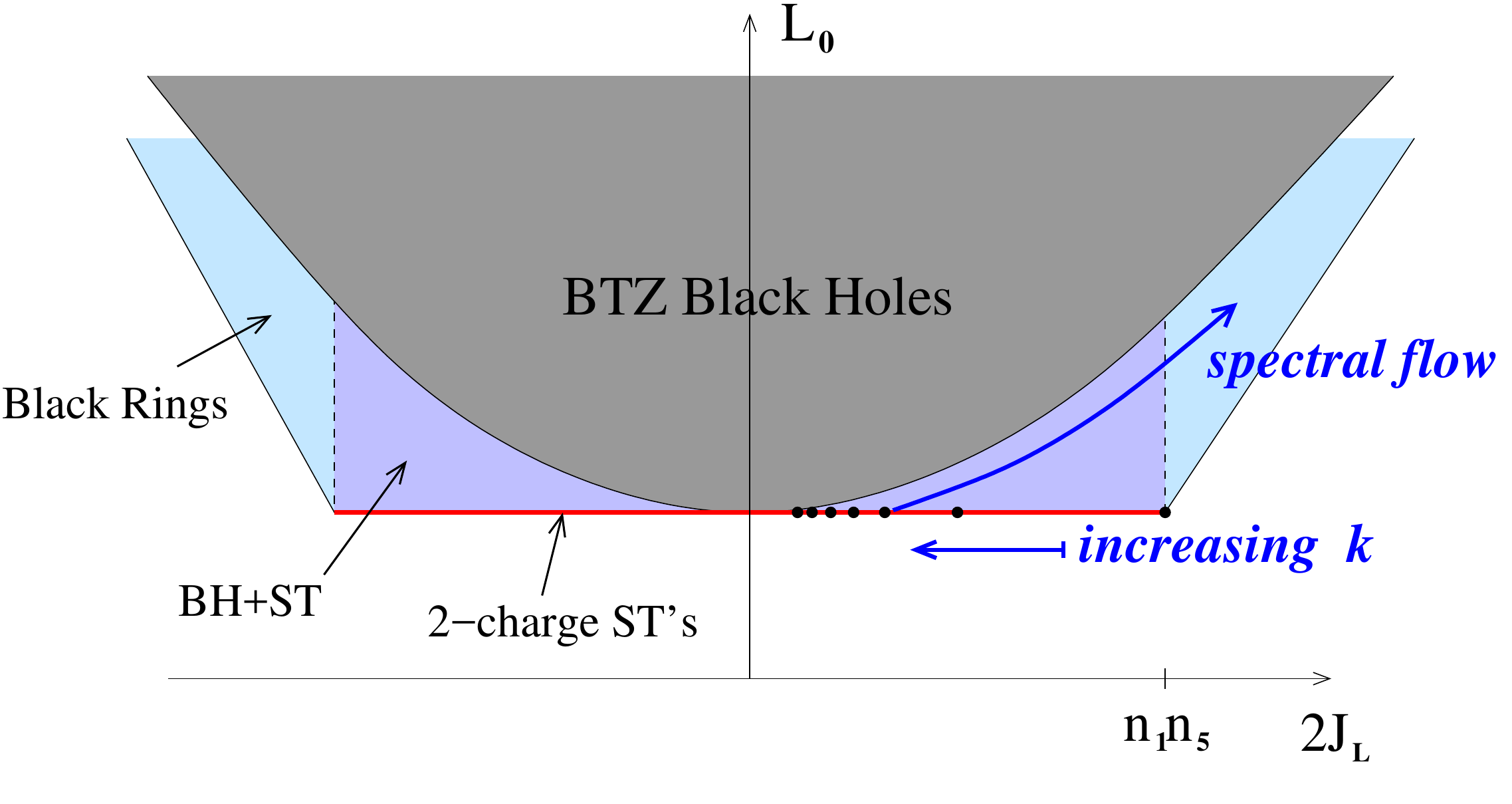}
\caption{\it 
Two-charge supertubes are BPS configurations at the lower bound in energy, and for angular momenta up to $J=n_1n_5/2$.   Between the unitarity bound and the BTZ threshold, there is a black hole plus supertube phase
at low angular momentum (purple), and a black ring phase at high angular momentum (light blue).  Spectral flow in the spacetime CFT takes one phase into the other.
}
\label{fig:Spectrum-moulting}
\end{figure}

The main unitarity bound is a polygon coming from integer spectral flow of the BPS bound (additional unitarity bounds come from the structure of the $\cN=4$ supersymmetric spacetime superconformal algebra~\cite{Boucher:1986bh}).  Outside this bound, there are no states in the spectrum.  Between the bound and the BTZ threshold, the density of states is dominated by a particular black object, depending on the angular momentum.
For $J<n_1n_5/2$, the system likes to carry all its angular momentum in a supertube surrounding a zero angular momentum (BMPV) black hole; for $J>n_1n_5/2$, the dominant configuration is a black ring.%
\footnote{One can check that the density of states of these two configurations flow into one another according to the standard spectral flow relations of the spacetime CFT, which on the gravity side of the duality is simply a large gauge transformation that mixes $AdS_3$ and $\bS^3$ angular coordinates. 
Both objects have horizon topology $\bS^3\times\bS^1$, so it is consistent that they flow into one another.}
This phase structure, worked out in~\cite{Bena:2011zw}, is depicted in Figure~\ref{fig:Spectrum-moulting}.

%%%%%%%%%%%%%%%%%%
%%%%%%%%%%%%%%%%%%

In bipolar coordinates of the sort used in this paper, the black hole plus supertube phase is characterized by a zero angular momentum (BMPV) black hole at $\rho=\theta=0$, and a supertube that carries the angular momentum at $\rho\tight=0$, $\theta\tight=\pi/2$; in the black ring phase, the ring is located at $\rho\tight=\theta\tight=\pi/2$, and $\rho=\theta=0$ is an ordinary smooth region of spacetime.

In the three-charge supertube, there can be orbifold structure at either or both locations $\rho\tight=0,\theta\tight=0$ and $\rho\tight=0,\theta\tight=\pi/2$, supporting W-branes in both places.  How much the orbifold structure coils at each location depends on the diophantine relations among $s$, $s+1$ and $k$ that determine ${\textit gcd}(s,k)\tight=\ell_1$ and ${\textit gcd}(s\tight+1,k)\tight=\ell_2$ (see the discussion around equation~(\ref{factorizations})); for instance the orbifold structure only exists at $\theta=0$ if $\ell_1=1$, or only at $\theta=\pi/2$ if $\ell_2=1$.  There are also potentially large dipole charges, for instance KK monopole charge of amounts $-s$ at $\theta=0$ and $s+1$ at $\theta=\pi/2$, as well as F1 and NS5 dipole charges of magnitude $n_{1,5}s(s+1)/k$ at these locations.  This does not directly fit the black object structure, which blackens only at $\theta=0$ for small angular momentum, and only at $\theta=\pi/2$ for large angular momentum, and in which 
%for the BMPV+tube 
the charge vectors at $\theta=0,\pi/2$ are quite different from the three-charge supertube.  All these features highlight the fact that the three-charge supertube is macroscopically different from an actual three-charge black object at the corresponding point in the phase diagram.  Nevertheless, it is encouraging that the central player in the black object entropy, namely the long string structure, is making an appearance in the cap of the three charge supertube, and in both the places that have the potential to blacken, depending on the route that thermodynamics favors.

%%%%%%%%%%%%%%%%%%%%%%%%%%%%%%%%%%%%
\subsection{W-brane excitations}

The structure of D-branes wrapping topology in the cap of two- and three-charge supertube backgrounds exhibits many features of the long string structure that characterizes the black hole phase of the spacetime CFT.  
In particular, excitations of these objects may teach us about the entropy-carrying degrees of freedom of the black hole phase, \ie\ the long string structure.
The open string spectrum characterizes the ways we can wiggle the W-brane.
While we defer a complete analysis of this spectrum to future work, we can make a few preliminary remarks.
In the open string sector the zero modes are restricted by the boundary conditions.  
Consider the string-like D2-brane in the NS5-P supertube of Figure~\ref{fig:WbraneFig}, for example. 
The brane is extended along the $\bS^2$ (twisted) conjugacy class in $\sutwo$ near $\theta=\pi/2$, and along the supertube helix with the slope~\eqref{helixpitch}, \ie\ $\nfive\, \delta\phi = -k\, \delta(\ytil/\Rytil)$.  In other words, we have a single fivebrane which winds $\nfive$ times around the $\ytil$ circle as it winds $k$ times around the $\phi$ circle, and the W-string tracks that structure.%
\footnote{Additionally the W-brane can wrap around $\bT^4$ as it winds along the supertube helix; or it can wrap only the $\bT^4$.  In the $AdS_3$ limit $\Ry^{-1}=\Rytil\to0$, the brane wrapping only the supertube helix is lightest.}  
A wave along the supertube helix thus has momentum fractionated by a factor $\nfive$ along $\ytil$ and by a factor $k$ along $\phi$.  More precisely, the total $\ytil$ and $\phi$ momenta are integral, but a fraction of each is carried by each coil of the supertube.
These waves can be polarized along any of the $\bT^4$ directions, exhibiting the hypermultiplet of excitations of the long string.  Dualizing to the NS5-F1 frame, the component of the momentum along $\ytil$ turns into fundamental string winding along $y$; the W-string now fractionates F1 charge by a factor $\nfive$.
Similarly, there are $k$ strips of W-brane lying vertically above a given point on the $\phi$ circle, and so open strings stretching between the $i^{\rm th}$ and $j^{th}$ strip stretch a fraction $\frac{|i-j|-1}{k} \Rytil$ around the $\ytil$ circle (for $i\ne j$).  In the T-dual picture one has $y$ momentum fractionated into amounts $\frac{|i-j|-1}{k\Ry}$.

The DBI effective action indicates that W-branes 
are rather heavy excitations compared to fundamental strings.  A rough estimate of the W-string DBI action in the NS5-P frame is given by the value of the dilaton in the cap% 
\footnote{Correcting a typo in~\cite{Martinec:2017ztd}.}
times the area of the $\ytil$-$\phi$ torus at $\rho=0,\theta=\pi/2$
\be
\int e^{-\Phi}\sqrt{\det(G+B+F)} \sim \sqrt{\frac{n_1 \lstr^4}{k^2 \nfive V_4}}\, \frac{\Rytil\sqrt{\nfive}}{\lstr^2} ~.
\ee
The $k$ dependence of this result comes from a slightly different route in the NS5-F1 frame~-- there the dilaton is the fixed scalar value, $\exp[2\Phi]=\nfive V_4/(n_1\lstr^4)$, but the vanishing cycle $\bS^2/\bZ_k$ of Figure~\ref{fig:NS5-vs-KKM} wrapped by the W-string has volume proportional to $1/k$.  The end result is that in the NS5-F1 frame the energy cost of a W-string is of order
\be
E \sim \sqrt{\frac{n_1 \lstr^4}{ V_4}}\, \frac{1}{k\Ry}~.
\ee
This has the appropriate scaling for the $AdS_3$ decoupling limit, where one holds $E\Ry$ fixed, but the deepest supertube throat has $k\sim n_1$ (since we have demanded that $k$ and $\nfive$ are relatively prime, and $k$ is bounded above by $n_1\nfive$).  Thus when we consider a supertube with a deep throat, the W-string is heavier than the lightest supergravity excitations in the bottom of the throat by a factor of order $\sqrt{n_1/V_4}$.  The volume $V_4/\lstr^4$ of the compactification is bounded by $n_1/n_5$ in order for the F1-NS5 description to be valid; beyond that, the valid weakly-coupled effective description switches to the S-dual D1-D5 frame (see for instance~\cite{Giveon:1998ns}).  Indeed we see that the W-string becomes lighter and lighter as we increase the torus volume and at the limit of validity of the NS5-F1 duality frame becomes as light as supergravity modes in the cap.  

This convergence of excitation energy scales is due to the fixed scalar condition in the cap of the NS5-F1 geometry,
\be
\exp[2\Phi] = g_s^2\frac{Q_5}{Q_1} = \frac{n_5 V_4}{n_1 \lstr^4} ~,
\ee
which strikes a balance between the onebrane and fivebrane charge radii.  Increasing the torus volume shifts the balance in favor of the fivebranes and increases the string coupling, making D-branes lighter.
Similarly, the radius $a$ of the supertube ring in the NS5-P frame (see Figure~\ref{fig:Supertube}),
\be
a^2 = \frac{Q_pQ_5\Rytil^2}{k^2\lstr^4} = \frac{n_5n_p g_s^2 \lstr^6}{k^2 V_4} ~,
\ee 
is determined by a balance between the tension of the fivebranes wanting to shrink the supertube radius and the angular momentum forcing the fivebranes to stretch; increasing the torus volume again makes the fivebranes heavier and shrinks the supertube radius, pushing the fivebranes closer together and making the W-string lighter.

Thus W-branes become light and compete with fundamental strings when the fivebranes approach one another, as expected from the picture painted in the introduction.  But when the fivebrane strands are well-separated and the W-branes are heavy, the supertube has a certain rigidity to it.  The fivebrane strands cannot typically come together unless either the supertube sheds its angular momentum~\cite{Marolf:2016nwu}, or two windings of the supertube cross through evolution of its transverse profile and start locally exciting W-branes.  Both processes require some energy to be supplied.%
\footnote{In the classical GR analysis of~\cite{Eperon:2016cdd} excitations can sit at the supertube locus where they cost zero energy, but as shown in~\cite{Martinec:2018nco} there is a gap in the spectrum.}
One of the excitations that costs rather little energy is to pry a fundamental string loose from the background.  In~\cite{Martinec:2018nco} it was shown that large gauge transformations in the $\sltwo$ factor of $\cG$ mediate processes by which F1 winding charge dissolved as background flux can be transferred to winding string excitations not bound to the cap.   The lightest perturbative string scattering states carry the same momentum and angular momentum per unit F1 charge as the background, and cost energy above the BPS bound (see~\cite[Eq.\;(4.63)]{Martinec:2018nco}),
\be
\varepsilon = E\Ry = \frac{\nfive\bigl[ s(s+1) +1 \bigr]w_y}{k^2}
~,~~~~~
n_y = P_y\Ry = \frac{\nfive s(s+1) w_y}{k^2}~,
\ee
where $w_y$ is the number of units of F1 winding carried by the string.  Thus as $k$ increases, these become very easy to excite, and their effect is to strip off the background F1 charge which is keeping the cap structure weakly coupled.  When enough energy is supplied so that enough charge is stripped away, the supertube shrinks to the point that the W-branes become competitive with elementary string excitations, and one starts to enter the black hole phase of thermally excited long/little strings.

Even if the W-brane is somewhat heavy, once one has paid the cost of creating it, further excitations are expected to be relatively light.  We hope in future work to analyze the spectrum of open strings on the W-branes constructed above, and estimate their contribution to the density of states near the black hole threshold.

%%%%%%%%%%%%%%%%%%%%%%%%%%%%%%%%%%%%
\subsection{The role of microstate geometries}

As we have emphasized, the primary role of microstate geometries in our considerations is to bring us near to the black hole threshold where we can study the most entropic degrees of freedom, which are excitations of the long/little string.  The round supertubes we have studied are quite close to the black hole threshold but have high curvature (an orbifold structure) in the cap.  The structure of topological bubbles here is thus nearly degenerate~-- the bubbles are sub-string scale in size so that the W-branes that wrap them are light.

There is an extensive zoology of smooth microstate geometries with large redshift to the bottom of the cap, and one might wonder how they fit into this picture.  There are two broad classes of such geometries that have been considered in the literature~-- bubbled geometries that carry all three background charges as fluxes~\cite{Bena:2007kg}, and so-called {\it superstrata}~\cite{Bena:2015bea} which decorate a two-charge supertube with a fully back-reacted (super)gravitational wave profile.  In both cases the background geometry is supported by angular momentum, and an issue is to determine how easily the background can shed that angular momentum and drive the configuration to a regime where other excitations become light.  In the case of bubbled geometries, it was argued in~\cite{Martinec:2015pfa} that (an admittedly crude) quiver quantum mechanics truncation of the dynamics of W-branes exhibits a finite fraction (a few percent) of the three-charge black hole entropy in the Higgs phase of the QM where the hypermultiplets of the quiver, whose quanta are the W-branes, have become light and condense.  This result suggests that once again the route to the black hole phase proceeds via a process in which the background sheds its angular momentum and the cap descends to a redshift where the entropic degrees of freedom become light enough to play a significant role.

Superstrata can support a deep $AdS_2$ throat and an approximate BTZ geometry via coherent supergravity waves on top of a two-charge supertube background~\cite{Bena:2017xbt}.  The fully nonlinear field equations are solved, with the wave profile tuned to avoid singularities and other pathologies such as closed timelike curves.  However, it has been shown~\cite{Tyukov:2017uig,Bena:2018mpb} that probes of this geometry experience large tidal forces, an indicator that the large blueshift experienced by the probe will result in processes well-approximated by the collision of gravitational shockwaves, resulting in a disruption of the delicately tuned superstratum structure.  There will be plenty of excitation energy available to transfer angular momentum away from the supertube, causing it to evolve toward the regime where W-branes become light.

We should mention that our proposal -- that W-brane excitations of supertubes are a precursor of the entropic degrees of freedom of black holes -- shares some similarity with earlier attempts to use supertube probes in microstate geometries for a similar purpose~\cite{Bena:2008nh,Bena:2008dw}.  In both cases, excitations of a string-like probe (or the U-dual of one) deep in the throat of a capped geometry are proposed as a way to account for black hole entropy in asymptotically $AdS_3$ spacetimes.  The backreaction of the probe is treated as a small correction.  There are however two significant differences.  First, we have taken pains to distinguish these configurations from generic black hole microstates, arguing that the three-charge microstate geometries constructed to date are not generic elements of the black hole phase, but rather particular coherent states.  Even their W-brane excitations should not be considered generic black hole microstates, but rather useful probes that exhibit certain features of the black hole phase in a context where we can apply perturbative string theory methods. The true black hole phase involves little string dynamics, and is inherently strongly coupled.%
\footnote{By which we mean having an effective coupling strength of order one -- not coupling far in excess of unity, which might lead to some other weak coupling approximation in a dual description.}
Second, the supertubes of~\cite{Bena:2008nh,Bena:2008dw} do not exhibit the charge fractionation that one expects of the entropic degrees of freedom; indeed, estimates of the entropy that could be achieved this way fell short of the parametric $\sqrt{\vphantom{t}n_5n_1n_p}$ growth of the BTZ black hole entropy, in large part because the fractionation coming from the third charge is absent in the two-charge supertube probes that were employed.  In our approach we have sought the source of this additional fractionation by finding where the fivebranes are hiding in the geometry.  The fact that the supertube winds $\nfive$ times around the $y$ circle means that the wiggles of the W-string will carry the necessary additional fractionation beyond the order $\sqrt{\vphantom{t}n_1n_p}$ entropy of excited fundamental string probes in the fivebrane throat~\cite{Giveon:2015raa}.

%%%%%%%%%%%%%%%%%%%%%%%%%%%%%%%%%%%%
\subsection{Speculations on the black hole phase}

A major difference between $AdS_3$ holography at large $\nfive$ and the perturbative string correspondence point studies described in the introduction is that there are now {\it two} string scales~-- the scale $\alpha'$ of the fundamental string and the scale $\apl=\nfive\alpha'$ of the little string, which are widely separated when $\nfive$ is large.  Physics that looks local and semiclassical to the fundamental string can look stringy and quantum to the little string.  For instance, the proper time from the bifurcation point of the BTZ horizon to the singularity is the curvature scale $R_{AdS}$ which is the inverse tension scale of the little string; the little string literally could not tell whether it is at the horizon or the singularity of the effective geometry. (Here we are assuming that effective geometry holds at the horizon -- that the effect of the little string is not felt through violent collisions but rather through soft momentum transfers of order the little string tension scale $(\apl)^{-1/2}=R_{AdS}^{-1}$, which perhaps not coincidentally is the scale of tidal forces in the effective geometry).  Of course, inside the horizon one expects that a probe fundamental string is fractionating into bits of little string, like a gauge theory meson shattering into partons upon entering a large nucleus; the picture of a localized fundamental string would be a description of the collective degrees of freedom of the underlying little string, much as the partons that compose the jet made by a probe meson entering the nucleus continue along the same center of mass trajectory of the original meson, but after entering the nucleus are spread out over a narrow cone in the transverse direction.%
\footnote{This picture has some affinity with the ``fuzzball complementarity'' scenario of~\cite{Mathur:2012jk,Mathur:2013gua} in that the physics of localized jets experiencing small transverse momentum interactions applies in a regime where the jet has momentum much larger than the string tension and is thus localized inside a narrow cone, while the evolution of probes of low momentum is strongly affected by parton interactions.}
To the extent that this spreading can be ignored, the center-of-mass trajectories of massless probes in localized wavepackets would reflect the causal structure of the {\it effective} geometry.  Whether one can regard this structure as ``real'' depends on the extent to which it can be decoupled from the spreading of the probe into the underlying little string degrees of freedom.  For the physics of outgoing trajectories near the effective horizon, this issue is almost certainly bound up with the maximally chaotic nature of dynamics there~\cite{Shenker:2013pqa,Maldacena:2015waa}.

The main difference between the interior and exterior is the deconfinement of the little string.  One imagines that a fundamental string probe entering the region behind the horizon will quickly fractionate.  To the extent that there is a geometrical picture of the black hole interior, it will be because one can still approximate the dynamics of probes by the local center-of-mass of their fractionated constituents, at least for a short time, until the fractionated constituents scramble with and thermalize into the little string ensemble.  In the effective description, this might amount to the probe hitting a singularity, signaling the breakdown of that description.  

From the present perspective, the question of whether something dramatic happens to an infalling probe when it crosses the black hole horizon is a question regarding the response function of the little string to probes made up of ordinary (F1) strings.  The tension of the latter is much larger, so geometry might look localized to F1's and completely different and fuzzy to little strings.  For there to be something akin to a firewall~\cite{Almheiri:2012rt} would require the little string to exhibit behavior dramatically different from that of fundamental strings, which do not transmit strong impulsive forces~\cite{Gross:1987ar}.
Instead, each string interaction typically transfers momentum on the order of the string scale, which for the little string is the curvature scale of the ambient geometry~-- little string interactions with the probe's collective modes might simply be seen as tidal forces from the perspective of the effective theory.

%%%%%%%%%%%%%%%%%%

A satisfactory resolution of the black hole information paradox should identify where and how Hawking's original calculation of black hole radiation makes a mistake.  
A similar issue arises in the phenomenon of ergoregion emission, as happens for instance in the JMaRT geometries~\cite{Jejjala:2005yu}, which are obtained from two-charge supertubes by non-supersymmetric spectral flow in the spacetime CFT~\cite{Chakrabarty:2015foa,Chowdhury:2007jx,Chowdhury:2008bd,Chowdhury:2008uj,Avery:2009xr}.  
In perturbative string theory, one sees in JMaRT backgrounds a process quite similar to Hawking pair creation~\cite{Cardoso:2005gj,Chowdhury:2007jx,Chowdhury:2008bd,Chowdhury:2008uj,Avery:2009xr,Chakrabarty:2015foa}.%
\footnote{However, the particle interpretation of the ergoregion emission process is quite subtle~\cite{Schiff:1940,Fulling:1989nb}.}
A probe scalar field in the background generates a pool of negative energy in the vicinity of the ergoregion at the same time that positive energy radiation escapes to infinity%
~\cite{Cardoso:2005gj,Chowdhury:2007jx,Chowdhury:2008bd,Chowdhury:2008uj,Avery:2009xr,Chakrabarty:2015foa,Martinec:2018nco} 
(where energy is defined in terms of the timelike Killing vector of the asymptotically flat region that the $AdS_3$ throat is joined to).  The backreaction of that radiation on the background will relax the latter by shrinking the ergoregion, though this backreaction has not been computed in detail.

In both cases -- ergoregion emission and Hawking radiation -- the probe approximation treats the background as fixed and one calculates the dynamics of field modes on that fixed background; the backreaction of the modes and in particular their entanglement with the background are ignored altogether.  In ergoregion emission one has the possibility to patch that up after the fact; the process by which the pool of negative energy modes in the cap relaxes the background doesn't necessarily violate cherished principles of local quantum field theory such as locality, causality or unitarity (though one faces puzzles regarding how the negative energy ``annihilates'' against the background, especially if the probe field carries some conserved quantum number, that are reminiscent of the sort of entanglement puzzles~\cite{Mathur:2009hf,Almheiri:2012rt} that have animated recent discussions of black hole radiance).  But in the Hawking process, one of the cherished principles of local QFT must be violated.

In the dual spacetime CFT, the decay of the JMaRT background is described as {\it de-excitation} of a particular collection of fermionic modes which have been macroscopically excited in order to create the JMaRT background from a two-charge ground state; each emitted quantum is directly responsible for relaxing a particular excitation of the underlying microstate, and so entanglement is properly looked after.  In fact, the mechanism by which the spacetime CFT relaxes is identical to the process by which Hawking quanta are emitted -- the only difference is in the initial state, which is a highly coherent excitation in the JMaRT states rather than the thermally excited structure of a typical black hole microstate.  In both the JMaRT background and in black hole microstates, the ``negative energy partner'' of the emitted radiation is simply a de-excitation of the underlying little strings, whose dynamics is causally connected to their surroundings; there is then no fundamental issue with the unitarity of the emission process and the preservation of quantum correlations among the constituents of the final state.

In the fuzzball paradigm, the black hole is simply a complicated bound state of fractionated brane constituents whose wavefunction extends out to the horizon scale; the black hole is yet another phase of matter, and radiates coherently from its surface.  
In the present context, Hawking quanta emerge from the Hagedorn gas of little strings when a fundamental (F1) string assembles itself out of its fractionated constituents and escapes the fuzzball.
One might then regard the effective field theory in which the Hawking calculation is performed, 
wherein field modes are continually drawn from some reservoir in the UV and stretched to macroscopic scales, 
as some approximation to the fuzzball dynamics that ignores much of its internal structure.  There are effective field theories in other contexts that similarly ignore correlations and entanglement between quanta of the effective theory and the underlying constituents.  In effective field theory of materials, the quasiparticles and quasiholes near the ground state are complicated, correlated excitations of the underlying constituents.  
In the Hawking process, the emitted quanta are correlated to ``negative energy'' partners behind the horizon that are akin to quasiholes in a material, whose lifetime is at most the scrambling time of the black hole.  But as in the material, exciting such a quantum amounts to the {\it absence} of an excitation in the underlying microscopics.  Isolating the effective field theory and ignoring the correlations of its excitations to the underlying substrate amounts to a sort of mean field theory which does not capture the correlations, leading to a seeming breakdown of unitarity which is simply an artifact of the approximation.

%%%%%%%%%%%%%%%%%%

%%%%%%%%%%%%%
%\vskip 0.7cm
\section*{Acknowledgements}
%%%%%%%%%%%%%

We thank 
Iosif Bena, Tristan Madeleine and Nicholas Warner for useful discussions.
%.
The work of EJM and SM is supported in part by DOE grant DE-SC0009924 and a UChicago FACCTS collaboration grant. 
EJM also thanks the Simons Foundation for support as a Fellow in Theoretical Physics for 2018-19, and the KITP for hospitality at its ChOrd workshop during the course of this work, supported in part by the National Science Foundation under Grant No. NSF PHY-1748958.
The work of DT is supported by a Royal Society Tata University Research Fellowship. For hospitality during the course of this work, we all thank the CEA Saclay; 
DT thanks the Enrico Fermi Institute at the University of Chicago, while EJM and SM thank the University of Southampton.

%%%%%%%%%%%%%%%%%%%%%%%%%%%%%%%%%%%%%
%%%%%%%%%%%%%%%%%%%%%%%%%%%%%%%%%%%%%

%\vskip 9mm
\appendix

%%%%%%%%%%%%%%%%%%%%%%%%%%%%%%%%%%%%
%%%%%%%%%%%%%%%%%%%%%%%%%%%%%%%%%%%%

%\newpage
\section{Conventions}
\label{app:conventions}

%%%%%%%%%%%%%%%%%%%%%%%%%%%%%%%%%%%%

In this appendix we record our group theory conventions.

\refstepcounter{subsection}
\subsection*{\thesubsection \quad SU(2)} \label{sec:sutwoapp}
%\subsection{SU(2)}\label{sec:sutwoapp}

We parametrize SU(2) via Euler angles
\be \label{eq:su2-param}
g_{\su} ~=~ e^{\frac{i}{2}(\psi-\phi)\sigma_3}e^{i \theta \sigma_1}e^{\frac{i}{2}(\psi + \phi)\sigma_3} 
~= ~
\begin{pmatrix} \cos\theta \;\! e^{i\psi}  &~~  i \:\!  \sin\theta\:\!  e^{-i\phi} \\ 
i \:\!  \sin\theta\:\!  e^{i\phi}  &~~ \cos\theta \;\! e^{-i\psi} \end{pmatrix}\,
\,.
\ee
These are the conventions used in~\cite{Martinec:2018nco} (note that~\cite{Martinec:2017ztd} had conventions related to these by $\phi \to - \phi$).
Here the $\sigma_a$ are the usual Pauli matrices, explicitly
\be
\sigma_1 = \left(\begin{array}{cc} 0 &1 \\ 1 &
    0 \end{array}\right), \qquad \sigma_2 = \left(\begin{array}{cc} 0 &-i \\ i &
    0  \end{array}\right), \qquad \sigma_3 = \left(\begin{array}{cc} 1 &0 \\ 0 &
    -1  \end{array}\right) \,.
\ee
The generators and structure constants of the Lie algebra $\su(2)$ are as usual
\be
T_a = \frac{1}{2}\sigma_a \,, \qquad \quad f_{ab}^{~\;c} = i\:\! \epsilon_{abc} \,, \qquad  \epsilon_{123}=1\,.
\ee
The group element parametrizes the unit sphere in $\bR^4$ through
\begin{equation}\label{SU2param1}
g_{\su} = 
\begin{pmatrix} X^0 + i X^3  &~~  iX^1 + X^2 \\ i X^1 -  X^2
&~~ X^0 - i X^3\end{pmatrix}\, ,
\end{equation}
where
\begin{equation} \label{eq:Xi}
X^0 + i X^3 = \cos \theta e^{i \psi} \, ,\quad X^1 + i X^2 = \sin\theta e^{i \phi} \, .
\end{equation}
Another parameterization is given by the unit quaternions
\begin{equation}
\label{SU2param2}
g = \cos \chi \, \One + i \sin\chi \Big( \sin \vartheta \cos\varphi \,\sigma_1
+\sin \vartheta \sin  \varphi \,\sigma_2 + \cos \vartheta \, \sigma_3\Big) \, ,
\end{equation}
for which
\begin{align}
X^0 = \cos \chi \, ,\quad X^1+ i X^2 = \sin\chi \sin \vartheta\, e^{ i
  \varphi}\, ,\quad X^3 = \sin \chi \cos \vartheta \, .
\end{align}
We refer to these as hyperspherical coordinates, as they describe the group manifold $\bS^3$ as the foliation of the three-sphere by the $\bS^2$'s parametrized by $\vartheta,\varphi$; these two-spheres are untwisted conjugacy classes of the group given by $\chi=const$.
The relation between the two parametrizations is given by 
\begin{equation}
\label{Euler-hyperspherical relation}
\cos \theta \cos \psi = \cos \chi \, ,\quad~ \cos \theta \sin \psi =
\sin \chi \cos \vartheta \, , \quad~  \sin \theta = \sin \chi\sin \vartheta   \,
, \quad~ \varphi = \phi \, . 
\end{equation}
The currents $J_3^{\su}$, $\bar{J}_3^{\su}$ are given by
\bea \label{eq:J3su}
J_3^{\su} &=& \nfive \tr \left[ (-i T_3 )\partial g_{\su} \;\! g_{\su}^{-1}  \right] ~=~  \nfive\bigl( \cos^2\!\theta \, \partial \psi - \sin^2 \!\theta \,\partial \phi \bigr) \,, \phantom{\Big\{}\cr
\bar{J}_3^{\su} &=&\nfive\tr \left[ (-i T_3) g_{\su}^{-1} \bar\partial g_{\su} \;\!   \right] ~=~ \nfive\bigl( \cos^2\!\theta \,\bar\partial \psi + \sin^2 \!\theta \, \bar\partial \phi \bigr) \,.
\eea

Defining in the usual way $T_\pm ~\;\equiv\; ~ T_1 \pm i T_2$, the left-invariant vector fields that correspond to the generators $T_a$, in the local coordinates in \eq{eq:su2-param}, are
\begin{align}
\begin{aligned}
(T_+)^L &~=~ -\frac{i}{2} e^{+i(\psi+\phi)} \bigl({\partial_\theta} - i \tan \theta \, {\partial_\psi} + i \cot \theta \,{\partial_\phi}\bigr), \\[3pt]
(T_-)^L &~=~  +\frac{i}{2} e^{-i(\psi+\phi)} \bigl({\partial_\theta} + i \tan \theta \, {\partial_\psi} - i \cot \theta\,{\partial_\phi}\bigr),\\[3pt]
(T_3)^L &~=~  -\frac{i}{2} \bigl({\partial_\psi} + {\partial_\phi}\bigr). 
\end{aligned}
\end{align}
Our conventions for the right-invariant vector fields follow those of \cite{Figueroa-OFarrill:2005vws}, as mentioned around \eq{eq:mc-1-forms}--\eq{theta-on-vec}. Thus we define the right-invariant vector fields to correspond to \emph{minus} the relevant Lie algebra element, i.e.
\be
 (T_3)^R  \quad \leftrightarrow \quad - T_3  \qquad \Rightarrow \qquad 
%\ee
%giving for instance
%\be
 (T_3)^R ~~=~~  \frac{i}{2} \left(\frac{\partial}{\partial \psi} - \frac{\partial}{\partial \phi}\right) . 
\ee

%%%%%%%%%%%%%%%%%%%%%%%%%%%%%%%%%%%%%%%

\refstepcounter{subsection}
\subsection*{\thesubsection \quad SL(2)} \label{sec:sltwoapp}
%\subsection{SL(2)}\label{sec:sltwoapp}

We parametrize $\sltwo$ as $SU(1,1)$ via
\be
g_{\sl} \;=\; e^{\frac{i}{2}(\tau-\sigma)\sigma_3}e^{\rho \sigma_1}e^{\frac{i}{2}(\tau + \sigma)\sigma_3} \,.
\ee
Again these are the same conventions we used in~\cite{Martinec:2018nco} (related to those of~\cite{Martinec:2017ztd} by $\sigma \to - \sigma$).
The generators and structure constants of the Lie algebra $\su(1,1)$ are 
\be
T_1^{\sl} \,=\, \frac{i}{2}\sigma_1 \,, \quad~ T_2^{\sl} \,=\, \frac{i}{2}\sigma_2 \,, \quad~ T_3^{\sl} \,=\, T_3 \,=\, \frac{1}{2}\sigma_3 \;; \qquad~~%
 f_{12}^{~\;3} \,=\, -i \,, \quad  f_{23}^{~\;1} \,=\, f_{31}^{~\;2} \,=\,  i \,.
\ee
The currents $J_3^{\sl}$, $\bar{J}_3^{\sl}$ are
\bea \label{eq:J3sl}
J_3^{\sl} &=& \nfive \tr \left[ (-i T_3 )\partial g_{\sl} \;\! g_{\sl}^{-1}  \right] ~=~ \nfive\bigl(\cosh^2\! \rho \,\partial\tau + \sinh^2 \!\rho \,\partial \sigma \bigr) \,,\nn\\[5pt]
\bar{J}_3^{\sl} &=& \nfive\tr \left[ (-i T_3) g_{\sl}^{-1} \bar\partial g_{\sl} \;\!   \right] ~=~ \nfive\bigl(\cosh^2\! \rho\, \bar\partial\tau - \sinh^2\! \rho \,\bar\partial\sigma \bigr) \,.
\eea
The left and right-invariant vector fields are similar to those given above for $\sutwo$, via the map
\be
\psi \to \tau \,, \qquad \phi \to \sigma \,, \qquad i\theta \to \rho \,.
\ee
In particular, we have
\be
(T_3^{\sl})^L ~=~  -\frac{i}{2} \left(\frac{\partial}{\partial \tau} + \frac{\partial}{\partial \sigma}\right) \,,
\qquad \quad
%\ee
%and 
%\be
 (T_3^{\sl})^R ~=~  \frac{i}{2} \left(\frac{\partial}{\partial \tau} - \frac{\partial}{\partial \sigma}\right) \,.
\ee

%%%%%%%%%%%%%%%%%%%%%%%%%%%%%%%%%%%%
%%%%%%%%%%%%%%%%%%%%%%%%%%%%%%%%%%%%

%%%%%%%%%%%%%%%%%%%%%%%%%%%%%%%%%%%%
%%%%%%%%%%%%%%%%%%%%%%%%%%%%%%%%%%%%

%\providecommand{\href}[2]{#2}\begingroup\raggedright\begin{thebibliography}{10}

\newpage
\vskip 1cm

\bibliographystyle{JHEP}      

\bibliography{microstates}

\providecommand{\href}[2]{#2}\begingroup\raggedright\begin{thebibliography}{100}

\bibitem{Seiberg:1997zk}
N.~Seiberg, \emph{{New theories in six-dimensions and matrix description of M
  theory on T**5 and T**5 / Z(2)}},
  \href{http://dx.doi.org/10.1016/S0370-2693(97)00805-8}{\emph{Phys. Lett.}
  {\bfseries B408} (1997) 98--104},
  [\href{https://arxiv.org/abs/hep-th/9705221}{{\ttfamily hep-th/9705221}}].

\bibitem{Aharony:1998ub}
O.~Aharony, M.~Berkooz, D.~Kutasov and N.~Seiberg, \emph{{Linear dilatons, NS
  five-branes and holography}},
  \href{http://dx.doi.org/10.1088/1126-6708/1998/10/004}{\emph{JHEP} {\bfseries
  10} (1998) 004}, [\href{https://arxiv.org/abs/hep-th/9808149}{{\ttfamily
  hep-th/9808149}}].

\bibitem{Aharony:1999ks}
O.~Aharony, \emph{{A Brief review of 'little string theories'}},
  \href{http://dx.doi.org/10.1088/0264-9381/17/5/302}{\emph{Class. Quant.
  Grav.} {\bfseries 17} (2000) 929--938},
  [\href{https://arxiv.org/abs/hep-th/9911147}{{\ttfamily hep-th/9911147}}].

\bibitem{Kutasov:2001uf}
D.~Kutasov, \emph{{Introduction to little string theory}}, {\emph{ICTP Lect.
  Notes Ser.} {\bfseries 7} (2002) 165--209}.

\bibitem{Maldacena:1996ya}
J.~M. Maldacena, \emph{{Statistical entropy of near extremal five-branes}},
  \href{http://dx.doi.org/10.1016/0550-3213(96)00368-9}{\emph{Nucl. Phys.}
  {\bfseries B477} (1996) 168--174},
  [\href{https://arxiv.org/abs/hep-th/9605016}{{\ttfamily hep-th/9605016}}].

\bibitem{Horowitz:1996nw}
G.~T. Horowitz and J.~Polchinski, \emph{{A correspondence principle for black
  holes and strings}},
  \href{http://dx.doi.org/10.1103/PhysRevD.55.6189}{\emph{Phys. Rev.}
  {\bfseries D55} (1997) 6189--6197},
  [\href{https://arxiv.org/abs/hep-th/9612146}{{\ttfamily hep-th/9612146}}].

\bibitem{Giveon:2005mi}
A.~Giveon, D.~Kutasov, E.~Rabinovici and A.~Sever, \emph{{Phases of quantum
  gravity in AdS(3) and linear dilaton backgrounds}},
  \href{http://dx.doi.org/10.1016/j.nuclphysb.2005.04.015}{\emph{Nucl. Phys.}
  {\bfseries B719} (2005) 3--34},
  [\href{https://arxiv.org/abs/hep-th/0503121}{{\ttfamily hep-th/0503121}}].

\bibitem{Kutasov:2005rr}
D.~Kutasov, \emph{{Accelerating branes and the string/black hole transition}},
  \href{https://arxiv.org/abs/hep-th/0509170}{{\ttfamily hep-th/0509170}}.

\bibitem{Mathur:2005zp}
S.~D. Mathur, \emph{{The fuzzball proposal for black holes: An elementary
  review}}, \href{http://dx.doi.org/10.1002/prop.200410203}{\emph{Fortsch.
  Phys.} {\bfseries 53} (2005) 793--827},
  [\href{https://arxiv.org/abs/hep-th/0502050}{{\ttfamily hep-th/0502050}}].

\bibitem{Bena:2007kg}
I.~Bena and N.~P. Warner, \emph{{Black holes, black rings and their
  microstates}}, \href{http://dx.doi.org/10.1007/978-3-540-79523-0}{\emph{Lect.
  Notes Phys.} {\bfseries 755} (2008) 1--92},
  [\href{https://arxiv.org/abs/hep-th/0701216}{{\ttfamily hep-th/0701216}}].

\bibitem{deBoer:2009un}
J.~de~Boer, S.~El-Showk, I.~Messamah and D.~Van~den Bleeken, \emph{{A bound on
  the entropy of supergravity?}},
  \href{http://dx.doi.org/10.1007/JHEP02(2010)062}{\emph{JHEP} {\bfseries 02}
  (2010) 062}, [\href{https://arxiv.org/abs/0906.0011}{{\ttfamily 0906.0011}}].

\bibitem{Bena:2012hf}
I.~Bena, M.~Berkooz, J.~de~Boer, S.~El-Showk and D.~Van~den Bleeken,
  \emph{{Scaling BPS Solutions and pure-Higgs States}},
  \href{http://dx.doi.org/10.1007/JHEP11(2012)171}{\emph{JHEP} {\bfseries 1211}
  (2012) 171}, [\href{https://arxiv.org/abs/1205.5023}{{\ttfamily 1205.5023}}].

\bibitem{Martinec:2015pfa}
E.~J. Martinec and B.~E. Niehoff, \emph{{Hair-brane Ideas on the Horizon}},
  \href{http://dx.doi.org/10.1007/JHEP11(2015)195}{\emph{JHEP} {\bfseries 11}
  (2015) 195}, [\href{https://arxiv.org/abs/1509.00044}{{\ttfamily
  1509.00044}}].

\bibitem{Eperon:2016cdd}
F.~C. Eperon, H.~S. Reall and J.~E. Santos, \emph{{Instability of
  supersymmetric microstate geometries}},
  \href{http://dx.doi.org/10.1007/JHEP10(2016)031}{\emph{JHEP} {\bfseries 10}
  (2016) 031}, [\href{https://arxiv.org/abs/1607.06828}{{\ttfamily
  1607.06828}}].

\bibitem{Raju:2018xue}
S.~Raju and P.~Shrivastava, \emph{{A Critique of the Fuzzball Program}},
  {\emph{Phys. Rev.} {\bfseries D99} (2019) 066009},
  [\href{https://arxiv.org/abs/1804.10616}{{\ttfamily 1804.10616}}].

\bibitem{Bena:2015bea}
I.~Bena, S.~Giusto, R.~Russo, M.~Shigemori and N.~P. Warner, \emph{{Habemus
  Superstratum! A constructive proof of the existence of superstrata}},
  \href{http://dx.doi.org/10.1007/JHEP05(2015)110}{\emph{JHEP} {\bfseries 05}
  (2015) 110}, [\href{https://arxiv.org/abs/1503.01463}{{\ttfamily
  1503.01463}}].

\bibitem{Bena:2016agb}
I.~Bena, E.~Martinec, D.~Turton and N.~P. Warner, \emph{{Momentum Fractionation
  on Superstrata}},
  \href{http://dx.doi.org/10.1007/JHEP05(2016)064}{\emph{JHEP} {\bfseries 05}
  (2016) 064}, [\href{https://arxiv.org/abs/1601.05805}{{\ttfamily
  1601.05805}}].

\bibitem{Bena:2016ypk}
I.~Bena, S.~Giusto, E.~J. Martinec, R.~Russo, M.~Shigemori, D.~Turton et~al.,
  \emph{{Smooth horizonless geometries deep inside the black-hole regime}},
  \href{http://dx.doi.org/10.1103/PhysRevLett.117.201601}{\emph{Phys. Rev.
  Lett.} {\bfseries 117} (2016) 201601},
  [\href{https://arxiv.org/abs/1607.03908}{{\ttfamily 1607.03908}}].

\bibitem{Bena:2017geu}
I.~Bena, E.~Martinec, D.~Turton and N.~P. Warner, \emph{{M-theory Superstrata
  and the MSW String}},
  \href{http://dx.doi.org/10.1007/JHEP06(2017)137}{\emph{JHEP} {\bfseries 06}
  (2017) 137}, [\href{https://arxiv.org/abs/1703.10171}{{\ttfamily
  1703.10171}}].

\bibitem{Bena:2017upb}
I.~Bena, D.~Turton, R.~Walker and N.~P. Warner, \emph{{Integrability and
  Black-Hole Microstate Geometries}},
  \href{http://dx.doi.org/10.1007/JHEP11(2017)021}{\emph{JHEP} {\bfseries 11}
  (2017) 021}, [\href{https://arxiv.org/abs/1709.01107}{{\ttfamily
  1709.01107}}].

\bibitem{Bena:2017xbt}
I.~Bena, S.~Giusto, E.~J. Martinec, R.~Russo, M.~Shigemori, D.~Turton et~al.,
  \emph{{Asymptotically-flat supergravity solutions deep inside the black-hole
  regime}}, \href{http://dx.doi.org/10.1007/JHEP02(2018)014}{\emph{JHEP}
  {\bfseries 02} (2018) 014},
  [\href{https://arxiv.org/abs/1711.10474}{{\ttfamily 1711.10474}}].

\bibitem{Bena:2018bbd}
I.~Bena, P.~Heidmann and D.~Turton, \emph{{AdS$_{2}$ holography: mind the
  cap}}, \href{http://dx.doi.org/10.1007/JHEP12(2018)028}{\emph{JHEP}
  {\bfseries 12} (2018) 028},
  [\href{https://arxiv.org/abs/1806.02834}{{\ttfamily 1806.02834}}].

\bibitem{Bakhshaei:2018vux}
E.~Bakhshaei and A.~Bombini, \emph{{Three-charge superstrata with internal
  excitations}}, \href{http://dx.doi.org/10.1088/1361-6382/ab01bc}{\emph{Class.
  Quant. Grav.} {\bfseries 36} (2019) 055001},
  [\href{https://arxiv.org/abs/1811.00067}{{\ttfamily 1811.00067}}].

\bibitem{Bena:2018mpb}
I.~Bena, E.~J. Martinec, R.~Walker and N.~P. Warner, \emph{{Early Scrambling
  and Capped BTZ Geometries}},
  \href{https://arxiv.org/abs/1812.05110}{{\ttfamily 1812.05110}}.

\bibitem{Ceplak:2018pws}
N.~Ceplak, R.~Russo and M.~Shigemori, \emph{{Supercharging Superstrata}},
  \href{http://dx.doi.org/10.1007/JHEP03(2019)095}{\emph{JHEP} {\bfseries 03}
  (2019) 095}, [\href{https://arxiv.org/abs/1812.08761}{{\ttfamily
  1812.08761}}].

\bibitem{Heidmann:2019zws}
P.~Heidmann and N.~P. Warner, \emph{{Superstratum Symbiosis}},
  \href{https://arxiv.org/abs/1903.07631}{{\ttfamily 1903.07631}}.

\bibitem{Bombini:2019vnc}
A.~Bombini and A.~Galliani, \emph{{AdS$_3$ four-point functions from
  $\frac{1}{8}$-BPS states}},
  \href{https://arxiv.org/abs/1904.02656}{{\ttfamily 1904.02656}}.

\bibitem{Bena:2019azk}
I.~Bena, P.~Heidmann, R.~Monten and N.~P. Warner, \emph{{Thermal Decay without
  Information Loss in Horizonless Microstate Geometries}},
  \href{https://arxiv.org/abs/1905.05194}{{\ttfamily 1905.05194}}.

\bibitem{Mathur:2011gz}
S.~D. Mathur and D.~Turton, \emph{{Microstates at the boundary of AdS}},
  \href{http://dx.doi.org/10.1007/JHEP05(2012)014}{\emph{JHEP} {\bfseries 05}
  (2012) 014}, [\href{https://arxiv.org/abs/1112.6413}{{\ttfamily 1112.6413}}].

\bibitem{Mathur:2012tj}
S.~D. Mathur and D.~Turton, \emph{{Momentum-carrying waves on D1-D5 microstate
  geometries}}, {\emph{Nucl.Phys.} {\bfseries B862} (2012) 764--780},
  [\href{https://arxiv.org/abs/1202.6421}{{\ttfamily 1202.6421}}].

\bibitem{Lunin:2012gp}
O.~Lunin, S.~D. Mathur and D.~Turton, \emph{{Adding momentum to supersymmetric
  geometries}},
  \href{http://dx.doi.org/10.1016/j.nuclphysb.2012.11.017}{\emph{Nucl.Phys.}
  {\bfseries B868} (2013) 383--415},
  [\href{https://arxiv.org/abs/1208.1770}{{\ttfamily 1208.1770}}].

\bibitem{Giusto:2013bda}
S.~Giusto and R.~Russo, \emph{{Superdescendants of the D1D5 CFT and their dual
  3-charge geometries}},
  \href{http://dx.doi.org/10.1007/JHEP03(2014)007}{\emph{JHEP} {\bfseries 1403}
  (2014) 007}, [\href{https://arxiv.org/abs/1311.5536}{{\ttfamily 1311.5536}}].

\bibitem{Skenderis:2008qn}
K.~Skenderis and M.~Taylor, \emph{{The fuzzball proposal for black holes}},
  \href{http://dx.doi.org/10.1016/j.physrep.2008.08.001}{\emph{Phys. Rept.}
  {\bfseries 467} (2008) 117--171},
  [\href{https://arxiv.org/abs/0804.0552}{{\ttfamily 0804.0552}}].

\bibitem{Mathur:2008nj}
S.~D. Mathur, \emph{{Fuzzballs and the information paradox: A Summary and
  conjectures}},  \href{https://arxiv.org/abs/0810.4525}{{\ttfamily
  0810.4525}}.

\bibitem{Bena:2013dka}
I.~Bena and N.~P. Warner, \emph{{Resolving the Structure of Black Holes:
  Philosophizing with a Hammer}},
  \href{https://arxiv.org/abs/1311.4538}{{\ttfamily 1311.4538}}.

\bibitem{Martinec:2014gka}
E.~J. Martinec, \emph{{The Cheshire Cap}},
  \href{http://dx.doi.org/10.1007/JHEP03(2015)112}{\emph{JHEP} {\bfseries 03}
  (2015) 112}, [\href{https://arxiv.org/abs/1409.6017}{{\ttfamily 1409.6017}}].

\bibitem{Mathur:2018tib}
S.~D. Mathur and D.~Turton, \emph{{The fuzzball nature of two-charge black hole
  microstates}},  \href{https://arxiv.org/abs/1811.09647}{{\ttfamily
  1811.09647}}.

\bibitem{Itzhaki:1998dd}
N.~Itzhaki, J.~M. Maldacena, J.~Sonnenschein and S.~Yankielowicz,
  \emph{{Supergravity and the large N limit of theories with sixteen
  supercharges}},
  \href{http://dx.doi.org/10.1103/PhysRevD.58.046004}{\emph{Phys. Rev.}
  {\bfseries D58} (1998) 046004},
  [\href{https://arxiv.org/abs/hep-th/9802042}{{\ttfamily hep-th/9802042}}].

\bibitem{Kraus:1998hv}
P.~Kraus, F.~Larsen and S.~P. Trivedi, \emph{{The Coulomb branch of gauge
  theory from rotating branes}},
  \href{http://dx.doi.org/10.1088/1126-6708/1999/03/003}{\emph{JHEP} {\bfseries
  03} (1999) 003}, [\href{https://arxiv.org/abs/hep-th/9811120}{{\ttfamily
  hep-th/9811120}}].

\bibitem{Giddings:1999zu}
S.~B. Giddings and S.~F. Ross, \emph{{D3-brane shells to black branes on the
  Coulomb branch}},
  \href{http://dx.doi.org/10.1103/PhysRevD.61.024036}{\emph{Phys. Rev.}
  {\bfseries D61} (2000) 024036},
  [\href{https://arxiv.org/abs/hep-th/9907204}{{\ttfamily hep-th/9907204}}].

\bibitem{Brandhuber:1999jr}
A.~Brandhuber and K.~Sfetsos, \emph{{Wilson loops from multicenter and rotating
  branes, mass gaps and phase structure in gauge theories}},
  \href{http://dx.doi.org/10.4310/ATMP.1999.v3.n4.a4}{\emph{Adv. Theor. Math.
  Phys.} {\bfseries 3} (1999) 851--887},
  [\href{https://arxiv.org/abs/hep-th/9906201}{{\ttfamily hep-th/9906201}}].

\bibitem{Danielsson:1999zt}
U.~H. Danielsson, E.~Keski-Vakkuri and M.~Kruczenski, \emph{{Spherically
  collapsing matter in AdS, holography, and shellons}},
  \href{http://dx.doi.org/10.1016/S0550-3213(99)00511-8}{\emph{Nucl. Phys.}
  {\bfseries B563} (1999) 279--292},
  [\href{https://arxiv.org/abs/hep-th/9905227}{{\ttfamily hep-th/9905227}}].

\bibitem{Horowitz:2006mr}
G.~T. Horowitz and E.~Silverstein, \emph{{The Inside story: Quasilocal tachyons
  and black holes}},
  \href{http://dx.doi.org/10.1103/PhysRevD.73.064016}{\emph{Phys. Rev.}
  {\bfseries D73} (2006) 064016},
  [\href{https://arxiv.org/abs/hep-th/0601032}{{\ttfamily hep-th/0601032}}].

\bibitem{tHooft:1981nnx}
G.~'t~Hooft, \emph{{Some Twisted Selfdual Solutions for the Yang-Mills
  Equations on a Hypertorus}},
  \href{http://dx.doi.org/10.1007/BF01208900}{\emph{Commun. Math. Phys.}
  {\bfseries 81} (1981) 267--275}.

\bibitem{Guralnik:1997sy}
Z.~Guralnik and S.~Ramgoolam, \emph{{Torons and D-brane bound states}},
  \href{http://dx.doi.org/10.1016/S0550-3213(97)00286-1}{\emph{Nucl. Phys.}
  {\bfseries B499} (1997) 241--252},
  [\href{https://arxiv.org/abs/hep-th/9702099}{{\ttfamily hep-th/9702099}}].

\bibitem{Dijkgraaf:1997ku}
R.~Dijkgraaf, E.~P. Verlinde and H.~L. Verlinde, \emph{{Notes on matrix and
  micro strings}}, \href{http://dx.doi.org/10.1142/9789814447287_0003,
  10.1016/S0920-5632(98)00138-8}{\emph{Nucl. Phys. Proc. Suppl.} {\bfseries 68}
  (1998) 28--54}, [\href{https://arxiv.org/abs/hep-th/9709107}{{\ttfamily
  hep-th/9709107}}].

\bibitem{Lunin:2001fv}
O.~Lunin and S.~D. Mathur, \emph{{Metric of the multiply wound rotating
  string}}, \href{http://dx.doi.org/10.1016/S0550-3213(01)00321-2}{\emph{Nucl.
  Phys.} {\bfseries B610} (2001) 49--76},
  [\href{https://arxiv.org/abs/hep-th/0105136}{{\ttfamily hep-th/0105136}}].

\bibitem{Lunin:2001jy}
O.~Lunin and S.~D. Mathur, \emph{{AdS/CFT duality and the black hole
  information paradox}},
  \href{http://dx.doi.org/10.1016/S0550-3213(01)00620-4}{\emph{Nucl. Phys.}
  {\bfseries B623} (2002) 342--394},
  [\href{https://arxiv.org/abs/hep-th/0109154}{{\ttfamily hep-th/0109154}}].

\bibitem{Lunin:2002iz}
O.~Lunin, J.~M. Maldacena and L.~Maoz, \emph{{Gravity solutions for the D1-D5
  system with angular momentum}},
  \href{https://arxiv.org/abs/hep-th/0212210}{{\ttfamily hep-th/0212210}}.

\bibitem{Kanitscheider:2007wq}
I.~Kanitscheider, K.~Skenderis and M.~Taylor, \emph{{Fuzzballs with internal
  excitations}}, {\emph{JHEP} {\bfseries 06} (2007) 056},
  [\href{https://arxiv.org/abs/0704.0690}{{\ttfamily 0704.0690}}].

\bibitem{Giusto:2019qig}
S.~Giusto, S.~Rawash and D.~Turton, \emph{{AdS$_3$ Holography at Dimension
  Two}},  \href{https://arxiv.org/abs/1904.12880}{{\ttfamily 1904.12880}}.

\bibitem{Rychkov:2005ji}
V.~S. Rychkov, \emph{{D1-D5 black hole microstate counting from supergravity}},
  \href{http://dx.doi.org/10.1088/1126-6708/2006/01/063}{\emph{JHEP} {\bfseries
  01} (2006) 063}, [\href{https://arxiv.org/abs/hep-th/0512053}{{\ttfamily
  hep-th/0512053}}].

\bibitem{Krishnan:2015vha}
C.~Krishnan and A.~Raju, \emph{{A Note on D1-D5 Entropy and Geometric
  Quantization}}, \href{http://dx.doi.org/10.1007/JHEP06(2015)054}{\emph{JHEP}
  {\bfseries 06} (2015) 054},
  [\href{https://arxiv.org/abs/1504.04330}{{\ttfamily 1504.04330}}].

\bibitem{Das:2005za}
S.~R. Das, S.~Giusto, S.~D. Mathur, Y.~Srivastava, X.~Wu and C.~Zhou,
  \emph{{Branes wrapping black holes}},
  \href{http://dx.doi.org/10.1016/j.nuclphysb.2005.11.011}{\emph{Nucl. Phys.}
  {\bfseries B733} (2006) 297--333},
  [\href{https://arxiv.org/abs/hep-th/0507080}{{\ttfamily hep-th/0507080}}].

\bibitem{Martinec:1999sa}
E.~J. Martinec and V.~Sahakian, \emph{{Black holes and five-brane
  thermodynamics}},
  \href{http://dx.doi.org/10.1103/PhysRevD.60.064002}{\emph{Phys. Rev.}
  {\bfseries D60} (1999) 064002},
  [\href{https://arxiv.org/abs/hep-th/9901135}{{\ttfamily hep-th/9901135}}].

\bibitem{Mathur:2005ai}
S.~D. Mathur, \emph{{The quantum structure of black holes}},
  \href{http://dx.doi.org/10.1088/0264-9381/23/11/R01}{\emph{Class. Quant.
  Grav.} {\bfseries 23} (2006) R115},
  [\href{https://arxiv.org/abs/hep-th/0510180}{{\ttfamily hep-th/0510180}}].

\bibitem{Chen:2014loa}
F.~Chen, B.~Michel, J.~Polchinski and A.~Puhm, \emph{{Journey to the Center of
  the Fuzzball}}, \href{http://dx.doi.org/10.1007/JHEP02(2015)081}{\emph{JHEP}
  {\bfseries 02} (2015) 081},
  [\href{https://arxiv.org/abs/1408.4798}{{\ttfamily 1408.4798}}].

\bibitem{Denef:2002ru}
F.~Denef, \emph{{Quantum quivers and Hall / hole halos}},
  \href{http://dx.doi.org/10.1088/1126-6708/2002/10/023}{\emph{JHEP} {\bfseries
  0210} (2002) 023}, [\href{https://arxiv.org/abs/hep-th/0206072}{{\ttfamily
  hep-th/0206072}}].

\bibitem{Denef:2007vg}
F.~Denef and G.~W. Moore, \emph{{Split states, entropy enigmas, holes and
  halos}}, \href{http://dx.doi.org/10.1007/JHEP11(2011)129}{\emph{JHEP}
  {\bfseries 11} (2011) 129},
  [\href{https://arxiv.org/abs/hep-th/0702146}{{\ttfamily hep-th/0702146}}].

\bibitem{deBoer:2008fk}
J.~de~Boer, F.~Denef, S.~El-Showk, I.~Messamah and D.~Van~den Bleeken,
  \emph{{Black hole bound states in AdS3 x S2}},
  \href{http://dx.doi.org/10.1088/1126-6708/2008/11/050}{\emph{JHEP} {\bfseries
  11} (2008) 050}, [\href{https://arxiv.org/abs/0802.2257}{{\ttfamily
  0802.2257}}].

\bibitem{deBoer:2008zn}
J.~de~Boer, S.~El-Showk, I.~Messamah and D.~Van~den Bleeken, \emph{{Quantizing
  N=2 Multicenter Solutions}},
  \href{http://dx.doi.org/10.1088/1126-6708/2009/05/002}{\emph{JHEP} {\bfseries
  05} (2009) 002}, [\href{https://arxiv.org/abs/0807.4556}{{\ttfamily
  0807.4556}}].

\bibitem{Sen:2009vz}
A.~Sen, \emph{{Arithmetic of Quantum Entropy Function}},
  \href{http://dx.doi.org/10.1088/1126-6708/2009/08/068}{\emph{JHEP} {\bfseries
  08} (2009) 068}, [\href{https://arxiv.org/abs/0903.1477}{{\ttfamily
  0903.1477}}].

\bibitem{Dabholkar:2010rm}
A.~Dabholkar, J.~Gomes, S.~Murthy and A.~Sen, \emph{{Supersymmetric Index from
  Black Hole Entropy}},
  \href{http://dx.doi.org/10.1007/JHEP04(2011)034}{\emph{JHEP} {\bfseries 04}
  (2011) 034}, [\href{https://arxiv.org/abs/1009.3226}{{\ttfamily 1009.3226}}].

\bibitem{Lee:2012sc}
S.-J. Lee, Z.-L. Wang and P.~Yi, \emph{{Quiver Invariants from Intrinsic Higgs
  States}}, \href{http://dx.doi.org/10.1007/JHEP07(2012)169}{\emph{JHEP}
  {\bfseries 07} (2012) 169},
  [\href{https://arxiv.org/abs/1205.6511}{{\ttfamily 1205.6511}}].

\bibitem{Manschot:2012rx}
J.~Manschot, B.~Pioline and A.~Sen, \emph{{From Black Holes to Quivers}},
  \href{http://dx.doi.org/10.1007/JHEP11(2012)023}{\emph{JHEP} {\bfseries 1211}
  (2012) 023}, [\href{https://arxiv.org/abs/1207.2230}{{\ttfamily 1207.2230}}].

\bibitem{Martinec:2017ztd}
E.~J. Martinec and S.~Massai, \emph{{String Theory of Supertubes}},
  \href{http://dx.doi.org/10.1007/JHEP07(2018)163}{\emph{JHEP} {\bfseries 07}
  (2018) 163}, [\href{https://arxiv.org/abs/1705.10844}{{\ttfamily
  1705.10844}}].

\bibitem{Martinec:2018nco}
E.~J. Martinec, S.~Massai and D.~Turton, \emph{{String dynamics in NS5-F1-P
  geometries}}, \href{http://dx.doi.org/10.1007/JHEP09(2018)031}{\emph{JHEP}
  {\bfseries 09} (2018) 031},
  [\href{https://arxiv.org/abs/1803.08505}{{\ttfamily 1803.08505}}].

\bibitem{Lunin:2004uu}
O.~Lunin, \emph{{Adding momentum to D1-D5 system}},
  \href{http://dx.doi.org/10.1088/1126-6708/2004/04/054}{\emph{JHEP} {\bfseries
  04} (2004) 054}, [\href{https://arxiv.org/abs/hep-th/0404006}{{\ttfamily
  hep-th/0404006}}].

\bibitem{Giusto:2004id}
S.~Giusto, S.~D. Mathur and A.~Saxena, \emph{{Dual geometries for a set of
  3-charge microstates}},
  \href{http://dx.doi.org/10.1016/j.nuclphysb.2004.09.001}{\emph{Nucl. Phys.}
  {\bfseries B701} (2004) 357--379},
  [\href{https://arxiv.org/abs/hep-th/0405017}{{\ttfamily hep-th/0405017}}].

\bibitem{Giusto:2004ip}
S.~Giusto, S.~D. Mathur and A.~Saxena, \emph{{3-charge geometries and their CFT
  duals}}, \href{http://dx.doi.org/10.1016/j.nuclphysb.2005.01.009}{\emph{Nucl.
  Phys.} {\bfseries B710} (2005) 425--463},
  [\href{https://arxiv.org/abs/hep-th/0406103}{{\ttfamily hep-th/0406103}}].

\bibitem{Jejjala:2005yu}
V.~Jejjala, O.~Madden, S.~F. Ross and G.~Titchener, \emph{{Non-supersymmetric
  smooth geometries and D1-D5-P bound states}},
  \href{http://dx.doi.org/10.1103/PhysRevD.71.124030}{\emph{Phys. Rev.}
  {\bfseries D71} (2005) 124030},
  [\href{https://arxiv.org/abs/hep-th/0504181}{{\ttfamily hep-th/0504181}}].

\bibitem{Giusto:2012yz}
S.~Giusto, O.~Lunin, S.~D. Mathur and D.~Turton, \emph{{D1-D5-P microstates at
  the cap}}, \href{http://dx.doi.org/10.1007/JHEP02(2013)050}{\emph{JHEP}
  {\bfseries 1302} (2013) 050},
  [\href{https://arxiv.org/abs/1211.0306}{{\ttfamily 1211.0306}}].

\bibitem{Alekseev:1998mc}
A.~{\relax Yu}. Alekseev and V.~Schomerus, \emph{{D-branes in the WZW model}},
  \href{http://dx.doi.org/10.1103/PhysRevD.60.061901}{\emph{Phys. Rev.}
  {\bfseries D60} (1999) 061901},
  [\href{https://arxiv.org/abs/hep-th/9812193}{{\ttfamily hep-th/9812193}}].

\bibitem{Alekseev:1999bs}
A.~{\relax Yu}. Alekseev, A.~Recknagel and V.~Schomerus, \emph{{Noncommutative
  world volume geometries: Branes on SU(2) and fuzzy spheres}},
  \href{http://dx.doi.org/10.1088/1126-6708/1999/09/023}{\emph{JHEP} {\bfseries
  09} (1999) 023}, [\href{https://arxiv.org/abs/hep-th/9908040}{{\ttfamily
  hep-th/9908040}}].

\bibitem{Stanciu:1999id}
S.~Stanciu, \emph{{D-branes in group manifolds}},
  \href{http://dx.doi.org/10.1088/1126-6708/2000/01/025}{\emph{JHEP} {\bfseries
  01} (2000) 025}, [\href{https://arxiv.org/abs/hep-th/9909163}{{\ttfamily
  hep-th/9909163}}].

\bibitem{Alekseev:2000fd}
A.~{\relax Yu}. Alekseev, A.~Recknagel and V.~Schomerus, \emph{{Brane dynamics
  in background fluxes and noncommutative geometry}},
  \href{http://dx.doi.org/10.1088/1126-6708/2000/05/010}{\emph{JHEP} {\bfseries
  05} (2000) 010}, [\href{https://arxiv.org/abs/hep-th/0003187}{{\ttfamily
  hep-th/0003187}}].

\bibitem{Bachas:2000fr}
C.~Bachas and M.~Petropoulos, \emph{{Anti-de Sitter D-branes}},
  \href{http://dx.doi.org/10.1088/1126-6708/2001/02/025}{\emph{JHEP} {\bfseries
  02} (2001) 025}, [\href{https://arxiv.org/abs/hep-th/0012234}{{\ttfamily
  hep-th/0012234}}].

\bibitem{Pawelczyk:2000hy}
J.~Pawelczyk and S.-J. Rey, \emph{{Ramond-ramond flux stabilization of
  D-branes}},
  \href{http://dx.doi.org/10.1016/S0370-2693(00)01159-X}{\emph{Phys. Lett.}
  {\bfseries B493} (2000) 395--401},
  [\href{https://arxiv.org/abs/hep-th/0007154}{{\ttfamily hep-th/0007154}}].

\bibitem{Giveon:2001uq}
A.~Giveon, D.~Kutasov and A.~Schwimmer, \emph{{Comments on D-branes in
  AdS(3)}}, \href{http://dx.doi.org/10.1016/S0550-3213(01)00438-2}{\emph{Nucl.
  Phys.} {\bfseries B615} (2001) 133--168},
  [\href{https://arxiv.org/abs/hep-th/0106005}{{\ttfamily hep-th/0106005}}].

\bibitem{Israel:2005ek}
D.~Israel, \emph{{D-branes in Lorentzian AdS(3)}},
  \href{http://dx.doi.org/10.1088/1126-6708/2005/06/008}{\emph{JHEP} {\bfseries
  06} (2005) 008}, [\href{https://arxiv.org/abs/hep-th/0502159}{{\ttfamily
  hep-th/0502159}}].

\bibitem{Maldacena:2001ky}
J.~M. Maldacena, G.~W. Moore and N.~Seiberg, \emph{{Geometrical interpretation
  of D-branes in gauged WZW models}},
  \href{http://dx.doi.org/10.1088/1126-6708/2001/07/046}{\emph{JHEP} {\bfseries
  07} (2001) 046}, [\href{https://arxiv.org/abs/hep-th/0105038}{{\ttfamily
  hep-th/0105038}}].

\bibitem{Gawedzki:2001ye}
K.~Gawedzki, \emph{{Boundary WZW, G / H, G / G and CS theories}},
  \href{http://dx.doi.org/10.1007/s00023-002-8639-0}{\emph{Annales Henri
  Poincare} {\bfseries 3} (2002) 847--881},
  [\href{https://arxiv.org/abs/hep-th/0108044}{{\ttfamily hep-th/0108044}}].

\bibitem{Elitzur:2001qd}
S.~Elitzur and G.~Sarkissian, \emph{{D branes on a gauged WZW model}},
  \href{http://dx.doi.org/10.1016/S0550-3213(02)00010-X}{\emph{Nucl. Phys.}
  {\bfseries B625} (2002) 166--178},
  [\href{https://arxiv.org/abs/hep-th/0108142}{{\ttfamily hep-th/0108142}}].

\bibitem{Fredenhagen:2001kw}
S.~Fredenhagen and V.~Schomerus, \emph{{D-branes in coset models}},
  \href{http://dx.doi.org/10.1088/1126-6708/2002/02/005}{\emph{JHEP} {\bfseries
  02} (2002) 005}, [\href{https://arxiv.org/abs/hep-th/0111189}{{\ttfamily
  hep-th/0111189}}].

\bibitem{Sarkissian:2002ie}
G.~Sarkissian, \emph{{Nonmaximally symmetric D-branes on group manifold in the
  Lagrangian approach}},
  \href{http://dx.doi.org/10.1088/1126-6708/2002/07/033}{\emph{JHEP} {\bfseries
  07} (2002) 033}, [\href{https://arxiv.org/abs/hep-th/0205097}{{\ttfamily
  hep-th/0205097}}].

\bibitem{Walton:2002db}
M.~A. Walton and J.-G. Zhou, \emph{{D-branes in asymmetrically gauged WZW
  models and axial vector duality}},
  \href{http://dx.doi.org/10.1016/S0550-3213(02)00996-3}{\emph{Nucl. Phys.}
  {\bfseries B648} (2003) 523--541},
  [\href{https://arxiv.org/abs/hep-th/0205161}{{\ttfamily hep-th/0205161}}].

\bibitem{Sarkissian:2002bg}
G.~Sarkissian, \emph{{On DBI action of the nonmaximally symmetric D-branes on
  SU(2)}}, \href{http://dx.doi.org/10.1088/1126-6708/2003/01/058}{\emph{JHEP}
  {\bfseries 01} (2003) 058},
  [\href{https://arxiv.org/abs/hep-th/0211038}{{\ttfamily hep-th/0211038}}].

\bibitem{Sarkissian:2002nq}
G.~Sarkissian, \emph{{On D branes in the Nappi-Witten and GMM models}},
  \href{http://dx.doi.org/10.1088/1126-6708/2003/01/059}{\emph{JHEP} {\bfseries
  01} (2003) 059}, [\href{https://arxiv.org/abs/hep-th/0211163}{{\ttfamily
  hep-th/0211163}}].

\bibitem{Quella:2002ns}
T.~Quella, \emph{{On the hierarchy of symmetry breaking D-branes in group
  manifolds}},
  \href{http://dx.doi.org/10.1088/1126-6708/2002/12/009}{\emph{JHEP} {\bfseries
  12} (2002) 009}, [\href{https://arxiv.org/abs/hep-th/0209157}{{\ttfamily
  hep-th/0209157}}].

\bibitem{Quella:2002fk}
T.~Quella and V.~Schomerus, \emph{{Asymmetric cosets}},
  \href{http://dx.doi.org/10.1088/1126-6708/2003/02/030}{\emph{JHEP} {\bfseries
  02} (2003) 030}, [\href{https://arxiv.org/abs/hep-th/0212119}{{\ttfamily
  hep-th/0212119}}].

\bibitem{Quella:2003kd}
T.~Quella, \emph{{Asymmetrically gauged coset theories and symmetry breaking
  D-branes: New boundary conditions in conformal field theory}}, Ph.D. thesis,
  Humboldt U., Berlin, 2003.

\bibitem{Israel:2005fn}
D.~Israel, A.~Pakman and J.~Troost, \emph{{D-branes in little string theory}},
  \href{http://dx.doi.org/10.1016/j.nuclphysb.2005.05.027}{\emph{Nucl. Phys.}
  {\bfseries B722} (2005) 3--64},
  [\href{https://arxiv.org/abs/hep-th/0502073}{{\ttfamily hep-th/0502073}}].

\bibitem{Sfetsos:1998xd}
K.~Sfetsos, \emph{{Branes for Higgs phases and exact conformal field
  theories}},
  \href{http://dx.doi.org/10.1088/1126-6708/1999/01/015}{\emph{JHEP} {\bfseries
  01} (1999) 015}, [\href{https://arxiv.org/abs/hep-th/9811167}{{\ttfamily
  hep-th/9811167}}].

\bibitem{Giveon:1999px}
A.~Giveon and D.~Kutasov, \emph{{Little string theory in a double scaling
  limit}}, \href{http://dx.doi.org/10.1088/1126-6708/1999/10/034}{\emph{JHEP}
  {\bfseries 10} (1999) 034},
  [\href{https://arxiv.org/abs/hep-th/9909110}{{\ttfamily hep-th/9909110}}].

\bibitem{Giveon:1999tq}
A.~Giveon and D.~Kutasov, \emph{{Comments on double scaled little string
  theory}}, \href{http://dx.doi.org/10.1088/1126-6708/2000/01/023}{\emph{JHEP}
  {\bfseries 01} (2000) 023},
  [\href{https://arxiv.org/abs/hep-th/9911039}{{\ttfamily hep-th/9911039}}].

\bibitem{Mateos:2001qs}
D.~Mateos and P.~K. Townsend, \emph{{Supertubes}},
  \href{http://dx.doi.org/10.1103/PhysRevLett.87.011602}{\emph{Phys. Rev.
  Lett.} {\bfseries 87} (2001) 011602},
  [\href{https://arxiv.org/abs/hep-th/0103030}{{\ttfamily hep-th/0103030}}].

\bibitem{Ooguri:1995wj}
H.~Ooguri and C.~Vafa, \emph{{Two-dimensional black hole and singularities of
  CY manifolds}},
  \href{http://dx.doi.org/10.1016/0550-3213(96)00008-9}{\emph{Nucl. Phys.}
  {\bfseries B463} (1996) 55--72},
  [\href{https://arxiv.org/abs/hep-th/9511164}{{\ttfamily hep-th/9511164}}].

\bibitem{Martinec:1988zu}
E.~J. Martinec, \emph{{Algebraic Geometry and Effective Lagrangians}},
  \href{http://dx.doi.org/10.1016/0370-2693(89)90074-9}{\emph{Phys. Lett.}
  {\bfseries B217} (1989) 431--437}.

\bibitem{Vafa:1988uu}
C.~Vafa and N.~P. Warner, \emph{{Catastrophes and the Classification of
  Conformal Theories}},
  \href{http://dx.doi.org/10.1016/0370-2693(89)90473-5}{\emph{Phys. Lett.}
  {\bfseries B218} (1989) 51--58}.

\bibitem{Israel:2004ir}
D.~Israel, C.~Kounnas, A.~Pakman and J.~Troost, \emph{{The Partition function
  of the supersymmetric two-dimensional black hole and little string theory}},
  \href{http://dx.doi.org/10.1088/1126-6708/2004/06/033}{\emph{JHEP} {\bfseries
  06} (2004) 033}, [\href{https://arxiv.org/abs/hep-th/0403237}{{\ttfamily
  hep-th/0403237}}].

\bibitem{Itzhaki:2005zr}
N.~Itzhaki, D.~Kutasov and N.~Seiberg, \emph{{Non-supersymmetric deformations
  of non-critical superstrings}},
  \href{http://dx.doi.org/10.1088/1126-6708/2005/12/035}{\emph{JHEP} {\bfseries
  12} (2005) 035}, [\href{https://arxiv.org/abs/hep-th/0510087}{{\ttfamily
  hep-th/0510087}}].

\bibitem{Hull:1989jk}
C.~M. Hull and B.~J. Spence, \emph{{The Gauged Nonlinear $\sigma$ Model With
  {Wess-Zumino} Term}},
  \href{http://dx.doi.org/10.1016/0370-2693(89)91688-2}{\emph{Phys. Lett.}
  {\bfseries B232} (1989) 204--210}.

\bibitem{Figueroa-OFarrill:2005vws}
J.~M. Figueroa-O'Farrill and N.~Mohammedi, \emph{{Gauging the Wess-Zumino term
  of a sigma model with boundary}},
  \href{http://dx.doi.org/10.1088/1126-6708/2005/08/086}{\emph{JHEP} {\bfseries
  08} (2005) 086}, [\href{https://arxiv.org/abs/hep-th/0506049}{{\ttfamily
  hep-th/0506049}}].

\bibitem{Callan:1991at}
C.~G. Callan, Jr., J.~A. Harvey and A.~Strominger, \emph{{Supersymmetric string
  solitons}},  \href{https://arxiv.org/abs/hep-th/9112030}{{\ttfamily
  hep-th/9112030}}.

\bibitem{Behrend:1999bn}
R.~E. Behrend, P.~A. Pearce, V.~B. Petkova and J.-B. Zuber, \emph{{Boundary
  conditions in rational conformal field theories}},
  \href{http://dx.doi.org/10.1016/S0550-3213(99)00592-1,
  10.1016/S0550-3213(00)00225-X}{\emph{Nucl. Phys.} {\bfseries B570} (2000)
  525--589}, [\href{https://arxiv.org/abs/hep-th/9908036}{{\ttfamily
  hep-th/9908036}}].

\bibitem{Gaberdiel:2001xm}
M.~R. Gaberdiel, A.~Recknagel and G.~M.~T. Watts, \emph{{The conformal boundary
  states for SU(2) at level 1}},
  \href{http://dx.doi.org/10.1016/S0550-3213(02)00033-0}{\emph{Nucl. Phys.}
  {\bfseries B626} (2002) 344--362},
  [\href{https://arxiv.org/abs/hep-th/0108102}{{\ttfamily hep-th/0108102}}].

\bibitem{Bars:1991pt}
I.~Bars and K.~Sfetsos, \emph{{Generalized duality and singular strings in
  higher dimensions}},
  \href{http://dx.doi.org/10.1142/S0217732392000963}{\emph{Mod. Phys. Lett.}
  {\bfseries A7} (1992) 1091--1104},
  [\href{https://arxiv.org/abs/hep-th/9110054}{{\ttfamily hep-th/9110054}}].

\bibitem{Driezen:2019ykp}
S.~Driezen, A.~Sevrin and D.~C. Thompson, \emph{{Integrable asymmetric
  $\lambda$-deformations}},
  \href{http://dx.doi.org/10.1007/JHEP04(2019)094}{\emph{JHEP} {\bfseries 04}
  (2019) 094}, [\href{https://arxiv.org/abs/1902.04142}{{\ttfamily
  1902.04142}}].

\bibitem{Driezen:2018glg}
S.~Driezen, A.~Sevrin and D.~C. Thompson, \emph{{D-branes in
  $\lambda$-deformations}},
  \href{http://dx.doi.org/10.1007/JHEP09(2018)015}{\emph{JHEP} {\bfseries 09}
  (2018) 015}, [\href{https://arxiv.org/abs/1806.10712}{{\ttfamily
  1806.10712}}].

\bibitem{Myers:1999ps}
R.~C. Myers, \emph{{Dielectric-branes}}, {\emph{JHEP} {\bfseries 12} (1999)
  022}, [\href{https://arxiv.org/abs/hep-th/9910053}{{\ttfamily
  hep-th/9910053}}].

\bibitem{Bachas:2000ik}
C.~Bachas, M.~R. Douglas and C.~Schweigert, \emph{{Flux stabilization of
  D-branes}},
  \href{http://dx.doi.org/10.1088/1126-6708/2000/05/048}{\emph{JHEP} {\bfseries
  05} (2000) 048}, [\href{https://arxiv.org/abs/hep-th/0003037}{{\ttfamily
  hep-th/0003037}}].

\bibitem{Fotopoulos:2004ut}
A.~Fotopoulos, V.~Niarchos and N.~Prezas, \emph{{D-branes and extended
  characters in SL(2,R) / U(1)}},
  \href{http://dx.doi.org/10.1016/j.nuclphysb.2004.12.030}{\emph{Nucl. Phys.}
  {\bfseries B710} (2005) 309--370},
  [\href{https://arxiv.org/abs/hep-th/0406017}{{\ttfamily hep-th/0406017}}].

\bibitem{Maldacena:2001ss}
J.~M. Maldacena, G.~W. Moore and N.~Seiberg, \emph{{D-brane charges in
  five-brane backgrounds}}, {\emph{JHEP} {\bfseries 10} (2001) 005},
  [\href{https://arxiv.org/abs/hep-th/0108152}{{\ttfamily hep-th/0108152}}].

\bibitem{Alekseev:2000jx}
A.~Alekseev and V.~Schomerus, \emph{{RR charges of D2-branes in the WZW
  model}},  \href{https://arxiv.org/abs/hep-th/0007096}{{\ttfamily
  hep-th/0007096}}.

\bibitem{Marolf:2000cb}
D.~Marolf, \emph{{Chern-Simons terms and the three notions of charge}},  in
  \emph{{Quantization, gauge theory, and strings. Proceedings, International
  Conference dedicated to the memory of Professor Efim Fradkin, Moscow, Russia,
  June 5-10, 2000. Vol. 1+2}}, pp.~312--320, 2000,
  \href{https://arxiv.org/abs/hep-th/0006117}{{\ttfamily hep-th/0006117}}.

\bibitem{Figueroa-OFarrill:2000lcd}
J.~M. Figueroa-O'Farrill and S.~Stanciu, \emph{{D-brane charge, flux
  quantization and relative (co)homology}},
  \href{http://dx.doi.org/10.1088/1126-6708/2001/01/006}{\emph{JHEP} {\bfseries
  01} (2001) 006}, [\href{https://arxiv.org/abs/hep-th/0008038}{{\ttfamily
  hep-th/0008038}}].

\bibitem{Witten:1991yr}
E.~Witten, \emph{{On string theory and black holes}},
  \href{http://dx.doi.org/10.1103/PhysRevD.44.314}{\emph{Phys. Rev.} {\bfseries
  D44} (1991) 314--324}.

\bibitem{Kiritsis:1991zt}
E.~B. Kiritsis, \emph{{Duality in gauged WZW models}},
  \href{http://dx.doi.org/10.1142/S0217732391003341}{\emph{Mod. Phys. Lett.}
  {\bfseries A6} (1991) 2871--2880}.

\bibitem{Dijkgraaf:1991ba}
R.~Dijkgraaf, H.~L. Verlinde and E.~P. Verlinde, \emph{{String propagation in a
  black hole geometry}},
  \href{http://dx.doi.org/10.1016/0550-3213(92)90237-6}{\emph{Nucl. Phys.}
  {\bfseries B371} (1992) 269--314}.

\bibitem{Douglas:1996xg}
M.~R. Douglas, \emph{{Enhanced gauge symmetry in M(atrix) theory}},
  \href{http://dx.doi.org/10.1088/1126-6708/1997/07/004}{\emph{JHEP} {\bfseries
  07} (1997) 004}, [\href{https://arxiv.org/abs/hep-th/9612126}{{\ttfamily
  hep-th/9612126}}].

\bibitem{Douglas:1997de}
M.~R. Douglas, B.~R. Greene and D.~R. Morrison, \emph{{Orbifold resolution by
  D-branes}},
  \href{http://dx.doi.org/10.1016/S0550-3213(97)00517-8}{\emph{Nucl. Phys.}
  {\bfseries B506} (1997) 84--106},
  [\href{https://arxiv.org/abs/hep-th/9704151}{{\ttfamily hep-th/9704151}}].

\bibitem{Hashimoto:1997gm}
A.~Hashimoto and W.~Taylor, \emph{{Fluctuation spectra of tilted and
  intersecting D-branes from the Born-Infeld action}},
  \href{http://dx.doi.org/10.1016/S0550-3213(97)00399-4}{\emph{Nucl. Phys.}
  {\bfseries B503} (1997) 193--219},
  [\href{https://arxiv.org/abs/hep-th/9703217}{{\ttfamily hep-th/9703217}}].

\bibitem{Tseytlin:1992ri}
A.~A. Tseytlin, \emph{{Effective action of gauged WZW model and exact string
  solutions}},
  \href{http://dx.doi.org/10.1016/0550-3213(93)90511-M}{\emph{Nucl. Phys.}
  {\bfseries B399} (1993) 601--622},
  [\href{https://arxiv.org/abs/hep-th/9301015}{{\ttfamily hep-th/9301015}}].

\bibitem{Buscher:1987qj}
T.~H. Buscher, \emph{{Path Integral Derivation of Quantum Duality in Nonlinear
  Sigma Models}},
  \href{http://dx.doi.org/10.1016/0370-2693(88)90602-8}{\emph{Phys. Lett.}
  {\bfseries B201} (1988) 466--472}.

\bibitem{Recknagel:2013uja}
A.~Recknagel and V.~Schomerus, \emph{{Boundary Conformal Field Theory and the
  Worldsheet Approach to D-Branes}}.
\newblock Cambridge Monographs on Mathematical Physics. Cambridge University
  Press, 2013,
  \href{http://dx.doi.org/10.1017/CBO9780511806476}{10.1017/CBO9780511806476}.

\bibitem{Freidel:2017wst}
L.~Freidel, R.~G. Leigh and D.~Minic, \emph{{Intrinsic non-commutativity of
  closed string theory}},
  \href{http://dx.doi.org/10.1007/JHEP09(2017)060}{\emph{JHEP} {\bfseries 09}
  (2017) 060}, [\href{https://arxiv.org/abs/1706.03305}{{\ttfamily
  1706.03305}}].

\bibitem{Freidel:2017nhg}
L.~Freidel, R.~G. Leigh and D.~Minic, \emph{{Noncommutativity of closed string
  zero modes}}, \href{http://dx.doi.org/10.1103/PhysRevD.96.066003}{\emph{Phys.
  Rev.} {\bfseries D96} (2017) 066003},
  [\href{https://arxiv.org/abs/1707.00312}{{\ttfamily 1707.00312}}].

\bibitem{Marolf:2016nwu}
D.~Marolf, B.~Michel and A.~Puhm, \emph{{A rough end for smooth microstate
  geometries}}, \href{http://dx.doi.org/10.1007/JHEP05(2017)021}{\emph{JHEP}
  {\bfseries 05} (2017) 021},
  [\href{https://arxiv.org/abs/1612.05235}{{\ttfamily 1612.05235}}].

\bibitem{Wald:1993nt}
R.~M. Wald, \emph{{Black hole entropy is the Noether charge}},
  \href{http://dx.doi.org/10.1103/PhysRevD.48.R3427}{\emph{Phys. Rev.}
  {\bfseries D48} (1993) 3427--3431},
  [\href{https://arxiv.org/abs/gr-qc/9307038}{{\ttfamily gr-qc/9307038}}].

\bibitem{Iyer:1994ys}
V.~Iyer and R.~M. Wald, \emph{{Some properties of Noether charge and a proposal
  for dynamical black hole entropy}},
  \href{http://dx.doi.org/10.1103/PhysRevD.50.846}{\emph{Phys. Rev.} {\bfseries
  D50} (1994) 846--864}, [\href{https://arxiv.org/abs/gr-qc/9403028}{{\ttfamily
  gr-qc/9403028}}].

\bibitem{Bena:2011zw}
I.~Bena, B.~D. Chowdhury, J.~de~Boer, S.~El-Showk and M.~Shigemori,
  \emph{{Moulting Black Holes}},
  \href{http://dx.doi.org/10.1007/JHEP03(2012)094}{\emph{JHEP} {\bfseries 1203}
  (2012) 094}, [\href{https://arxiv.org/abs/1108.0411}{{\ttfamily 1108.0411}}].

\bibitem{Boucher:1986bh}
W.~Boucher, D.~Friedan and A.~Kent, \emph{{Determinant Formulae and Unitarity
  for the N=2 Superconformal Algebras in Two-Dimensions or Exact Results on
  String Compactification}},
  \href{http://dx.doi.org/10.1016/0370-2693(86)90260-1}{\emph{Phys. Lett.}
  {\bfseries B172} (1986) 316}.

\bibitem{Giveon:1998ns}
A.~Giveon, D.~Kutasov and N.~Seiberg, \emph{{Comments on string theory on
  AdS(3)}}, {\emph{Adv. Theor. Math. Phys.} {\bfseries 2} (1998) 733--780},
  [\href{https://arxiv.org/abs/hep-th/9806194}{{\ttfamily hep-th/9806194}}].

\bibitem{Tyukov:2017uig}
A.~Tyukov, R.~Walker and N.~P. Warner, \emph{{Tidal Stresses and Energy Gaps in
  Microstate Geometries}},
  \href{http://dx.doi.org/10.1007/JHEP02(2018)122}{\emph{JHEP} {\bfseries 02}
  (2018) 122}, [\href{https://arxiv.org/abs/1710.09006}{{\ttfamily
  1710.09006}}].

\bibitem{Bena:2008nh}
I.~Bena, N.~Bobev, C.~Ruef and N.~P. Warner, \emph{{Entropy Enhancement and
  Black Hole Microstates}},
  \href{http://dx.doi.org/10.1103/PhysRevLett.105.231301}{\emph{Phys. Rev.
  Lett.} {\bfseries 105} (2010) 231301},
  [\href{https://arxiv.org/abs/0804.4487}{{\ttfamily 0804.4487}}].

\bibitem{Bena:2008dw}
I.~Bena, N.~Bobev, C.~Ruef and N.~P. Warner, \emph{{Supertubes in Bubbling
  Backgrounds: Born-Infeld Meets Supergravity}},
  \href{http://dx.doi.org/10.1088/1126-6708/2009/07/106}{\emph{JHEP} {\bfseries
  07} (2009) 106}, [\href{https://arxiv.org/abs/0812.2942}{{\ttfamily
  0812.2942}}].

\bibitem{Giveon:2015raa}
A.~Giveon, J.~Harvey, D.~Kutasov and S.~Lee, \emph{{Three-Charge Black Holes
  and Quarter BPS States in Little String Theory}},
  \href{http://dx.doi.org/10.1007/JHEP12(2015)145}{\emph{JHEP} {\bfseries 12}
  (2015) 145}, [\href{https://arxiv.org/abs/1508.04437}{{\ttfamily
  1508.04437}}].

\bibitem{Mathur:2012jk}
S.~D. Mathur and D.~Turton, \emph{{Comments on black holes I: The possibility
  of complementarity}},
  \href{http://dx.doi.org/10.1007/JHEP01(2014)034}{\emph{JHEP} {\bfseries 1401}
  (2014) 034}, [\href{https://arxiv.org/abs/1208.2005}{{\ttfamily 1208.2005}}].

\bibitem{Mathur:2013gua}
S.~D. Mathur and D.~Turton, \emph{{The flaw in the firewall argument}},
  \href{http://dx.doi.org/10.1016/j.nuclphysb.2014.05.012}{\emph{Nucl.Phys.}
  {\bfseries B884} (2014) 566--611},
  [\href{https://arxiv.org/abs/1306.5488}{{\ttfamily 1306.5488}}].

\bibitem{Shenker:2013pqa}
S.~H. Shenker and D.~Stanford, \emph{{Black holes and the butterfly effect}},
  \href{http://dx.doi.org/10.1007/JHEP03(2014)067}{\emph{JHEP} {\bfseries 03}
  (2014) 067}, [\href{https://arxiv.org/abs/1306.0622}{{\ttfamily 1306.0622}}].

\bibitem{Maldacena:2015waa}
J.~Maldacena, S.~H. Shenker and D.~Stanford, \emph{{A bound on chaos}},
  \href{http://dx.doi.org/10.1007/JHEP08(2016)106}{\emph{JHEP} {\bfseries 08}
  (2016) 106}, [\href{https://arxiv.org/abs/1503.01409}{{\ttfamily
  1503.01409}}].

\bibitem{Almheiri:2012rt}
A.~Almheiri, D.~Marolf, J.~Polchinski and J.~Sully, \emph{{Black Holes:
  Complementarity or Firewalls?}},
  \href{http://dx.doi.org/10.1007/JHEP02(2013)062}{\emph{JHEP} {\bfseries 1302}
  (2013) 062}, [\href{https://arxiv.org/abs/1207.3123}{{\ttfamily 1207.3123}}].

\bibitem{Gross:1987ar}
D.~J. Gross and P.~F. Mende, \emph{{String Theory Beyond the Planck Scale}},
  \href{http://dx.doi.org/10.1016/0550-3213(88)90390-2}{\emph{Nucl.Phys.}
  {\bfseries B303} (1988) 407}.

\bibitem{Chakrabarty:2015foa}
B.~Chakrabarty, D.~Turton and A.~Virmani, \emph{{Holographic description of
  non-supersymmetric orbifolded D1-D5-P solutions}},
  \href{http://dx.doi.org/10.1007/JHEP11(2015)063}{\emph{JHEP} {\bfseries 11}
  (2015) 063}, [\href{https://arxiv.org/abs/1508.01231}{{\ttfamily
  1508.01231}}].

\bibitem{Chowdhury:2007jx}
B.~D. Chowdhury and S.~D. Mathur, \emph{{Radiation from the non-extremal
  fuzzball}},
  \href{http://dx.doi.org/10.1088/0264-9381/25/13/135005}{\emph{Class. Quant.
  Grav.} {\bfseries 25} (2008) 135005},
  [\href{https://arxiv.org/abs/0711.4817}{{\ttfamily 0711.4817}}].

\bibitem{Chowdhury:2008bd}
B.~D. Chowdhury and S.~D. Mathur, \emph{{Pair creation in non-extremal fuzzball
  geometries}},
  \href{http://dx.doi.org/10.1088/0264-9381/25/22/225021}{\emph{Class. Quant.
  Grav.} {\bfseries 25} (2008) 225021},
  [\href{https://arxiv.org/abs/0806.2309}{{\ttfamily 0806.2309}}].

\bibitem{Chowdhury:2008uj}
B.~D. Chowdhury and S.~D. Mathur, \emph{{Non-extremal fuzzballs and ergoregion
  emission}},
  \href{http://dx.doi.org/10.1088/0264-9381/26/3/035006}{\emph{Class. Quant.
  Grav.} {\bfseries 26} (2009) 035006},
  [\href{https://arxiv.org/abs/0810.2951}{{\ttfamily 0810.2951}}].

\bibitem{Avery:2009xr}
S.~G. Avery and B.~D. Chowdhury, \emph{{Emission from the D1D5 CFT: Higher
  Twists}}, \href{http://dx.doi.org/10.1007/JHEP01(2010)087}{\emph{JHEP}
  {\bfseries 1001} (2010) 087},
  [\href{https://arxiv.org/abs/0907.1663}{{\ttfamily 0907.1663}}].

\bibitem{Cardoso:2005gj}
V.~Cardoso, O.~J.~C. Dias, J.~L. Hovdebo and R.~C. Myers, \emph{{Instability of
  non-supersymmetric smooth geometries}},
  \href{http://dx.doi.org/10.1103/PhysRevD.73.064031}{\emph{Phys. Rev.}
  {\bfseries D73} (2006) 064031},
  [\href{https://arxiv.org/abs/hep-th/0512277}{{\ttfamily hep-th/0512277}}].

\bibitem{Schiff:1940}
L.~I. Schiff, H.~Snyder and J.~Weinberg, \emph{{On The Existence of Stationary
  States of the Mesotron Field}}, {\emph{Phys. Rev.} {\bfseries 57} (1940)
  315}.

\bibitem{Fulling:1989nb}
S.~A. Fulling, \emph{{Aspects of Quantum Field Theory in Curved Space-time}},
  {\emph{London Math. Soc. Student Texts} {\bfseries 17} (1989) 1--315}.

\bibitem{Mathur:2009hf}
S.~D. Mathur, \emph{{The information paradox: A pedagogical introduction}},
  \href{http://dx.doi.org/10.1088/0264-9381/26/22/224001}{\emph{Class. Quant.
  Grav.} {\bfseries 26} (2009) 224001},
  [\href{https://arxiv.org/abs/0909.1038}{{\ttfamily 0909.1038}}].

\end{thebibliography}\endgroup

%\input{supertube}

%%%%%%%%%%%%%%%%%%%%%%%%%%%%%%%%%%%%
%\end{thebibliography}

\end{document}